%% file: main.tex
\documentclass[a4paper, oneside, english,12pt]{report}
\usepackage[utf8]{inputenc}
\usepackage[T1]{fontenc}
\usepackage{graphicx}
\usepackage{hyperref}
\usepackage{booktabs,caption}
\usepackage[normalem]{ulem}
\usepackage{placeins}
\usepackage{amsmath}
\usepackage{amssymb}
\usepackage{amsfonts}
\usepackage{textcomp}
\usepackage{mathpazo}
\usepackage{bbm}

\usepackage[left=2.5cm, right=2.5cm, top=3.cm, bottom=3.cm, head=15pt, headsep=.5in]{geometry} 
\usepackage{setspace}
\usepackage{pdfpages}
\usepackage{fancyhdr}
\usepackage[numbib,notlof,notlot,nottoc]{tocbibind}

\usepackage{mathptmx}
\usepackage{subcaption}
\usepackage[backend=bibtex,style=ieee,natbib=true,isbn=false,doi=false]{biblatex}
\usepackage{adjustbox}
\usepackage[autostyle=true]{csquotes} 

\DeclareFieldFormat{pages}{#1}

\usepackage{floatrow}
\usepackage{amsmath}	
\usepackage{amssymb}	
\usepackage{multicol}        
\usepackage{bm}		
\usepackage{pdflscape}	
\usepackage{multirow}
\usepackage{verbatim}
\usepackage[outdir=./]{epstopdf}
\usepackage{braket}

\usepackage{xcolor}
\definecolor{blue(ryb)}{rgb}{0.01, 0.28, 1.0}
\definecolor{cobalt}{rgb}{0.0, 0.28, 0.67}
\definecolor{brickred}{rgb}{0.8, 0.25, 0.33}
\usepackage{ae,aecompl}
\hypersetup{pdfpagemode={UseOutlines},
bookmarksopen=true,
bookmarksopenlevel=0,
hypertexnames=true,
colorlinks=true,
citecolor=magenta,
pdfstartview={FitV},
unicode,
breaklinks=true,
filecolor=cobalt,
linkcolor=blue(ryb),
urlcolor=brickred
}
\renewbibmacro{in:}{%
  \ifboolexpr{%
     test {\ifentrytype{article}}%
     or
     test {\ifentrytype{inproceedings}}%
  }{}{\printtext{\bibstring{in}\intitlepunct}}%
}

\usepackage{amsthm}

\newtheorem{theorem}{Theorem}

\usepackage{pifont}

\usepackage{tikz}

\usepackage{mathtools}

\usepackage{array}
\newcommand{\PreserveBackslash}[1]{\let\temp=\\#1\let\\=\temp}
\newcolumntype{C}[1]{>{\PreserveBackslash\centering}p{#1}}
\newcolumntype{R}[1]{>{\PreserveBackslash\raggedleft}p{#1}}
\newcolumntype{L}[1]{>{\PreserveBackslash\raggedright}p{#1}}

\usepackage{makecell}

\usepackage[toc,page]{appendix}

\usepackage{pbox}

\usepackage{diagbox}

\renewcommand\bra[1]{{\langle{#1}|}}
\renewcommand\ket[1]{{|{#1}\rangle}}

\usetikzlibrary{quotes,angles}

\usepackage{CJKutf8}

\usepackage{centernot}
\usepackage{mathtools}
\usepackage{ stmaryrd }

\usetikzlibrary{arrows}

\def\blankpage{%
      \clearpage%
      \thispagestyle{empty}%
      \addtocounter{page}{-1}%
      \null%
      \clearpage}
\usepackage[autostyle=true]{csquotes} 

\newcommand{\beq}[0]{\begin{equation}}
\newcommand{\eeq}[0]{\end{equation}}

\def\ra{\rangle}
\def\la{\langle}

\newcommand{\one}{\leavevmode\hbox{\small1\normalsize\kern-.33em1}}
\def\tr{\mbox{tr}}

\font\Bbb =msbm10 
 \def\C{{\hbox {\Bbb C}}}

\def\be{\begin{equation}}
\def\ee{\end{equation}}
\def\ben{\begin{eqnarray}}
\def\een{\end{eqnarray}}
\def\eea{\end{array}}
\def\bea{\begin{array}}

\newcommand{\Tr}[1]{\mathrm{Tr}#1}
\newcommand{\bei}{\begin{itemize}}
\newcommand{\eei}{\end{itemize}}

\newcommand{\proj}[1]{\ket{#1}\!\bra{#1}}

\def\ra{\rangle}
\def\la{\langle}

\newcommand{\N}{\mathbb{N}}
\newcommand{\Q}{\mathbb{Q}}

\newcommand{\Id}{\mathbb{1}}
\newcommand{\w}{\omega}
\newcommand{\p}{\Vec{\textbf{p}}}

\renewcommand{\emph}[1]{\textbf{#1}}

\newcommand{\eqnref}[1]{(\ref{#1})}

\makeatletter
\newtheorem*{rep@theorem}{\rep@title}
\newcommand{\newreptheorem}[2]{%
\newenvironment{rep#1}[1]{%
 \def\rep@title{#2 \ref{##1}}%
 \begin{rep@theorem}}%
 {\end{rep@theorem}}}
\makeatother
\usepackage{amsthm}

\newtheorem{innercustomgeneric}{\customgenericname}
\providecommand{\customgenericname}{}
\newcommand{\newcustomtheorem}[2]{%
  \newenvironment{#1}[1]
  {%
   \renewcommand\customgenericname{#2}%
   \renewcommand\theinnercustomgeneric{##1}%
   \innercustomgeneric
  }
  {\endinnercustomgeneric}
}

\newcustomtheorem{customthm}{Theorem}
\newcustomtheorem{customlemma}{Lemma}
\newcustomtheorem{customobs}{Observation}
\theoremstyle{plain}
\newtheorem{thm}{Theorem}
\newtheorem*{thm*}{Theorem}
\newreptheorem{thm}{Theorem}

\newtheorem{fakt}{Fact}

\newtheorem*{ucor}{Corollary}
\newtheorem{defn}[thm]{Definition}

\theoremstyle{definition}

\theoremstyle{remark}

\usepackage[T1]{fontenc}
\title{Certification of entangled quantum states and quantum measurements
in Hilbert spaces of arbitrary dimension}

\date{September, 2020}
\newcommand{\thesisauthor}{Shubhayan Sarkar
} 
\newcommand{\supervisor}{Dr. Remigiusz Augusiak}

\begin{document}

\input{Template/Title.tex}
\blankpage{}
\thispagestyle{empty}%
\topskip0pt
\vspace*{\fill}
\begin{center}
    \textit{Dedicated to my parents, my mother Swapna Sarkar and father Soumen Sarkar,}\\
  \textit{for making me who I am today.}
\end{center}
\vspace*{\fill}
\blankpage{}
\pagenumbering{roman}
\input{Template/Abstract.tex}
\pagebreak
\blankpage{}

\input{Template/polish_abstract.tex}
\pagebreak
\blankpage{}

\input{Template/Declaration}
\pagebreak
\blankpage{}
\input{Template/Acknowledgement.tex}
\pagebreak
\clearpage
\thispagestyle{empty}%
\tableofcontents


\pagenumbering{gobble}
\pagebreak
\clearpage
\thispagestyle{empty}%
\pagebreak
\input{Sections/chapter_1}
\input{Sections/chapter_2}

\input{Sections/chapter_3}
\input{Sections/chapter_4}

\input{Sections/chapter_5}

\input{Sections/conclusions}
\clearpage

\printbibliography
\clearpage

\input{Sections/appendices}
\clearpage



\end{document}

%% file: Template/Title.tex
\makeatletter
\begin{titlepage}
    \begin{center}
        \hrule
        \vspace*{1cm}
        \Huge
        \textbf{\@title}
        \vspace{1cm}
        \hrule
        \vspace{0.5cm}
        \begin{figure}[htbp]
             \centering
             \includegraphics[width=.3\linewidth]{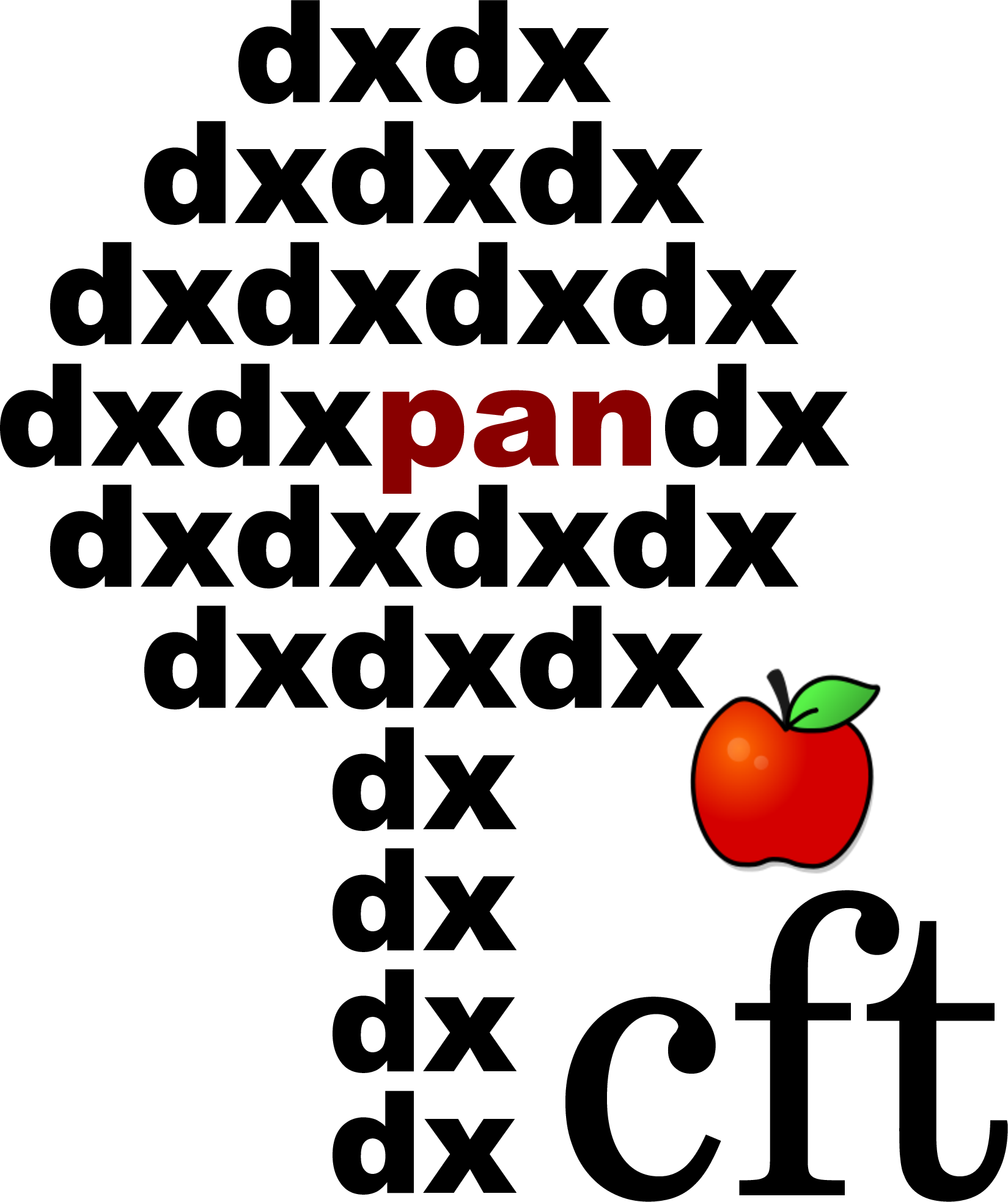}
        \end{figure}

        \vspace{1.5cm}

        \large
        \textbf{Author}: \Large{\thesisauthor{}}\\
        \large
        \textbf{Supervisor}: \Large{\supervisor{}}\\

        \vspace{1cm}
        \Large{\textbf {Centrum Fizyki Teoretycznej Polskiej Akademii Nauk}}\\
        \vspace{1cm}
       \small{\textit{A thesis submitted in partial fulfillment of the requirements for the degree of}}\\
       \textit{Doctor of Philosophy in Physics}\\
       \textit{}\\
        
        \vspace{1cm}
        {\normalsize September 2022}

    \end{center}
\end{titlepage}
\makeatother

%% file: Template/Abstract.tex
\vspace{-1cm}
\begin{center}
    \section*{Abstract}
\end{center}
\label{sec:abstract}

The emergence of quantum theory at the beginning of 20$-th$ century has changed our view of the microscopic world and has led to applications such as quantum teleportation, quantum random number generation and quantum computation to name a few, that could never have been realised using classical systems. One such application that has attracted considerable attention lately is device-independent (DI) certification of composite quantum systems. The basic idea behind it is to treat a given device as a black box that given some input generates an output, and then to verify whether it works as expected by only studying the statistics generated by this device. The novelty of these certification schemes lies in the fact that one can almost completely characterise the device (up to certain equivalences) under minimal physically well-motivated assumptions such as that the device is described using quantum theory. The resource required in most of these certification schemes is quantum non-locality.

A lot of work has recently been put into finding DI certification schemes for composite quantum systems. Most of them are however restricted to lower-dimensional systems, in particular two-qubit states. In this thesis, we consider the problem of designing general DI schemes that apply to composite quantum systems of arbitrary local dimensions. First, we construct a fully DI certification scheme, also known as self-testing, that allows us to certify generalised Greenberger-Horne-Zeilinger (GHZ) states of arbitrary local dimension shared among any number of parties from the maximal violation of a certain family of Bell inequalities, for two parties, the generalised GHZ state represents the two-qudit maximally entangled state. Importantly, this is the first instance where such states can be certified using only two measurements per party which is in fact the minimal number of measurements required to observe quantum non-locality. 


While a substantial progress has been recently made in designing device-independent certification schemes, most of these schemes are concerned with entangled quantum states. 
At the same time the problem of certification of quantum measurements remains largely unexplored. In particular, a general scheme allowing one to certify any set of incompatible quantum measurements has not been proposed so far. As designing such a scheme within the DI setting is certainly a difficult task, here we consider a relaxation of the Bell scenario
%
known as the one-sided device-independent (1SDI) scenario. In this scenario, we have an additional assumption that one of the parties is trusted and the measurements performed by this party are known. We propose a scheme for certification of a general class of projective measurements, termed here ``genuinely incompatible". To this end, we construct a family of steering inequalities that are maximally violated by any set of genuinely incompatible measurements. Interestingly,  mutually unbiased bases belong to this class of measurements. Finally, in the 1SDI scenario, we construct a family of steering inequalities, whose maximal violation can be used to certify any pure entangled bipartite state using the minimal number of two measurements per observer. Building on this result, we then provide a method to certify any rank-one extremal measurement, including non-projective measurements on the untrusted side.

Interestingly, self-testing of entangled states and measurements can be harnessed to propose schemes to certify that the outcomes of measurements performed on quantum states are perfectly random in the sense that they can not be predicted by an external party. This makes our scheme suitable for quantum cryptographic tasks. We first show that one can generate randomness in a fully DI way using projective measurements from composite quantum states of arbitrary local dimension. Later in the 1SDI scenario, we construct a scheme to certify the optimal amount of randomness that can be generated using a quantum system of any dimension and non-projective extremal measurements, which is twice the amount one can generate using projective measurements. 

%% file: Template/polish_abstract.tex
\vspace{-1cm}
\begin{center}
    \section*{Streszczenie}
\end{center}    

Powstanie teorii kwantów na początku XX wieku zmieniło nasz pogląd na świat mikroskopowy i doprowadziło do powstania takich zastosowań efektów kwantowych jak teleportacja kwantowa, kwantowe generowanie liczb losowych i obliczenia kwantowe. Żadne z nich nie mogło by zostać zrealizowane w ramach fizyki klasycznej. Jednym z takich zastosowań, które przyciągnęło ostatnio wiele uwagi, jest certyfikacja złożonych układów kwantowych w wersji niezależnej od urządzeń (ang. \textit{device-independent}). Podstawową jej ideą jest traktowanie danego urządzenia jak czarną skrzynkę, która po podaniu danych wejściowych generuje dane wyjściowe, a następnie wykorzystane obserwowanych statystyk do sprawdzenia, czy urządzenie to działa zgodnie z oczekiwaniami. Nowatorskość schematów certyfikacji tego typu polega na tym, że można prawie całkowicie scharakteryzować urządzenie (z dokładnością do pewnych równoważności) przy minimalnych, dobrze umotywowanych fizycznie założeniach, takich jak to, że urządzenie działa zgodnie z zasadami fizyki kwantowej. Zasobem kwantowym wykorzystywanym przez większość tych schematów certyfikacji jest nielokalność Bella.

Wiele pracy włożono ostatnio w stworzenie schematów certyfikacji w wersji niezależnej od urządzeń dla złożonych układów kwantowych. Większość z nich stosuje się jednak do układów o relatywnie niskich wymiarach lokalnych, w szczególności do stanów dwukubitowych. W tej pracy rozważamy problem projektowania ogólnych schematów, które mają zastosowanie do układów kwantowych o dowolnych wymiarach lokalnych. Najpierw konstruujemy w schemat certyfikacji, znany również jako samotestowanie, który pozwala certyfikować uogólnione stany Greenbergera-Horne'a-Zeilingera (GHZ) o dowolnym wymiarze lokalnym
dzielony przez dowolną liczbę obserwatorów na podstawie maksymalnego łamania pewnej rodziny nierówności Bella; w szczególnym przypadku dwóch podukładów stan GHZ sprowadza się do stanu maksymalnie splątanego dwóch kuditów. Co ważne, jest to pierwszy przypadek, w którym takie stany kwantowe mogą być certyfikowane przy użyciu tylko dwóch pomiarów przez każdego z obserwatorów, co jest w rzeczywistości minimalną liczbą pomiarów wymaganą do zaobserwowania nielokalności Bella.

Choć w ostatnich latach dokonano znacznego postępu w projektowaniu schematów certyfikacji w wersji niezależnej od urządzeń, większość z nich dotyczy splątanych stanów kwantowych. Jednocześnie problem certyfikacji pomiarów kwantowych pozostaje w dużej mierze niezbadany.
Brakuje w szczególności ogólnego schematu pozwalającego na certyfikację dowolnego zestawu niekompatybilnych pomiarów kwantowych. Ponieważ stworzenie takiego schematu w scenariuszu niezależnym od urządzeń jest trudnym zadaniem, w rozprawie rozważamy pewien uproszczony scenariusz, znany jako scenariusz jednostronnie niezależny od urządzeń (1SDI). W tym scenariuszu czynimy dodatkowe założenie, że jedno z urządzeń pomiarowych jest w pełni scharakteryzowane i wykonuje znane pomiary kwantowe. Proponujemy schemat certyfikacji ogólnej klasy pomiarów rzutowych, określanych tu jako prawdziwie niekompatybilne. W tym celu konstruujemy rodzinę nierówności sterowania (ang. \textit{steering inequalities}), których maksymalna wartość kwantowa osiągana jest przez dowolny zbiór prawdziwie niekompatybilnych pomiarów. Co ciekawe, do tej klasy pomiarów kwantowych zaliczają się te, które odpowiadają bazom wzajemnie niejednoznacznym. Wreszcie, w scenariuszu 1SDI, konstruujemy rodzinę nie-\linebreak równości sterowania, których maksymalne łamanie może być wykorzystane do certyfikacji dowolnego czystego, splątanego stanu dwucząstkowego przy użyciu minimalnej liczby dwóch pomiarów wykonywanych przez obu obserwatorów. Opierając się na tym wyniku, podajemy następnie metodę certyfikacji dowolnego ekstremalnego pomiaru kwantowego rzędu jeden, włączając w to pomiary nierzutowe.

Co ciekawe, metody samotestowania stanów oraz pomiarów kwantowych mogą być wykorzystane do stworzenia schematów poświadczania, że wyniki pomiarów wykonywanych na stanach kwantowych są losowe w tym sensie, że nie mogą być przewidziane przez żadnego zewnętrznego obserwatora. To sprawia, że nasze wyniki stają się użyteczne w zadaniach kryptograficznych. Najpierw pokazujemy, że ze splątanych stanów kwantowych o dowolnym wymiarze lokalnym można generować losowość w scenariuszu niezależnym od urządzeń używając pomiarów rzutowych. Następnie, w scenariuszu 1SDI, konstruujemy schemat poświadczający optymalną ilość losowości, która może być wygenerowana przy użyciu układu kwantowego o dowolnym wymiarze i nierzutowych pomiarów ekstremalnych, która równa jest dwukrotności maksymalnej ilości losowości, którą można wygenerować przy użyciu pomiarów rzutowych. 
    

\label{sec:pol_abstract}

%% file: Template/Declaration.tex
\vspace{-1cm}
\begin{center}
    \section*{Declaration}
\end{center}
\label{sec:declaration}
The work described in this thesis was undertaken between October 2018 and April 2022 while the author was a doctoral student under the supervision of Prof. Remigiusz Augusiak at the Center for Theoretical Physics, Polish Academy of Sciences. All the required coursework was completed between October 2018 and July 2020 at the Institute of Physics, Polish Academy of Sciences. 
No part of this thesis has been submitted for any other degree at the Center for Theoretical Physics, Polish Academy of Sciences or any other scientific institution.
\vspace{0.02\textwidth}

This thesis is based upon four different research works carried on during the doctoral studies all of which are listed below. 
\vspace{0.01\textwidth}

     \begin{enumerate}

                \item \textbf{Shubhayan Sarkar}, Remigiusz Augusiak, \textit{Self-testing of multipartite GHZ states of arbitrary local dimension with arbitrary number of measurements per party}, \href{https://doi.org/10.1103/PhysRevA.105.032416}{Phys. Rev. A \textbf{105}, 032416 (2022) \cite{sarkar4}.}
                
                \item \textbf{Shubhayan Sarkar}, Jakub J. Borkała, Chellasamy Jebarathinam, Owidiusz Makuta, Debashis Saha, Remigiusz Augusiak, \textit{Self-testing of any pure entangled state with minimal number of measurements and optimal randomness certification in one-sided device-independent scenario}, \href{https://arxiv.org/abs/2110.15176}{arXiv:2110.15176 (2021) \cite{sarkar3}.}

                \item  \textbf{Shubhayan Sarkar}, Debashis Saha, Remigiusz Augusiak, \textit{Certification of incompatible measurements using quantum steering},
               \href{https://arxiv.org/abs/2107.02937  }{arXiv:2107.02937   (2021) \cite{sarkar2}.}
               
                \item \textbf{Shubhayan Sarkar}, Debashis Saha, Jędrzej Kaniewski, Remigiusz Augusiak, \textit{Self-testing quantum systems of arbitrary local dimension with minimal number of measurements}, 
               \href{https://www.nature.com/articles/s41534-021-00490-3}{npj Quantum Information \textbf{7}, 151 (2021) \cite{sarkar}. }
               
\end{enumerate}
In addition to the work presented in this thesis, the author has also carried out the following research tasks all of which are listed below.

\begin{enumerate}
                \item \textbf{Shubhayan Sarkar}, Chandan Datta, Saronath Halder, Remigiusz Augusiak, Self-testing composite measurements and bound entangled state in a unified framework, \href{https://arxiv.org/abs/2301.11409}{	arXiv:2301.11409 [quant-ph] (2023) \cite{sarkar8}}.
                
                \item \textbf{Shubhayan Sarkar}, Certification of the maximally entangled state using non-projective measurements, \href{https://arxiv.org/abs/2210.14099}{	arXiv:2210.14099 [quant-ph] \cite{sarkar7}}.
                
                \item  Jakub Jan Borkała, Chellasamy Jebarathinam, \textbf{Shubhayan Sarkar}, Remigiusz Augusiak, \textit{Device-independent certification of maximal randomness from pure entangled two-qutrit states using non-projective measurements}, 
               \href{https://doi.org/10.3390/e24030350}{Entropy \textbf{24(3)}, 350 (2022) \cite{sarkar5}}.

                \item \textbf{Shubhayan Sarkar}, Debashis Saha, \textit{Probing measurement problem of quantum theory with an operational approach}, \href{https://arxiv.org/abs/2107.08447 }{arXiv:2107.08447  (2021) \cite{sarkar10}.}

                \item \textbf{Shubhayan Sarkar}, \textit{Universal notion of classicality based on ontological framework}, \href{https://arxiv.org/abs/2104.14355  }{arXiv:2104.14355 (2021) \cite{sarkar9}.}
                
                \item Colin Benjamin, \textbf{Shubhayan Sarkar}, \textit{Emergence of Cooperation in the thermodynamic limit}, \href{https://doi.org/10.1016/j.chaos.2020.109762}{Chaos, Solitons \& Fractals \textbf{135}, 109762 (2020) \cite{sarkar11}.}
                
                \item Colin Benjamin, \textbf{Shubhayan Sarkar}, \textit{Triggers for cooperative behavior in the thermodynamic limit: a case study in Public goods game}, \href{https://doi.org/10.1063/1.5085076}{Chaos \textbf{29}, 053131 (2019) \cite{sarkar12}.}
                
                \item \textbf{Shubhayan Sarkar}, Colin Benjamin, \textit{Quantum Nash equilibrium in the thermodynamic limit}, \href{https://doi.org/10.1007/s11128-019-2237-2}{Quantum
Inf. Process. \textbf{18}, 122 (2019) \cite{sarkar13}.}

\item \textbf{Shubhayan Sarkar}, Colin Benjamin, \textit{Entanglement makes free riding redundant in the thermodynamic limit}, \href{https://doi.org/10.1016/j.physa.2019.01.085}{Physica A: Statistical Mechanics and its Applications \textbf{521}, 607 (2019) \cite{sarkar14}.}

\item \textbf{Shubhayan Sarkar}, \textit{Entropy as a bound for expectation values and variances of a general quantum
mechanical observable}, \href{https://doi.org/10.1142/S0219749918500363}{International Journal of Quantum Information, \textbf{16}
1850036 (2018) \cite{sarkar15}.}

\item \textbf{Shubhayan Sarkar}, Chandan
Datta, \textit{Can quantum correlations increase in a quantum communication task?}, \href{https://doi.org/10.1007/s11128-018-2019-2}{Quantum Inf Process \textbf{17}, 248 (2018) \cite{sarkar16}.}
\end{enumerate}

%% file: Template/Acknowledgement.tex
\begin{center}
\section*{Acknowledgements}    
\end{center}
\label{sec:acknowledgement}

First and foremost, I would like to express my deepest gratitude to my supervisor, Remigiusz Augusiak. This thesis would not have been completed without his support and constant encouragement. He always helped me whenever I was stuck in academic as well as administrative problems. At all times, he was understanding and willing to help.

Furthermore, I feel indebted to thank Center for Theoretical Physics of the Polish Academy of Sciences for providing me with the research environment and also to the administrative team, whom I troubled a lot during my studies. Not to mention, they were a constant support and helped me even with my personal difficulties.

This thesis would not have been possible without the help of my group members Debashis Saha, Gautam Sharma, Jakub J. Borkała, Chellasamy Jebarathinam and Owidiusz Makuta and also Jędrzej Kaniewski, who were involved in the research tasks that finally led to this thesis. I would like to thank my friends Chandan Datta, Rubina Ghosh, Kaustav Sengupta, Sudeep Sarkar, Saubhik Sarkar, Suparna Saha, Ishika Palit, Suhani Gupta, Kiran Saba, Saikruba Krishnan, Rafael Santos, Julius Serbenta, Grzegorz Rajchel-Mieldzioc and Michele Grasso.

I would like to thank my family specially my parents who believed in me and were with me in my ups and downs. Also special thanks to Bohnishikha Ghosh for being a constant support.
\vspace{1 cm}

Finally, I would also like to thank the Foundation for Polish Science for their support through the First Team project
(No First TEAM/2017-4/31) co-financed by the European Union under the European
Regional Development Fund. The grant not just financed my stay in  Poland but also provided opportunities to visit various institutes across Europe to share my research works and learn from the leading experts in the field. 

%% file: Sections/chapter_1.tex
\pagenumbering{arabic}
\chapter{Introduction}
\label{}
\vspace{-1cm}
\rule[0.5ex]{1.0\columnwidth}{1pt} \\[0.2\baselineskip]
\definecolor{airforceblue}{rgb}{0.36, 0.54, 0.66}
\definecolor{bleudefrance}{rgb}{0.19, 0.55, 0.91}

The advent of quanta by Planck in 1900 for describing black-body radiation marked as the beginning of ``quantum era of theoretical physics". With the increasingly precise experiments on microscopic objects, it was soon realised that the structure of atom was not describable with classical physics but required a new non-classical description. In this pursuit, Schr$\mathrm{\ddot{{o}}}$dinger in 1923, using the ideas of Bohr and De-Broglie that microscopic objects might possess both wave and particle characteristics, came up with a wave equation describing the microscopic world, which is now known as the Schr$\mathrm{\ddot{{o}}}$dinger's equation. It is excellently successful in predicting the results of the experiments performed till date. However, the Schr$\mathrm{\ddot{{o}}}$dinger equation involves an object known as a wavefunction, whose meaning was hugely debated and was believed to be just a mathematical object without any physical significance. It was the work of Max Born that established that the wavefunction contains information about the probability of the system being at a particular position in space. This can be considered as one of the biggest paradigm shifts in the human understanding of nature, as the microscopic world might be inherently unpredictable, and the maximum information that can be gained is the probability of the system being at a particular state. Many prominent physicists did not buy this idea and even led Einstein to one of his famous quotes, ``God does not play dice". In fact, the unpredictable nature of the microscopic world can be considered as the foundational philosophy behind ``quantum theory". 

With even more experiments probing the microscopic regime, it was soon realised that quantum theory is the best description of this regime, even when the theory leads to counter-intuitive phenomena and paradoxes. With later works of pioneers like Heisenberg, Bohr, Von-Neumann and Dirac, to name a few, we arrive at four postulates on which quantum theory is based on [cf. \cite{Preskill, niel}].
\begin{enumerate}
    \item Any state of the system is represented by a density matrix $\rho$ acting on some Hilbert space $\mathcal{H}$.
    \item The time evolution of the quantum system is generated by completely positive and trace preserving (CPTP) maps. 
    \item A system composed of two subsystems $A$ and $B$ is described using a quantum state belonging to the Hilbert space $\mathcal{H}_A\otimes\mathcal{H}_B$ where $\mathcal{H}_{A}, \mathcal{H}_{B}$ are the Hilbert spaces associated with the subsystems $A,B$ respectively.
    \item Any measurement to observe the state of the system $\rho$ is represented by a set of positive operators $M=\{E_k\}$ that act on the Hilbert space $\mathcal{H}$. Here $k$ represents the outcome of the measurement and $\sum_kE_k=\mathbb{1}_{\mathcal{H}}$. The probability of observing the the outcome $k$ is denoted by $p(k|M,\rho)$ and can be computed as,
    \begin{eqnarray}
    p(k|M,\rho)=\Tr(E_k\rho).
    \end{eqnarray}
    \end{enumerate}
All the notions presented here will later be revisited again in details.

An important step towards the understanding of quantum theory was put forward by Einstein, Podolski and Rosen in 1935 in their seminal paper \cite{EPR35}, where it was claimed that quantum theory is incomplete. 
It was later speculated by physicists like Bohm and Von-Nuemann that an underlying classical-like theory would be able to predict all the quantum effects and would remove all the counter-intuitiveness of the theory. This remained only a theoretical idea until Bell's pioneering work in 1964 \cite{bel64}. He put forward a way to test whether quantum theory is inherently different from classical physics. It was later confirmed by experiments that any classical description is insufficient to explain some quantum phenomena \cite{Aspect, Aspect1, Bellex1, Bellex2}.  

This led to the understanding that there can exist tasks in which strategies involving quantum states and measurements would perform much better than classical strategies, giving birth to the field of quantum information theory. In this thesis, we explore one of the recently contrived areas in quantum information theory, namely device-independent quantum protocols \cite{Acin2} focusing particularly on device-independent certification of quantum states, measurements and intrinsic randomness that can be generated using quantum systems. 

The idea of device-independent certification has gained much attention lately, mainly due to their implications in quantum cryptography as well as foundations of quantum theory. 
Let us say that two spatially separated labs are given a device whose inner mechanism is unknown and thus they can be treated as black boxes. If an input is provided to this device, it produces an output. Based on various inputs and outputs, one can construct the statistics.  
Assuming that these devices satisfy the rules of quantum theory, any such statistics needs to be explained via quantum measurements acting on some quantum state. The basic idea behind device-independent certification is that using only the statistics obtained by performing local measurements on a composite quantum system, one can characterise this system and also the measurements up to certain equivalences. 

The strongest device-independent certification scheme is known as self-testing where no assumptions on the devices are made apart from the fact that they are governed by quantum theory. The idea of self-testing was first introduced by Mayers and Yao in 1998 \cite{Mayers, Yao} in the cryptographic context as a way of certifying that the parties share a desired state. Since then, numerous self-testing schemes for composite quantum systems and measurements have been introduced. For instance Refs. \cite{qubitst1, qubitst2, qubitst3, qubitst4, qubitst5, qubitst6, qubitst7, qubitst8, qubitst9, qubitst10} provide schemes for certification of pure bipartite entangled states that are locally qubits. Then, Refs. \cite{multist1, multist2, multist3, multist4, multist5, multist6, multist7, multist8, multist9} provide methods to certify multipartite states of local dimension two. A few works Refs. \cite{Jed1, ALLst1, ALLst2, quditst1, quditst2} provide certification schemes of entangled states of local dimension higher than two. In \cite{ALLst1} (see also \cite{ALLst2}) a strategy for certification of every pure bipartite entangled state was introduced. 

In Chapter \ref{chapter_3} of this thesis, we provide a general scheme that self-tests the generalised Greenberger-Horne-Zeilenger (GHZ) state \cite{GHZ} of arbitrary local dimension shared among arbitrary number of parties. Restricting to two parties, the above state is, in fact, the maximally entangled state of arbitrary local dimension. Unlike the previous self-testing scheme \cite{ALLst1} that allows for certification of arbitrary dimensional states, we use the minimum possible number of measurements per party, that is, two. This is particularly useful from an application point of view, as it reduces the experimental effort necessary to implement our scheme. With the aim to self-test measurements, we generalise this scheme to arbitrary number of measurements per party. An important application of our self-testing scheme is towards device-independent certification of genuine randomness. Sources generating genuine randomness are useful in many areas such as numerical computation or cryptography. We provide a scheme for certifying the maximum amount of randomness that can be extracted from a quantum system of arbitrary local dimension using projective measurements. This chapter is based on two of our works, \cite{sarkar} and \cite{sarkar4}. The first one proves self-testing statement for the maximally entangled state of two qudits based on the maximal violation of the SATWAP Bell inequality introduced in \cite{SATWAP}. The second paper generalises this result to the GHZ state shared among arbitrary number of parties based on the maximal violation of ASTA Bell inequality introduced in \cite{ACTA}.

Any quantum device consists mainly of two parts: a source that generates a quantum state and a measuring device that performs a measurement on this state. While there has been considerable progress in designing self-testing schemes for quantum states, the avenue for certification of quantum measurements remains largely unexplored. Recently, in \cite{random1} and \cite{sarkar5}, the authors introduced a way of self-testing the bases of the set of two-outcome and three-outcome observables respectively. Apart from a few results \cite{parallelst1, ALLst1, parallelst2, parallelst3, parallelst4, Jed1, mea1, mea2, qubitst8}, there is no general method allowing to certify any set of incompatible measurements.

To simplify the problem, depending on the physical scenario, one can make some assumptions about the quantum devices. In this thesis, we consider a scheme in which one of the parties is trusted, known as one-sided device-independent (1SDI) scenario. The resource that one needs here to certify quantum states or measurements is known as quantum steering, which is  another form of quantum non-locality \cite{Sch35, Wiseman}. The quantum steering scenario is similar to the Bell scenario, apart from the fact that one of the parties is trusted, or equivalently, that the measurement device of the trusted party performs known measurements. The first 1SDI certification scheme was introduced in 2016 \cite{Supic} (see also \cite{Alex}). Their scheme allows for certification of the two-qudit maximally entangled state and two binary outcome measurements. 

In Chapter \ref{chapter_4} of this thesis, we provide the first certification scheme that applies to a general family of incompatible projective measurements which we termed  ``genuinely incompatible". An important class of projective measurements that are well-studied in quantum information theory are those corresponding to mutually unbiased bases. For instance, they are crucial for the security of quantum cryptographic tasks \cite{MUB1, MUB2, MUB3, MUB4, MUB5}. Also, there is a close link between mutually unbiased bases and quantum cloning \cite{Cloning1, Cloning2, Cloning3} (see \cite{MUBreview}, for a review on mutually unbiased bases). As a matter of fact, we show that mutually unbiased bases are also genuinely incompatible, and thus we provide a certification scheme of mutually unbiased bases of arbitrary dimension. For two observables corresponding to mutually unbiased bases, we also find the robustness of our scheme against experimental errors, which makes it relevant for practical applications.
This chapter is based on our work \cite{sarkar2}.

Recently, quantum non-locality has been realised as an effective way to generate genuine randomness and device-independently verify that there is no intruder who has access to it \cite{Pironio1, Colbeck1, di4}. The maximum amount of randomness that one can, in principle, obtain from a quantum system of a dimension $d$ is $2\log_2 d$ bits. A long-standing question in quantum information theory is whether one can find protocols that can be used to certify this maximal amount of randomness. 

In Chapter \ref{chapter_5} of this thesis, we again consider the one-sided device-independent scenario. We begin by devising a 1SDI scheme that can be used to certify any pure bipartite entangled state. 
Importantly, our scheme utilises only two measurements per party, which is the minimal number of measurements necessary to observe quantum steering. Moreover, these measurements are independent of the state to be certified. This makes our scheme much easier to implement in experiments. Using these results, we provide a scheme for certification of any rank-one extremal generalised measurements. This allows us to finally certify the maximal amount of randomness that one can generate from a quantum system of any dimension. We further show for some dimensions that the amount of randomness is independent of the amount of entanglement in the quantum system.
This chapter is based on our work \cite{sarkar3}.

Before presenting the main results of this thesis, in Chapter \ref{chapter_2} we introduce the relevant technical concepts and notions which will be required throughout this thesis. The thesis ends with concluding remarks and a list of open problems in Chapter \ref{chapter_6}.

%% file: Sections/chapter_2.tex
\chapter{Technical introduction}
\label{chapter_2}
\vspace{-1cm}
\rule[0.5ex]{1.0\columnwidth}{1pt} \\[0.2\baselineskip]
\definecolor{airforceblue}{rgb}{0.36, 0.54, 0.66}
\definecolor{bleudefrance}{rgb}{0.19, 0.55, 0.91}


\section{Basics of quantum information theory} \label{2.1.1}

Let us begin by elaborating on the four postulates of quantum theory presented in Introduction. 
\subsection{States in quantum theory}
\begin{defn}[Pure quantum states] Pure quantum states are normalised vectors belonging to a Hilbert space $\mathcal{H}$ and are denoted by $\ket{\psi}$. The normalisation condition ensures that
\begin{eqnarray}\label{norm}
\langle\psi\ket{\psi}=1.
\end{eqnarray}
\end{defn}
In general, the state $\ket{\psi}$ can belong to Hilbert space of infinite dimension. However in this work, we consider only finite-dimensional Hilbert spaces. Any pure quantum state $\ket{\psi}\in\mathcal{H}$ such that the dimension of the Hilbert space is $d$, can be expressed using a set of $d$ linearly independent vectors, $\mathcal{B}=\{\ket{e_i}\}_{i=0}^{d-1}$ known as a basis. Now, any quantum system belonging to a $d-$dimensional Hilbert space is known as qudit. A two dimensional quantum system is known as qubit.
An important basis that is widely used in quantum information theory and will be extensively used in this work is known as the computational or standard basis, represented by $\mathcal{B}_c=\{\ket{i}\}_i$, where 
\begin{eqnarray}
\ket{0}=\begin{pmatrix}
1\\
0\\
0\\
\vdots\\0
\end{pmatrix},\quad
\ket{1}=\begin{pmatrix}
0\\
1\\
0\\
\vdots\\0
\end{pmatrix},\quad \ldots\quad
\ket{d-1}=\begin{pmatrix}
0\\
0\\
0\\
\vdots\\
1
\end{pmatrix}.
\end{eqnarray}
Note that the elements in the basis $\mathcal{B}_c$ are  orthogonal, that is, $\langle i\ket{j}=\delta_{i,j}$ where 
\begin{eqnarray}\label{compbasis}
\delta_{i,j}=
\begin{cases}
1\quad \text{if}\quad i=j\\
0\quad \text{if}\quad i\ne j
\end{cases}.
\end{eqnarray}
In certain situations, considered also in this thesis, one needs to use density matrices denoted by $\rho$ to describe a quantum system. These are matrices acting on the Hilbert space $\mathcal{H}$ that satisfy the following properties 
\begin{eqnarray}
\Tr(\rho)=1,\quad \rho\geq0,\quad \text{and}\quad \rho=\rho^{\dagger},
\end{eqnarray}
that is, they are normalised and positive semi-definite.
For any pure state $\ket{\psi}$, the corresponding density matrix is given by $\rho_p=\ket{\psi}\!\bra{\psi}$.
\begin{defn}[Mixed states]\label{mixedstate}
Quantum states that can be written as convex combination of other quantum states, are known as mixed states, that is, 
\begin{eqnarray}\label{mixedstateeq}
\rho_m=\sum_i p_i\ket{\psi_i}\!\bra{\psi_i}.
\end{eqnarray}
\end{defn}
\noindent For a note, a particular decomposition of $\rho_m$ \eqref{mixedstateeq} is not unique. 

Let us now consider the scenario when a system consists of two subsystems $A$ and $B$. Let $\mathcal{H}_A$ and $\mathcal{H}_B$ denote the Hilbert spaces of $A$ and $B$ respectively. According to the postulates of quantum theory the Hilbert space of their joint system is given by $\mathcal{H}_{AB}=\mathcal{H}_A\otimes\mathcal{H}_B$. In such composite systems, we can have two major classes of states, separable and entangled. Let us begin with pure separable states also known as product states.

\begin{defn}[Product states] Consider a pure state $\ket{\psi_{AB}}\in\mathcal{H}_{A}\otimes\mathcal{H}_B$. We call it a product state if it can be written as $\ket{\psi}_{AB}=\ket{\psi}_A\otimes\ket{\psi}_B$ where $\ket{\psi}_A\in\mathcal{H}_{A}$ and $\ket{\psi}_B\in\mathcal{H}_{B}$.
\end{defn}
 
\begin{defn}[Separable States \cite{Werner}]\label{sepstate} A mixed state $\rho_{sep}$ acting on $\mathcal{H}_{A}\otimes\mathcal{H}_B$ is called separable if it can be written as convex mixture of product states.
\end{defn}
Thus, any separable state can be written as
\begin{eqnarray}\label{sepstateeq}
\rho_{sep}=\sum_ip_i\ket{\psi_i}_A\bra{\psi_i}\otimes \ket{\phi_i}_B\bra{\phi_i}
\end{eqnarray}
where $\ket{\psi_i}_A\in \mathcal{H}_A$ and $\ket{\phi_i}_B\in \mathcal{H}_B$.
Let us now look at a class of states that do not exist in classical physics.
\begin{defn}[Entangled States] Quantum states that are not separable as defined in Def. \ref{sepstate} are classified as entangled states. 
\end{defn}
\noindent For example, when both the Hilbert spaces of $A$ and $B$ is $\mathbbm{C}^2$, then the quantum state
\begin{eqnarray}
     \ket{\psi_{AB}}=\frac{1}{\sqrt{2}}\left(\ket{0}_A\ket{0}_B+\ket{1}_A\ket{1}_B\right)
\end{eqnarray}
is entangled. 

A convenient way to express any pure bipartite state, that is quantum systems consisting of only two subsystems,  $\ket{\psi}_{AB}\in\mathcal{H}_{AB}$ is by using the Schmidt decomposition \cite{niel}. Let us say Hilbert space of subsystem $A$, denoted by $\mathcal{H}_A$ is of dimension $m$ and Hilbert space of subsystem $B$, denoted by $\mathcal{H}_B$ is of dimension $n$, then any state $\ket{\psi}_{AB}$ can be written as
\begin{eqnarray}\label{genent}
\ket{\psi}_{AB}=\sum_{i=1}^{\min\{m,n\}}\lambda_i\ket{e_i}\otimes\ket{f_i}
\end{eqnarray}
such that $\lambda_i\geq 0$ with $\sum_i\lambda_i^2=1$ and $\{\ket{e_i}\},\ \{\ket{f_i}\}$ are set of orthonormal vectors defined on $\mathcal{H}_A$ and $\mathcal{H}_B$ respectively\footnote{For ease of mathematical notation, most of the times the symbol ``$\otimes$" will be dropped.}. 

Let us now consider quantum systems consisting of $N$ subsystems where $N$ is any positive integer greater than two such that the local Hilbert spaces are denoted by $\mathcal{H}_i$ for $i=1,2,\ldots,N$. Then such a state is described by a density matrix  $\rho_N$ acting on $\mathcal{H}_1\otimes\mathcal{H}_2\otimes\ldots\otimes\mathcal{H}_N$.  We can straightforwardly generalise the notion of separability and entanglement to the multipartite states as done above. 
This completes the classification of quantum states. We now move on to characterising dynamics in quantum theory. 

\subsection{Dynamics in quantum theory}
Within quantum theory, any evolution in general is represented by completely positive and trace preserving (CPTP) maps. However, in this work we restrict ourselves to a family of maps that are known as unitary transformations or unitary matrices.

\begin{defn}[Unitary transformation]\label{unitary} A unitary transformation $U$ is a mapping from a Hilbert space $\mathcal{H}$ to $\mathcal{H}$ that preserves the distance between any two vectors belonging to $\mathcal{H}$.

\end{defn}
A unitary matrix $U$ acting on $\mathcal{H}$ is characterised by the following properties,
\begin{eqnarray}
UU^{\dagger}=\Id\quad \text{or}\quad U^{-1}=U^{\dagger},
\end{eqnarray}
that is, the inverse of any unitary matrix is equal to its conjugate transpose.
Interestingly, any CPTP map can be realised as a unitary map acting on some higher dimensional system \cite{stine}. In fact, the dynamics of closed quantum systems are reproduced by unitary maps. Another class of CPTP maps that is relevant for this work is known as isometry \cite{iso}. 
\begin{defn}[Isometry]\label{isometry} An isometry is generalisation of a unitary matrix, mapping one Hilbert space $\mathcal{H}$ of lower dimension to some other Hilbert space $\mathcal{H'}$ of higher dimension in a way that preserves the distance between any two vectors belonging the Hilbert space $\mathcal{H}$. 
\end{defn}
We now move on to characterising measurements in quantum theory.   

\subsection{Measurements in quantum theory}\label{2.1.3}

Any measurement $M$ with $m$ outcomes in quantum theory is defined by the set of positive operators $\{E_k\}$ that act on some Hilbert space $\mathcal{H}$ for $k=0,1,2,\ldots,m-1$. These positive operators are hermitian, that is, $E_k=E_k^{\dagger}$ and sum up to identity $\sum_k E_k=\Id_{\mathcal{H}}$  which is a consequence of the fact that the probabilities of outcomes must sum up to $1$. Due to this, measurements in quantum theory are also referred to as positive-operator valued measures or simply POVM's. The probability of observing the outcome $k$ if the measurement $M$ has been performed on a state $\rho\in\mathcal{H}$ is given by,
\begin{eqnarray}
p(k|M,\rho)=\Tr(E_k\rho).
\end{eqnarray}
It is clear to see from the above expression that $\sum_kp(k|M,\rho)=1$. Let us now try to understand the idea of normalisation of the state as discussed in the previous subsection \ref{2.1.1}. Suppose that state of the quantum system is $\ket{\psi}$. Now, we consider a two-outcome measurement $M_2$ with measurement elements
\begin{eqnarray}
     E_0=\ket{\psi}\!\bra{\psi},\quad\text{and}\quad E_1=\Id-\ket{\psi}\!\bra{\psi}.
\end{eqnarray}
Then the probability of observing the system in the quantum state $\ket{\psi}$ must be $1$ or equivalently, the measurement must always output the $0^{th}$ outcome. Thus, we have that 
\begin{eqnarray}
     p(0|M_2,\ket{\psi}\!\bra{\psi})=|\langle\psi\ket{\psi}|^2=1.
\end{eqnarray}
Consequently, the quantum state must be normalised so that the total probability of finding the system in any quantum state is one.

Quantum measurements for a given Hilbert space form a convex set.
Now, we look at a special class of POVM's that lie at the boundary of this set.
\begin{defn}[Extremal POVM's]
Any POVM that can not be written as a convex combination of other POVM's is called extremal.

\end{defn}
As was proven in \cite{APP05}, the measurement operators of any rank-one extremal POVM can be written as $E_k=\lambda_k\Pi_k$, where
\begin{eqnarray}
\Pi_k=\ket{\psi_k}\!\bra{\psi_k}
\end{eqnarray}
and $\lambda_k\geq 0$. Moreover, $E_k's$ are linearly independent for all $k$. It was further proven in \cite{APP05} that for POVM's that act on a Hilbert space $\mathcal{H}$ of dimension $d$, an extremal POVM can have atmost $d^2$ outcomes. Any other POVM can be written as convex combination of extremal POVM's.  
A special class of extremal POVM's, are known as projective measurements.
\begin{defn}[Projective measurements]
Projective measurements are POVM's where the measurement elements are represented by projectors $\Pi_i$ such that $\Pi_i\Pi_j=\delta_{i,j}\Pi_i$. 
\end{defn}
For any rank-one projective measurement, the corresponding measurement elements are given by $E_k=\Pi_k=\ket{\psi_k}\!\bra{\psi_k}$. Along with the condition that $\sum_k\Pi_k=\Id_\mathcal{H}$ and $\Pi_i\Pi_j=\delta_{i,j}\Pi_i$, it can be concluded that $\{\ket{\psi_{k}}\}$ for all $k's$ form a complete basis of the Hilbert space $\mathcal{H}$ which the measurement acts on. 
The state of the system after the measurement is performed on it is known as the post-measurement state. For instance, let us consider the  measurement $\{E_k\}$ and a quantum state $\rho$, then the post-measurement state $\rho_{pm}^a$ after the outcome $a$ is observed is given by,
\begin{eqnarray}
\rho_{pm}^a=\frac{\sqrt{E_a}\rho\sqrt{E_a}}{\Tr(E_a\rho)}\qquad \forall a.
\end{eqnarray}

An equivalent way to represent measurements in quantum theory is by using quantum observables. Let us first consider the simplest scenario where the measurements have only two outcomes and are projective. The quantum observable $\mathbb{A}$ corresponding to such a measurement $M=\{\Pi_0,\Pi_1\}$ is represented by,
\begin{eqnarray}
\mathbb{A}=\Pi_0-\Pi_1.
\end{eqnarray}
From the above expression, we can conclude that such quantum observables are hermitian and unitary,
\begin{eqnarray}
\mathbb{A}=\mathbb{A}^{\dagger},\quad \text{and}\quad \mathbb{A}\mathbb{A}^{\dagger}=\Id.
\end{eqnarray}
One can define the expectation value of $\mathbb{A}$ in terms of the probabilities of obtaining outcome $0,1$ as 
\begin{eqnarray}
\langle\mathbb{A}\rangle_{\rho}=\Tr(\mathbb{A}\rho)=p(0|M,\rho)-p(1|M,\rho).
\end{eqnarray}
The above construction of quantum observables can also be generalised to two-outcome POVM's where $\mathbb{A}=E_0-E_1$. However, such an observable is not unitary.
This construction of quantum observables was generalised to arbitrary outcome POVM's in \cite{Jed1}. We refer them here as generalised observables. Consider a $d-$outcome measurement. One defines generalised expectation values $\langle\mathbb{A}^{(l)}\rangle_{\rho}$ for $l=0,1,\ldots,d-1$ as the Fourier transform of the probabilities $p(k|M,\rho)$ as,
\begin{eqnarray}\label{obs0}
\langle\mathbb{A}^{(l)}\rangle_{\rho}=\sum_{k=0}^{d-1}\omega^{lk}p(k|M,\rho)\quad \text{for}\quad l=\{0,1,\ldots,d-1\}
\end{eqnarray}
where $\omega=e^{2\pi i/d}$ is the $d-th$ root of unity. Notice that using inverse Fourier transform, we can obtain the probabilities from the expectation values as,
\begin{eqnarray}\label{obs0.1}
p(k|M,\rho)=\frac{1}{d}\sum_{l=0}^{d-1}\omega^{-lk}\langle\mathbb{A}^{(l)}\rangle_{\rho}\quad \text{for}\quad k=\{0,1,\ldots,d-1\}
\end{eqnarray}
Using the above definition \eqref{obs0}, analogous to the two-outcome case, one can define the following expression for the corresponding observables $\mathbb{A}^{(l)}$ in terms of the measurement elements $E_k$
\begin{eqnarray}\label{obs1}
\mathbb{A}^{(l)}=\sum_{k=0}^{d-1} \omega^{lk}E_k\quad \text{for}\quad l=\{0,1,\ldots,d-1\}.
\end{eqnarray}
Equivalently the measurement elements $E_k$ can be obtained from the quantum observables $\mathbb{A}^{(l)}$ by considering the inverse Fourier transform,
\begin{eqnarray}\label{obs21}
E_k=\frac{1}{d}\sum_{l=0}^{d-1} \omega^{-lk}\mathbb{A}^{(l)}\quad \text{for}\quad k=\{0,1,\ldots,d-1\}.
\end{eqnarray}
Some relevant properties about the observable $\mathbb{A}^{(l)}$ can be obtained from Eq. \eqref{obs1} such as,
\begin{eqnarray}\label{obs2}
\mathbb{A}^{(d)}=\Id, \quad\text{and}\quad \mathbb{A}^{(d-l)}=\mathbb{A}^{(-l)}=\mathbb{A}^{(l){\dagger}}
\end{eqnarray}
along with
\begin{eqnarray}\label{obs3}
\mathbb{A}^{(l)}\mathbb{A}^{(l){\dagger}}\leq\Id,\qquad\text{and}\qquad \mathbb{A}^{(l){\dagger}}\mathbb{A}^{(l)}\leq\Id,
\end{eqnarray}
where to obtain the relations \eqref{obs2} we used the fact that $\sum_kE_k=\Id$ and $E_k=E_k^{\dagger}$. The relation \eqref{obs3} was proven in \cite{Jed1}. 
An important result about generalised observables that are unitary was also proven in \cite{Jed1}.

\begin{fakt}\label{factgenobs1}
Consider the generalised observables $\mathbb{A}^{(l)}$ as defined in \eqref{obs1}. These observables are unitary, that is,
\begin{eqnarray}
\mathbb{A}^{(l)}\mathbb{A}^{(l)\dagger}=\Id,\quad \text{and}\quad  \mathbb{A}^{(l)}=\mathbb{A}^l \quad \forall l
\end{eqnarray}
if and only if the corresponding measurement is projective, that is, $E_kE_{k'}=\delta_{k,k'}E_k$ for every $k,k'$.  
\end{fakt}
This fact will be used extensively throughout this thesis. Often at times, the generalised observables $\mathbb{A}^{(l)}$ would be referred to as measurements as they contain all the information to reconstruct the actual measurement. Consider now a multipartite system and on each subsystem a local measurement $M_i=\{E_{i,k}\}$ is performed where $i=1,2,\ldots,N$ denotes the subsystems. The probability to obtain outcomes $k_1,k_2,\ldots k_N$ when the measurements are performed on a state $\rho_N$ acting on $\mathcal{H}_1\otimes\mathcal{H}_2\otimes \ldots\otimes\mathcal{H}_N$,
\begin{eqnarray}
p(k_1,k_2,\ldots,k_N|\rho_N)=\Tr(E_{1,k_1}\otimes E_{2,k_2}\otimes\ldots E_{N,k_N}\rho_N)
\end{eqnarray}
The notion for the observables can also be generalised to the multipartite case
\begin{eqnarray}
\mathbb{A}_{1}^{(l_1)}\otimes\mathbb{A}_{2}^{(l_2)}\otimes\ldots\otimes\mathbb{A}_{N}^{(l_N)}=\sum_{k_1,k_2,\ldots,k_N=0}^{d-1}\omega^{l_1k_1+l_2k_2+\ldots l_Nk_N}E_{1,k_1}\otimes E_{2,k_2}\otimes\ldots \otimes E_{N,k_N}.
\end{eqnarray}
for every $l_1,l_2,\ldots,l_N$. 

An important distinction between any classical and quantum theories is the existence of quantum measurements that are mutually incompatible. 
Two projective measurements $M_1$ and $M_2$ are mutually incompatible, if these two measurements do not commute. For example, consider the Pauli observables $\sigma_z$ and $\sigma_x$ corresponding to  projective measurements that acts on qubits in $z$ and $x$ direction respectively
\begin{eqnarray}\label{pauliz,x}
\sigma_z=\begin{pmatrix}
1 & 0\\
0 &-1 
\end{pmatrix},\quad\text{and}\quad
\sigma_x=\begin{pmatrix}
0 & 1\\
1 & 0 
\end{pmatrix}
\end{eqnarray}
are incompatible. For a review of incompatibility of quantum measurements refer to \cite{Busch, Guhne1}.

\subsection{Purification of quantum states and measurements}
Another important concept for multipartite quantum systems is the extraction of the local quantum state for each subsystem given some global quantum state. For simplicity, we review it here for bipartite quantum systems but the concept can be straightforwardly generalised to the multipartite case. Let us suppose the global quantum state is given by $\rho_{AB}$, then the local quantum states $\rho_A$ and $\rho_B$ are given by
\begin{eqnarray}
\rho_A=\Tr_B(\rho_{AB})\quad \text{and}\quad \rho_B=\Tr_A(\rho_{AB})
\end{eqnarray}
where $\Tr_B$ and $\Tr_A$ denote partial trace over the subsystem $B$ and $A$ respectively and are computed as,
\begin{eqnarray}
\Tr_x(\rho_{AB})=\sum_j\bra{j_x}\rho_{AB}\ket{j_x}\qquad (x=A,B),
\end{eqnarray}
where $\{\ket{j_x}\}$ is any orthonormal basis than spans the Hilbert space $\mathcal{H}_x$ of the subsystem $x$ with $x=A,B$. For any separable state $\rho_{sep}$ [cf. Def. \ref{sepstate}] using Eq. \eqref{sepstateeq}, the local quantum states of the subsystem $A$ can be simply computed as
\begin{eqnarray}
\rho_A=\Tr_B(\rho_{sep})=\sum_ip_i\ket{\psi_i}_A\bra{\psi_i}\otimes \Tr\left(\ket{\phi_i}_B\bra{\phi_i}\right).
\end{eqnarray}
Using the fact that $\Tr\left(\ket{\phi_i}_B\bra{\phi_i}\right)=1$, we arrive at
\begin{eqnarray}
\rho_A=\Tr_B(\rho_{sep})=\sum_ip_i\ket{\psi_i}_A\bra{\psi_i}.
\end{eqnarray}
Analogously, the local quantum state of the subsystem $B$ is given by,
\begin{equation}
    \rho_B=\Tr_A(\rho_{sep})=\sum_ip_i\ket{\phi_i}_B\bra{\phi_i}.
\end{equation}
Computing the local quantum states of subsystem $A$ and $B$ when they are entangled is not as straightforward as separable states. For an example, let us consider a pure bipartite entangled state of local dimension $d$ given by,
\begin{eqnarray}\label{entsta1}
\ket{\psi}_{AB}=\sum_{i=0}^{d-1}\lambda_i\ket{i}_A\otimes\ket{i}_B
\end{eqnarray}
where $\{\ket{i}\}$ is the computational basis \eqref{compbasis} such that $\lambda_i> 0$ and $\sum_i\lambda_i^2=1$. For a note, any pure entangled bipartite state of local dimension $d$ can be written as \eqref{entsta1} up to some local unitary transformation. The local quantum state corresponding to the subsystem $\rho_A$ and $\rho_B$ are given by
\begin{eqnarray}
\rho_A=\sum_i\lambda_i^2\ket{i}\!\bra{i}_A,\quad \text{and}\quad \rho_B=\sum_i\lambda_i^2\ket{i}\!\bra{i}_B.
\end{eqnarray}
Any mixed quantum state [cf. Def. \ref{mixedstate}] can be realised as a pure state by adding an additional ancillary system to this quantum state. This is known as Stinespring's dilation \cite{stine}. A possible purification of any mixed quantum state $\rho$ admitting the form given in \eqref{mixedstateeq} can be written as,
\begin{eqnarray}
\ket{\psi_{mA}}=\sum_i\sqrt{p_i}\ket{\psi_i}_m\otimes\ket{e_i}_A
\end{eqnarray}
where $\{\ket{e_i}_A\}$ is an orthonormal basis over the Hilbert space of the ancillary system $A$. The mixed state $\rho_m$ can be extracted from this purified state as
\begin{eqnarray}
\rho_m=\Tr_A\left(\ket{\psi_{mA}}\!\bra{\psi_{mA}}\right).
\end{eqnarray}
On similar lines, any POVM can be realised as a projective measurement that acts on some higher dimensional system, known as Naimark's dilation \cite{paul}. Any $n$-outcome POVM $M=\{E_k\}$ that acts on the Hilbert space $\mathcal{H}$ can be realised using a projective measurement $\Pi_k$ and an isometry [cf. Def. \ref{isometry}] $V_{iso}:\mathcal{H}\rightarrow\mathcal{H}\otimes\mathcal{H}_A$ where $\mathcal{H}_A$ is some finite dimensional Hilbert space corresponding to some ancillary system $A$ \cite{Preskill, wat}, as
\begin{eqnarray}
 E_k=V_{iso}\Pi_k V_{iso}^{\dagger} \quad \forall k.
\end{eqnarray}
Here, the projectors $\Pi_k$ are given by
\begin{eqnarray}
\Pi_i=\Id_{\mathcal{H}}\otimes\ket{i}\!\bra{i}_A\quad \forall i
\end{eqnarray}
such that $\{\ket{i}\}$ is the computational basis and the isometry $V_{iso}$ is given by,
\begin{equation}
V_{iso}=\sum_{k=1}^n\sqrt{E_k}\otimes\ket{i}_A.  
\end{equation}

This completes the introduction of basic formalism of quantum information theory that is relevant to this work. We now move on to introducing concepts more central to this thesis.
\section{Quantum non-locality}
In the seminal work \cite{EPR35} published in 1935, Einstein, Podolsky and Rosen (EPR) pointed out to one of the key features of quantum theory which makes it inherently different from classical physics. In this work, two spatially separated quantum systems were considered on which two parties named Alice, and Bob can perform local measurements. The joint quantum state of both the systems were considered to be entangled. 
It was concluded in this work that quantum theory is incomplete. It was later hypothesised that there could exist some hidden variables that can not be directly observed but the knowledge of which would remove the paradoxes within quantum theory. In other words, these hidden variables would provide a classical explanation of the observed quantum phenomenon. In the subsequent years it was further noted by Schr$\mathrm{\ddot{o}}$dinger \cite{Sch35} from the EPR work, that Alice can in fact affect the quantum state of the system which is spatially separated from her by just performing measurements on her part of the system. The problem was debated upon for decades but mostly from a philosophical perspective without much consensus. We would elaborate on these ideas with details in the subsequent subsections.

The problem was revisited again by Bell in 1964 \cite{bel64, Bell66}. With the improved tools of quantum theory, Bell was able to devise a way that can put an end to the debate whether there exist any hidden variables that can provide a classical description of quantum theory. It turned out that there exist some set of joint probabilities that Alice and Bob can observe that can not be reproduced by local hidden variables [see below]. It was later confirmed in numerous experiments such as \cite{Aspect,Aspect1, Bellex1, Bellex2}, that it is indeed the case that quantum theory is intrinsically different from classical physics. This not just led to paradigm shift in foundation of physics but also led to enormous development in the newly born field of quantum information theory. We now proceed towards understanding Bell's solution to the EPR problem and the discovery of quantum non-locality also known as ``Bell non-locality".  

\subsection{Bell non-locality}
\begin{figure}
    \centering
    \includegraphics[scale=.75]{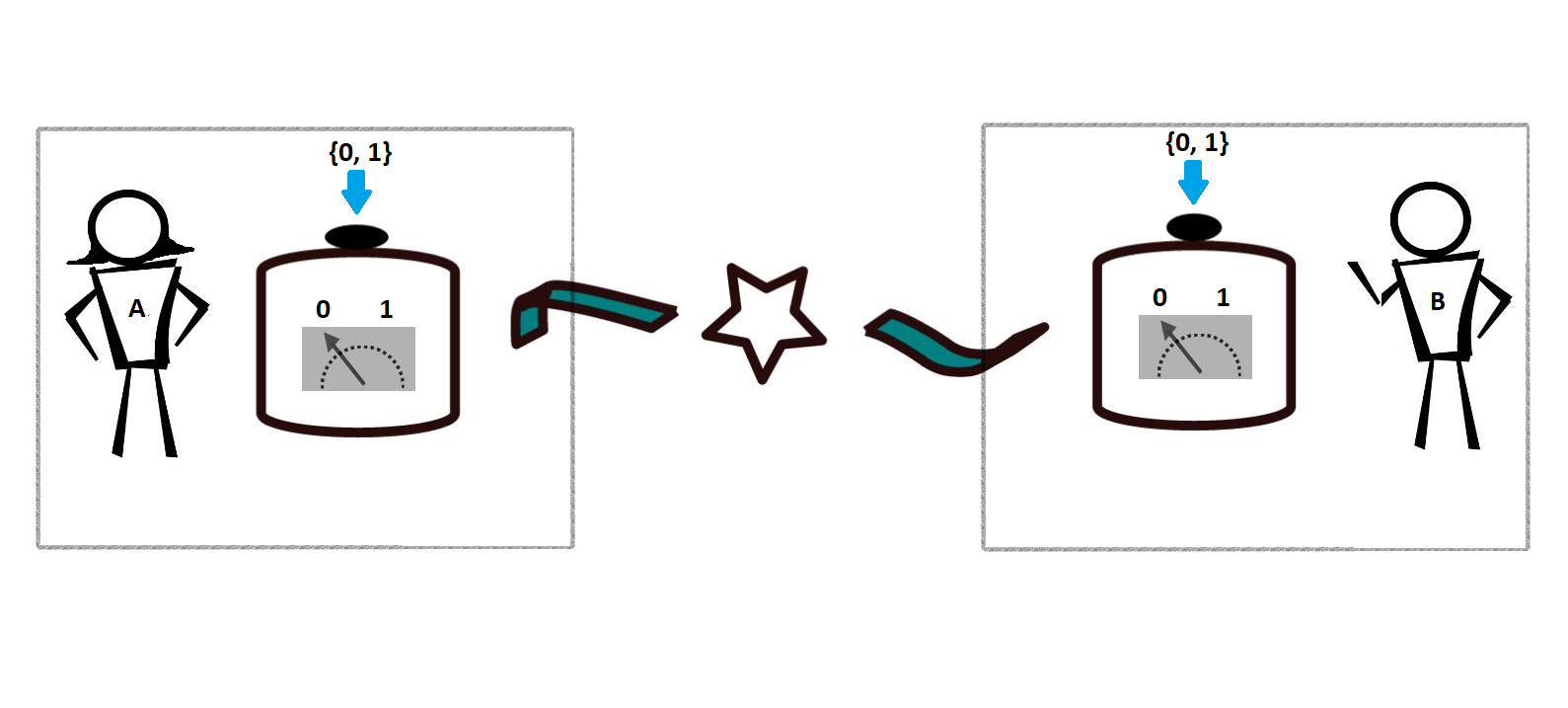}
    \caption{Simplest Bell scenario: Two parties Alice and Bob are located at separated labs. A source sends one subsystem to Alice and the other to Bob. Inside their labs, Alice and Bob perform two two-outcome measurements each on the received subsystem. After the experiment is complete, they compare their results to construct the joint probability distribution $p(a,b|x,y)$. }
    \label{fig1}
\end{figure}
We begin by considering the arguments by Einstein, Podolsky and Rosen in details. Consider two parties, namely Alice and Bob in two different spatially separated labs. Each of them receive a subsystem corresponding to an entangled quantum state such that knowing the position or momentum of either of the subsystem gives the information about the position and momentum of the counterpart. Now, Alice measures the position of her subsystem and Bob measures the momentum of his subsystem. EPR argued that for each of the subsystem one has information about its position as well as momentum which was forbidden in quantum theory as position and momentum are non-commuting measurements. This led them to conclude that there is an inherent inconsistency in quantum theory and it is incomplete. They argued that such a phenomenon should not exist which seems ``non-local" in the sense that one can know the state of a far away system based on the experiment done locally on its counterpart even when there is no interaction between them. Due to this, in the subsequent years it was hypothesised by physicists like Bohm and Von-Neumann that there might exist some hidden variables that would remove this inconsistency and provide a local description of quantum theory.

Building on this idea, Bell in his original work \cite{bel64} described a similar scenario. There are two parties Alice and Bob in two different labs that are spatially separated from each other. Both of them receive a quantum system from the preparation device. In their respective labs, Alice and Bob can perform two local measurements on their part of the system. During this operation they are not allowed to communicate among each other or equivalently, they are not supposed to know each others measurement choice or the outcomes of the measurements. The scenario is schematically depicted in Fig. \ref{fig1}. The experiment is repeated enough number of times to gather statistics corresponding to the joint probability distributions $\vec{p}=\{p(a,b|x,y)\}$ where $p(a,b|x,y)$ denotes the probability of obtaining outcome $a,b$ when Alice and Bob perform $x,y$ measurement respectively. One usually refers $\vec{p}$ as ``correlations".

Let us now again consider the hypothesis that there might exist some hidden variables that would provide a consistent local description of quantum theory. Let us denote those hidden variables by $\lambda$. In presence of such hidden variables the joint probability distribution can be expressed as
\begin{eqnarray}
p(a,b|x,y)=\sum_{\lambda\in\Lambda}p(a,b|x,y,\lambda)p(\lambda)
\end{eqnarray}
where $\Lambda$ denotes the set of the hidden variables $\lambda$, and $p(\lambda)$ denotes the probability with which a particular $\lambda$ occurs. Here, $p(a,b|x,y,\lambda)$ denotes the probability of occurrence of outcome $a,b$ given the input $x,y$ and the hidden variable $\lambda$. The above statement points to the fact that we do not have access to $\lambda$ and can only observe the probabilities averaged over it.

Let us now concentrate on $p(a,b|x,y,\lambda)$. From rule of conditional probabilities $p(a,b)=p(b|a)p(a)$ where $p(b|a)$ denotes the the conditional probability of occurrence of $b$ given $a$, we have
\begin{eqnarray}
p(a,b|x,y,\lambda)=p(a|x,\lambda)p(b|a,x,y,\lambda).
\end{eqnarray}
Now, the assumption of ``locality" or ``local realism" states that outcome of Bob does not depend on the outcome of Alice or her measurement choice
\begin{eqnarray}
p(b|a,x,y,\lambda)=p(b|y,\lambda).
\end{eqnarray}
Thus, for any local hidden variable, we have that
\begin{eqnarray}
p(a,b|x,y,\lambda)=p(a|x,\lambda)p(b|y,\lambda)\qquad \forall \lambda.
\end{eqnarray}
Consequently, any joint probability distribution admitting a local hidden variable model must be of the form
\begin{eqnarray}\label{locality}
p(a,b|x,y)=\sum_{\lambda\in\Lambda}p(a|x,\lambda)p(b|y,\lambda)p(\lambda).
\end{eqnarray}
Now, using these joint probabilities, one can construct a functional of the form
\begin{eqnarray}\label{genBell}
\mathcal{B}=\sum_{a,b,x,y}c_{a,b,x,y}p(a,b|x,y)
\end{eqnarray}
where $c_{a,b,x,y}$ are some real numbers. For ease of understanding we refer here to a functional presented by Clauser, Horne, Shimony and Holt (CHSH) \cite{CHSH} in which the cofficients in \eqref{genBell} are chosen as
\begin{eqnarray}
c(a,b,x,y)=\begin{cases}
1\quad \text{if}\quad a\oplus_2 b=x.y\\
-1\quad \text{otherwise}
\end{cases}
\end{eqnarray}
where $a\oplus_2 b$ represents $a+b$ modulo $2$. Usually the CHSH Bell functional is represented in the expectation value picture as
\begin{eqnarray}\label{CHSH}
\mathcal{B}_{CHSH}=\langle A_0\otimes B_0\rangle+\langle A_0\otimes B_1\rangle+\langle A_1\otimes B_0\rangle-\langle A_1\otimes B_1\rangle.
\end{eqnarray}
Assuming that the joint probabilities can be described by a local hidden variable model \eqref{locality}, one arrives at an upper bound of the CHSH expression \eqref{CHSH} given by
\begin{eqnarray}
\mathcal{B}_{CHSH}\leq 2.
\end{eqnarray}
The maximum value that one can achieve for a Bell expression using local hidden variable models is usually referred to as the ``classical bound" or the ``local bound" and is denoted by $\beta_L$. It turns out that there exist some states and measurements in quantum theory that violate this bound. To give an example let us consider that the preparation device in Fig. \ref{fig1} prepares the two-qubit maximally entangled state
\begin{eqnarray}\label{maxentstate}
\ket{\psi}_{AB}=\frac{1}{\sqrt{2}}\left(\ket{0}_A\ket{0}_B+\ket{1}_A\ket{1}_B\right).
\end{eqnarray}
Assume then that on her share of their state Alice and Bob perform the following measurements 
\begin{eqnarray}\label{1.49}
A_0=\sigma_z\qquad\text{and}\qquad A_1=\sigma_x,
\end{eqnarray}
and,
\begin{eqnarray}\label{1.50}
B_0=\frac{\sigma_z+\sigma_x}{\sqrt{2}}\qquad\text{and}\qquad B_1=\frac{\sigma_z-\sigma_x}{\sqrt{2}}.
\end{eqnarray}
 respectively. It follows that for this choice of the state and measurements, the value of the CHSH expression \eqref{CHSH} is
\begin{eqnarray}\label{CHSHquant}
\mathcal{B}_{CHSH}= 2\sqrt{2}.
\end{eqnarray}
One thus concludes that there exist joint probability distributions in quantum theory that violate the notion of local realism. Contrary to EPR, Bell showed that even if there exists some hidden variables that can not be observed directly, they are insufficient to give a local explanation of some predictions of quantum theory. This is known as ``Bell non-locality". 

As a matter of fact, the value $2\sqrt{2}$ in Eq. \eqref{CHSHquant} is the maximum value attainable using quantum state and measurements. This was proven by Tsirelson in \cite{Tsirel}, and is referred as ``Tsirelson bound" or simply ``quantum bound" and is denoted by $\beta_Q$.
 
An additional constraint in the above experiment as mentioned before is that there is no communication between Alice and Bob during the experiment. This restricts the joint probability distributions to be ``no-signalling" \cite{Sandu}, that is,
\begin{eqnarray}\label{NS1}
p(a|x)=\sum_bp(a,b|x,y)=\sum_bp(a,b|x,y')\qquad \forall y,y'
\end{eqnarray}
and
\begin{eqnarray}\label{NS2}
p(b|y)=\sum_ap(a,b|x,y)=\sum_ap(a,b|x',y)\qquad \forall x,x'.
\end{eqnarray}
Assuming that the joint probability distributions are no-signalling, one can even outperform quantum theory, that is,
\begin{eqnarray}
\mathcal{B}_{CHSH}= 4
\end{eqnarray}
which is also the algebraic bound of the Bell functional $\mathcal{B}_{CHSH}$. This is also known as the ``no-signalling bound" and is denoted by $\beta_{NS}$.

\begin{figure}[t]
    \centering
    \includegraphics[scale=.4]{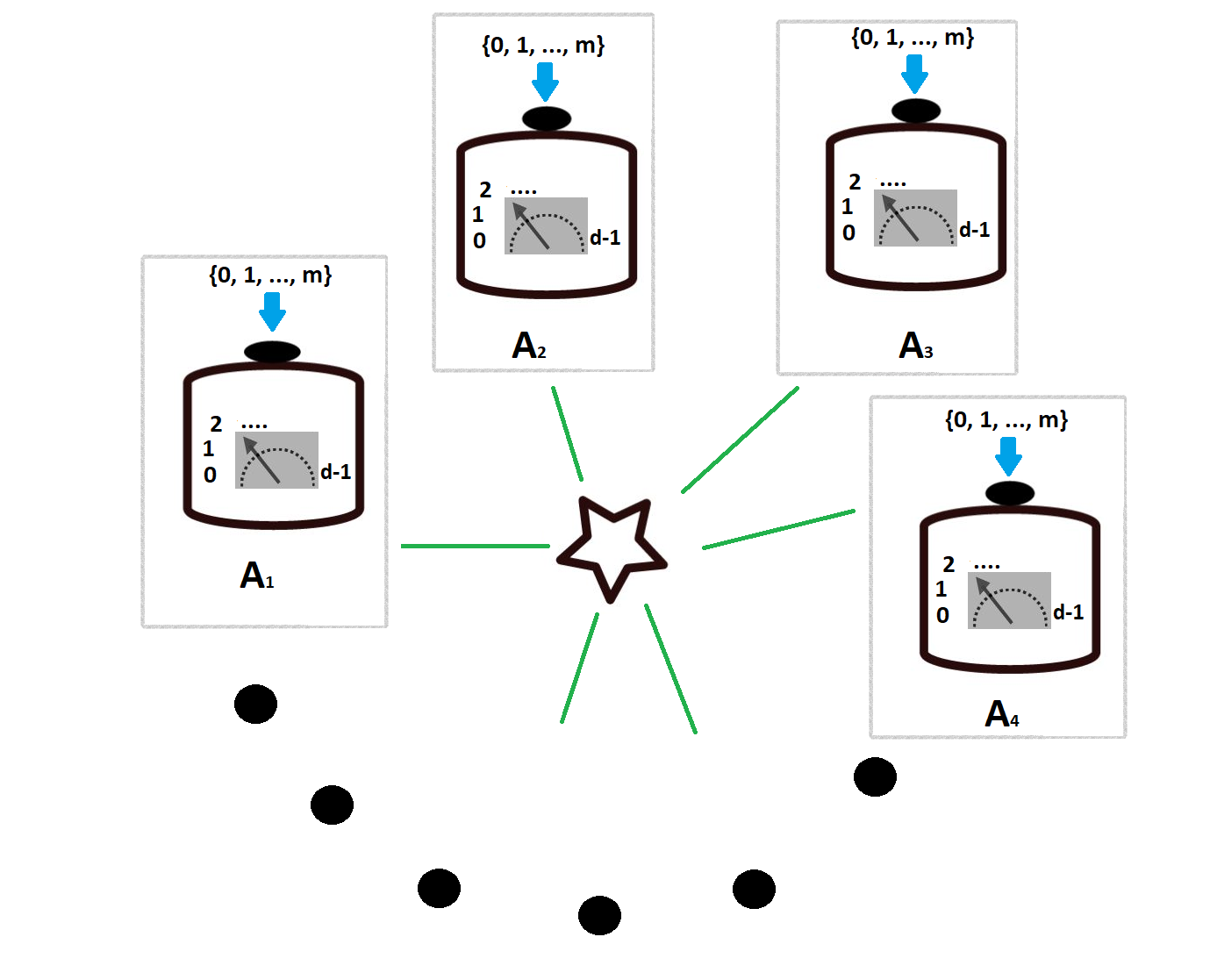}
    \caption{Generalised Bell scenario: $N$ parties denoted by $A_i$ for $i=1,2,\ldots,N$ are located at separated labs. A source sends one subsystem to each of the labs. Inside their labs, each party performs $m$ number of $d-$outcome measurements on the received subsystem. After the experiment is complete, they compare their results to construct the joint probability distribution $p(a_1,a_2,\ldots, a_N|x_1,x_2,\ldots,x_N)$.}
    \label{fig2}
\end{figure}

The above scenario can be straightforwadly generalised to the multipartite scenario where there is a preparation device, that sends subsystems to $N$ spatially separated parties denoted by $A_i$ for $i=1,2,,\ldots,N$. Each of the parties can now perform $m-$measurements each of which are $d-$outcome where $m,d$ are arbitrary positive integers strictly greater than $1$. The scenario is depicted in Fig. \ref{fig2}. 

The joint probability distributions in such a scenario is denoted by \begin{equation}
    \vec{p}=\{p(a_1,a_2,\ldots,a_N|x_1,x_2,\ldots,x_N)\},
\end{equation} where $a_k=\{0,1,\ldots,d-1\}$ denotes the outcome when $A_k$ performs the measurement $x_k=\{1,2,\ldots,m\}$. It is important to note here that
\begin{eqnarray}
\vec{p}\in\mathbb{R}^{(md)^N},
\end{eqnarray}
where $\mathbb{R}^{(md)^N}$ denotes the real vector space of dimension $(md)^N$. The Bell functional in the multipartite case can be written as
\begin{eqnarray}\label{genBell2}
\mathcal{B}_{m,d,N}=\vec{c}\cdot\vec{p}
\end{eqnarray}
where $\vec{c}=\{c_{a_1,a_2\ldots a_N,x_1,x_2,\ldots,x_N}\}$ and $\vec{c}\in\mathbb{R}^{(md)^N}$.

\begin{figure}[t]
    \centering
    \includegraphics{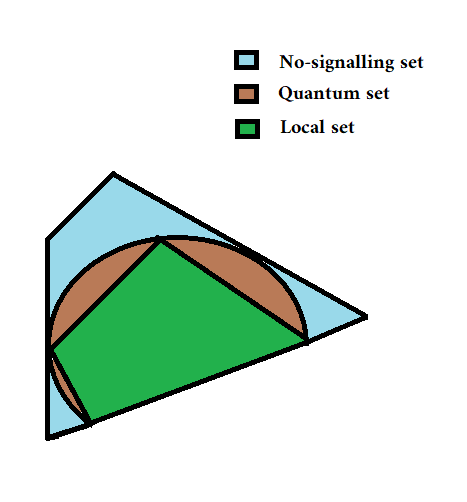}
    \caption{A two-dimensional representation of the set of correlations. The local set is a subset of the quantum set which is a subset of the no-signalling set. The no-signalling and local set are polytopes, that is, every point inside these sets can be written as a convex combination of the vertices of the polytope. On the other hand, the quantum set is not a polytope but convex.}
    \label{fig3}
\end{figure}

The collection of joint probability distributions that admit a local hidden variable are known as the ``set of local correlations" or simply ``local set" denoted by $\mathcal{L}_{m,d,N}$. Similarly, one can define ``set of quantum correlations" or simply ``quantum set" denoted by $\mathcal{Q}_{m,d,N}$ and ``set of no-signalling correlations" or ``no-signalling set" denoted by $\mathcal{N}_{m,d,N}$ referring to joint probability distributions that admit a quantum and no-signalling models respectively. One can understand Bell's non-locality arising from the fact that the local set lies strictly inside the quantum set. All these sets are convex in nature, that is, convex combination of different elements of the set is also an element belonging to this set. Elements of the set that can not be written as a convex combination of other elements are known as "extremal points" of the set and any other element in the set can be written as a convex combination of these extremal points. By definition, the local set lies inside the quantum set that lies inside the no-signalling set \cite{Bellreview},

\begin{eqnarray}
\mathcal{L}_{m,d,N}\subseteq\mathcal{Q}_{m,d,N}\subseteq\mathcal{N}_{m,d,N}.
\end{eqnarray}
It follows from the work of Bell \cite{bel64} and then Popescu and Rohrlich \cite{Sandu} that these inclusions are strict in the minimal scenario [m=d=N=2]. This is schematically represented in Fig. \ref{fig3}.
Now, Bell inequalities can be understood as a hyper plane that cuts the quantum set into two different parts. The first part consists of correlations that admit a local model and the second part consists the rest. We now move on to a different notion of non-locality known as quantum steering.

\subsection{Quantum steering}

\begin{figure}[t]
    \centering
    \includegraphics[scale=.35]{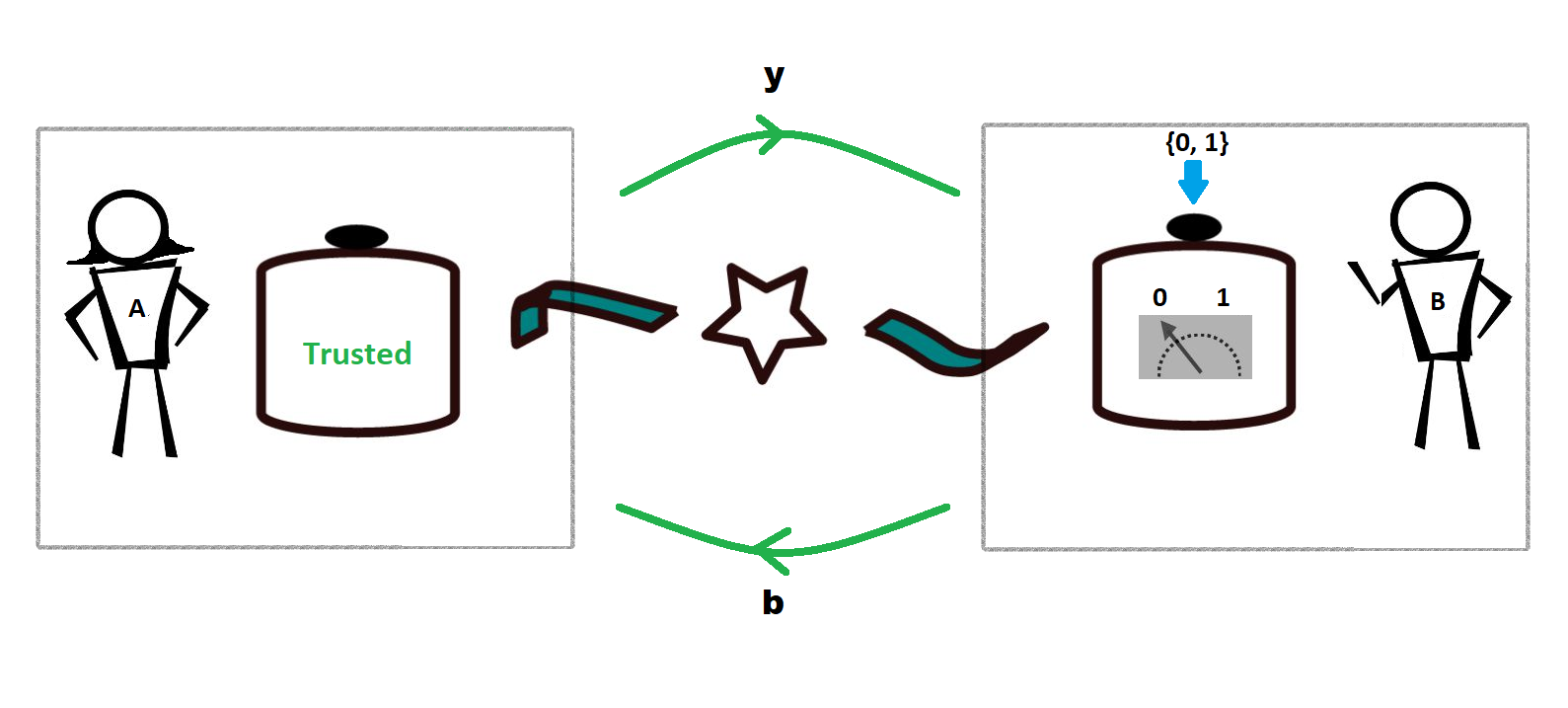}
    \caption{Simplest quantum steering scenario: Two parties Alice and Bob are located at separated labs. A source sends one subsystem to Alice and the other to Bob. Bob performs two two-outcome measurements on the received subsystem which might change the state of the subsystem of Alice. Alice is trusted and can perform a tomography on her subsystem and reconstruct the assemblage (the set of unnormalised states) and can figure out whether the assemblage is steerable or not.}
    \label{fig4}
\end{figure}

The idea of quantum steering was conceived by Schr$\mathrm{\ddot{o}}$dinger in 1935 but was formalised after almost $70$ years in 2007 by Wisemen et. al. \cite{Wiseman}. We begin by describing the simplest quantum steering scenario described in \cite{CJWR09} that consits of two spatially separated parties Alice and Bob. A preparation device sends one subsystem to Alice and another subsystem to Bob. Alice sends input $y=0,1$ to Bob based on which Bob performs a measurement on his subsystem and gets an outcome $b=0,1$ which is sent back to Alice. Contrary to Bell scenario, Alice is trusted which means that Alice has full control of her lab and can perform a tomography on her subsystem. The experiment is repeated enough number of times such that Alice can reconstruct the state of the subsystem $\sigma_{b|y}$ that acts on Alice's Hilbert space $\mathcal{H}_A$ for every $y,b$. As a matter of fact these states are un-normalised and they form a set known as an assemblage, denoted by $\{\sigma_{b|y}\}$. The scenario is schematically represented in Fig. \ref{fig4}.

Similar to the idea of local hidden variables in Bell non-locality, one can define a notion which every classical description of the experiment must abide. In the quantum steering scenario presented above, the classical notion corresponds to the assemblage $\{\sigma_{b|y}\}$ admitting a local hidden state model. One can understand this as, the assemblage observed by Alice only depends on some local state $\rho_\lambda$ that might be classically correlated with Bob, 
\begin{eqnarray}\label{LHS1}
\sigma_{b|y}=\sum_{\lambda\in\Lambda}p(\lambda)p(b|y,\lambda)\rho_\lambda. 
\end{eqnarray}
\noindent Here $p(b|y,\lambda)$ represents the probability of obtaining outcome $b$ by Bob given input $y$ and hidden variable $\lambda$ and $p(\lambda)$ denotes the probability distribution of the hidden variables. If the assemblage admits a local hidden state model then it is non-steerable from Bob to Alice. On the other hand, the assemblage is called steerable from Bob to Alice if no hidden state model can be constructed to express $\{\sigma_{b|y}\}$ as in \eqref{LHS1}. The way to detect whether the assemblage is steerable or not, is to use a steering functional of the form
\begin{eqnarray}\label{stefn}
W=\sum_{b}\sum_{y}\Tr(F_{b|y}\sigma_{b|y}).
\end{eqnarray}
where the coefficients $F_{b|y}$ are some positive
semi-definite operators acting on $\mathcal{H}_A$. 

Let us now consider the first steering inequality proposed in \cite{CJWR09}. The scenario considered there is minimal in the sense that there are only two parties and Bob performs only two measurements. The matrices $F_{b|y}$ of the steering functional were chosen as,
\begin{eqnarray}
F_{0|0}=\ket{0}\!\bra{0},\qquad F_{1|0}=\ket{1}\!\bra{1}
\end{eqnarray}
and
\begin{eqnarray}
F_{0|1}=\ket{+}\!\bra{+},\qquad F_{1|0}=\ket{-}\!\bra{-},
\end{eqnarray}
where $\ket{+}=\frac{\ket{0}+\ket{1}}{\sqrt{2}}$ and $\ket{-}=\frac{\ket{0}-\ket{1}}{\sqrt{2}}$. Consequently, the simplest steering functional can be represented as
\begin{eqnarray}
     W_{2,2,2}=\Tr\left(\ket{0}\!\bra{0}\sigma_{0|0}\right)+\Tr\left(\ket{1}\!\bra{1}\sigma_{1|0}\right)+\Tr\left(\ket{+}\!\bra{+}\sigma_{0|1}\right)+\Tr\left(\ket{-}\!\bra{-}\sigma_{1|1}\right).
\end{eqnarray}
For assemblages that admit a local hidden state model, one can find an upper bound to the steering functional \eqref{stefn} as
\begin{eqnarray}
W_{2,2,2}\leq1+\frac{1}{\sqrt{2}}.
\end{eqnarray}

This is known as ``local hidden state bound" or simply ``local bound" and is again denoted by $\beta_L$. The quantum bound $\beta_Q$ of the steering inequality $W_{2,2,2}$ is $2$ and for instance can be achieved when the preparation device prepares the maximally entangled state \eqref{maxentstate} and Bob performs the Pauli z and Pauli x measurements
\begin{eqnarray}
B_0=\sigma_z,\qquad B_1=\sigma_x.
\end{eqnarray}
The scenario can be straightforwardly generalised to the case when Bob performs arbitrary number of measurements $m$ with arbitrary number of outcomes $d$. The steering functional in this case admits a form
\begin{eqnarray}\label{stefn1}
W_{m,d,2}=\sum_{y=1}^{m}\sum_{b=0}^{d-1}\Tr(F_{b|y}\sigma_{b|y}).
\end{eqnarray}
Comparing Bell scenario to quantum steering scenario, in Bell scenario one obtains joint probability distributions from the experiment and constructs the Bell functional by suitably choosing the real coefficients $c_{a,b,x,y}$. In quantum steering scenario the experiment generates the assemblage and the steering functional is constructed by suitably choosing the positive semi-definite matrices $F_{b|y}$. However, it is always possible to map quantum steering to the Bell scenario and express the steering functional in terms of expectation values of joint observables or equivalently joint probability distributions $\vec{p}$ similar to Bell functional. This idea will be particularly useful for this thesis.

Let us consider the steering functional $W_{m,d,2}$ \eqref{stefn1} with the coefficients $F_{b|y}$ being positive, Hermitian and summing up to the identity for all $y$, that is, 
\begin{equation}
  F_{b|y}\geq 0,\qquad  F_{b|y}=F_{b|y}^{\dagger},\qquad\sum_{b}F_{b|y}=\Id.
\end{equation}
Consequently, $\{F_{b|y}\}$ is a valid quantum measurement. As discussed in subsection \ref{2.1.3}, one can equivalently represent a quantum measurement by using $d$ generalised observables $A_{k|y}$ \eqref{obs21}, that is, 
\begin{equation}\label{Akx}
    F_{b|y}=\frac{1}{d}\sum_{b=0}^{d-1}\omega^{-bk} A_{k|y}
\end{equation}
with $k=0,\ldots,d-1$ and $\omega=\mathrm{exp}(2\pi i/d)$. Let us now consider that Bob performs the measurements $\{N_{b|y}\}$ on some state $\rho_{AB}$. Then, the assemblage $\sigma_{b|y}$ can be expressed as
\begin{equation}\label{Akx1}
    \sigma_{b|y}=\Tr_B[(\Id_A\otimes N_{b|y})\rho_{AB}].
\end{equation}
Again, we represent the measurements $\{N_{b|y}\}$ using $d$ generalised observables $B_{l|y}$ as
\begin{equation}\label{Akx2}
    N_{b|y}=\frac{1}{d}\sum_{b=0}^{d-1}\omega^{-bl} B_{l|y}.
\end{equation}
Now, we are ready to express the steering functional \eqref{stefn1} in terms of expectation values. First using the observation $\eqref{Akx1}$, the steering functional \eqref{stefn1} can be written as
\begin{eqnarray}
W_{m,d,2}&=&\sum_{b}\sum_{y}\Tr(F_{b|y}\sigma_{b|y})\nonumber\\
        &=&\sum_{b}\sum_{y}\Tr(F_{b|y}\otimes N_{b|y}\rho_{AB}).
\end{eqnarray}
Now, using observations \eqref{Akx} and \eqref{Akx2}, we arrive at
\begin{eqnarray}\label{stefn2}
\sum_{b=0}^{d-1}\sum_{y=1}^{N}\Tr(F_{b|y}\sigma_{b|y})=\frac{1}{d}
   \sum_{k=0}^{d-1} \sum_{y=1}^N\left\langle A_{k|y}\otimes B_{-k|y}\right\rangle.
\end{eqnarray}
For a detailed review on quantum steering, refer to \cite{Guhne2}. In the next part, we briefly discuss the applications of quantum non-locality.

\subsection{Applications of quantum non-locality}
Apart from the foundational aspects, quantum non-locality has has given rise to enormous number of applications in computation, communication and information theory. Here we first give a brief account of various applications of Bell non-locality and then quantum steering.

Suppose, Alice and Bob are spatially separated and both of them receive $n-$bit strings denoted by $x,y$ and Bob wants to compute a function $f(x,y)$. The minimum amount of information needed by Bob from Alice to compute $f(x,y)$ is known as communication complexity of $f(x,y)$. It has been shown that for certain functions $f(x,y)$, Bell non-locality reduces the communication complexity when compared to classical communication between Alice and Bob \cite{Bellapp1}. Another important application of Bell-nonlocality is in the field of quantum cryptography. The earliest connection between quantum non-locality and cryptography was realised by Herbert in 1975 \cite{Herbert}. However, the breakthrough was achieved by Ekert in his seminal paper in 1991 \cite{Ekert} that showed that Bell-nonlocality serves as a way to generate secure key among two spatially separated parties. The protocol was based on the maximal violation of CHSH inequality which makes it physically impossible for some external attacker to know the key shared between Alice and Bob. As a consequence this protocol can be realised in a device-independent way, that is, without assuming any details about the inner working of the device apart from the fact that it is governed by quantum theory. In fact device-independent quantum key distribution (DIQKD) has been one of the key applications of quantum non-locality \cite{Acin1, Acin2, Acin3, Masanes1, Masanes2, Masanes3, barrett1, Barrett2, Pironio1, Scarani1, hangi}. Using Bell inequalities one can even find the minimum dimension of a quantum system required for obtaining certain correlations \cite{dimwit1}. Since Bell inequalities can be violated using only entangled states, they serve as an important tool to detect entanglement \cite{entwit1, entwit2, entwit3, entwit4}. Bell non-locality has been identified as a way to generate genuine randomness \cite{AM16, di4, Pironio2, Colbeck1, Colbeck2, Colbeck3, Hall, Gallego, Fehr, Koh, rand1, rand2, rand3, random0, random1}.

Mayers and Yao in \cite{Mayers, Yao} realised that correlations obtained by performing local measurements on spatially separated systems can used for device-independent certification of quantum states and measurements. This was termed as self-testing. It was later realised that in fact Bell-nonlocality allows one to self-test quantum states and measurements \cite{qubitst1, qubitst2, qubitst5, multist1, multist2, parallelst1, parallelst3, ALLst1}. In the subsequent sections, we discuss device-independent certification and randomness generation in details.

Quantum steering is also useful in quantum cryptography as was realised in \cite{Crypto1, Crypto2}. In a practical scenario where one of the parties is trusted, for instance, in a bank-client relationship where the bank can be considered to be trusted, one can construct one-sided device-independent protocols. Since, then there have been numerous protocols for quantum key distribution using quantum steering \cite{steecrypt1, steecrypt2, steecrypt3, steecrypt4, steecrypt5}. Quantum steering is also useful in tasks like secret sharing, where a referee sends an encrypted message to multiple players which can be decoded only if the players work together \cite{secsha1, secsha2, secsha3, secsha4}. Also, randomness can be certified using quantum steering to be secure and genuine \cite{steerand1, steerand2, steerand3, steerand4}. Quantum steering has also shown advantage in tasks like subchannel discrimination, that is, how well one can distinguish different branches of a quantum channel \cite{piani1, uola1}. Further, quantum steering can be used for certification of quantum states and measurements \cite{Supic, Alex, suchet, Bharti}. 

This completes the analysis of quantum non-locality relevant to this thesis. In the next section we introduce the idea of device-independent certification and in particular self-testing and one-sided device certification. 

\section{Device-independent certification}
\begin{figure}[h!]
\centering
\begin{subfigure}{.5\textwidth}
  \centering
  \includegraphics[width=.63\linewidth]{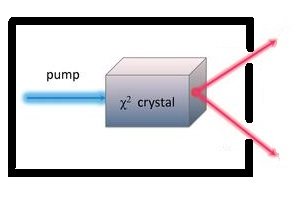}
  \caption{}
  \label{fig5:sub1}
\end{subfigure}%
\begin{subfigure}{.5\textwidth}
  \centering
  \includegraphics[width=\linewidth]{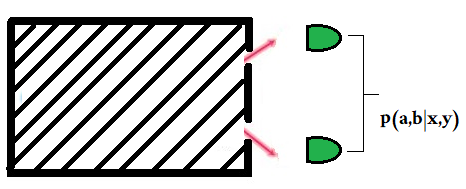}
  \caption{}
  \label{fig5:sub2}
\end{subfigure}
\caption{(a) Device-dependent scenario: To verify the device, one has to assume the inner workings of the device. Here, if it is known that the device consists a photon source that passes through a $\chi^2$ crystal, then the device can produce entangled photons. (b) Device-independent scenario: The inner workings of the device are unknown. To verify whether the device generates entangled photons, one needs to perform a Bell test by measuring the joint correlation statistics $p(a,b|x,y)$. }
\label{fig5}
\end{figure}

In recent years, the field of device-independent schemes has gained a lot of interest. The novelty of such schemes lies in the fact that one can predict some properties about the system by looking only at the statistics generated by this system. For instance, consider that we are given a device that is promised to produce entangled photons. The natural question is how we verify that the device works as promised and whether we should trust the manufacturer. One way is to break the device and check the source. Another way is not to break the device but perform a quantum tomography on the state generated by the device. But one needs the knowledge of quantum optics and also would have to trust his measurement device to infer the state. However, using Bell inequalities, we can verify the device without breaking it or without trusting any other device. For this, as depicted in Fig. \ref{fig5}, the entangled subsystems are allowed to be far enough such that they are spatially separated. Now, two local measurements are performed on each of these arms. Here the verifiers do not have any knowledge about the measurement device but choose two inputs freely corresponding to the two measurements and record their outputs. The experiment is repeated enough number of times to collect joint probability distribution for different inputs and outputs. If this distribution violates some Bell inequality, it can be concluded that the device generates entangled photons. 

Thus, quantum nonlocality serves as a way to infer properties about a device without knowing the inner workings of it apart from the fact that it is governed by quantum theory. Such device-independent schemes provide maximum security in cryptographic scenarios where there might be some hidden mechanisms that are uncontrollable or the device leaks out some information that can be used by some external attacker. However, in practical scenarios one can make some well justified assumptions on the device. In this thesis, we focus on two different types of device-independent certification, first, fully device-independent certification or self-testing and second, one-sided device independent certification. 

\subsection{Self-testing}
Let us now consider the following task: Instead of detecting some property about the state such as entanglement, we want to characterise the state and measurement which generates the observed statistics with the device being treated as a black box. This is the strongest form of device-independent certification and is known as self-testing. The idea of fully device-independent certification or self-testing was first introduced by Mayers and Yao in 1998 \cite{Mayers, Yao}.  The scenario for self-testing is same as Bell scenario Fig. \ref{fig1}. There are two independent observers Alice and Bob who are spatially separated from each other and can freely choose their inputs. Also, they are not allowed to communicate among each other during the experiment. However, the natural assumption here is that the device behaves according to quantum theory, that is, every statistics is generated by some quantum measurement acting on some quantum state. Another important assumption here is that the preparation device always prepares the same state and there is no correlation between the measurement device and the preparation device.
From the experiment Alice and Bob obtain the joint probability distribution $\{p(a,b|x,y)\}$ where $a,b$ denote the outcomes when they choose the measurements labelled by $x,y$, respectively. 

Now, consider that the preparation device generates the state $\rho$ and the measurements performed by Alice and Bob in the observable picture as $A_{a|x}$ and $B_{b|y}$ [see Sec. \ref{2.1.3}] respectively for all $a,b,x,y$. For simplicity, from here on we refer the observables $A_{a|x}$ and $B_{b|y}$ as measurements. Using the joint probability distribution  $\{p(a,b|x,y)\}$, Alice and Bob want to characterise the state $\rho$ sent by the preparation device and also their measurements represented in the observable picture as $A_{a|x}$ and $B_{b|y}$ for all $a,b,x,y$. It is important here to note that one can identify such quantum realisations only up to the equivalences under which the probability distribution remains invariant. Also, notice that in the scenario presented above, one can not certify the source to be producing a unique mixed state. The reason being that any mixed state can be decomposed in terms of pure states. Let us say that the target or the ideal state that is expected to be produced by the source is denoted by $\ket{\tilde{\psi}}$ and the target measurements that are expected to be performed are denoted in the observable form as $\Tilde{A}_{a|x}$ for Alice and $\Tilde{B}_{b|y}$ for Bob. Then, there are two major equivalences under which the probability remains invariant,
\begin{enumerate}
    \item An additional state might be attached to the target state on which the measurements act trivially, that is, the actual state and measurements in the device can be 
    \begin{eqnarray}
    \rho=\proj{\tilde{\psi}}\otimes\sigma
    \end{eqnarray}
    such that $\sigma$ acts on some unknown but finite dimensional Hilbert space $\mathcal{H}''$ and,
    \begin{eqnarray}
    A_{a|x}=\tilde{A}_{a|x}\otimes\Id''\quad B_{b|y}=\tilde{B}_{b|y}\otimes \Id''.
    \end{eqnarray}

    \item The actual state and measurements might be unitarily rotated with respect to the target state, that is,
    \begin{eqnarray}\label{2.76}
    \rho=U_A\otimes U_B\ \proj{\tilde{\psi}}\  U_A^{\dagger}\otimes U_B^{\dagger},
    \end{eqnarray}
    and
    \begin{eqnarray}\label{2.77}
    A_{a|x}=U_A\tilde{A}_{a|x}U_A^{\dagger},\quad B_{b|y}=U_B\tilde{B}_{b|y}U_B^{\dagger}
        \end{eqnarray}
    where $U_A:\mathcal{H}_A\rightarrow\mathcal{H}_A,\ U_B:\mathcal{H}_B\rightarrow\mathcal{H}_B$ are local unitary transformations on the subsystem of Alice and Bob respectively.

\end{enumerate}
Given these two equivalences, we now present the self-testing definition that is relevant for this thesis. 
\begin{defn}[Self-testing]\label{st} Consider the above Bell experiment with Alice and Bob performing measurements $A_{a|x}$ and $B_{b|y}$ on a state $\rho_{AB}$ and observing correlations $\{p(a,b|x,y)\}$. 
The state $\rho_{AB}$ and measurements $A_{a|x}$ and $B_{b|y}$ are certified to be the target state $\ket{\tilde{\psi}}$ and target measurements $\tilde{A}_{a|x}$ and $\tilde{B}_{b|y}$ from $\{p(a,b|x,y)\}$ if:
\begin{enumerate}
    \item The Hilbert space of Alice and Bob decompose as 
\begin{eqnarray}
\mathcal{H}_A=\mathcal{H}_{A'}\otimes\mathcal{H}_{A''},\qquad \mathcal{H}_B=\mathcal{H}_{B'}\otimes\mathcal{H}_{B''}
\end{eqnarray}
where the target state $\ket{\tilde{\psi}}$ belongs to $\mathcal{H}_{A'}\otimes\mathcal{H}_{B'}$ and $\mathcal{H}_{A''}$ and $\mathcal{H}_{B''}$ represents the auxiliary Hilbert space of Alice and Bob respectively.

\item There exists a unitary $U_B:\mathcal{H}_B\to \mathcal{H}_B$ and $U_A:\mathcal{H}_A\to \mathcal{H}_A$ such that the state is

\begin{equation}
 U_A\otimes U_B\ \rho_{AB}\ U_A^{\dagger}\otimes U_B^{\dagger}=\proj{\tilde{\psi}}_{A'B'}\otimes\sigma_{A''B''},
\end{equation}
where $\sigma_{A''B''}$ is some state acting on the Hilbert space $\mathcal{H}_{A''}\otimes\mathcal{H}_{B''}$.
\item The measurements are certified as
\begin{equation}\label{stmea1}
    U_A\,A_{a|x}\,U_A^{\dagger}=\tilde{A}_{a|x}\otimes\Id_{A''},\qquad U_B\,B_{b|y}\,U_B^{\dagger}=\tilde{B}_{b|y}\otimes\Id_{B''}.
\end{equation}
for all $a,b,x,y$. Here, $\Id_{A''}, \Id_{B''}$ are identities acting on the Hilbert space  $\mathcal{H}_{A''}$ and $\mathcal{H}_{B''}$ respectively. 
\end{enumerate}
\end{defn}
 One natural assumption that we make in the above definition is that the local states of Alice and Bob are full-rank. This comes from the fact that Alice and Bob can only characterise the part of the quantum measurements that act on the quantum state. Notice, that in general the unitaries appearing in the second equivalence can be replaced by isometries [see Def. \ref{isometry}]. Another important equivalence pointed out in \cite{conjugate}, that is not stated above is that the actual quantum state and measurements can not be distinguished from the conjugate of the target state and measurements, that is,
 \begin{eqnarray}
    \rho=\proj{\tilde{\psi}}^*,\quad A_{a|x}=\tilde{A}_{a|x}^*,\quad B_{b|y}=\tilde{B}_{b|y}^*.
    \end{eqnarray}
However, this equivalence is not relevant for this thesis (it will be clarified when analysing the results of this thesis). For other definitions of self-testing refer to \cite{SB20}.

As discussed before, the set of joint probability distributions is convex. It is important to note here that one can only certify quantum states and measurements from probability distribution that lie at the boundary of this set, that is, extremal probability distributions \cite{Jed4}. The reason is that any point inside this set can be represented as convex combination of the extremal points which does not correspond to a unique probability distribution and thus can not be achieved by only a particular state and measurements. 

\begin{figure}[t]
    \centering
    \includegraphics{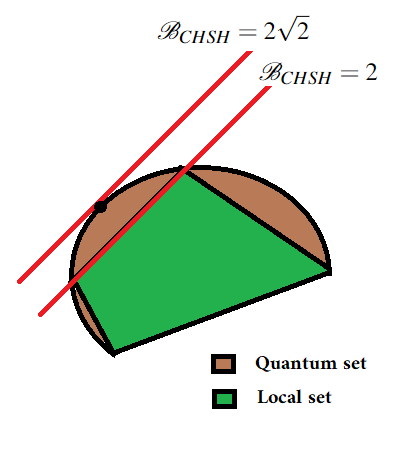}
    \caption{The CHSH functional $\mathcal{B}_{CHSH}=2$ represents a facet of the local set of correlations $\mathcal{L}_{2,2,2}$. In other words, CHSH functional is a hyperplane that represents one of the face of the local polytope. Interestingly, the CHSH functional $\mathcal{B}_{CHSH}=2\sqrt{2}$ touches the quantum set at a single extremal point making it particularly useful for self-testing.}
    \label{fig6}
\end{figure}

Now, It turns out that if one utilises the maximal violation of the Bell inequalities, then the desired self-testing of quantum state and measurements can be achieved by collecting less statistics compared to the case when one uses tomography. For instance, both Alice and Bob are required to choose at least four inputs corresponding to a tomographically complete set of measurements to certify any two qubit state. However, employing Bell inequalities one can certify two qubit states using only two measurements per party. The maximal violation of a Bell inequality is achieved by the joint probability distributions lying at the boundary of the quantum set as shown in Fig. \ref{fig6}. However, it might happen that the Bell functional  touches the boundary of the quantum set at more than one point and one can have some weaker self-testing statement where one certifies a family of states or measurements \cite{weakst1, KST+19, weakst2}. To certify some particular quantum state and measurements, the Bell violation should point to a unique probability distribution. As a consequence, to self-test any quantum realisation, we need to ensure the maximum violation of the Bell inequality and then we need to show that the probability distribution that gives the maximum violation is generated by unique quantum state and measurements up to the equivalences as suggested before. Thus, in the definition of self-testing [cf. Def. \ref{st}], we can replace the statement ``observing correlations $\{p(a,b|x,y)\}$" with ``observing the maximal violation of a Bell inequality $\mathcal{B}$". 

In a physical experiment, we can never achieve the maximal violation of a Bell inequality but some value which is close to this violation. From an experimental perspective, it is thus necessary to understand self-testing in the presence of noise and whether the proposed certification schemes are robust against experimental imperfections.

\begin{defn}[Robust self-testing]\label{rst} Consider the above Bell experiment with Alice and Bob performing the measurements $A_{a|x}$ and $B_{b|y}$ respectively on a quantum state $\rho_{AB}$. Assume that the value of a given Bell expression $\mathcal{B}$ for the observed correlations satisfy
\begin{eqnarray}
\mathcal{B}\geq \beta_Q-\epsilon.
\end{eqnarray}
Alice and Bob can robustly certify a target state $\ket{\tilde{\psi}}$ and target measurements $\tilde{A}_{a|x}$ and $\tilde{B}_{b|y}$ from the observed Bell value, if there exists a unitary $U_B:\mathcal{H}_B\to \mathcal{H}_B$ and $U_A:\mathcal{H}_A\to \mathcal{H}_A$ such that the state is
\begin{equation}
 \left|\left|U_A\otimes U_B\rho_{AB}U_A^{\dagger}\otimes U_B^{\dagger}-\proj{\tilde{\psi}}_{A'B'}\otimes\sigma_{A''B''}\right|\right|_2\leq f_1(\epsilon),
\end{equation}
where $\sigma_{A''B''}$ is some state acting on the Hilbert space $\mathcal{H}_{A''}\otimes\mathcal{H}_{B''}$ and the measurements are
\begin{eqnarray}
    ||U_A\,A_{a|x}U_A^{\dagger}-\tilde{A}_{a|x}\otimes\Id_{A''}||_2&\leq& f_2(\epsilon), \quad\text{and}\nonumber\\ ||U_B\,B_{b|y}\,U_B^{\dagger}-\tilde{B}_{b|y}\otimes\Id_{B''}||_2&\leq& f_3(\epsilon).
\end{eqnarray}
for all $a,b,x,y$ and $\Id_{A''}, \Id_{B''}$ are identities acting on to the support of Alice's and Bob's subsystem $\rho_A, \rho_B$ respectively. Further, when $\epsilon$ goes to $0$, the functions $f_1(\epsilon), f_2(\epsilon)$ and $f_3(\epsilon)$ must vanish.
\end{defn}
For a note, the $||X||_2$ denotes the Hilbert-Schmidt norm of an operator $X$, and is defined as $||X||_2=\sqrt{\Tr(X^{\dagger}X)}$. For two unitary matrices $A$ and $\tilde{A}$, we have that
\begin{eqnarray}
    ||A-\tilde{A}||_2=2\left[\Tr(\Id)-\mathrm{Re}\Tr(A\tilde{A}^{\dagger})\right].
\end{eqnarray}

The above statement can be understood as, if one observes a value $\epsilon$ lower than the maximal violation in a Bell experiment, then overlap between the state shared between the parties and the target state up to some equivalences is bounded from below by a function of $\epsilon$. One can arrive at a similar conclusion for the measurements. 

\subsubsection{Ways to do self-testing}

There are several works that provide self-testing statements using numerical approaches based on semi-definite programs \cite{multist6, multist7, quditst1, quditst2}. However, these methods can not be used to provide self-testing statements for higher dimensional states due to computational requirements. In other words, there does not exist any numerical scheme that can certify entangled states of arbitrary local dimension in an efficient way. Due to this, in this work we focus on analytical methods of self-testing.

There are a few analytical techniques that have been explored for the task of self-testing but most of these techniques work when the state to be certified is a two-qubit state. For instance, the initial self-testing schemes \cite{qubitst10, multist2, multist9} were based on Jordan's lemma \cite{Pironio1, JL2} that says that if two Hermitian matrices with eigenvalues $\pm 1$ act on a Hilbert space $\mathcal{H}$, then these matrices decompose as a direct sum of matrices acting on Hilbert space of dimension less than or equal to two. Another such method is using the swap gate that serves as an isometry that maps the non-ideal state to the target state \cite{qubitst1, qubitst2, qubitst4, qubitst5, ALLst1, multist4}. 

In this thesis, we follow another approach to derive self-testing statements that is based on ``sum of squares (SOS) decomposition" of the Bell operator. This involves decomposing the Bell operator in terms of some positive operators. This method was first explored for self-testing of any pure entangled two-qubit state in \cite{qubitst3}. Based on this method, some self-testing schemes were provided in \cite{multist3, weakst1, random1} that can be used to certify multi-qubit states and subspaces. A scheme to self-test two-qutrit maximally entangled state was proposed in \cite{Jed1} that employed the method of SOS decomposition of the Bell operator. In \cite{Jed1, SATWAP} family of Bell inequalities were constructed using this technique, the maximal violation of which was achieved by maximally entangled state of arbitrary local dimension. 

Let us now elaborate on the SOS decomposition of a Bell operator. Any Bell operator \eqref{genBell} can be represented as
\begin{eqnarray}\label{2.84}
    \hat{\mathcal{B}}=\sum_{a,b,x,y}c_{a,b,x,y}M_{a|x}\otimes N_{b|y}
\end{eqnarray}
where $M_{a|x},  N_{b|y}$ are measurement operators corresponding to the input $x$ and output $a$ of Alice and input $y$ and output $b$ of Bob respectively. If $\beta_Q$ is the quantum bound of the Bell expression $\langle\hat{\mathcal{B}}\rangle$ where $\hat{\mathcal{B}}$ is given in \eqref{2.84}, then let us assume that the positive semi-definite operator $\beta_Q\Id-\hat{\mathcal{B}}$ can be decomposed in the following way
\begin{eqnarray}\label{genSOS}
    \beta_Q\Id-\hat{\mathcal{B}}=\sum_iP_i^{\dagger}P_i.
\end{eqnarray}
Note that the right hand side of the above equation consists of positive operators $P_i^{\dagger}P_i$ composed of the operators $M_{a|x},  N_{b|y}$. Given a generic Bell operator, it is not that straightforward to find such decompositions. One can numerically find it using the Navascu\'es-Pironio-Ac\'in (NPA) hierarchy \cite{NPA1, NPA2, NPA3} which is based on semi-definite programming. Let us now consider that the state $\ket{\psi}$ achieves the maximal violation of a Bell inequality, then the left hand side of \eqref{genSOS} vanishes and thus
\begin{eqnarray}
    \sum_i||P_i\ket{\psi}||^2=0.
\end{eqnarray}
All the terms are positive in the above sum which allows to conclude that $||P_i\ket{\psi}||=0$ for all $i$. Thus,
\begin{eqnarray}
   \forall i\qquad  P_i\ket{\psi}=0.
\end{eqnarray}
This simple expression contains all the information about the state and measurements that give rise to the maximal violation of the corresponding Bell inequality. Often, these relations can be used to derive self-testing results. 

Since, this technique is central to the results presented in this thesis we would provide a simple example of self-testing based on SOS decomposition. Before proceeding let us state an important fact that was proven in Ref. \cite{Jed1} which is central to some of the results presented in this thesis.

\begin{fakt}\label{fact2}
Consider two unitary matrices $R_0, R_1$ that act on a finite-dimensional Hilbert space $\mathcal{H}$ satisfying $R_0^d = R_1^d = \Id$. If $R_0$ and $R_1$ satisfy the relation $R_0R_1 =\omega R_1R_0$, then $dim(\mathcal{H}) = d.t$ for some positive integer $t$ and there exists a unitary $U:\mathcal{H}\rightarrow\mathcal{H}$ such that
\begin{eqnarray}
    UR_0 U^{\dagger}=Z_d\otimes\Id, \quad \text{and}\quad  UR_1 U^{\dagger}=X_d\otimes\Id,
\end{eqnarray}
where $Z_d, X_d$ are the $d-$dimensional generalisation of the Pauli matrices $\sigma_z,\sigma_x$ \eqref{pauliz,x} given by,
\begin{eqnarray}\label{X,Z}
Z_d=\sum_{i=0}^{d-1}\omega^i\ket{i}\!\bra{i},\qquad X_d=\sum_{i=0}^{d-1}\ket{i+1}\!\bra{i}.
\end{eqnarray}
\end{fakt}

Let us now consider the CHSH inequality \eqref{CHSH}, which for our convenience is scaled down by the factor $\frac{1}{\sqrt{2}}$, and consider its operator form
\begin{eqnarray}\label{CHSHop}
    \hat{\mathcal{B}}_{CHSH}=\frac{1}{\sqrt{2}}\left(A_0\otimes B_0+A_0\otimes B_1+ A_1\otimes B_0- A_1\otimes B_1\right).
\end{eqnarray}
We assume here that the measurements of Alice and Bob are projective and thus, $A_x^2=B_y^2=\Id$ for all $x,y$. The local and quantum bound of this Bell inequality is $\sqrt{2}$ and $2$ respectively \eqref{CHSHquant} and thus the SOS decomposition of the corresponding shifted Bell operator is given by
\begin{eqnarray}
    2\Id-\hat{\mathcal{B}}_{CHSH}=\frac{1}{2}\left(P_0^{\dagger}P_0+P_1^{\dagger}P_1\right)
\end{eqnarray}
where,
\begin{eqnarray}\label{SOSCHSH}
    P_0=\Id-A_0\otimes\frac{B_0+B_1}{\sqrt{2}}\quad\text{and}\quad P_1=\Id-A_1\otimes\frac{(B_0-B_1)}{\sqrt{2}}.
\end{eqnarray}
Let us now suppose that a state $\rho_{AB}$ attains the quantum bound of the above Bell operator. Then, we can always add an ancillary system $E$ to purify this state. Let us denote this purified state by $\ket{\psi_{ABE}}$ such that $\rho_{AB}=\Tr_E(\proj{\psi}_{ABE})$. Then, from Eq. \eqref{SOSCHSH} the following relations must hold,
\begin{eqnarray}\label{introSOS1}
   \left[ A_0\otimes\frac{B_0+B_1}{\sqrt{2}}\otimes\Id_E\right]\ket{\psi_{ABE}}=\ket{\psi_{ABE}}
\end{eqnarray}
and,
\begin{eqnarray}\label{introSOS2}
     \left[A_1\otimes\frac{B_0-B_1}{\sqrt{2}}\otimes\Id_E\right]\ket{\psi_{ABE}}=\ket{\psi_{ABE}}
\end{eqnarray}
Using the fact that $A_0$ are unitary and Hermitian, we have from Eq. \eqref{introSOS1} that
\begin{eqnarray}\label{introSOS3}
     \frac{B_0+B_1}{\sqrt{2}}\otimes\Id_{AE}\ket{\psi_{ABE}}=A_0\otimes\Id_{BE}\ket{\psi_{ABE}}
\end{eqnarray}
Let us recall that the measurements can be characterised only on the support of the  local reduced states $\rho_A$ and $\rho_B$ and consequently, in the proof we assume that the local states of Alice and Bob are full-rank. Also,
From here on, for convenience we drop the term $\Id_E$.
Now, multiplying by $A_0\otimes\Id_B$ on both the sides of Eq. \eqref{introSOS3} , we have that
\begin{eqnarray}
     \left[\Id_A\otimes\frac{B_0+B_1}{\sqrt{2}}\right]A_0\otimes\Id_B\ket{\psi_{ABE}}=A_0^2\otimes\Id_B\ket{\psi_{ABE}}
\end{eqnarray}
where we used the fact that measurements acting on subsystem $A$ and $B$ commute. Using \eqref{introSOS3} and the fact that $A_0^2=\Id$, we have that
\begin{eqnarray}\label{1.99}
     \Id\otimes\left[\frac{B_0+B_1}{\sqrt{2}}\right]^2\ket{\psi_{ABE}}=\ket{\psi_{ABE}}.
\end{eqnarray}
Taking a partial trace over the subsystems $A,E$, we have that
\begin{eqnarray}
     \left[\frac{B_0+B_1}{\sqrt{2}}\right]^2\rho_B=\rho_B.
\end{eqnarray}
The fact that $\rho_B$ is full-rank and thus invertible allows us to conclude from the above equation that
\begin{eqnarray}\label{1.102}
      \left[\frac{B_0+B_1}{\sqrt{2}}\right]^2=\Id_B.
\end{eqnarray}
Similarly, one obtains from \eqref{introSOS2} that
\begin{equation}\label{1.103}
    \left[\frac{B_0-B_1}{\sqrt{2}}\right]^2=\Id_B.
\end{equation}
Expanding the the above two equations \eqref{1.102} and \eqref{1.103}, we have that
\begin{equation}\label{1.104}
     B_0^2+B_1^2+\big\{B_0,B_1\big\}=2\Id_B, \quad \text{and}\quad B_0^2+B_1^2-\big\{B_0,B_1\big\}=2\Id_B
\end{equation}
where $\big\{B_0,B_1\big\}=B_0B_1+B_1B_0$ denotes the anti-commutator of $B_0, B_1$. Now, using the fact that $B_0^2=B_1^2=\Id_B$, we obtain that
\begin{eqnarray}\label{1.106}
      \big\{B_0,B_1\big\}=0.
\end{eqnarray}
Now, as stated in Fact \ref{fact2}, for two unitary matrices $B_0, B_1$ with the additional property that $ B_i^2=\Id_B$ for $i=0,1$ and satisfy the condition \eqref{1.106}, there exist local unitary transformation $V_B:\mathcal{H}_B\rightarrow\mathcal{H}_B$ such that
\begin{equation}\label{1.113}
    V_B\ B_0\ V_B^{\dagger}=\sigma_z\otimes\Id_{B''}, \quad \text{and}\quad  V_B\ B_1\  V_B^{\dagger}=\sigma_x\otimes\Id_{B''}
\end{equation}
where $\Id_{B''}$ acts on the auxiliary Hilbert space $\mathcal{H}_{B''}$ of some finite dimension and $\sigma_z,\sigma_x$ are Pauli matrices \eqref{pauliz,x}. This also shows that the Hilbert space of Bob decomposes as  $\mathcal{H}_B=\mathbb{C}^2\otimes\mathcal{H}_{B''}$.
Now, consider a unitary matrix $W=W'\otimes\Id_{B''}$ such that $W'$ is also unitary and can be expressed as a matrix written in the $2-$dimensional computational basis $\{\ket{0},\ket{1}\}$ as
\begin{eqnarray}
     W'=\begin{pmatrix}
\cos\left(\frac{\pi}{8}\right) & \sin\left(\frac{\pi}{8}\right)\\
\sin\left(\frac{\pi}{8}\right)& -\cos\left(\frac{\pi}{8}\right)
\end{pmatrix}.
\end{eqnarray}
The unitary matrix $W'$ transforms the Pauli observables $\sigma_z, \sigma_x$ to the ideal ones \eqref{1.50} and thus from Eq. \eqref{1.113} we arrive at
\begin{equation}\label{certmeas1}
    U_BB_0U_B^{\dagger}=\frac{\sigma_z+\sigma_x}{\sqrt{2}}\otimes\Id_{B''}\qquad\text{and}\qquad U_BB_1U_B^{\dagger}=\frac{\sigma_z-\sigma_x}{\sqrt{2}}\otimes\Id_{B''},
\end{equation}
where $U_B=WV_B$. Now, to find Alice's observables $A_i$ we consider an analogous SOS decomposition of the CHSH operator given by
\begin{eqnarray}
    2\Id-\hat{\mathcal{B}}_{CHSH}=\frac{1}{2}\left(Q_0^{\dagger}Q_0+Q_1^{\dagger}Q_1\right)
\end{eqnarray}
where,
\begin{eqnarray}\label{SOSCHSH2}
    Q_0=\Id-\frac{A_0+A_1}{\sqrt{2}}\otimes B_0\quad\text{and}\quad Q_1=\Id-\frac{A_0-A_1}{\sqrt{2}}\otimes B_1.
\end{eqnarray}
Again, for any state $\ket{\psi_{ABE}}$ that maximally violates the CHSH inequality, from Eq. \eqref{SOSCHSH2} the following relations must hold,
\begin{eqnarray}\label{introSOS12}
   \frac{A_0+A_1}{\sqrt{2}}\otimes B_0\otimes\Id_E\ket{\psi_{ABE}}=\ket{\psi_{ABE}}
\end{eqnarray}
and
\begin{eqnarray}\label{introSOS22}
     \frac{A_0-A_1}{\sqrt{2}}\otimes B_1\otimes\Id_E\ket{\psi_{ABE}}=\ket{\psi_{ABE}}.
\end{eqnarray}
As concluded for Bob's observables, the Hilbert space of Alice decomposes as $\mathcal{H}_A=\mathbb{C}^2\otimes\mathcal{H}_{A''}$ and there exist local unitary transformation $U_A:\mathcal{H}_A\rightarrow\mathcal{H}_A$ such that
\begin{equation}\label{certmeas2}
    U_A\ A_0\ U_A^{\dagger}=\sigma_z\otimes\Id_{A''}, \quad \text{and}\quad  U_A\ A_1\  U_A^{\dagger}=\sigma_x\otimes\Id_{A''}
\end{equation}
where $\Id_{A''}$ acts on the auxiliary Hilbert space $\mathcal{H}_{A''}$.

After determining the form of the measurements, we can finally certify the quantum state $\ket{\psi_{ABE}}$ that achieves the maximum violation of the CHSH Bell inequality. Since, $\mathcal{H}_A=\mathbb{C}^2\otimes\mathcal{H}_{A''}$ and $\mathcal{H}_B=\mathbb{C}^2\otimes\mathcal{H}_{B''}$, we can decompose the state $\ket{\psi_{ABE}}$ as
\begin{equation}\label{genstate1}
   U_A\otimes U_B \ket{\psi_{ABE}}=\ket{\tilde{\psi}_{ABE}}=\sum_{i,j=0,1}\ket{i}_{A'}\ket{j}_{B'}\ket{\psi_{ij}}_{A''B''E}
\end{equation}
where $\ket{\psi_{ij}}_{A''B''E}\in\mathcal{H}_{A''}\otimes\mathcal{H}_{B''}\otimes\mathcal{H}_E$. For convenience, we drop the subscripts from the state. Now, plugging in the certified measurements \eqref{certmeas1} and \eqref{certmeas2} in the SOS relation \eqref{introSOS1}
\begin{equation}
    \sigma_z\otimes\sigma_z\otimes\Id_{A''B''E}\ket{\tilde{\psi}_{ABE}}=\ket{\tilde{\psi}_{ABE}}.
\end{equation}
Substituting the general form of the state \eqref{genstate1} and simplifying, we obtain that
\begin{eqnarray}\label{2.120}
     \sum_{i,j=0,1}(-1)^{ij}\ket{i}_{A'}\ket{j}_{B'}\ket{\psi_{ij}}_{A''B''E}=\sum_{i,j=0,1}\ket{i}_{A'}\ket{j}_{B'}\ket{\psi_{ij}}_{A''B''E}.
\end{eqnarray}
By projecting the above equation on to $\bra{i}_{A'}\!\bra{j}_{B'}$ for all $i,j$ such that $i\ne j$, we arrive at the following condition
\begin{eqnarray}
     \ket{\psi_{ij}}_{A''B''E}=0\quad s.t.\quad i\ne j
\end{eqnarray}
and thus the state that achieves the maximal violation of the CHSH inequality simplifies to
\begin{equation}\label{genstate2}
    \ket{\tilde{\psi}_{ABE}}=\ket{0}_{A'}\ket{0}_{B'}\ket{\psi_{00}}_{A''B''E}+\ket{1}_{A'}\ket{1}_{B'}\ket{\psi_{11}}_{A''B''E}.
\end{equation}
Now, we consider the other SOS relation \eqref{introSOS2} and plug into it the certified measurements  \eqref{certmeas1} and \eqref{certmeas2}. This gives us
\begin{eqnarray}
     \sigma_x\otimes\sigma_x\otimes\Id_{A''B''E}\ket{\tilde{\psi}_{ABE}}=\ket{\tilde{\psi}_{ABE}}.
\end{eqnarray}
Now, substituting the simplified state \eqref{genstate2}, we obtain that
\begin{equation}
     \ket{1}_{A'}\ket{1}_{B'}\ket{\psi_{00}}_{A''B''E}+\ket{0}_{A'}\ket{0}_{B'}\ket{\psi_{11}}_{A''B''E}=\ket{0}_{A'}\ket{0}_{B'}\ket{\psi_{00}}_{A''B''E}+\ket{1}_{A'}\ket{1}_{B'}\ket{\psi_{11}}_{A''B''E}.
     \end{equation}
Again projecting the above equation on to $\bra{0}_A'\bra{0}_B'$, we arrive at the condition that
\begin{eqnarray}
     \ket{\psi_{00}}_{A''B''E}=\ket{\psi_{11}}_{A''B''E}.
\end{eqnarray}
Thus, the state which achieves the maximal violation of the CHSH inequality is given by
\begin{eqnarray}\label{2.126}
      U_A\otimes U_B\ket{\psi_{ABE}}=\frac{1}{\sqrt{2}}(\ket{0}_{A'}\ket{0}_{B'}+\ket{1}_{A'}\ket{1}_{B'})\otimes\ket{aux}_{A''B''E}
\end{eqnarray}
where $\ket{aux}_{A''B''E}=\sqrt{2}\ket{\psi_{00}}_{A''B''E}$. This completes the proof that the maximal violation of CHSH Bell inequality can be used to certify the two-qubit maxmally entangled state. In Chapter \ref{chapter_3}, we generalise this proof to self-test maximally entangled state of arbitrary local dimension. We now move on to presenting the idea of  one-sided device independent certification.

\subsection{One-sided device independent certification}
Let us now consider the following task inspired from a real-world scenario consisting of a bank that wants to securely send information to its clients. The security of the Bank can be considered to be strong and chances of attack directly on the bank is much lower. However, individual clients are prone to intruders who want to steal their information. In such a situation, bank can be considered to be trusted and the clients are untrusted. 

Based on the above example, let us consider a scenario where there are two parties Alice and Bob and a preparation device which sends one system to Alice and other to Bob. Here Alice is trusted, that is, she can perform a full tomography on her subsystem. Bob now performs measurements on his subsystem which might steer or affect Alice's subsystem. This scenario is in fact the quantum steering scenario. Observing some specific joint probability distribution $\{p(a,b|x,y)\}$ allows one to certify the preparation device as well the measurements performed by the untrusted Bob up to the equivalences as discussed in the previous subsection.  

\begin{defn}[One-sided device independent certification (1-SDI certification)]\label{1SDI} Consider the above experiment with Alice and Bob performing measurements $A_{a|x}$ and $B_{b|y}$ on a state $\rho_{AB}$ and observing correlations $\{p(a,b|x,y)\}$ along with the fact that Alice's measurements $A_{a|x}$ are known and act on Hilbert space $\mathcal{H}_A$. 
The state $\rho_{AB}$ and measurements $B_{b|y}$ are certified to be the target state $\ket{\tilde{\psi}}$ and target measurements $\tilde{B}_{b|y}$ from $\{p(a,b|x,y)\}$ if:
\begin{enumerate}
    \item The Hilbert space of Bob decomposes as 
\begin{eqnarray} \mathcal{H}_B=\mathcal{H}_{B'}\otimes\mathcal{H}_{B''}
\end{eqnarray}
where the target state $\ket{\tilde{\psi}}$ belongs to $\mathcal{H}_{A}\otimes\mathcal{H}_{B'}$ and $\mathcal{H}_{B''}$ represents the auxiliary Hilbert space of Bob.

\item There exists a unitary $U_B:\mathcal{H}_B\to \mathcal{H}_B$ such that the state is

\begin{equation}
 \Big(\Id_A\otimes U_B\Big)\rho_{AB}\left(\Id\otimes U_B^{\dagger}\right)=\proj{\tilde{\psi}}_{AB'}\otimes\sigma_{B''},
\end{equation}
where $\sigma_{B''}$ is some state acting on the Hilbert space $\mathcal{H}_{B''}$.
\item The measurements are certified as
\begin{equation}\label{ststeemea1}
    U_B\,B_{b|y}\,U_B^{\dagger}=\tilde{B}_{b|y}\otimes\Id_{B''}.
\end{equation}
where $\Id_{B''}$ is identity acting on the Hilbert space  $\mathcal{H}_{B''}$. 
\end{enumerate}
\end{defn}

Analogous to self-testing, any correlation that can be used for 1SDI certification must belong to the boundary of the quantum set. As discussed before, the quantum correlations that achieve the maximal violation of a steering inequality are extremal or lie at the boundary of the quantum set. If one employs the maximum violation of a  steering inequality to certify the quantum realisations, one needs to measure much lesser number of correlations than the standard procedure of observing at least the full tomographically complete set of correlations. One can thus provide an analogous definition of 1SDI certification by replacing the line ``observing correlations $\{p(a,b|x,y)\}$" by ``observing the maximal violation of a steering inequality" in Def. \ref{1SDI}. Again, we can define robust 1SDI analogously to Def. \ref{rst} imposing that Alice is trusted.

\subsubsection{Ways to do 1SDI certification} 
The first results on 1SDI certification were derived in \cite{Supic, Alex} where the idea of swap isometry was employed to certify the maximally entangled state and Pauli observables. In \cite{Bharti} any pure bipartite entangled state was certified by using the subspace method as was done in \cite{ALLst1}. There is also a numerical method \cite{Jordi} that aims to characterise the steering assemblages but again the limitation being that higher dimensional assemblages can not be certified in an efficient way using this method.

In chapter \ref{chapter_4} in this thesis, we develop an analytical method for 1SDI certification. We explore the fact that the algebraic bound of the steering inequality gives some relations which can be solved to find the desired certification. Here we give a short proof to highlight this idea using the simplest steering inequality \eqref{stefn}.

Let us consider the simplest steering functional \eqref{stefn} and express it in the correlation picture \eqref{stefn2} as discussed above 
\begin{eqnarray}\label{stefn2.3}
    W_{2,2,2}=\langle\sigma_z\otimes B_0\rangle+\langle\sigma_x\otimes B_1\rangle.
\end{eqnarray}
Note that for convenience, we scaled the inequality \eqref{stefn} by a factor of $2$ and also removed the term corresponding to $\Id_A\otimes\Id_B$. Achieving the maximal quantum value implies that
\begin{eqnarray}\label{2.125}
    \langle\sigma_z\otimes B_0\rangle+\langle\sigma_x\otimes B_1\rangle=2.
\end{eqnarray}
Since, $B_0$ and $B_1$ are valid observables, both the terms in the above expression are upper bounded by $1$. Consequently, one can achieve the quantum bound iff each of the expectation value in the above expression \eqref{2.125} is $1$, 
\begin{eqnarray}\label{relstee1}
    \sigma_z\otimes B_0\otimes\Id_E\ket{\psi_{ABE}}=\ket{\psi_{ABE}}
\end{eqnarray}
and,
\begin{eqnarray}\label{relstee2}
    \sigma_x\otimes B_1\otimes\Id_E\ket{\psi_{ABE}}=\ket{\psi_{ABE}}.
\end{eqnarray}
Here, as done for self-testing, we introduce an external system $E$ to purify the state shared between Alice and Bob. For convenience, we will drop $\Id_E$ from further calculations. Let us recall that the measurements can be characterised only on the support of the  local reduced states $\rho_A$ and $\rho_B$. Thus, without loss of generality we assume that these local states are full-rank. 
Now multiplying \eqref{relstee1} with $\sigma_x\otimes B_1$ on both the sides and then using \eqref{relstee2}, we arrive at
\begin{eqnarray}\label{relstee3}
    \sigma_x\sigma_z\otimes B_1B_0\ket{\psi_{ABE}}=\sigma_x\otimes B_1\ket{\psi_{ABE}}=\ket{\psi_{ABE}}.
\end{eqnarray}
Similarly, multiplying \eqref{relstee2} with $\sigma_z\otimes B_0$ on both the sides and then using \eqref{relstee1}, we arrive at
\begin{eqnarray}
    \sigma_z\sigma_x\otimes B_0B_1\ket{\psi_{ABE}}=\sigma_z\otimes B_0\ket{\psi_{ABE}}=\ket{\psi_{ABE}}.
\end{eqnarray}
Now, using the fact that $\sigma_z\sigma_x+\sigma_x\sigma_z=0$ in the above equation, we have that
\begin{eqnarray}\label{relstee4}
    \sigma_x\sigma_z\otimes B_0B_1\ket{\psi_{ABE}}=-\ket{\psi_{ABE}}.
\end{eqnarray}
Now, adding \eqref{relstee3} and \eqref{relstee4}, we have that
\begin{eqnarray}
    \sigma_x\sigma_z\otimes\left(B_0B_1+B_1B_0\right)\ket{\psi_{ABE}}=0.
\end{eqnarray}
Since, $\sigma_x, \sigma_z$ are invertible we get
\begin{eqnarray}
    \Id_A\otimes\left(B_0B_1+B_1B_0\right)\ket{\psi_{ABE}}=0.
\end{eqnarray}
Now, tracing over the subsystems $A, E$ we have
\begin{eqnarray}
    \left(B_0B_1+B_1B_0\right)\rho_B=0.
\end{eqnarray}
Again, $\rho_B$ is full-rank and thus invertible which finally gives us the desired relation
\begin{eqnarray}
    B_0B_1+B_1B_0=0.
\end{eqnarray}
Now, using Fact \ref{fact2} we can conclude that the Hilbert space of Bob decomposes as $\mathcal{H}_B=\mathbb{C}^2\otimes\mathcal{H}_{B''}$ and there exist a local unitary transformation $U_B:\mathcal{H}_B\rightarrow\mathcal{H}_B$ such that
\begin{eqnarray}\label{certmea2}
     U_BB_0 U_B^{\dagger}=\sigma_z\otimes\Id_{B''}, \quad \text{and}\quad  U_BB_1 U_B^{\dagger}=\sigma_x\otimes\Id_{B''}.
\end{eqnarray}
Now, we find the state $\ket{\psi_{ABE}}$ that gives the maximum violation of the steering inequality \eqref{stefn2.3}. As was done for self-testing, we consider a general state of the form remembering that Alice is trusted
\begin{equation}\label{genstatestee1}
     \Id_A\otimes U_B\ket{\psi_{ABE}}=\ket{\tilde{\psi}_{ABE}}=\sum_{i,j=0,1}\ket{i}_{A'}\ket{j}_{B'}\ket{\psi_{ij}}_{B''E}.
\end{equation}
where $\ket{\psi_{ij}}_{B''E}\in\mathcal{H}_{B''}\otimes\mathcal{H}_E$.
Now, using this state and the certified measurements \eqref{certmea2}, we solve the relations \eqref{relstee1} and \eqref{relstee2}, that is,
\begin{eqnarray}\label{relstee5}
    \sigma_z\otimes \sigma_z\otimes\Id_{B''}\otimes\Id_E\ket{\tilde{\psi}_{ABE}}=\ket{\tilde{\psi}_{ABE}}
\end{eqnarray}
and,
\begin{eqnarray}\label{relstee6}
    \sigma_x\otimes \sigma_x\otimes\Id_{B''}\otimes\Id_E\ket{\tilde{\psi}_{ABE}}=\ket{\tilde{\psi}_{ABE}}.
\end{eqnarray}
Following exactly the same approach, as the one used for self-testing from Eq. \eqref{2.120} to Eq. \eqref{2.126}, we find that up to a local unitary transformation $U_B$ we have that
\begin{eqnarray}
    \Id_A\otimes U_B\ket{\psi_{ABE}}=\frac{1}{\sqrt{2}}(\ket{00}+\ket{11})\otimes\ket{aux}_{B''E}
\end{eqnarray}
This completes our analysis of one-sided device independent certification. Now, we proceed towards another important avenue for device-independent protocols namely device-independent certification of genuine randomness.

\subsection{Randomness certification}
\begin{figure}
    \centering
    \includegraphics[scale=.475]{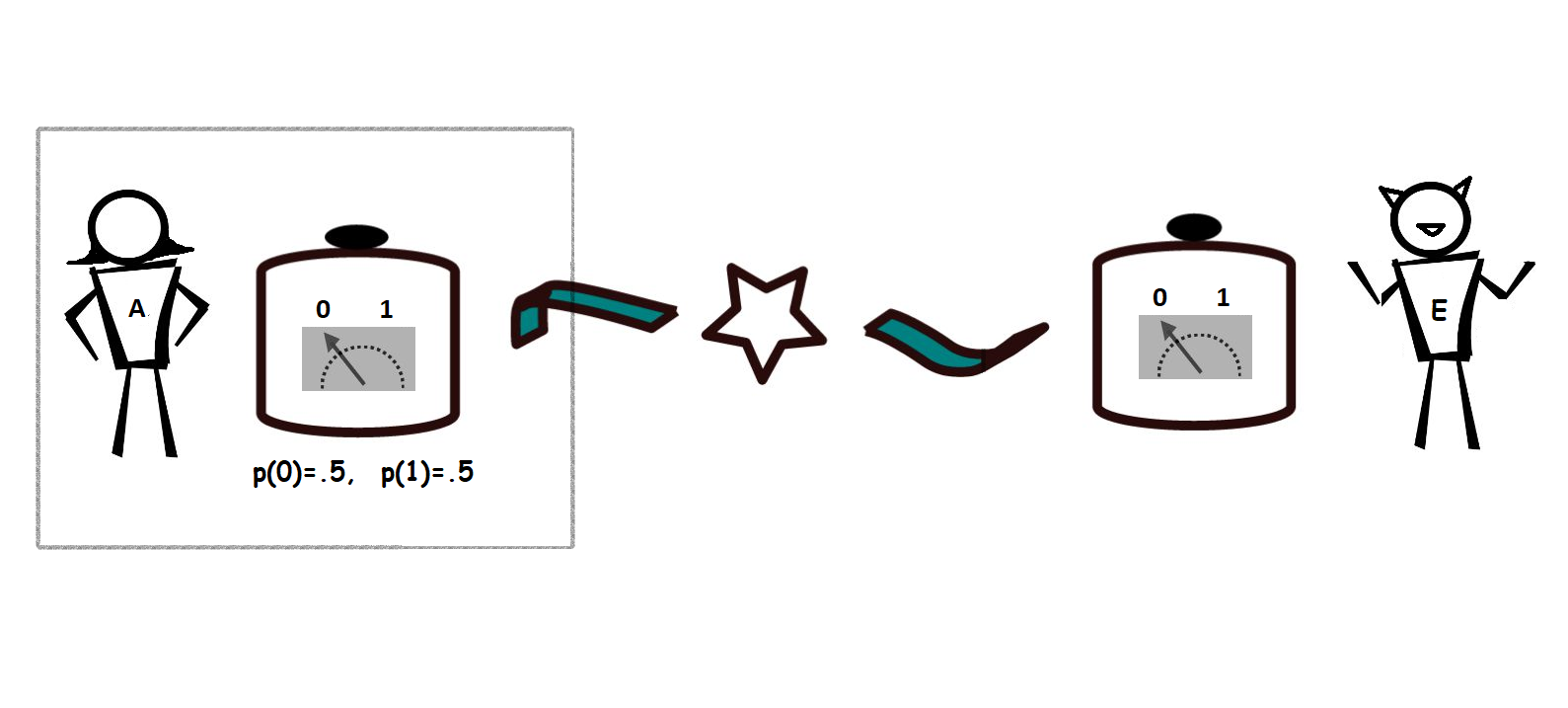}
    \caption{Alice locally in her lab gets perfectly random outcomes. However, her subsystem can be correlated with another subsystem possessed by an intruder Eve. Thus, locally to Alice her outcomes seem perfectly random but Eve can perfectly guess her outcome based on her results.}
    \label{fig7}
\end{figure}

One of the central requirements for any cryptographic task is access to devices that can generate outputs not predictable by any attacker. Within classical framework, the security of such cryptographic tasks relies majorly on the idea that some functions can not be efficiently computed by a classical computer. For instance, the security of the RSA protocol \cite{RSA} relies on the fact that large numbers can not be factored efficiently using classical computers. It was shown by Shor in 1994 \cite{shor} that one could efficiently factorise large numbers using quantum resources. Consequently, with the advent of quantum technologies, many such classical protocols have been shown to be prone to attackers possessing quantum computers. Thus, it is a matter of time when the classical cryptosystems would fail and we need better protocols that are secure against quantum attacks. This is the basis for the origin of the field of quantum cryptography.
\newpage
One of the inferences that one can make about quantum theory is that it is inherently unpredictable. For instance, consider a source that generates the quantum state 

\begin{eqnarray}\label{2.140}
    \ket{\psi_A}=\frac{1}{\sqrt{2}}\left(\ket{0}+\ket{1}\right).
\end{eqnarray}
Now, if a measurement is performed by Alice on this state in the $z-$basis, that is, $M=\{\ket{0}\!\bra{0},\ket{1}\!\bra{1}\}$, then the outcomes occur with equal probability and are unpredictable. However, here we require that the state and measurements are exactly same as the ideal ones. This is device-dependent generation of randomness, that is, one has to completely trust his devices to generate this randomness. However, let us say that the source generates the maximally entangled state
\begin{eqnarray}
     \ket{\psi_{AE}}=\frac{1}{\sqrt{2}}\left(\ket{0}_{A}\ket{0}_E+\ket{1}_A\ket{1}_E\right)
\end{eqnarray}
instead of the state \eqref{2.140} such that some attacker Eve has access to subsystem $E$. Again, the measurement is performed on the subsystem $A$ in the $z-$basis. Even if the outcomes seems unpredictable to Alice, if Eve also performs a measurement in the $z-$basis then she is able to guess the outcome of Alice with certainty as depicted in Fig. \ref{fig7}. Further, Eve might also have access to Alice's measurement device. Thus, it is necessary to find protocols for randomness generation in which one does not has to trust the device used to produce randomness. This is known as device-independent certification of randomness.  

The idea of DI randomness certification schemes relies on the premise that there is some attacker who has access to the devices that are used to generate randomness. However, the attacker can only guess the outcome of the devices randomly. For instance, if the device generates two outputs, then we say that these outputs are perfectly random if chances that the attacker can guess them is never more than fifty percent. On the other hand, if the attacker can guess the outcome with certainty then the outputs are not random. 

Let us now elaborate on this problem in a more formal way. Let us say that Alice performs a measurement $M$ on a system $S$ and observes an outcome $a$. Now, the attacker Eve can perform some measurement $\{Z_e\}$ on her subsystem and gets an outcome $e$. The outcome $e$ is the guess that Eve makes about Alice's outcome. Then the best average probability of guessing by Eve $p_{guess}(E|S)$ has to be maximised over all possible measurements $z$, and thus we arrive at the following expression as suggested in Ref. \cite{Bellreview},
\begin{eqnarray}
     p_{guess}(E|S)=\max_z\sum_ep(e|z)p(e=a|a,z)
\end{eqnarray}
where the above quantity $p(e=a|a,z)$ denotes the probability of Eve's outcome $e$ to be the same as Alice's outcome $a$ and $S$ denotes the system of Alice. When Alice and Eve posses quantum resources, that is, let us say that the state shared between Alice and Eve is given by $\rho_{AE}$ and Alice performs a measurement $\{M_a\}$ and Eve performs a measurement $\{Z_e\}$ then the above guessing probability can be written as
\begin{eqnarray}\label{guessprob}
      p_{guess}(E|\rho_A)=\max_Z\sum_a\Tr(\rho_{AE}M_a\otimes Z_a)
\end{eqnarray}
where $Z=\{Z_a\}$ and $\rho_A=\Tr_E(\rho_{AE})$. The amount of randomness that can be generated by Alice is quantified by the min-entropy of the of the guessing probability \cite{Konig}, that is,
\begin{eqnarray}
     H_{min}(E|S)=-\log_2p_{guess}(E|S).
\end{eqnarray}
Note that the above quantity is $0$ if Eve can perfectly guess Alice's outcome. 
Interestingly, the maximum amount of randomness that in principle can be generated from a $d-$dimensional quantum system is of amount $2\log_2d$ bits. 

Let us now discuss how quantum non-locality, in particular, self-testing serves as a tool in designing methods for randomness certification.
\begin{figure}[t]
    \centering
    \includegraphics[scale=.475]{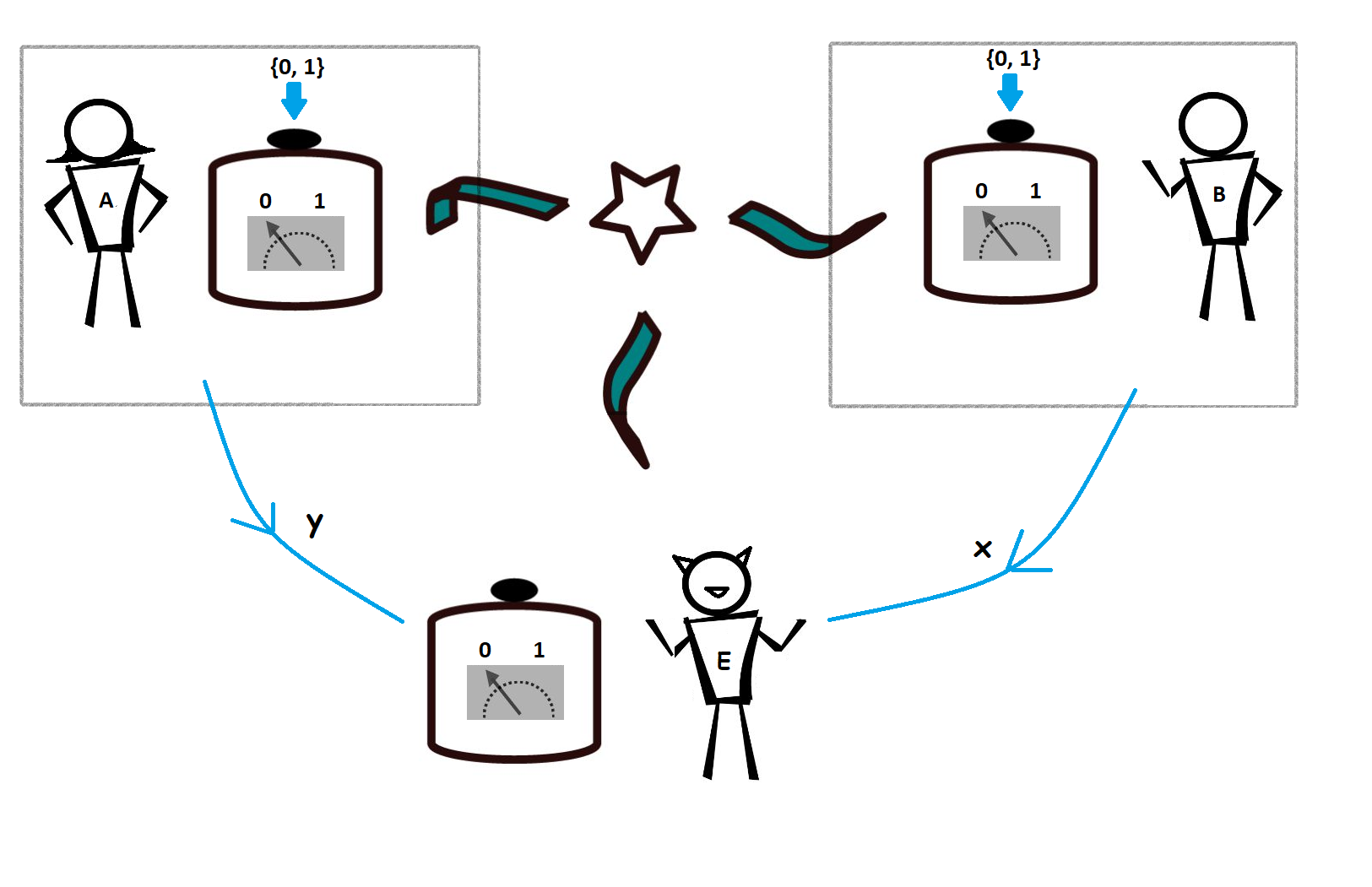}
    \caption{Bell scenario with an intruder Eve: In the simplest Bell scenario, we consider an additional party Eve who represents an attacker who wants to guess Alice's and Bob's outputs. She can receive some subsystem from the source correlated with the subsystem of Alice and Bob. She also might know the inputs of Alice and Bob. She uses all these information to guess their outputs by performing measurements on her received subsystem.}
    \label{fig8}
\end{figure}
Consider again the Bell scenario but now with an additional party Eve (the intruder) who has access to the preparation device as well as the measurement devices of Alice and Bob. The scenario is depicted in Fig. \ref{fig8}. 
Now, Eve wants to guess the  measurements outcome of one of the parties, let us say, Alice. In this case, Eve has the following information or strategies to guess the outcomes:
\begin{enumerate}
    \item   Eve can posses some subsystem $E$ that is correlated with the state sent by the preparation device. As a consequence, the state shared among the parties is defined by $\rho_{AB}=\Tr_E(\rho_{ABE})$, where $\rho_{ABE}\in \mathcal{H}_A\otimes \mathcal{H}_B\otimes \mathcal{H}_E $ denotes the state shared between Alice, Bob and Eve. There can not be any restriction on the dimension of the local Hilbert spaces of any of the parties. This also allows one to consider that the state shared among the parties to be pure as we can always add additional ancillary systems with Eve and purify the state.
    \item Eve might have control over Alice's and Bob's measurement devices in the sense that the measurements might be performed on some auxiliary system of Alice and Bob which can interfere with their observations. 
    \item Eve might perform a measurement on her subsystem given by POVM $Z=\{Z_a\}$ that acts on $\mathcal{H}_E$. As discussed above, the probability of obtaining outcome $a$ from measurement performed by Eve on her share of the joint state $\rho_{ABE}$ is the best guess of Alice's outcome $a$.
\end{enumerate}
It is also necessary for Eve to remain undetected during the attack otherwise Alice and Bob can discard those runs and start over again. Thus, the joint as well as local statistics of Alice and Bob should remain equivalent to the ideal statistics as expected by them, that is, 
\begin{eqnarray}\label{randomcons}
      \bra{\psi_{ABE}}A_{a|x}\otimes B_{b|y}\otimes\Id_E\ket{\psi_{ABE}}=\bra{\tilde{\psi}_{AB}}\tilde{A}_{a|x}\otimes \tilde{B}_{b|y}\ket{\tilde{\psi}_{AB}}
\end{eqnarray}
where $\ket{\tilde{\psi}_{AB}}$ is the ideal state that is expected to be be sent by the preparation device and $\tilde{A}_{a|x}$ and $\tilde{B}_{b|y}$ represent the ideal measurement effects that are expected to be performed by the respective parties. Note that there is no constraint on Eve's measurement as Alice and Bob do not know about Eve and the statistics are averaged over Eve's outcomes. Let us denote the set of strategies as $S_p$ that are employed by Eve such the constraint \eqref{randomcons} is satisfied. Thus, the probability of Eve to guess Alice's outcome for some input $x$ is given by,
\begin{eqnarray}\label{guessEve2}
      G(x,P)=\sup_{S_p}\sum_a\bra{\psi_{ABE}}A_{a|x}\otimes \Id_B\otimes Z_a\ket{\psi_{ABE}}.
\end{eqnarray}
This is also known as local guessing probability \cite{AAP12, APP13, APP14, random0}. Here $P$ denotes the set of observed correlations.

Now, suppose that Alice wants to generate randomness in her lab by performing the measurements $A_x$ on the state $\rho_{A}=\Tr_{BE}(\ket{\psi_{ABE}}\!\bra{\psi_{ABE}})$. Now,
Alice and Bob perform a Bell state using the state $\ket{\psi_{ABE}}$ and measurements $A_{x}$ and $B_{y}$ respectively, and observe the maximal violation of a Bell inequality. For simplicity of argument, we assume here that the measurements are projective. Let us say that the state and measurements can be self-tested from this violation to be the ideal ones, that is,
\begin{eqnarray}
     U_A\otimes U_B \ket{\psi_{ABE}}=\ket{\tilde{\psi}_{A'B'}}\otimes\ket{aux}_{A''B''E}
\end{eqnarray}
and the measurements on the local support of the state are certified to be the ideal measurements as
\begin{eqnarray}
     U_A\,A_{x}\,U_A^{\dagger}=\tilde{A}_{x}\otimes\Id_{A''},\qquad U_B\,B_{y}\,U_B^{\dagger}=\tilde{B}_{y}\otimes\Id_{B''}.
\end{eqnarray}
Then the guessing probability of Eve \eqref{guessEve2} is given by
\begin{eqnarray}
     G(0,E)=\sup_{\rho_E,Z_a}\sum_a\Tr(\tilde{A}_{a|0}\rho_{A'})\Tr( Z_a\rho_E).
\end{eqnarray}
where $\rho_E=\Tr_{AB}(\ket{\psi_{ABE}}\!\bra{\psi_{ABE}})$. Using the fact $\sum_aZ_a=\Id$, we conclude that
\begin{eqnarray}
     G(0,E)\leq \max_a\Tr(\tilde{A}_{a|0}\rho_{A'}).
\end{eqnarray}
Thus, Alice can securely generate at least
\begin{eqnarray}
     H_{min}(E)\geq-\log_2(\max_a\Tr(\tilde{A}_{a|0}\rho_{A'}))
\end{eqnarray}
amount of randomness. Notice that this bound is independent of any strategy of Eve. Thus, self-testing allows one to locally generate randomness in a highly secure way even in the presence of an intruder. This completes the subsection on randomness and the section on technical introduction. We now move onto the main results of this thesis.

%% file: Sections/chapter_3.tex
\chapter{Certification of multipartite entangled states of arbitrary local dimension}

\label{chapter_3}
\vspace{-1cm}
\rule[0.5ex]{1.0\columnwidth}{1pt} \\[0.2\baselineskip]
\definecolor{airforceblue}{rgb}{0.36, 0.54, 0.66}
\definecolor{bleudefrance}{rgb}{0.19, 0.55, 0.91}

\section{Introduction}
The strongest form of device-independent certification or self-testing has attracted considerable attention lately as it allows to obtain maximal information up to certain equivalences about a concerned quantum system with minimal assumptions. For instance, in Refs.  \cite{qubitst1, qubitst2, qubitst3, qubitst4, qubitst5, qubitst6, qubitst7, qubitst8, qubitst9, qubitst10}, self-testing schemes have been proposed for certification of pure bipartite entangled states that are locally qubits. There are a few results that allow for certification pure bipartite entangled states of local dimension three \cite{quditst1, quditst2, Jed1}. A scheme for self-testing of pure bipartite states of arbitrary local dimension was proposed in \cite{ALLst1}. However this scheme relies on self-testing of two-qubit states \cite{qubitst1, qubitst2, qubitst3} by considering different two-dimensional subspaces of the local state of both the parties. In their scheme, one party has to perform three and the other has to perform four measurements on their respective subsystems. As far as the experimental implementation of self-testing schemes is concerned, it is thus a question of utmost importance as to whether one can design schemes exploiting the minimal number of measurements necessary to observe non-locality which is two per observer. 

Self-testing schemes were also designed for multipartite states, in particular the tripartite ones, which are states shared among three parties and are locally qubits  \cite{ multist5, multist6, multist7, multist8, multist9}. The problem to certify states shared among arbitrary number of parties has not been much explored  with a few exceptions \cite{multist1, multist2, multist3} that provide a scheme to self-test $N-$qubit graph states and the Dicke states. A scheme that applies to states of arbitrary local dimension and shared among arbitrary number of parties was proposed in \cite{multist4}. This scheme again adapts the procedure of Ref. \cite{ALLst1} to the multiparty scenario that relies on the self-testing technique for the two-qubit states. Here again, one party has to perform three and the other parties have to perform four measurements. In Ref. \cite{everything}, the authors show a proof-of-principle demonstration that every pure entangled state can be certified. However, in this scheme all the parties have to perform at least $d^2$ measurements where $d$ is the local dimension of the state. Each of the measurements have $2^s$ outcomes where $s$ is chosen such that $2^s$ is the smallest integer larger than $d$. All these limitations make this scheme practically difficult to implement in experiments.

It is thus a problem of key interest to find device-independent certification schemes that are experimentally friendly in the sense that they require minimal resources and effort for their practical implementation. We consider the general problem to design a scheme for self-testing of pure entangled states of arbitrary local dimension shared among arbitrary number of parties using Bell inequalities composed of truly $d-$outcome measurements where $d$ is any positive integer. Further, each party must perform the minimal number of measurements possible to observe a Bell violation, which is, two. 

In this Chapter, we provide a method to self-test one of most studied multipartite states, namely the $N$-partite Greenberger–Horne–Zeilinger (GHZ) states of local dimension $d$, or simply, generalised GHZ state
\begin{eqnarray}\label{GHZ}
\ket{\text{GHZ}_{N,d}}=\frac{1}{\sqrt{d}}\sum_{i=0}^{d-1}\ket{i}^{\otimes N}
\end{eqnarray}
with $N$ and $d$ being arbitrary integers such that $N,d\geq 2$. Notice that the generalised GHZ state \eqref{GHZ} reduces to the two-qudit maximally entangled state \eqref{maxent} when $N=2$.
For this, we utilise the Bell inequalities proposed \cite{ACTA} which is a generalisation of the Bell inequalities proposed in \cite{SATWAP}. In this scheme, every party needs to perform only two measurements. With the view to self-test measurements we generalise this result when each party performs arbitrary number of measurements. The results presented below are based on our works \cite{sarkar, sarkar4}.

\section{Family of Bell inequalities}
We consider the most general Bell scenario as depicted in Fig.  \ref{fig2}. It involves $N$ parties in spatially separated labs. Each of them receives a subsystem from a source and performs one of $m$ $d-$outcome measurements on it. After the experiment is complete, they combine their results to reconstruct the joint probability distribution. 
Let us now say that we want to certify that the source generates the desired state and the parties perform the desired quantum measurements. Then, the first step to achieve this goal is to find a Bell inequality that is maximally violated by these quantum realisations. For this, we consider two recently derived Bell inequalities.  

\subsection{SATWAP Bell inequalities}
Let us first recall the Salavrakos-Augusiak-Tura-Wittek-Acín-Pironio (SATWAP) Bell inequality \cite{SATWAP} in the simplest scenario where each party performs only two measurements. When expressed in the observable picture (cf. Chapter \ref{chapter_2}) SATWAP Bell inequality reads as
 \begin{equation}\label{SATWAP}
\mathcal{B}_{2,d,2}=\sum_{k=1}^{d-1} \left( a_k \langle A_1^{k}B_1^{d-k}\rangle +a_k^*\omega^k \langle A_1^{k}B_2^{d-k} \rangle 
+ a_k^* \langle A_2^{k} B_1^{d-k} \rangle +  a_k \langle A_2^{k} B_2^{d-k} \rangle \right) \leq \beta_{C}^{2,d,2},
\end{equation}
where 
\begin{equation}\label{ak}
     a_k=\frac{1}{\sqrt{2}}\omega^{\frac{2k-d}{8}}=\frac{1-\mathbbm{i}}{2}\omega^{k/4}=\frac{1-\mathbbm{i}}{2}\mathrm{e}^{\frac{\pi \mathbbm{i}k}{2d}},
\end{equation}
where $\omega=e^{2\pi i/d}$ is the $d-th$ root of unity. The classical bound of the SATWAP Bell inequality $\beta_{C}^{2,d,2}$ was computed in \cite{SATWAP} to be 
\begin{equation}\label{SATWAPclass}
    \beta_{C}^{2,d,2}=\frac{1}{2}\left[3\cot\left(\frac{\pi}{4d}\right)
    -\cot\left(3\frac{\pi}{4d}\right)\right]-2.
\end{equation}
It is worth noting that for $d=2$ the SATWAP Bell inequality reduces to the well-known CHSH inequality \eqref{CHSH} introduced in Chapter \ref{chapter_2}.

\subsubsection{Sum of Squares (SOS) decomposition}
The maximal quantum value of $\mathcal{B}_{2,2,d}$ turns out to be $\beta_Q^{2,d,2}=2(d-1)$ and can be attained by the two-qudit maximally entangled state
\begin{equation}\label{maxent}
    \ket{\phi_{d}^+}=\frac{1}{\sqrt{d}}\sum_{i=0}^{d-1}\ket{ii}
\end{equation}
and $d$-outcome measurements referred as Collins-Gisin-Linden-Massar-Popescu (CGLMP) measurements \cite{CGLMP,BKP}. These measurements have been experimentally realised in \cite{expCGLMP}. In the next subsection, we provide the explicit form of these measurements.
As discussed in Chapter \ref{chapter_2}, to compute the quantum bound one can find the SOS decomposition of the corresponding Bell operator of SATWAP Bell inequality \eqref{SATWAP}. To prove that $\beta_{Q}^{2,d,2}$ is indeed the maximum quantum value of the SATWAP Bell inequality, the following SOS decomposition was derived in \cite{SATWAP}:
\begin{equation}
    \beta_{Q}^{2,d,2}\Id-\mathcal{\hat{B}}_{2,2,d}=\frac{1}{2}\sum_{k=1}^{d-1}\left(P_{1,k}^{\dagger}P_{1,k}+P_{2,k}^{\dagger}P_{2,k}\right),
\end{equation}
with
\begin{equation}
    P_{i,k}=\Id-A_{i}^{k}\otimes C_{i}^{(k)}
\end{equation}
for $i=1,2$ and $k=1,\ldots,d-1$ such that
\begin{equation}\label{Coperators}
C_1^{(k)}=a_k B_1^{-k} + a_{k}^{*}\omega^k B_2^{-k}, \qquad 
C_2^{(k)}=a_{k}^{*}B_1^{-k} + a_k B_2^{-k}.
\end{equation}
Note from \eqref{ak} that $a_{d-k}=a_{k}^{*}$ and thus $C_i^{(d-k)}=[C_{i}^{(k)}]^{\dagger}$ for 
all $i, k$.

\subsection{ASTA Bell inequalities}
The SATWAP Bell inequalities were later generalised to an arbitrary number of measurements per party and also to arbitrary number of parties in  \cite{ACTA}, which from here on will be referred to as Augusiak-Salavrakos-Tura-Acín (ASTA) Bell inequalities. As a matter of fact, ASTA Bell inequalities are the tilted version of Bell inequality proposed in \cite{aolita}. The ASTA Bell inequalities can be expressed in the observable picture (cf. Chapter \ref{chapter_2}) as
\begin{equation}\label{ASTA}
\mathcal{B}_{m,d,N}:=\sum_{\alpha_1,\ldots,\alpha_{N-1}=1}^m\sum_{k=1}^{d-1}\left\langle\left(a_kA_{1,\alpha_1}^k+a_k^*A_{1,\alpha_1+1}^k\right)\otimes\bigotimes_{i=2}^NA^{(-1)^{i-1}k}_{i,\alpha_{i-1}+\alpha_i-1}\right\rangle\geq\beta_C^{m,d,N}
\end{equation}
where,
\begin{equation}\label{app:ak}
     a_k=\frac{\omega^{\frac{2k-d}{4m}}}{2\cos(\pi/2m)}.
\end{equation}
and $A_{i,x}$ is the $x-$th measurement of $A_i-$th party, where $\alpha_N=1$. The classical bound of the above Bell inequality when $N=m=2$ is given in \eqref{SATWAPclass}. For some cases, such as $N=3,4 $ and $m=2,3$ the classical bound was computed numerically in \cite{ACTA}.  For all the other cases it is difficult to obtain the classical bound due to computational constraints. However, it was proven in \cite{ACTA} that the Svetlichny bound \cite{Svet} of the above Bell inequalities is strictly lower than quantum bound (written below). At the same time the Svetlichny bound is known to be an upper bound to the classical bound which implies that ASTA Bell inequalities are non-trivial for any $N,m,d$. Recall that the Svetlichny bound is the maximal value of a Bell expression over all correlations that are convex combination of correlators that are local with respect to all possible non-trivial bipartitions.

\subsubsection{Sum of Squares (SOS) decomposition}
The maximal quantum value of $\mathcal{B}_{m,d,N}$ turns out to be $\beta_Q^{m,d,N}=m^N(d-1)$ and can be achieved by the generalised GHZ state \eqref{GHZ} and the following measurements written in the observable form as,  
\begin{equation}\label{measurements}
\mathcal{O}_{1,x}=U_{x}F_d\,\Omega_dF_d^{\dagger}U_{x}^{\dagger},\qquad
\mathcal{O}_{2,x}=V_{x}F_d^{\dagger}\Omega_dF_dV_{x}^{\dagger},
\end{equation}
for the first two parties, and for parties indexed by odd numbers as
\begin{equation}\label{measurements2}
\mathcal{O}_{\mathrm{odd},x}=W_{x}F_d\,\Omega_dF_d^{\dagger}W_{x}^{\dagger}
\end{equation}
and parties indexed by even numbers as
\begin{equation}\label{measurements3}
\mathcal{O}_{\mathrm{ev},x}=W_{x}^{\dagger}F_d^{\dagger}\Omega_dF_d W_{x}    
\end{equation}
for all other parties $A_i$ such that $(i=3,\ldots,N)$. 
%
In the above formulas
\begin{equation}
F_d=\frac{1}{\sqrt{d}}\sum_{i,j=0}^{d-1}\omega^{ij}\ket{i}\!\bra{j},\qquad
\Omega_d=\mathrm{diag}[1,\omega,\ldots,\omega^{d-1}]
\end{equation}
with $\omega=\mathrm{exp}(2\pi\mathbbm{i}/d)$. Then, the unitary operations 
$U_x$, $V_x$ and $W_x$ are defined as
\begin{equation}
U_{x}=\sum_{j=0}^{d-1}\omega^{-j \gamma_{m}(x)}\proj{j},\qquad
V_x=\sum_{j=0}^{d-1}\omega^{j \zeta_{m}(x)}\proj{j},\qquad
W_x=\sum_{j=0}^{d-1}\omega^{-j \theta_{m}(x)}\proj{j},
\end{equation}
where
\begin{eqnarray}\label{measupara}
\gamma_m(x)=\frac{x}{m}-\frac{1}{2m},\qquad \zeta_m(x)=\frac{x}{m},\qquad \text{and}\qquad \theta_m(x)=\frac{x-1}{m}. 
\end{eqnarray}
The above observables can also be written in the matrix form as
\begin{eqnarray}\label{Obs12}
\mathcal{O}_{1,x}&=&\sum_{i=0}^{d-2}\omega^{\gamma_m(\alpha)}\ket{i}\!\bra{i+1}+\omega^{(1-d)\gamma_m(\alpha)}\ket{d-1}\!\bra{0},\nonumber\\  \mathcal{O}_{2,x}&=&\sum_{i=0}^{d-2}\omega^{\zeta_m(\alpha)}\ket{i+1}\!\bra{i}+\omega^{(1-d)\zeta_m(\alpha)}\ket{0}\!\bra{d-1}
\end{eqnarray}
for the first two parties, and 
\begin{eqnarray}\label{Obsoddeven}
\mathcal{O}_{\mathrm{odd},x}&=&\sum_{i=0}^{d-2}\omega^{\theta_m(\alpha)}\ket{i}\!\bra{i+1}+\omega^{(1-d)\theta_m(\alpha)}\ket{d-1}\!\bra{0}, \nonumber\\ \mathcal{O}_{\mathrm{ev},x}&=&\sum_{i=0}^{d-2}\omega^{\theta_m(\alpha)}\ket{i+1}\!\bra{i}+\omega^{(1-d)\theta_m(\alpha)}\ket{0}\!\bra{d-1}
\end{eqnarray}
for the remaining parties. When $m=2$, the measurement of the first two parties are the previously discussed CGLMP measurements. Further on, we would refer to the above measurements as generalised CGLMP measurements. 

To prove that $\beta_Q^{m,d,N}$ is maximal quantum value of the ASTA Bell expression, the authors of \cite{ACTA} constructed the followig sum-of-squares decomposition of the Bell operator $\mathcal{\hat{B}}_{m,d,N}$,
\begin{equation}\label{SOSAp}
    \beta_{Q}^{m,d,N}\Id-\mathcal{\hat{B}}_{m,d,N}=\frac{1}{2}\sum_{\alpha_1,\ldots,\alpha_{N-1}=1}^m\sum_{k=1}^{d-1}\left(P_{{\alpha_1,\ldots,\alpha_{N-1}}}^{(k)}\right)^{\dagger}P_{{\alpha_1,\ldots,\alpha_{N-1}}}^{(k)}+\frac{m^{N-2}}{2}\sum_{\alpha=1}^{m-2}\sum_{k=1}^{d-1}\left(R_{\alpha}^{(k)}\right)^{\dagger}R_{\alpha}^{(k)}
\end{equation}
with
\begin{equation}\label{SOSR1}
   P_{\alpha_1,\ldots,\alpha_{N-1}}^{(k)}=\Id-\overline{A}_{1,\alpha_1}^{(k)}\otimes\bigotimes_{i=2}^NA^{(-1)^{i-1}k}_{i,\alpha_{i-1}+\alpha_i-1} 
\end{equation}
and,
\begin{equation}\label{3.21}
    R_{\alpha}^{(k)}=\mu_{\alpha,k}^*A_{1,2}^k+\nu_{\alpha,k}^*A_{1,\alpha+2}^k+\tau_{\alpha,k}A_{1,\alpha+3}^k
\end{equation}
for $k=1,\ldots,d-1$ and all $\alpha_1,\ldots,\alpha_N$ and $\alpha=1,2,\ldots,m-2$, where  
\begin{equation}\label{barA1}
\overline{A}_{1,\alpha_1}^{(k)}=a_kA_{1,\alpha_1}^k+a_k^*A_{1,\alpha_1+1}^k.
\end{equation}
The coefficients $\mu_{\alpha,k}, \nu_{\alpha,k}$ and $\tau_{\alpha,k}$ are given by
\begin{eqnarray}\label{Rcoef1}
\mu_{\alpha,k}&=&\frac{\omega^{(\alpha+1)(d-2k)/2m}}{2\cos(\pi/2m)}\frac{\sin(\pi/m)}{\sqrt{\sin(\pi\alpha/m)\sin[\pi(\alpha+1)/m]}},\nonumber\\
\nu_{\alpha,k}&=&-\frac{\omega^{(d-2k)/2m}}{2\cos(\pi/2m)}\frac{\sqrt{\sin[\pi(\alpha+1)/m]}}{\sqrt{\sin(\pi\alpha/m)}},\nonumber\\
\tau_{\alpha,k}&=&\frac{1}{2\cos(\pi/2m)}\frac{\sqrt{\sin(\pi\alpha/m)}}{\sqrt{\sin[\pi(\alpha+1)/m]}}
\end{eqnarray}
for all $k$ and $\alpha=1,2,\dots,m-3$. For $\alpha=m-2$, we have
\begin{eqnarray}\label{Rcoef2}
\mu_{m-2,k}&=&-\frac{\w^{-k}\omega^{-(d-2k)/2m}}{2\cos(\pi/2m)\sqrt{2\cos(\pi/m)}},\nonumber\\
\nu_{m-2,k}&=&-\frac{\omega^{(d-2k)/2m}}{2\cos(\pi/2m)\sqrt{2\cos(\pi/m)}},\nonumber\\
\tau_{m-2,k}&=&\frac{\sqrt{2\cos(\pi/m)}}{2\cos(\pi/2m)}.
\end{eqnarray}
Note from \eqref{ak} that $a_{d-k}=a_{k}^{*}$ and thus $\overline{A}_i^{(d-k)}=[\overline{A}_{i}^{(k)}]^{\dagger}$ for 
all $i, k$. 
For our considerations, we also construct several analogous sum-of-squares decomposition parametrized of the Bell operator $\mathcal{\hat{B}}_{m,d,N}$ given by

\begin{eqnarray}\label{extraSOS}
\beta_Q\Id-\mathcal{\hat{B}}_{N,m,d}=\frac{1}{2}\sum_{\alpha_1,\ldots\alpha_N=1,2}\sum_{k=1}^{d-1}P_{n,\alpha_1,\ldots\alpha_N}^{(k)}P_{n,\alpha_1,\ldots\alpha_N}^{(k)\dagger}+\frac{m^{N-2}}{2}\sum_{\alpha=1}^{m-2}\sum_{k=1}^{d-1}\left(R_{n,\alpha}^{(k)}\right)^{\dagger}R_{n,\alpha}^{(k)},
\end{eqnarray}
where $n=2,3,\ldots,N$. In the above decomposition \eqref{extraSOS}, we have that
\begin{eqnarray}\label{SOS27}
P_{n,\alpha_1,\ldots,\alpha_N}^{(k)}=\Id-A_{1,\alpha_1}^{(k)}\otimes\overline{A}_{n,\alpha_{n-1}+\alpha_n-1}^{(k)}\otimes\bigotimes_{\substack{i=2\\i\ne n}}^N A^{(-1)^{i-1}k}_{i,\alpha_{i-1}+\alpha_i-1}.
\end{eqnarray}
For odd $n$, $\overline{A}_{n,\alpha_{n-1}+\alpha_n-1}^{(k)}$ and $R_{n,\alpha}^{(k)}$ are defined as
\begin{eqnarray}\label{SOS23}
\overline{A}_{n,\alpha_{n-1}+\alpha_n-1}^{(k)}=a_kA_{n,\alpha_{n-1}+\alpha_n-1}^{k}+a_k^*A_{n,\alpha_{n-1}+\alpha_n}^{k}
\end{eqnarray}
and,
\begin{eqnarray}\label{SOS25}
R_{n,\alpha}^{(k)}=\mu_{\alpha,k}^*A_{n,2}^k+\nu_{\alpha,k}^*A_{n,\alpha+2}^k+\tau_{\alpha,k}A_{n,\alpha+3}^k,
\end{eqnarray}
whereas, if $n$ is even then
\begin{eqnarray}\label{SOS24}
\overline{A}_{n,\alpha_{n-1}+\alpha_n-1}^{(k)}=a_kA_{n,\alpha_{n-1}+\alpha_n-1}^{-k}+a_k^*A_{n,\alpha_{n-1}+\alpha_n-2}^{-k}
\end{eqnarray}
and,
\begin{eqnarray}\label{SOS26}
R_{n,\alpha}^{(k)}=\mu_{\alpha,k}A_{n,2}^{-k}+\nu_{\alpha,k}A_{n,\alpha+2}^{-k}+\tau_{\alpha,k}A_{n,\alpha+3}^{-k}.
\end{eqnarray}
It is important to note here that $A_{n,\alpha+m}=\omega A_{n,\alpha}$ and $A_{n,0}=\omega^{-1}A_{n,m}$ for all $n,\alpha$. Further, the coefficients $\mu_{\alpha,k}$, $\nu_{\alpha,k}$ and $\tau_{\alpha,k}$ appearing in conditions \eqref{SOS25} and \eqref{SOS26} are given in Eq. \eqref{Rcoef1} and Eq. \eqref{Rcoef2}. 

Before moving on to the proof of self-testing, let us introduce the following unitary matrices which are essential for our considerations. The first matrix is the $d-$dimensional generalisation of the Pauli-$z$ matrix
\begin{equation}\label{Zd}
    Z_d=\sum_{i=0}^{d-1}\omega^i\proj{i}
\end{equation}
and second,
\begin{equation}\label{ZdTd}
 T_{d,m}=\sum_{i=0}^{d-1}\omega^{i+\frac{1}{m}}\proj{i}-\frac{2\mathbbm{i}}{d}\sin\left(\frac{\pi}{m}\right)\sum_{i,j=0}^{d-1}(-1)^{\delta_{i,0}+\delta_{j,0}}\omega^{\frac{i+j}{2}-\frac{d-2}{2m}}|i\rangle\!\langle j|.
\end{equation}
Notice that the above matrices represent valid observabales, that is, they are unitaries with eigenvalues $1,\omega,\ldots,\omega^{d-1}$. Also notice that when $d=m=2$, the second matrix is proportional to Pauli-$x$ matrix, that is, $T_{2,2}=-\sigma_x$. 

\section{Self-testing}\label{sec:selftesting}

Here, we demonstrate how achieving the maximal value of the ASTA Bell inequalities \eqref{ASTA}, that is, $\mathcal{B}_{m,d,N}=\beta_Q$, can be used for self-testing of generalised GHZ states \eqref{GHZ} and the generalised CGLMP measurements \eqref{measurements},\eqref{measurements2} and \eqref{measurements3}. Let us first recall that we can only characterise the observables on the support of the local states $\rho_{A_i}$. Thus, without loss of generality we can assume them to be of full rank. The main result comprises of one long proof and is very technical. To make it easier to follow, we shift some parts of the proof to Appendix and refer them here as Lemmas.

\renewcommand{\thetheorem}{3.\arabic{theorem}}

\setcounter{thm}{0}

\begin{theorem}\label{Theo3.1} 
Assume that all the parties $A_i$ perform the Bell experiment and observe that the Bell inequality \eqref{ASTA} 
is maximally violated, that is, $\beta_Q^{m,d,N}=m^N(d-1)$ where $N$ is number of parties, $d$ denotes the number of outcomes of each measurement and $m$ is the number of measurements performed by each party. Let us now say that the maximal quantum bound is achieved using the state $\rho_{N}$ acting on $\mathcal{H}_{A_1}\otimes\ldots\otimes\mathcal{H}_{A_N}$ and unitary observables $A_{i,\alpha}$ for all $i$ and $\alpha\in \{1,2,\ldots,m\}$ acting on $\mathcal{H}_{A_i}$. Then, the following statements hold true: 
\begin{enumerate}
    \item The Hilbert space $\mathcal{H}_{A_i}$ of all the parties $A_i$ admits a decomposition into a $d-$dimensional Hilbert space $\mathbbm{C}^d$ and an auxiliary Hilbert space of unkown but finite dimension $\mathcal{H}_{A_i''}$,
    \begin{eqnarray}\label{lem1.0}
    \mathcal{H}_{A_i}=(\mathbbm{C}^d)_{A_i'}\otimes \mathcal{H}_{A_i''}.
    \end{eqnarray}
    
    \item A local unitary transformation $U_{A_i}:\mathcal{H}_{A_i}\rightarrow\mathcal{H}_{A_i}$ can be applied on each side, such that
\begin{eqnarray}\label{lem1.2.1}
\left(U_{1}\otimes \ldots \otimes U_{N}\right)\rho_{N}\left(U_{1}^{\dagger}\otimes \ldots \otimes U_{N}^{\dagger}\right)=\ket{\mathrm{GHZ}_{N,d}}\!\bra{\mathrm{GHZ}_{N,d}}_{A_1'A_2'\ldots A_N'} \otimes\rho^\mathrm{aux}_{A_1''A_2''\ldots A_N''} 
\end{eqnarray}
where $\ket{\mathrm{GHZ}_{N,d}}$ is the generalised GHZ state \eqref{GHZ} and
\begin{eqnarray}\label{lem1.1.1}
\forall i,\alpha, \qquad 
U_{i}\,A_{i,\alpha}\, U^{\dagger}_{i} = \mathcal{O}_{i,\alpha}\otimes \Id_{i}''
\end{eqnarray}  
where $A_i''$ denotes the auxiliary system of every party on which the measurements act trivially and $\Id_{i}''$ acts on the Hilbert space $\mathcal{H}_{A_i''}$.
\end{enumerate}

\end{theorem}
\begin{proof} The proof consists of two major steps. The first one is divided into two sub-steps. In the first of them, we concentrate on the first party and prove that in $\mathcal{H}_{A_1}$ one can identify a qudit in the sense of Eq. \eqref{lem1.0}. Then, the observables $A_{1,x}$ can be certified up to local unitary to be the generalised CGLMP measurement \eqref{measurements}. In the next sub-step, we extend the above proofs to the remaining parties. For our convinience, in the proof we consider a purification of the state $\rho_N$ by adding an ancillary system $E$ and $\ket{\psi_N}\in\mathcal{H}_{1}\otimes\ldots\otimes\mathcal{H}_{N}\otimes\mathcal{H}_E$ such that $\rho_N=\Tr_E(\ket{\psi}\!\bra{\psi}_N)$.

In the second major step of the proof, we use the obtained observables to certify that $\ket{\psi_N}$ is unitarily equivalent to the generalised GHZ state \eqref{GHZ}.

\subsubsection{The Hilbert space structure and characterization of observables} 

\subsubsection{The first party}
To begin, as discussed in Chapter \ref{chapter_2}, we notice that any state $\ket{\psi}_N\in\mathcal{H}_{A_1}\otimes\ldots\otimes\mathcal{H}_{A_N}\otimes\mathcal{H}_E$ that maximally violates the Bell inequalities \eqref{ASTA} must satisfy the following relation due to the SOS decomposition (\ref{SOSAp}),
%
\begin{equation}
 P_{\alpha_1,\ldots,\alpha_N}^{(k)}\otimes\Id_E\ket{\psi_{N}}=0
\end{equation}
for $k=1,2,\ldots,d-1$ and $\alpha_i=1,2,\ldots,m$ for $i=1,2,\ldots,N-1$ and $\alpha_N=1$. Expanding the above term with the aid of Eq. (\ref{SOSR1}), this implies that
\begin{eqnarray}\label{SOSrel2}
\overline{A}_{1,\alpha_1}^{(k)}\otimes\bigotimes_{i=2}^NA^{(-1)^{i-1}k}_{i,\alpha_{i-1}+\alpha_i-1}\otimes\Id_E\ket{\psi_{N}}=\ket{\psi_{N}}
\end{eqnarray}
for all $k$ and $\alpha_i$. From here on, for simplicity we drop the term $\Id_E$. Since $A_{i,\alpha_{i}}$ are unitary for all $i$ and $\alpha_i$, we have that
\begin{equation}\label{3.32}
    \overline{A}_{1,\alpha_1}^{(k)}\otimes\Id_{A_2A_3\ldots A_N}\ket{\psi_{N}}=\Id_{A_1}\otimes\bigotimes_{i=2}^NA^{(-1)^{i}k}_{i,\alpha_{i-1}+\alpha_i-1}\ket{\psi_{N}}.
\end{equation}
After applying $\Id_{A_1}\otimes\bigotimes_{i=2}^NA^{(-1)^{i-1}k}_{i,\alpha_{i-1}+\alpha_i-1}$ to the above condition and taking into account Eq. \eqref{3.32} for $k\rightarrow d-k$, we arrive at 
\begin{eqnarray}
 \overline{A}_{1,\alpha_1}^{(k)}\overline{A}_{1,\alpha_1}^{(d-k)}\otimes\Id_{A_2A_3\ldots A_N}\ket{\psi_{N}}=\ket{\psi_{N}}.
\end{eqnarray}
Taking then the partial trace over the subsystems $A_2, A_3,\ldots ,A_N, E$ we obtain
\begin{eqnarray}
\overline{A}_{1,\alpha_1}^{(k)}\overline{A}_{1,\alpha_1}^{(d-k)}\rho_{A_1}=\rho_{A_1},
\end{eqnarray}
where $\rho_{A_1}=\Tr_{A_2, A_3,\ldots ,A_N,E}(\ket{\psi}\!\bra{\psi}_N)$. Since, $\rho_{A_1}$ is full-rank and thus invertible, we conclude from the above formula that
\begin{eqnarray}\label{SOSrela}
\overline{A}_{1,\alpha_1}^{(k)}\overline{A}_{1,\alpha_1}^{(d-k)}=\Id.
\end{eqnarray}
Now, using the relation \eqref{3.32} for $k=1$, and then applying $\overline{A}_{1,\alpha_1}^{(1)}$ recursively to it, we also obtain that
\begin{equation}\label{SOSrelb}
    \overline{A}_{1,\alpha_1}^{(k)}=\left[\overline{A}_{1,\alpha_1}^{(1)}\right]^k
\end{equation}
for all $k, \alpha_1$. Recall that $\overline{A}_{1,\alpha_1}^{(d-k)}=\overline{A}_{1,\alpha_1}^{(k)\dagger}$ for any $k=1,\ldots,d-1$. The other term in the SOS decomposition (\ref{SOSAp}) yields the following relation,
\begin{eqnarray}
R_{\alpha}^{(k)}\ket{\psi_{N}}=0\qquad \forall k,\alpha.
\end{eqnarray}
Note that $R_{\alpha}^{(k)}$ is composed of only $A_1's$ observables and thus acts only on the first party's subsystem  $\rho_{A_1}$. Taking a partial trace over the subsystems $A_2, A_3,\ldots ,A_N, E$, the above condition is equivalent to $R_{\alpha}^{(k)}\rho_{A_1}=0$. Again, taking into account that $\rho_{A_1}$ is full-rank and thus inevertible, we have that
\begin{eqnarray}\label{SOSrelc}
R_{\alpha}^{(k)}=0
\end{eqnarray}
for all $k$ and $\alpha$. 
The conditions \eqref{SOSrela}, \eqref{SOSrelb} and \eqref{SOSrelc} are solely composed of the $A_1's$ observables and as a matter of fact, enough to characterise the observables $A_{1,\alpha_1}$ for all $\alpha_1$ up to local unitary operations. 

From here on, for simplicity we denote $\mathcal{H}_{A_1}$ as $\mathcal{H}_{1}$ and $\mathcal{H}_{A_1''}$ as $\mathcal{H}_{1}''$. Let us first give short overview of the idea behind the proof. First, we show that the Hilbert space of first party can be written as a direct sum of $d-$dimensional Hilbert space, that is, 
\begin{equation}\label{hstruc}
\mathcal{H}_{1}=\mathbb{C}^d\otimes\mathcal{H}_{1}'',    
\end{equation}
where $\mathcal{H}_{1}''$ is a Hilbert of unknown but finite dimension. For this purpose, using the conditions (\ref{SOSrela}) and (\ref{SOSrelb}) for $\alpha_1=2$, we show in Lemma
\ref{le:traceless} which is stated below that 
\begin{equation}\label{eq17}
    \Tr(A_{1,2}^n)=\Tr(A_{1,3}^n)=0
\end{equation}
for any $n$ that is a divisor of $d$ such that $n<d$. Then using (\ref{SOSrelc}), we can also conclude the same result for the rest of the observbales, that is, $\Tr(A_{1,\alpha}^n)=0$ for all $\alpha=1,2,\ldots,m$ and any $n$ that is a divisor of $d$ such that $n<d$.
Notice that for a unitary matrix $M$ with eigenvalues $m_i$, we have that $\Tr(M^n)=\sum_{i=0}^{d-1}m_i^n$ for any $n$. Thus from \eqref{eq17}, we can conclude that
\begin{eqnarray}
\Tr(A_{1,2}^n)=\sum_i\lambda_i\omega^{in}=0
\end{eqnarray}
for any $n$ that is a divisor of $d$, where $\lambda_i$ is the multiplicity of the eigenvalue $\omega^i$. Now using Fact \ref{le:pol} stated below, whose proof can be found in \cite{sarkar}, we can conclude that the multiplicities of these eigenvalues are equal, that is, $\lambda_0=\lambda_1=\ldots=\lambda_{d-1}$. 

\begin{fakt}\label{le:pol}
Consider a real polynomial 
\begin{equation}\label{WX}
    W(x)=\sum_{i=0}^{d-1}\lambda_ix^i
\end{equation}
with rational coefficients $\lambda_i \in \Q$. Assume that $\omega^n$ with $\w=\mathrm{e}^{2\pi \mathbbm{i}/d}$ is a root of $W(x)$ for any $n$ being a proper divisor of $d$, i.e., $n\neq d$ such that $d/n\in \N$. Then, $\lambda_0=\lambda_1=\ldots=\lambda_{d-1}$.
\end{fakt}

This allows us to conclude that the observables of $A_{1}$ act on a Hilbert space $\mathcal{H}_{1}$ of dimension $d\times D$ where $D$ is positive integer.
Moreover, one can always rotate one of $A_1's$ observables to some observable that acts on $\mathbbm{C}^d$ tensored with identity acting on $\mathcal{H}_1''$ \eqref{hstruc}. Precisely one can find a unitary transformation $V_{1}:\mathcal{H}_{1}\rightarrow \mathcal{H}_{1}$ such that 
\begin{equation}
    V_{1}\,A_{1,2}\,V_{1}^{\dagger}=Z_d\otimes\Id_{1}'',
\end{equation}
where $Z_d$ is defined in Eq. (\ref{ZdTd}). Then in Lemma \ref{le:Fij}, using the above form of $A_{1,2}$, we show that $A_{1,3}$ is also unitarily equivalent to an observable acting on a $\mathbbm{C}^d$ tensored with identity acting on $\mathcal{H}_1''$, that is,  
\begin{equation}
   V_{1} A_{1,3} V_{1}^{\dagger}=T_{d,m}\otimes \Id_{1}''
\end{equation}
with $T_{d,m}$ defined in Eq. (\ref{ZdTd}). Next, we find a unitary transformation $ U_{1}:\mathcal{H}_{1}\rightarrow \mathcal{H}_{1}$ stated in Fact \ref{fact:1} of Appendix \ref{chap3}, that rotate these observables to the ideal ones, that is,
\begin{eqnarray}\label{3.50}
 U_{1}\,A_{1,\alpha_1}\,U_{1}^{\dagger}=\mathcal{O}_{1,\alpha_1}\otimes\Id_{1}''\quad \text{for}\quad\alpha_1=2,3.
\end{eqnarray}
Finally, using the derived observables and the condition \eqref{SOSrelc} we find the rest of the observables $A_{1,\alpha_1}$. For this, we first consider the relation \eqref{SOSrelc} for $k=1$ and $\alpha=1$. After plugging the explicit form of $R_1^{(1)}$ from \eqref{3.21} we have that
\begin{eqnarray}\label{3.51}
\mu_{1,1}^*A_{1,2}+\nu_{1,1}^*A_{1,3}+\tau_{1,1}A_{1,4}=0.
\end{eqnarray}
Plugging the observables $A_{1,2}$ and $A_{1,3}$ from \eqref{3.50}, we find that 
\begin{eqnarray}
 U_{1}\,A_{1,4}\,U_{1}^{\dagger}=\mathcal{O}_{1,4}\otimes\Id_{1}''.
\end{eqnarray}
 The above statement is easy to conclude based on the fact that the ideal observables satisfy the condition \eqref{SOSrelc} derived from the maximal violation of the Bell inequalities. Continuing this procedure recursively for the remaining values of $\alpha_1$, we see that 
 \begin{eqnarray}\label{3.53}
 U_{1}\,A_{1,\alpha_1}\,U_{1}^{\dagger}=\mathcal{O}_{1,\alpha_1}\otimes\Id_{1}''\qquad\forall\alpha_1.
\end{eqnarray}

Now let us prove the two lemmas which were used to arrive at the desired form of the observables \eqref{3.53}. 
\begin{customlemma}{3.1}\label{le:traceless}
Consider two unitary observables $A_{1,2}$ and $A_{1,3}$ with eigenvalues $\{1,\omega,\ldots,\omega^{d-1}\}$ that act on a finite-dimensional Hilbert space and satisfy the conditions \eqref{SOSrela} and \eqref{SOSrelb}. Then for any $n$ that is a divisor of $d$ such that $n<d$ we have that,
\be \label{traceB}
\Tr(A_{1,2}^n)= 0,\quad\text{and}\qquad\Tr(A_{1,3}^n)= 0.
\ee 
\end{customlemma}
\begin{proof} To begin the proof, let us consider the relations \eqref{SOSrela} for $\alpha_1=2$, in which we substitute the explicit form of $\overline{A}_{1,2}$ \eqref{barA1},
\be 
\left(a_kA_{1,2}^k+a_k^*A_{1,3}^k\right)  \left(a_k^*A_{1,2}^{-k}+a_kA_{1,3}^{-k}\right) = \Id.
\ee
Simplifying the above expression and then plugging in the value of $a_k$ from \eqref{app:ak} leads us to the following condition,
\begin{equation}\label{Obs22}
\omega^{\frac{2k-d}{2m}}A_{1,2}^{k}A_{1,3}^{-k}+\omega^{-\frac{2k-d}{2m}}A_{1,3}^{k}A_{1,2}^{-k}=
2\cos\left(\frac{\pi}{m}\right)\Id.
\end{equation} 
We multiply the above equation \eqref{Obs22} by $A_{1,3}^{k}$ and then take the trace to obtain
\begin{eqnarray}\label{Obs25}
\omega^{\frac{2k-d}{2m}}\Tr(A_{1,2}^{k})+\omega^{-\frac{2k-d}{2m}}\Tr(A_{1,3}^{2k}A_{1,2}^{-k})=
2\cos\left(\frac{\pi}{m}\right)\Tr(A_{1,3}^{k}).
\end{eqnarray}
Let us again consider the second relation \eqref{SOSrelb} for $\alpha_1=2$ where we substitute $\overline{A}_{1,2}$ and $a_k$ from \eqref{barA1} and \eqref{app:ak} respectively, to obtain,
\begin{eqnarray}
a_{2k}A_{1,2}^{2k}+a_{2k}^*A_{1,3}^{2k} =\left(a_k A_{1,2}^k+a_k^*A_{1,3}^k\right)\left(a_k A_{1,2}^k+a_k^*A_{1,3}^k\right)
\end{eqnarray}
for $k = 1,\dots, \left\lfloor{\frac{d}{2}}\right\rfloor$ where $\left\lfloor x \right\rfloor$ is the largest integer smaller than $x$. A simple calculation using the explicit form of $a_k$ and some trigonometric identities leads us to
\begin{eqnarray}\label{Obs23}
\omega^{k/m}A_{1,2}^{2k}+\omega^{-k/m}A_{1,3}^{2k}=A_{1,2}^kA_{1,3}^k+A_{1,3}^kA_{1,2}^{k}.
\end{eqnarray}
We then multiply the above equation \eqref{Obs23} by $A_{1,2}^{-k}$ and take the trace to get
\begin{eqnarray}\label{Obs24}
\omega^{k/m}\Tr(A_{1,2}^{k})+\omega^{-k/m}\Tr(A_{1,3}^{2k}A_{1,2}^{-k})=2\Tr(A_{1,3}^k).
\end{eqnarray}
After substituting the term $\Tr(A_{1,3}^{2k}A_{1,2}^{-k})$ from \eqref{Obs24} to the above equation \eqref{Obs25}, we obtain that
\begin{eqnarray}\label{Obs261}
\Tr(A_{1,2}^{k})=2\omega^{-k/m}\frac{1-\cos(\pi/m)\omega^{-d/2m}}{1-\omega^{-d/m}}\Tr(A_{1,3}^{k}).
\end{eqnarray}
Using the fact that 
\begin{eqnarray}
\cos\left(\frac{\pi}{m}\right)=\frac{1}{2}\left(\omega^{-\frac{d}{2m}}+\omega^{\frac{d}{2m}}\right),
\end{eqnarray}
we can simplify the relation \eqref{Obs261} to the following form
\begin{eqnarray}\label{Obs26}
\Tr(A_{1,2}^{k})=\omega^{-k/m}\Tr(A_{1,3}^{k})\qquad k = 1,\dots, \left\lfloor{\frac{d}{2}}\right\rfloor.
\end{eqnarray}
%
To prove that the traces of these observables vanish, we use the following observation whose proof is deferred to Appendix \ref{chap3}. 
\begin{customobs}{3.1} \label{fact:4s} Consider two unitary observables $A_{1,2}$ and $A_{1,3}$ with eigenvalues $\{1,\omega,\ldots,$ $\omega^{d-1}\}$ that act on a finite-dimensional Hilbert space and satisfy the conditions \eqref{SOSrela} and \eqref{SOSrelb}. Then for any $n$ that is a divisor of $d$ such that $n<d$ we have that,
\begin{equation}\label{NewId1}
    \Tr(A_{1,2}^x)=\omega^{\frac{2tx}{m}}\,\Tr\left(A_{1,2}^{(2t+1)x}A_{1,3}^{-2tx}\right).
\end{equation}
for any non-negative integer $t\in \mathbb{N} \cup \{0\}$ and $x=1,\ldots,\lfloor d/2\rfloor$
\end{customobs}
%
Now, consider a positive integer $n$ that is a divisor of $d$, that is, $d/n \in \mathbb{N}$ where $\mathbb{N}$ denotes the set of natural numbers\footnote{As a matter of fact, any non-trivial divisor of $d$ is always less than or equal to $d/2$.}. Now, $d/n$ can be either even or odd. Let us first consider the case, when $d/n$ is even. This implies that there is an integer $t$ such that $2t = d/n$. Substituting $x= n = d/2t$ in the conditon \eqref{NewId1} we obtain that
\begin{eqnarray}\label{traces22}
\Tr(A_{1,2}^{n})=\omega^{d/m}\,\Tr(A_{1,2}^{d+n}A_{1,3}^{-d}).
\end{eqnarray}
The above relation can be simplified to
\be 
\Tr(A_{1,2}^{n})= \omega^{d/m}\Tr(A_{1,2}^{n}).
\ee 
where we used the fact that $A_{1,2}^d=A_{1,3}^{-d}=\Id$.
Thus, for any $m\geq2$, the only possible solution of the above condition is $\Tr(A_{1,2}^{n})=0$ for any $n$ such that $d/n$ is even. Using then Eq. \eqref{Obs26} one can similarly conclude that $\Tr(A_{1,3}^{n})=0$. 
Now, consider the second case, that is, $n$ is a divisor of $d$ such that $d/n$ is odd. This implies that there is an integer $t$ such that $2t+1 = d/n$. Substituting $x = n = d/(2t+1)$ again in condition \eqref{NewId1}, we obtain that
\be \label{even-1}
\Tr(A_{1,2}^{n})= \w^{d/m} \w^{-n/m}\, \Tr\left(A_{1,3}^{n}\right).
\ee
Comparing the above expression with Eq. \eqref{Obs26}, one directly concludes that $\Tr(A_{1,\alpha})=0$ for any $n$ such that $d/n$ is odd 
and $n\leq d/2$. Thus, we have shown that for any $n$ which is a divisor of $d$, $\Tr(A_{1,\alpha}^{n})=0$ for $\alpha=2,3$. This completes the proof of Lemma \ref{le:traceless}.
\end{proof}
%


%
%
%

Now, we move onto finding the explicit form of the measurements $A_{1,\alpha}$ for $\alpha=2,3$.

\begin{customlemma}{3.2}\label{le:Fij}
Consider two unitary observables $A_{1,2}$ and $A_{1,3}$ with eigenvalues $\{1,\omega,\ldots,\omega^{d-1}\}$ that act on a $\C^d\otimes \mathcal{H}_{1}''$ such that $\mathcal{H}_{1}''$ is a finite-dimensional Hilbert space and satisfy the conditions \eqref{SOSrela} and \eqref{SOSrelb}. 
%
Then, we can find a unitary $V_1:\mathcal{H}_{1}\rightarrow\mathcal{H}_{1}$ that transforms $A_{1,2}$ and $A_{1,3}$ as
\begin{eqnarray}
V_1\,A_{1,2}\, V_1^\dagger = Z_d\otimes\Id_{1}''
\end{eqnarray}
and 
\begin{eqnarray}
V_1\,A_{1,3}\,V_1^{\dagger}=T_{d,m}\otimes\Id_{{1}}''
\end{eqnarray}
where $Z_d, T_{d,m}$ are defined in \eqref{Zd} and \eqref{ZdTd}.
\end{customlemma}
\begin{proof}
Let us begin by proving a relation between $A_{1,2}$ and $A_{1,3}$ given by,
\begin{equation}\label{FijEq1}
A_{1,3}^k=-(k-1)\omega^{\frac{k}{m}}A_{1,2}^k+\omega^{\frac{k-1}{m}}\sum_{t=0}^{k-1}A_{1,2}^tA_{1,3}A_{1,2}^{k-1-t}.
\end{equation}
for any $k=1,\ldots,d$. To prove the above relation, we use the technique of mathematical induction. We can easily check that this relation \eqref{FijEq1} holds trivially for $k=1$. Now, let us assume that this relation (\ref{FijEq1}) holds for some $k=s$. To prove that it also holds for $k=s+1$, we need to examine \eqref{SOSrelb} for $\alpha_1=2$ and $k=s+1$
\begin{equation}
 \overline{A}_{1,2}^{(s+1)}=\left[\overline{A}_{1,2}^{(1)}\right]^{(s)}\overline{A}_{1,2}^{(1)} .
\end{equation}
for $s=1,\dots,d-1$. Again, using \eqref{SOSrelb} for $\alpha_1=2$ and $k=s$ on the right hand side of the above equation, we have that
\begin{eqnarray}
 \overline{A}_{1,2}^{(s+1)}=\overline{A}_{1,2}^{(s)}\overline{A}_{1,2}^{(1)}.
\end{eqnarray}
Expanding $\overline{A}_{1,2}^{(s)}$ using \eqref{barA1} and then simplifying the obtained expression with the aid of the formula $a_{s+1}-a_{s}a_{1}=\omega^{\frac{s+1}{2m}}/{2\cos^2(\pi/2m)}$, we obtain that
\begin{eqnarray}
   A_{1,3}^{s+1}=-\omega^{\frac{s+1}{m}}A_{1,2}^{s+1}+\omega^{\frac{s}{m}}A_{1,2}^sA_{1,3}+\omega^{\frac{1}{m}}A_{1,3}^sA_{1,2}.
\end{eqnarray}
Replacing $A_{1,3}^s$ using the relation \eqref{FijEq1} into the above equation, we arrive at
\begin{equation}
   A_{1,3}^{s+1} = -\omega^{\frac{s+1}{m}}A_{1,2}^{s+1}+\omega^{\frac{s}{m}}A_{1,2}^sA_{1,3}+\omega^{\frac{1}{m}}\left[-(s-1)\omega^{\frac{s}{m}}A_{1,2}^s+\omega^{\frac{s-1}{m}}\sum_{t=0}^{s-1}A_{1,2}^tA_{1,3}A_{1,2}^{s-1-t}\right]A_{1,2}
\end{equation} 
which on simplification gives us the above relation \eqref{FijEq1} for $k=s+1$,
\begin{eqnarray}
A_{1,3}^{s+1}
= -s\omega^{\frac{s+1}{m}}A_{1,2}^{s+1}+\omega^{\frac{s}{m}}\sum_{t=0}^{s}A_{1,2}^tA_{1,3}A_{1,2}^{s-t}.
\end{eqnarray}
As discussed before, $A_{1,2}$ and $A_{1,3}$ act on $\mathbbm{C}^d\otimes\mathcal{H}_{1}''$ and therefore we can always find a a unitary $\overline{V}_1:\mathcal{H}_{1}\rightarrow\mathcal{H}_{1}$ such that $\overline{V}_1\,A_{1,2}\,\overline{V}_1^{\dagger}=Z_d\otimes\Id_{1}''$. Let us then decompose $A_{1,3}$ under the action of  $\overline{V}_1$ as
\be  \label{B2form}
\overline{V}_1\,A_{1,3}\,\overline{V}_1^{\dagger}=\sum_{i,j=0}^{d-1}\ket{i}\!\bra{j}\otimes F_{ij},
\ee  
where $F_{ij}$ for $i,j=0,1,\ldots,d-1$ are matrices acting on $\mathcal{H}_{1}''$. Notice that any matrix acting on a  Hilbert space $\mathbbm{C}^d\otimes\mathcal{H}_{1}''$ can be decomposed in this way. From here on, for simplicity we drop the unitary $V_1$ and recall it back at the end of the proof of this lemma. Also, we simplify the notation by replacing $\Id_1''$ by $\Id$ and use the correct notation at the end of the proof. 

For characterising $A_{1,3}$, it is now enough to find the matrices $F_{ij}$. To do this we first determine the matrices $F_{ii}$ for all $i$ using the relations \eqref{FijEq1}. Using then the derived $F_{ii}$, we then proceed to determine $F_{ij}$ for $i\neq j$. Let us now consider the relation \eqref{FijEq1} for $k=d-1$,
\begin{eqnarray}\label{FijEq3}
    A_{1,3}^\dagger=-(d-2)\omega^{\frac{d-1}{m}}A_{1,2}^{\dagger}+\omega^{\frac{d-2}{m}}\sum_{t=0}^{d-2}A_{1,2}^t A_{1,3}A_{1,2}^{d-t-2}.
\end{eqnarray}
Substituting $A_{1,2}=Z_d\otimes \Id$ and $A_{1,3}$ from (\ref{B2form}) in the above expression \eqref{FijEq3} and then simplifying it, we arrive at
\begin{equation}\label{FijEq9}
 \sum\limits_{i,j=0}^{d-1}\ket{j}\!\bra{i}\otimes F_{ij}^\dagger =
-(d-2)\omega^{\frac{d-1}{m}}\sum_{i=0}^{d-1}\omega^{-i}\ket{i}\!\bra{i}\otimes \Id + \omega^{\frac{d-2}{m}} \sum_{i,j=0}^{d-1}\sum_{t=0}^{d-2}\omega^{-2j+t(i-j)}\ket{i}\!\bra{j}\otimes F_{ij}.
\end{equation}
Sandwiching the above equation with $\bra{i}.\ket{i}$, we get the following expression
\begin{eqnarray}\label{FijEq8}
  F_{ii}^\dagger =-(d-2)\omega^{\frac{d-1}{m}} \omega^{-i}\Id +(d-1)\omega^{\frac{d-2}{m}} \omega^{-2i}F_{ii}.
\end{eqnarray}
We can get another relation by its Hermitian conjugation,
\begin{eqnarray}\label{FijEq85}
     F_{ii}=-(d-2)\omega^{-\frac{d-1}{m}}\omega^{i}\ \Id +(d-1)\omega^{-\frac{d-2}{m}}\omega^{2i} F_{ii}^\dagger.
\end{eqnarray}
Substituting $F_{ii}^\dagger$ from the first relation \eqref{FijEq8} into the above formula \eqref{FijEq85} we arrive at
\begin{equation} \label{Fijfact1}
F_{ii}=-(d-2)\omega^{-\frac{d-1}{m}}\omega^{i} \ \Id -(d-2)(d-1)\omega^{\frac{1}{m}+i} \Id +(d-1)^2F_{ii}, 
\end{equation}
which after rearranging the terms simplifies to,
\begin{equation}\label{TintoDeVerano}
    F_{ii}=\omega^{i+\frac{1}{m}} \left(\frac{d-1+\w^{-\frac{d}{m}}}{d}\right)\Id=\omega^{i+\frac{1}{m}} \left(1-\frac{2\mathbbm{i}\sin(\pi/m)}{d}\w^{-\frac{d}{2m}}\right)\Id.
\end{equation}
where we used the trigonometric identity 
\begin{eqnarray}\label{sin}
\sin \left(\frac{\pi}{m}\right)=\frac{1}{2\mathbbm{i}}\left(\omega^{\frac{d}{2m}}-\omega^{\frac{d}{2m}}\right).
\end{eqnarray}

After determining $F_{ii}$ we proceed towards finding the matrices $F_{ij}$ for $i\neq j$. 
The first observation can be simply made by considering the relation \eqref{FijEq9} and sandwiching it with $\bra{i}.\ket{j}$ for $i\neq j$, to get
\begin{eqnarray}
F_{ji}^\dagger=\omega^{\frac{d-2}{m}}\omega^{-2j}\sum_{t=0}^{d-2}\omega^{t(i-j)} F_{ij}.
\end{eqnarray}
Using the fact that $\sum_{t=0}^{d-2}\omega^{t(i-j)}=-\omega^{-(i-j)}$ for $i\ne j$, the above equation reduces to 
\be \label{fijEq12}
F_{ij}=-\w^{-\frac{d-2}{m}}\omega^{i+j}F_{ji}^\dagger.
\ee 
Finding the exact form of $F_{ij}$ requires another observation that involves higher order terms in $F_{ij}$. Due to highly technical nature of the proof, we defer it to Appendix \ref{chap3}.

\begin{customobs}{3.2}
The following conditions hold true for any $k=1,\ldots,d-1$ and $m\geq 2$,
 \begin{eqnarray} \label{FijObs3}
  -(k-1)\sum_{i,j=0}^{d-1}\omega^{ki}\ket{i}\!\bra{j}\otimes F_{ij}+\omega^{-\frac{1}{m}}\sum_{i,j=0}^{d-1}\ket{i}\!\bra{j}\otimes\left[\sum_{\substack{l=0\\l\ne i}}^{d-1}\left(\frac{\omega^{ki}-\omega^{kl}}{\omega^{i}-\omega^{l}}\right) F_{il}F_{lj}+k\omega^{(k-1)i}F_{ii}F_{ij}\right] \nonumber\\
  = -k\omega^{\frac{1}{m}}\sum_{i=0}^{d-1}\omega^{(k+1)i}\ket{i}\!\bra{i}\otimes \Id+\sum_{i,j=0}^{d-1}\ket{i}\!\bra{j}\otimes\sum_{t=0}^{k}\omega^{k j+t(i-j)} F_{ij}.\nonumber\\
\end{eqnarray}
\end{customobs}
We sandwich the relation (\ref{FijObs3}) with $\bra{i}.\ket{i}$ to get
\begin{eqnarray}
-(k-1)\sum_{i=0}^{d-1}\omega^{ki} F_{ii}+\omega^{-\frac{1}{m}}\sum_{i=0}^{d-1}\left[\sum_{\substack{l=0\\l\ne i}}^{d-1}\left(\frac{\omega^{ki}-\omega^{kl}}{\omega^{i}-\omega^{l}}\right) F_{il}F_{li}+k\omega^{(k-1)i}F_{ii}^2\right] \nonumber\\
  = -k\omega^{\frac{1}{m}}\sum_{i=0}^{d-1}\omega^{(k+1)i} \Id+\sum_{i=0}^{d-1}\sum_{t=0}^{k}\omega^{k i} F_{ii}
\end{eqnarray}
which after some simple rearrangement of the terms gives us
\begin{eqnarray}
 \sum_{\substack{l=0\\l\ne i}}^{d-1}\left(\frac{\omega^{ki}-\omega^{kl}}{\omega^{i}-\omega^{l}}\right) F_{il}F_{li} =
 k\omega^{ki}\left[2\omega^{\frac{1}{m}}F_{ii}-\omega^{-i}F_{ii}^2-\omega^{i+\frac{2}{m}}\Id\right].
 %
\end{eqnarray}
Now replacing $F_{ii}$ from Eq. \eqref{TintoDeVerano} and evaluating the above relation, we arrive at
\begin{equation}
 \sum_{\substack{l=0\\l\ne i}}^{d-1}\left(\frac{\omega^{ki}-\omega^{kl}}{\omega^{i}-\omega^{l}}\right) F_{il}F_{li} =-\frac{k}{d^2}\omega^{i(k+1)+\frac{2}{m}}(1-\w^{-d/m})^2\Id,
 \end{equation}
which can also be rewritten in the following form
\begin{equation}
\sum_{\substack{l=0\\l\ne i}}^{d-1}\left(\frac{1-\omega^{k(l-i)}}{1-\omega^{i-l}}\right)F_{il}F_{li}\omega^{-\left(i+l+\frac{2}{m}\right)}\w^{\frac{d}{m}}=\frac{k}{d^2}\w^{\frac{d}{m}}(1-\w^{-d/m})^2 \Id.
\end{equation}
Without loss of generality, we can replace the index $l$ with $j$. Now, substituting $F_{ji}$ from \eqref{fijEq12} and also using the identity \eqref{sin} we obtain
\begin{eqnarray}\label{FijFact2}
\sum_{\substack{j=0\\j\ne i}}^{d-1}\left(\frac{1-\omega^{k(j-i)}}{1-\omega^{i-j}}\right)F_{ij}F_{ij}^{\dagger}=\frac{4k}{d^2}\sin^2\left(\frac{\pi}{m}\right) \Id,
\end{eqnarray}
for all $k=0,\ldots,d-1$ and $i=0,\ldots,d-1$. 
Multiplying the above equation (\ref{FijFact2}) with $\omega^{kn}$ such that
$n=1,\ldots,d-1,$ then summing over $k$ and using the identity $\sum_{k=0}^{d-1}\omega^{kn}=0$ that holds true for any $n=1,2,\ldots,d-1$, we arrive at 
\begin{equation}\label{formulaTwo}
-\sum_{\substack{j=0\\j\ne i}}^{d-1}\frac{1}{1-\omega^{i-j}}F_{ij}F_{ij}^\dagger\sum_{k=0}^{d-1}\omega^{k(j-i+n)}=\frac{4}{d^2}\sin^2\left(\frac{\pi}{m}\right)\sum_{k=0}^{d-1}k\omega^{kn}\Id .
\end{equation}
Let us now consider some simple identities
\begin{eqnarray}
 \sum_{k=0}^{d-1}\omega^{kn}=0,\quad \text{and}\quad \sum_{k=0}^{d-1}\omega^{k(j-i)}=d\delta_{j,i}
\end{eqnarray}
which can be simply computed using the formula for the sum of geometric sequence, and,
\begin{eqnarray} \label{FijIden3} 
  \sum_{k=0}^{d-1}k\omega^{kn}=\frac{d}{\omega^n-1}
\end{eqnarray}
which has been proven in Fact \ref{iden} in Appendix \ref{chapano}
for any $n=1,\ldots,d-1$. After applying these identities to equation \eqref{formulaTwo} we arrive at the following relation
\begin{eqnarray}
F_{i(i-n \mod d)}F_{i(i-n\mod d)}^\dagger=\frac{4}{d^2}\sin^2\left(\frac{\pi}{m}\right)\Id.
\end{eqnarray}
Let us notice that for a fixed
$i$, $i-n\mod d$ covers all numbers from $\{0,\ldots,d-1\}$ except $i$ by varying $n$ from $1$ to $d-1$. Thus, we can simply represent the above expression as

\begin{eqnarray}\label{FijEq25}
F_{ij}F_{ij}^\dagger=\frac{4}{d^2}\sin^2\left(\frac{\pi}{m}\right)\Id.
\end{eqnarray}
Let us now consider a unitary transformation $\widetilde{V}:\mathcal{H}_1\to\mathcal{H}_1$ 
of the following form 
\begin{eqnarray}
\widetilde{V} = \sum^{d-1}_{i=0} \ket{i}\!\bra{i} \otimes \widetilde{V}_i,
\end{eqnarray} 
where $\widetilde{V}_i$ are unitary matrices acting on $\mathcal{H}_1''$ defined as
%
\begin{equation}
    \widetilde{V}_0=\Id,\qquad \widetilde{V}_i=-\frac{d\mathbbm{i}}{2\sin(\pi/m)}\omega^{-\frac{i}{2}+\frac{d-2}{2m}}F_{0i}
\end{equation}
for $i=1,\ldots,d-1.$ Notice that $A_{1,2}$ remains invariant under application of $\widetilde{V}$,
\begin{eqnarray}
\widetilde{V}A_{1,2}\widetilde{V}^{\dagger}=\widetilde{V}\left[Z_{d}\otimes\Id\right]\widetilde{V}^{\dagger}=Z_{d}\otimes\Id,
\end{eqnarray}
which a consequence of the fact that $Z_d\otimes\Id$ commutes with $\widetilde{V}$. Applying $\widetilde{V}$ to $A_{1,3}$, we obtain
\begin{eqnarray}
 \widetilde{V}\,A_{1,3}\,\widetilde{V}^\dagger  = \sum^{d-1}_{i,j=0} \ket{i}\!\bra{j} \otimes \tilde{F}_{ij},
\end{eqnarray}
where we denoted $\tilde{F}_{ij}=\widetilde{V}_i\,F_{ij}\,\widetilde{V}_j^\dagger$.
Notice that all the algebraic relations obtained till now for $F_{ij}$ are equally valid for $\tilde{F}_{ij}$, and $\tilde{F}_{ii} = F_{ii}$. Employing the relation (\ref{FijEq25}) for $i=0$, we obtain that
\begin{eqnarray}\label{FijEq28}
  \tilde{F}_{0j}=\widetilde{V}_0\,F_{0j}\,\widetilde{V}_j^\dagger
  =\frac{d}{2}\omega^{\frac{2j+d}{4}+\frac{2-d}{2m}}F_{0j}F_{0j}^\dagger
  =\frac{2\mathbbm{i}}{d}\sin\left(\frac{\pi}{m}\right)\omega^{\frac{j}{2}+\frac{2-d}{2m}}\Id,
\end{eqnarray} 
Then employing the relation between $\tilde{F}_{ij}$ and $\tilde{F}_{ji}^{\dagger}$ from Eq. (\ref{fijEq12}) for $i=0$, we obtain that $\tilde{F}_{j0}=\tilde{F}_{0j}$.

To determine the remaining matrices $F_{ij}'s$, the above relations are not enough and we also need to look at the off-diagonal elements of \eqref{FijFact2}. For this, we again sandwich the relation \eqref{FijFact2} with $\bra{i}.\ket{j}$ such that $i\ne j$, which leads us to
\begin{equation}
 -(k-1)\omega^{ki}F_{ij}+\omega^{-\frac{1}{m}}\sum_{\substack{l=0\\l\ne i}}^{d-1}\left(\frac{\omega^{ki}-\omega^{kl}}{\omega^{i}-\omega^{l}}\right) F_{il}F_{lj}+k\omega^{(k-1)i}\omega^{-\frac{1}{m}}F_{ii}F_{ij}=\frac{\omega^{(k+1)i}-\omega^{(k+1)j}}{\omega^{i}-\omega^{j}}F_{ij}. 
\end{equation}
Plugging $F_{ii}$ from Eq. \eqref{TintoDeVerano} and then using some simple algebra, we arrive at
 \ben 
 \sum_{\substack{l=0\\l\ne i}}^{d-1}\frac{\omega^{ki}-\omega^{kl}}{\omega^{i}-\omega^{l}} F_{il}F_{lj}=\omega^{\frac{1}{m}}\left[\frac{\omega^{(k+1)i}-\omega^{(k+1)j}}{\omega^{i}-\omega^{j}}+\left(\frac{(1-\w^{-d/m})k}{d}-1\right)\omega^{ki}\right]F_{ij}.
\end{eqnarray}
Now, let us consider the above relation for $i=0$, 
\begin{eqnarray}
\sum_{l=1}^{d-1}\frac{1-\omega^{kl}}{1-\omega^{l}} F_{0l}F_{lj}=\omega^{\frac{1}{m}}\left(\frac{1-\omega^{(k+1)j}}{1-\omega^{j}}+\frac{(1-\w^{-d/m})k}{d}-1\right)F_{0j}.
\end{eqnarray}
Replacing ${F}_{0j}$ as derived in \eqref{FijEq28},
\begin{equation}\label{VihnoVerde}
    \sum_{l=1}^{d-1}\frac{1-\omega^{kl}}{1-\omega^{l}}\omega^{\frac{l}{2}} F_{lj}=\omega^{\frac{j}{2}+\frac{1}{m}}\left(\frac{1-\omega^{(k+1)j}}{1-\omega^{j}}+\frac{(1-\w^{-d/m})k}{d}-1\right)\Id.
\end{equation}
As the matrix $F_{jj}$ was derived in \eqref{TintoDeVerano}, we separate it out from the sum on the left hand side of the above equation and replace the index $l$ with $i$ to finally obtain
\begin{equation}\label{FijFact3}
\sum_{\substack{l=1\\l\ne j}}^{d-1}\left(\frac{1-\omega^{kl}}{1-\omega^{l}}\right) \omega^{\frac{l}{2}}F_{lj}=
\frac{1-\w^{-d/m}}{d}\omega^{\frac{j}{2}+\frac{1}{m}}\left(k+\frac{1-\omega^{kj}}{1-\omega^{j}}\omega^{j}\right) \Id.
\end{equation}
We now multiply the above relation \eqref{FijFact3} by $\omega^{-kn}$ with $n=1,\ldots,d-1$ and $n\neq j$. Next, we sum the resulting relation over all $k$, which yields
\begin{eqnarray}
\sum_{\substack{i=1\\i\ne j}}^{d-1}\frac{\omega^{i/2}}{1-\omega^{i}}F_{ij}\sum_{k=0}^{d-1}\left(\omega^{-kn}-\omega^{k(i-n)}\right)&&\nonumber\\=
\frac{1-\w^{-d/m}}{d}&\omega^{\frac{j}{2}+\frac{1}{m}}&\left[\sum_{k=0}^{d-1}k\omega^{-kn}+\frac{\omega^{j}}{1-\omega^{j}}
\sum_{k=0}^{d-1}\left(\omega^{-kn}-\omega^{k(j-n)}\right)\right]\Id.\nonumber\\
\end{eqnarray}
%
Now, exploiting the fact that $\sum_{k=0}^{d-1}\omega^{k(n-i)}=d\delta_{n,i}$ and the identity \eqref{FijIden3}, we arrive at
%
%
\begin{equation}
-d\frac{\omega^{n/2}}{1-\omega^{n}}F_{nj}=\frac{1-\w^{-d/m}}{d}\omega^{\frac{j}{2}+\frac{1}{m}}\left(\frac{d}{\omega^{-n}-1}\right)\I,
\end{equation}
which after rearranging some terms and replacing the index $n$ with $i$, gives the matrices $F_{ij}$ for $i\ne j$,
\begin{eqnarray}\label{FijEq34}
F_{ij}&=&-\frac{1-\w^{-d/m}}{d}\omega^{\frac{i+j}{2}+\frac{1}{m}}\Id\nonumber\\&=&
-\frac{2\mathbbm{i}\sin(\pi/m)}{d}\omega^{\frac{i+j}{2}+\frac{2-d}{2m}}\Id, \qquad i, j=1,\ldots,d-1,\ \ \  i\ne j
\end{eqnarray}
where to obtain the second equality, we used the identity \eqref{sin}.
Combining all the derived identities, (first, $F_{ii}$ from \eqref{TintoDeVerano} for all $i$, second, $F_{0j}$ and $F_{j0}$ from \eqref{FijEq28} for all $j$ and finally, the rest of the matrices $F_{ij}$ from \eqref{FijEq34}) into the form of $A_{1,3}$ (\ref{B2form}), we conclude that the unitary $V_1=\widetilde{V}\overline{V}_1$,
transforms $A_{1,2}$ and $A_{1,3}$ as:
\begin{eqnarray}
V_1\,A_{1,2}\,V_1^{\dagger}=Z_d\otimes\Id,
\end{eqnarray}
and
\be 
V_1\, A_{1,3}\, V_1^\dagger = T_{d,m} \otimes \Id
\ee 
with $T_{d,m}$ given by
\begin{eqnarray}\label{Td}
T_{d,m}=\sum_{i=0}^{d-1}\omega^{i+\frac{1}{m}}\proj{i}-\frac{2\mathbbm{i}}{d}\sin\left(\frac{\pi}{m}\right)\sum_{i,j=0}^{d-1}(-1)^{\delta_{i,0}+\delta_{j,0}}\omega^{\frac{i+j}{2}-\frac{d-2}{2m}}|i\rangle\!\langle j|
\end{eqnarray}
This completes the characterisation of $A_{1,2}$ and $A_{1,3}$.
\end{proof}

%
\subsubsection{Rest of the parties}

To find the observables for rest of the parties, we follow the exact same lines as were used for finding the observables of the first party $A_{1,\alpha_1}$. Let us now focus on the SOS decomposition of the Bell operator $\mathcal{\hat{B}}_{m,d,N}$ given in Eq. \eqref{extraSOS}. As discussed before in Chapter \ref{chapter_2} any state $\ket{\psi_N}$ belonging to the Hilbert space $\mathcal{H}_{A_1''}\otimes\ldots\otimes\mathcal{H}_{A_N''}\otimes\mathcal{H}_E$ and observables acting on it that maximally violate the Bell inequalities \eqref{ASTA} must satisfy the following relations due to the SOS decomposition \eqref{extraSOS},
%
\begin{equation}\label{3113}
 P_{n,\alpha_1,\ldots,\alpha_N}^{(k)}\ket{\psi_{N}}=0.
\end{equation}
After expanding the above expession \eqref{3113} with the aid of Eq. \eqref{SOS27}, we obtain the following relation
\begin{eqnarray}\label{3114}
A_{1,\alpha_1}^{(k)}\otimes\overline{A}_{n,\alpha_{n-1}+\alpha_n-1}^{(k)}\otimes\bigotimes_{\substack{i=2\\i\ne n}}^N A^{(-1)^{i-1}k}_{i,\alpha_{i-1}+\alpha_i-1}\ket{\psi_N}=\ket{\psi_N},
\end{eqnarray}
and
\begin{eqnarray}\label{3115}
R_{n,\alpha}^{(k)}\ket{\psi_{N}}=0
\end{eqnarray}
for $k=1,2,\ldots,d-1$, $\alpha_i=1,2,\ldots,m$ for $i=1,2,\ldots,N-1$, $\alpha_N=1$ and $n=2,3,\ldots,N$. The expressions $\overline{A}_{n,\alpha_{n-1}+\alpha_n-1}^{(k)}$ and $R_{n,\alpha}^{(k)}$ are given in Eqs. \eqref{SOS23} and \eqref{SOS25} when $n$ is odd and in Eqs. \eqref{SOS24} and \eqref{SOS26} when $n$ is even.
Taking into account that the local states of all the parties are full-rank, we get similar relations among the observables for every party as those in Eqs. \eqref{SOSrela} and \eqref{SOSrelb} from the above expressions \eqref{3114} and \eqref{3115},
\begin{eqnarray}\label{SOSrela1}
\overline{A}_{n,\alpha_{n-1}+\alpha_n-1}\ \overline{A}_{n,\alpha_{n-1}+\alpha_n-1}^{\dagger}=\Id,
\end{eqnarray}
and 
\begin{eqnarray}\label{SOSrela2}
\overline{A}_{n,\alpha_{n-1}+\alpha_n-1}^{(k)}=\left[\overline{A}_{n,\alpha_{n-1}+\alpha_n-1}^{(1)}\right]^k,
\end{eqnarray}
and
\begin{eqnarray}\label{SOSrelb1}
R_{n,\alpha}^{(k)}=0,
\end{eqnarray}
for $k=1,2,\ldots,d-1$, $\alpha_i=1,2,\ldots,m$ for $i=1,2,\ldots,N-1$, $\alpha_N=1$ and $n=2,3,\ldots,N$.

Let us notice that the forms of $ \overline{A}_{n,\alpha_{n-1}+\alpha_n-1}^{(k)}$ in \eqref{SOS23} and \eqref{SOS24} are identical to $\overline{A}_{1,\alpha_1}$. Further, $R_{n,\alpha}^{(k)}$ in \eqref{SOS25} and \eqref{SOS26} are identical to $R_{1,\alpha}^{(k)}$. Thus, we employ the same technique to find observables $A_{n,\alpha_n}$ for all $n$ and $\alpha_n$. We first use the relations \eqref{SOSrela1} and \eqref{SOSrela2} for $\alpha_{n-1}=2, \alpha_{n}=1$ when $n$ is odd and   $\alpha_{n-1}=2, \alpha_{n}=2$ when $n$ is even, yielding the exactly same equations as \eqref{SOSrela} and \eqref{SOSrelb} for $\alpha_1=2$. Thus, from Lemma \ref{le:traceless} we can conclude that $\Tr(A_{n,2}^s)=\Tr(A_{n,3}^s)=0$ for any $s$ that is a divisor of $d$. Now using Fact \ref{le:pol} stated in Appendix \ref{chap3} we can conclude that the multiplicity of the eigenvalues are equal. 
This allows us to deduce that the observables of $A_{n}$ act on a Hilbert space $\mathcal{H}_{n}$ of dimension $d\times D_n$ where $D_n$ is a positive integer, that is,
\begin{eqnarray}
\mathcal{H}_{n}=\mathbbm{C}^d\otimes\mathcal{H}_{n}''.
\end{eqnarray}
As a consequence, one can always rotate one of $A_n's$ observables to some observable that acts on a $\mathbbm{C}^d$ tensored with identity acting on $\mathcal{H}_n''$.
Moreover, we can find a unitary transformation $V_{n}:\mathcal{H}_{n}\rightarrow \mathcal{H}_{n}$ such that 
\begin{equation}
    V_{n}\,A_{n,2}\,V_{n}^{\dagger}=Z_d\otimes\Id_{n}''
    \end{equation}
where $Z_d$ is defined in Eq. (\ref{ZdTd}). Then, using Lemma \ref{le:Fij}, using the above form of $A_{n,2}$ we show that $A_{n,3}$ is also unitarily equivalent to a an observable acting on a $\mathbbm{C}^d$ tensored with identity acting on $\mathcal{H}_1''$, that is,  
\begin{equation}
   V_{n}\ A_{n,3} V_{n}^{\dagger}=T_{d,m}\otimes \Id_{n}''
\end{equation}
with $T_{d,m}$ defined in Eq. (\ref{ZdTd}). Next, we show that there exist unitary transformations $ U_{n}:\mathcal{H}_{n}\rightarrow \mathcal{H}_{n}$ such that
\begin{eqnarray}\label{3.50r}
 U_{n}\,A_{n,\alpha_n}\,U_{n}^{\dagger}=\mathcal{O}_{n,\alpha_n}\otimes\Id_{n}''\quad \text{for}\quad\alpha_1=2,3.
\end{eqnarray}
These unitaries $U_n$ are explicitly calculated in Observation \ref{fact:1} stated in Appendix \ref{chap3}.
Finally, using the derived observables and the condition \eqref{3115} we find the rest of the observables of $A_{n,\alpha_n}$. For this, we first consider the relation \eqref{3115} for $k=1$ and $\alpha=1$. After plugging the explicit form of $R_n^{(1)}$ from \eqref{SOS26} and \eqref{SOS27} and then plugging the observables $A_{n,2}$ and $A_{n,3}$ from \eqref{3.50r}, we can easily compute that 
\begin{eqnarray}
 U_{n}\,A_{n,4}\,U_{n}^{\dagger}=\mathcal{O}_{n,4}\otimes\Id_{n }''.
\end{eqnarray}
 The above statement can also be checked based on the fact that the ideal observables satisfy all the above conditions derived from the maximal violation of the Bell inequalities. Continuing this procedure recursively for the remaining values of $\alpha$, we see that 
 \begin{eqnarray}
 U_{n}\,A_{n,\alpha_n}\,U_{n}^{\dagger}=\mathcal{O}_{n,\alpha_1}\otimes\Id_{n}''\qquad\forall\alpha_n.
\end{eqnarray}
This completes the characterisation of all the observables of every party. We have shown that the maximal violation of ASTA Bell inequalities \eqref{ASTA} is attained only by observables that up to local unitary transformations and addditional degrees of freedom are the ideal observables \eqref{measurements}, \eqref{measurements2} and \eqref{measurements3}.

\subsubsection{The state} 

We finally have all the tools required to determine the state that maximally violates the Bell inequalities \eqref{ASTA}. For this, we only need to consider the relation \eqref{SOSrela}. As was derived in the previous subsection that up to a local unitary all the observables are the ideal ones. Thus, we can rewrite the relation \eqref{SOSrela} for $k=1$ by expanding $\overline{A}_{1,\alpha_1}^{(1)}$ as
\begin{eqnarray}\label{3125}
\left[\left(a_1\mathcal{O}_{1,\alpha_1}+a_1^*\mathcal{O}_{1,\alpha_1+1}\right)\otimes\bigotimes_{i=2}^N\mathcal{O}^{(-1)^{i-1}}_{i,\alpha_{i-1}+\alpha_i-1}\otimes\Id''\right]\ket{\tilde{\psi}_{N}}=\ket{\tilde{\psi}_{N}}
\end{eqnarray}
for all $\alpha_i=1,2,\ldots,m$ such that $i=1,2,\ldots,N-1$ and $\alpha_N=1$. Also, in the above condition $\Id''$ acts on the Hilbert space $\mathcal{H}_1''\otimes\mathcal{H}_2''\otimes\ldots\otimes\mathcal{H}_N''\otimes\mathcal{H}_E$ and
\begin{eqnarray}
\ket{\tilde{\psi}_{N}}=U_1\otimes U_2\otimes\ldots\otimes U_N\otimes\Id_E\ket{\psi_N}.
\end{eqnarray}
Let us simplify the term $a_1\mathcal{O}_{1,\alpha_1}+a_1^*\mathcal{O}_{1,\alpha_1+1}$ by expanding it using \eqref{Obs12} and \eqref{Obsoddeven},
\begin{eqnarray}\label{3127}
a_1\mathcal{O}_{1,\alpha}&+&a_1^*\mathcal{O}_{1,\alpha+1}\nonumber\\
&=&\left[\sum_{i=0}^{d-2}\omega^{\gamma_m(\alpha)}\left(a_1+a_1^*\omega^{\frac{1}{m}}\right)\ket{i}\!\bra{i+1}+\omega^{(1-d)\gamma_2(\alpha)}\left(a_1+a_1^*\omega^{-\frac{d-1}{m}}\right)\ket{d-1}\!\bra{0}\right]\otimes\Id_{n}''\nonumber\\
&=&\left[\sum_{i=0}^{d-2}\omega^{\zeta_m(\alpha)}\ket{i}\!\bra{i+1}+\omega^{-(d-1)\zeta_m(\alpha)}\ket{d-1}\!\bra{0}\right]\otimes\Id_{n}'',
\end{eqnarray}
where to arrive at the third line of the above equation, we  use the fact that $a_1+a_1^*\omega^{1/m}=\w^{1/2m}$ and $a_1+a_1^*\omega^{-(d-1)/m}=\w^{-(d-1)/2m}$ and also that $\gamma_m(x)+1/2m=\zeta_m(x)$ [cf. \eqref{measupara}]. 
Let us now notice the action of the ideal observables on vectors belonging to the computational basis $\ket{j}$ of Hilbert space $\mathbbm{C}^d$, 
\begin{eqnarray}\label{matrel1}
(a_1O_{1,x}+a_1^*O_{1,x+1})\ket{j}&=&\omega^{(1-d\delta_{j,0})(x/m)}\ket{j-1},
\nonumber\\
\mathcal{O}_{2,x}^{-1}\ket{j}&=&\omega^{-(1-d\delta_{j,0})(x/m)}\ket{j-1},
\end{eqnarray}
where the first equation follows from Eq. \eqref{3127} and the second equation follows from \eqref{Obs12}. Now, when $n$ is odd or even
\begin{eqnarray}\label{matrel2}
\mathcal{O}_{n_{\mathrm{odd}},x}\ket{j}&=&\omega^{(1-d\delta_{j,0})\theta_m(x)}\ket{j-1},\nonumber\\
\mathcal{O}_{n_{\mathrm{ev}},x}^{-1}\ket{j}&=&\omega^{-(1-d\delta_{j,0})\theta_m(x)}\ket{j-1},
\end{eqnarray}
where we employed \eqref{Obsoddeven}. Here, the natural convention is $\ket{-1}\equiv \ket{d-1}$.

As the local Hilbert spaces of all the parties are of the form $\mathcal{H}_i=\mathbbm{C}^d\otimes\mathcal{H}_{i}''$, we can express the state as
\begin{eqnarray}\label{genstate}
\ket{\tilde{\psi}_{N}}=\sum_{i_1,\ldots, i_N=0}^{d-1}\ket{i_1,\ldots, i_N}\ket{\psi_{i_1,\ldots, i_N,E}}
\end{eqnarray}
where the vectors $\ket{\psi_{i_1,\ldots ,i_N,E}}\in\mathcal{H}_{1}''\otimes\ldots\otimes\mathcal{H}_N''\otimes \mathcal{H}_E$ are in general unnormalised.
Let us plug this state into the relation \eqref{3125} for $\alpha_1=\alpha_2=\ldots=\alpha_N=1$ and $k=1$.
Noting moreover that $\theta_m(1)=0$, gives us
\begin{equation}\label{stateeq1.0}
\sum_{i_1,\ldots, i_N=0}^{d-1}\w^{\frac{d}{m}\left(\delta_{i_2,0}-\delta_{i_1,0}\right)}\ket{i_1-1}\ldots \ket{i_N-1}\ket{\psi_{i_1,\ldots, i_N,E}}=\sum_{i_1,\ldots, i_N=0}^{d-1}\ket{i_1,\ldots, i_N}\ket{\psi_{i_1,\ldots, i_N,E}},
\end{equation}
where to arrive at the above expression we have also used the relations \eqref{matrel1} and \eqref{matrel2}. Multiplying the above expression with $\bra{i_1-1}\ldots \bra{i_N-1}$, we obtain
\begin{eqnarray}\label{state1.1}
\w^{\frac{d}{m}\left(\delta_{i_2,0}-\delta_{i_1,0}\right)}\ket{\psi_{i_1,\ldots, i_N,E}}=\ket{\psi_{i_1-1,\ldots, i_N-1,E}}
\end{eqnarray}
for all $i_1,\ldots,i_N$. Again, in the relation \eqref{3125}, we set $\alpha_1=2$ and $\alpha_2=\ldots=\alpha_N=1$ to obtain  for all $i_1,\ldots,i_N$ that
\begin{eqnarray}\label{state1.2}
 \w^{\frac{2d}{m}\left(\delta_{i_2,0}-\delta_{i_1,0}\right)}\ket{\psi_{i_1,\ldots, i_N,E}}=\ket{\psi_{i_1-1,\ldots, i_N-1,E}}.
\end{eqnarray}
The relations \eqref{state1.1} and \eqref{state1.2} can be divided into two possible cases. The first case is $\delta_{i_2,0}=\delta_{i_1,0}$ which holds true when and $i_1,i_2=1,2,\ldots,d-1\ \ \text{or}\ \  i_1=i_2=0$ for which
\begin{eqnarray}\label{state1.3}
\ket{\psi_{i_1,i_2\ldots, i_N,E}}=\ket{\psi_{i_1-1,i_2-1,\ldots, i_N-1,E}}
\end{eqnarray}
and for all $i_3,i_4,\ldots,i_N$. The second case is $\delta_{i_1,0}\ne\delta_{i_2,0}$ which holds true if $i_1=0$ and $i_2=1,\ldots,d-1$ or $i_2=0$ and $i_1=1,\ldots,d-1$, for which Eq. \eqref{state1.1} and Eq. \eqref{state1.2} can be simultaneously satisfied if and only if
\begin{eqnarray}\label{state1.4}
\ket{\psi_{i_1,0,i_3,\ldots, i_N,E}}=0, \quad \ket{\psi_{0,i_2,\ldots, i_N,E}}=0 
\end{eqnarray}
and for all $i_3,i_4,\ldots,i_N$. Considering \eqref{state1.3} for $i_2=1$ and $i_1\ne 1$, we get $\ket{\psi_{i_1,1\ldots, i_N,E}}=\ket{\psi_{i_1-1,0,\ldots, i_N-1,E}}=0$. Again, considering \eqref{state1.3} for $i_2=2$ and $i_1\ne 2$, we get $\ket{\psi_{i_2,2,\ldots, i_N,E}}=\ket{\psi_{i_1-1,1,\ldots, i_N-1,E}}=0$. Considering all the assignments of $i_2$ from $3$ to $d-1$ in \eqref{state1.3} and $i_1\ne i_2$, we can similarly obtain that
\begin{eqnarray}\label{state1.51}
\ket{\psi_{i_1,i_2,\ldots, i_N,E}}=0\quad \quad \forall i_1,i_2,\ldots,i_N\ \ \text{s.t.}\ \ i_1\ne i_2
\end{eqnarray}
and,
\begin{eqnarray}\label{state1.5}
\ket{\psi_{i_2-1,i_2-1,i_3-1,\ldots, i_N-1,E}}=\ket{\psi_{i_2,i_2,i_3,\ldots, i_N,E}}\qquad \forall i_2,i_3,\ldots,i_N.
\end{eqnarray}
We again consider the relations \eqref{3125} for $\alpha_1=\alpha_3=\ldots=\alpha_N=1$ and $\alpha_2=2$. Using the above derived condition \eqref{state1.51}, we can focus only on the cases when $i_1=i_2$ as rest of the terms are $0$. Due to this we arrive at the following condition for all $i_2,\ldots,i_N$,
\begin{eqnarray}\label{eq132}
 \w^{\frac{d}{m}\left(\delta_{i_2,0}-\delta_{i_3,0}\right)}\ket{\psi_{i_2,i_2,i_3\ldots, i_N,E}}=\ket{\psi_{i_2-1,i_2-1,i_3-1,\ldots, i_N-1,E}}.
\end{eqnarray}
Again, there are two posible solutions when simultaneously solving Eq. \eqref{state1.5} and the above Eq. \eqref{eq132}. The first solution is that $\delta_{i_2,0}=\delta_{i_3,0}$ which holds true for $i_2,i_3=1,2,\ldots,d-1\ \ \text{or}\ \  i_2=i_3=0$ for which 
\begin{eqnarray}\label{state1.6}
\ket{\psi_{i_2,i_2,i_3,\ldots, i_N,E}}=\ket{\psi_{i_2-1,i_2-1,i_3-1,\ldots, i_N-1,E}} 
\end{eqnarray}
and for all $i_4,i_5\ldots,i_N$. The second solution is that $\delta_{i_3,0}\ne\delta_{i_2,0}$ which holds true if $i_3=0$ and $i_2=1,\ldots,d-1$ or $i_2=0$ and $i_3=1,\ldots,d-1$, for which 
\begin{eqnarray}\label{state133}
\ket{\psi_{0,0,i_3,\ldots, i_N,E}}=0, \quad \ket{\psi_{i_2,i_2,0,\ldots, i_N,E}}=0 
\end{eqnarray}
and for all $i_4,i_5\ldots,i_N$. Considering the above equation \eqref{state1.6} for $i_2=1$ and $i_3\ne 1$, we obtain that $\ket{\psi_{1,1,i_3,\ldots, i_N,E}}=\ket{\psi_{0,0,i_3,\ldots, i_N-1,E}}=0$, Again, considering the above equation \eqref{state1.6} for $i_2=2$ and $i_3\ne 2$, we obtain that $\ket{\psi_{2,2,i_3,\ldots, i_N,E}}=\ket{\psi_{1,1,i_3,\ldots, i_N-1,E}}=0$. Considering all the assignments of $i_2$ from $3$ to $d-1$ in \eqref{state1.6} and $i_3\ne i_2$, we can similarly obtain that
\begin{eqnarray}\label{state1.7}
\ket{\psi_{i_2,i_2,i_3,\ldots, i_N,E}}=0\qquad \forall i_2,i_3,\ldots,i_N\ \ \text{s.t.}\ \ i_2\ne i_3
\end{eqnarray}
and,
\begin{eqnarray}\label{state1.71}
\ket{\psi_{i_2-1,i_2-1,i_2-1,\ldots, i_N-1,E}}=\ket{\psi_{i_2,i_2,i_2,\ldots ,i_N,E}}\qquad \forall i_2,i_4,\ldots,i_N.
\end{eqnarray}
Proceeding in a similar manner, we assign $\alpha_n=2$ for any $n=3,4,\ldots,N-1$ with the rest of coefficients as $\alpha_1=\alpha_2=\alpha_3=\ldots=\alpha_N=1$ in Eq. \eqref{3125}, we obtain $N-3$  different conditions. We solve them exactly the same way as was done for $n=2$, and  can finally conclude that the only non-zero terms among $\ket{\psi_{i_1,i_2,i_3,\ldots, i_N,E}}$ are related as,
\begin{eqnarray}
\ket{\psi_{i-1,i-1,i-1,\ldots, i-1,E}}=\ket{\psi_{i,i,i,\ldots, i,E}}\quad \forall i.
\end{eqnarray}
Thus, we can conclude from \eqref{genstate} that
\begin{eqnarray}
\ket{\tilde{\psi}_N}=\sum_{i=0}^{d-1}\ket{ii\ldots i}\otimes\ket{\psi_{0,0,\ldots,0,E}}.
\end{eqnarray}
Normalising the state, we can rewrite it as
\begin{eqnarray}\label{genstate3}
U_1\otimes\ldots \otimes U_N\ket{\psi_{N}}=\left(\frac{1}{\sqrt{d}}\sum_{i=0}^{d-1}\ket{i}^{\otimes N}\right)\otimes\ket{\tilde{\psi}_{0,0,\ldots,0,E}}
\end{eqnarray}
where $\ket{\tilde{\psi}_{0,0,\ldots,0}}=1/\sqrt{d}\ket{\psi_{0,0,\ldots,0,E}}$. Thus, the state that maximally violates the Bell inequalities \eqref{ASTA} up to some local unitaries is infact the generalised GHZ state \eqref{GHZ} along with some uncorrelated auxiliary state on which the measurements act trivially. This finally completes the proof of our self-testing scheme.
\end{proof}

\section{Randomness certification}
As was discussed before in introduction Chapter \ref{chapter_2}, self-testing of quantum states and measurements can be used to design methods of certification of genuine randomness that can be generated using the outcomes of the measurement device. Even if some external attacker Eve has access to the measurement devices and the state, self-testing restricts the maximum probability by which Eve can guess the generated outputs. 

For this, let us consider the scenario in which one of the observers, say the first party $A_1$, wishes to generate randomness using the outcomes of their measurements. As discussed before, Eve might supply the measurement devices which might give some pre-determined outputs that are known to her. 
Interestingly, the self-testing scheme presented in this Chapter can be used to 
certify $\log_2 d$ bits of perfect 
randomness in the outcomes of any measurement of any party. Let us now focus on $A_1$'s measurements, keeping in mind that the results apply to any party. 
We compute the probability of Eve to guess the measurement outcomes of $A_1$'s. For this, we refer to the local guessing probability \eqref{guessEve2} introduced in Chapter \ref{chapter_2} which can be straightforwardly extended to the multipartite scenario,  
\begin{eqnarray}\label{lgp}
G(\alpha_1,\p) = \sup_{S_{p}} \sum_b\bra{\psi_N}Q_
{\alpha_1}^{(b)}\otimes\Id_{A_2}\otimes\ldots\otimes\Id_N\otimes
E^{(b)}\ket{\psi_N},
\end{eqnarray}
where $\ket{\psi_N}\in\mathcal{H}_1\otimes\mathcal{H}_2\otimes\ldots\otimes\mathcal{H}_N\otimes\mathcal{H}_E$ is the $N$-partite state shared by all the parties as well as Eve. Here
$Q_{\alpha_1}^{(b)}$ is the $b-$th projector corresponding to the $\alpha_1-$th measurement performed by $A_1$, and $E^{(b)}$ corresponds to the $b-$th outcome of a $d$-outcome measurement performed by Eve on her share of the state. This is Eve's best guess of $A_1$'s outcome. Finally, $S_{p}$ is the set of all possible strategies that Eve can use to guess of $A_1$'s measurement outputs. 

Let us consider that all the parties perform the Bell test on a state $\ket{\psi}$ and observe the maximal violation of the Bell inequalities (\ref{ASTA}). As concluded in the previous section, up to local unitary operations, the quantum state is given by \eqref{genstate3} and the measurement operators of $A_1$'s can be expressed as $Q_{\alpha_1}^{(b)}=\bar{Q}_{\alpha_1}^{(b)}\otimes\Id_{A''}$, where $\bar{Q}_{\alpha_1}^{(b)}$ are eigenprojectors of the ideal observables \eqref{measurements}. Going back to the local guessing probability \eqref{lgp}, which can be rewritten as
\begin{eqnarray}\label{lgp1}
G(\alpha_1,\p) = \sup_{S_{p}}\sum_b\Tr\left(Q_{\alpha_1}^{(b)}\otimes E^{(b)}\rho_{A_1E}\right)
\end{eqnarray}
where $\rho_{A_1E}=\Tr_{A_2A_3\ldots A_N}(\ket{\psi_N}\!\bra{\psi_N})$. From \eqref{genstate3}, we obtain the following density matrix
\begin{eqnarray}
\rho_{A_1E}=\frac{1}{d}\sum_{i=0}^{d-1}\ket{i}\!\bra{i}\otimes\rho_{A_1''E}.
\end{eqnarray}
Plugging the above state and measurement $Q_{\alpha_1}^{(b)}$ to the guessing probability \eqref{lgp1}, we arrive at
\begin{eqnarray}
G(\alpha_1,\p) = \frac{1}{d}\sup_{S_{p}}\sum_b\sum_i\bra{i}Q_{\alpha_1}^{(b)}\ket{i}\Tr\left(\Id_{A_1''}\otimes E^{(b)}\rho_{A_1''E}\right)
\end{eqnarray}
Notice that $\sum_i\bra{i}Q_{\alpha_1}^{(b)}\ket{i}=1$ for any $b$, which allows us to finally arrive at
\begin{eqnarray}
G(\alpha_1,\p)&=&\frac{1}{d}\sup_{S_{p}}\sum_b\Tr\left(\Id_{A_1''}\otimes E^{(b)}\rho_{A_1''E}\right)\nonumber\\
&=& \frac{1}{d}
\end{eqnarray}
where we used the fact that Eve performs a valid measurement and thus $\sum_bE^{(b)}=\Id_E$ and $\Tr(\rho_{A_1''E})=1$.
Thus,  we can certify $-\log_2 G(\alpha_1,\p)=\log_2d$ bits of randomness from the maximal violation of the Bell inequalities \eqref{ASTA}. The same analysis can be extended to any party. 

\section{Conclusions and discussions}

We proposed the first self-testing scheme for the certification of generalised GHZ state that relies on violation of a single Bell inequality and requires only two measurements per observer. Apart from this our approach relies on the maximum violation of a Bell inequality that involves $d-$outcome measurements. The previous approach to self-test the generalised GHZ state in Ref. \cite{multist4} extends the scheme of Ref. \cite{ALLst1} by utilising the maximum violation of the tilted CHSH inequality \cite{AAP12} by considering two dimensional subspaces among two parties. Here the first party needs to perform three and the rest of the parties need to perform four measurements each. Unlike this approach, our method does not rely on self-testing results for two-dimensional systems. We propose a novel mathematical approach to derive self-testing statements. Moreover, our scheme is experimentally friendly as we can self-test generalised GHZ states using the minimal number of measurements required to observe Bell nonlocality, that is, two. This makes our scheme more practical and easier to implement experimentally as compared to \cite{ALLst1} because it reduces the amount of data necessary to certify the devices. As a matter of fact,  violation of the SATWAP Bell inequalities has been experimentally demonstrated in Ref. \cite{Science} for $d=3$. Our scheme can also be considered as a generalisation of the self-testing scheme based on chained Bell inequalities \cite{qubitst5} to quantum systems of an arbitrary local dimension as well as arbitrary number of parties.

We also showed that our scheme can be used to securely generate the maximum amount of randomness using projective measurements with arbitrary number of outcomes. 
This result is also interesting from a foundational point of view as this is the first instance, where a single measurement can be used to generate genuine unbounded randomness with the highest possible security. Our self-testing scheme has also been exploited in Ref. \cite{qs5} to show that the set of quantum correlations in a certain Bell scenario is not closed.  
An interesting follow-up of our work was presented in \cite{Fu, mancin} where correlations of constant size where enough to self-test any two-qudit maximally entangled state.

Our work provokes some follow up problems. The first interesting problem would be to devise an analytic technique that can also give information about the state and measurements even when one does not obtain the exact quantum bound but a value slightly lower than it. The only analytical method to derive such robustness bounds is restricted to scenarios where the parties perform two-outcome measurements \cite{qubitst1, qubitst2,qubitst8,qubitst9}. For the SATWAP Bell inequalities, in Ref. \cite{SATWAP} the robustness was derived for $d=3$ and $m=2$ using the numerical approach based on semi-definite programming. Such results would be particularly important for experimental implementations. 

Another interesting problem would be to derive genuinely $d-$outcome Bell inequalities that are maximally violated by partially entangled states in the bipartite case and other classes of multipartite entangled states such as $W-$states and then explore whether these inequalities can be used for self-testing. As was discussed in introduction, the maximum amount of randomness that one can generate using a $d-$dimensional system is of amount $2\log_2 d$ bits using non-projective measurements. Thus, it would be interesting to see whether our self-testing scheme can be used to certify this optimal randomness along the lines of Refs. \cite{random1, sarkar5} which consider qubit and qutrit states respectively. 

%% file: Sections/chapter_4.tex
\chapter{Certification of incompatible measurements}
\label{chapter_4}
\vspace{-1cm}
\rule[0.5ex]{1.0\columnwidth}{1pt} \\[0.2\baselineskip]
\definecolor{airforceblue}{rgb}{0.36, 0.54, 0.66}
\definecolor{bleudefrance}{rgb}{0.19, 0.55, 0.91}
\section{Introduction}

Most of the self-testing schemes aims to certify quantum states without much emphasis on measurements even when one of the necessary conditions for existence of non-classical correlations can be attributed to the presence of incompatible measurements in quantum theory. Moreover, self-testing of quantum measurements in the Bell scenario has restrictions as was shown in \cite{Brunner, Vertesi}. Specifically, there exist pair of incompatible measurements that do not violate any Bell inequality. As a consequence, it might not be possible to certify every pair of incompatible measurements in a fully device-independent way and therefire it is reasonable to look for scenarios that are weaker than the Bell scenario. One such possibility is to assume that one of the parties in the Bell experiment is fully characterised and performs known measurements. This is the well-known quantum steering scenario. As a matter a fact, it was recently proven that there is a one-to-one correspondence between quantum steering and measurement incompatibility \cite{Guhne,quintino, Caval1},  suggesting that every pair of incompatible measurements can be certified in a steering scenario. This makes quantum steering well suited for our task of certifying incompatible measurements. Certification using quantum steering was proposed recently \cite{Supic, Alex} but only for qubits and for a specific pair of 2-outcome measurements. It is worth noting that, the technique used in these schemes \cite{Supic, Alex} cannot be used to certify arbitrary pair of 2-outcome incompatible measurements. 

We provide here a simple scheme for certification
of $d$-outcome incompatible projective 
measurements and the maximally entangled state of local dimension $d$. Our scheme can be used to certify a family of
quantum observables termed here "genuinely incompatible". Roughly speaking, genuinely incompatible observables are those that do not share a common invariant proper subspace [see below for a precise definition]. For instance, mutually
unbiased bases (MUBs) acting on $d$-dimensional Hilbert spaces are genuinely incompatible. Inspired by the inequalities presented in Refs. \cite{Horodecki,Stein}, we introduce a family of steering inequalities that are maximally violated by the maximally entangled state and any set of genuinely incompatible measurements. We analyse the case when this inequality can be maximally violated by measurements that are not genuinely incompatible. We also study the robustness of our 
certification scheme towards experimental imperfections when trusted Alice chooses a pair of mutually unbiased bases to measure her subsystem.

\section{Family of steering inequalities}\label{AppA}

\begin{figure}[h!]
    \centering
    \includegraphics[scale=.5]{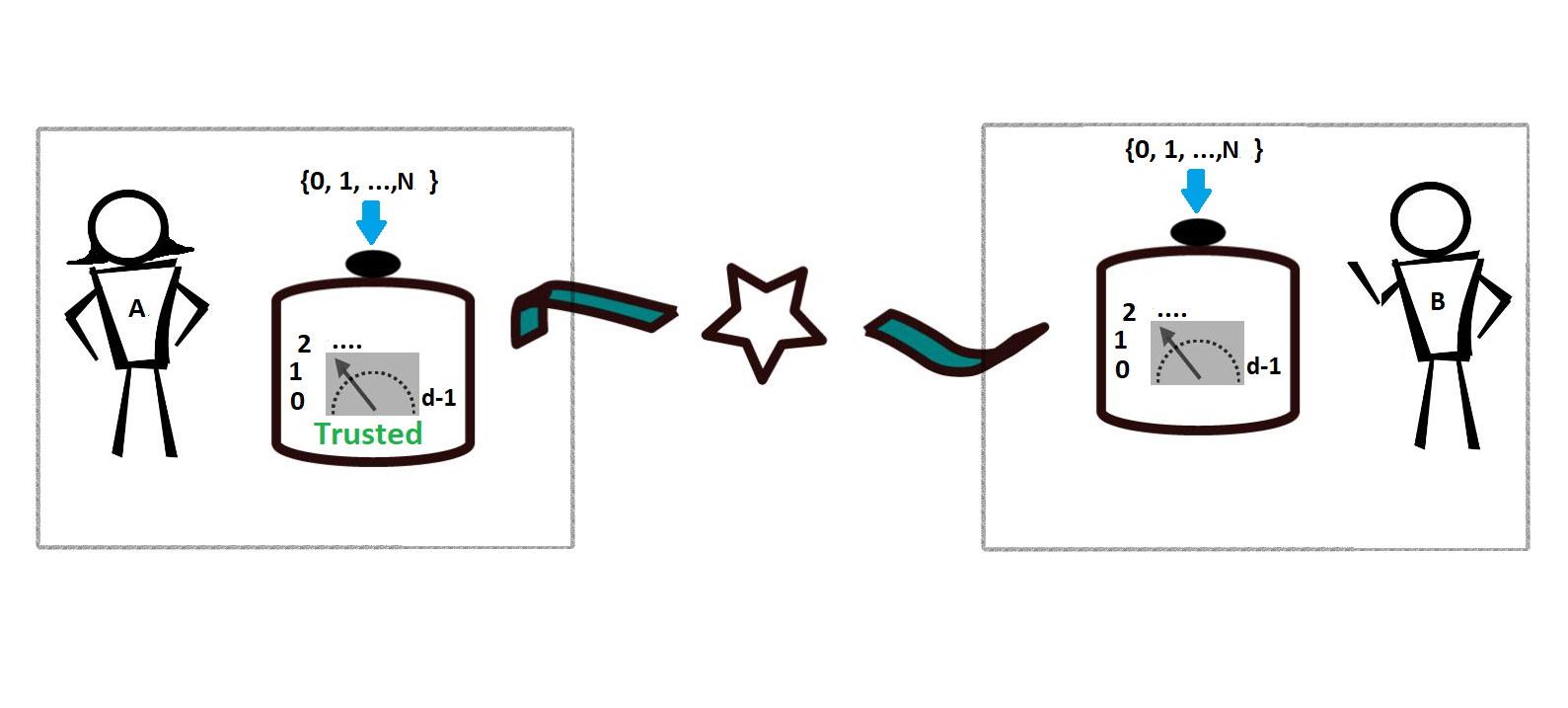}
    \caption{Quantum steering scenario: Alice and Bob both receive a system from the preparation device on which each of them perform $N$ $d-$outcome measurements such that $N,d$ are integers greater than or equal to two. The experiment is repeated enough number of times to generate the relevant joint probability distribution. The key difference between this and Bell scenario is that Alice is trusted and her measurements are known.}
    \label{fig3.1}
\end{figure}

Let us first recall the quantum steering scenario introduced in Chapter \ref{chapter_2} in an analogous way to Bell scenario. Alice and Bob are located in spatially separated labs. Both of them receive two unknown subsystems from a preparation device. Alice is trusted and performs $N$ known $d-$outcome measurements on the received subsystem. On the other hand, Bob also performs $N$ $d-$outcome measurements on his subsystem but these measurements are unknown. The measurements of both Alice and Bob are labelled by $x,y=1,2,\ldots,N$ whereas the outcomes as $a,b=0,1,\ldots,d-1$. They collect enough statistics to construct the joint probability distribution $\{p(a,b|x,y)\}$. The scenario is depicted in Fig. \ref{fig3.1}.

Inspired by \cite{Horodecki,Stein}, we construct the following family of steering inequalities written using generalised observables [cf. Sec. \ref{2.1.3}] as
\begin{equation}\label{SteeringFunctional21}
    W_{2,d,N}=
   \sum_{k=1}^{d-1} \sum_{y=1}^N\left\langle A_{y}^k\otimes B_{k|y}\right\rangle\leq \beta_L.
\end{equation}
Note that the steering functional in the above expression looks similar to \eqref{stefn2}, here however we assume that the observables $A_{k|x}$ are unitary, that is, $A_{k|x}=A_x^k$. We also moved the term for $k=0$ to the classical bound, as it simply reduces to the expectation value of $\Id$ which is one. Alice is trusted and thus $A_{y}$ are known to act on Hilbert space of dimension $d$. Recall that Bob's measurements act on a Hilbert space of unknown but finite dimension. The above steering functional \eqnref{SteeringFunctional21} can also be expressed in terms of joint probabilities as
\begin{equation}\label{SteeringFunctional23}
W_{2,d,N} = d\sum^N_{x=1} \sum^{d-1}_{a,b=0} c_{a,b} \ p(a,b|x,y=x)-N, 
\end{equation}
where,
\begin{eqnarray}\label{c_steer}
c(a,b,x,y)=\begin{cases}
1\quad \text{if}\quad a\oplus_d b=0\\
0\quad \text{otherwise}
\end{cases}
\end{eqnarray}
where $a\oplus_d b$ represents $a+b$ modulo $d$. To get from the observable picture \eqref{SteeringFunctional21} to the joint probability picture \eqref{SteeringFunctional23}, we used the relation \eqref{obs0} from Chapter \ref{chapter_2}, that is,
\begin{eqnarray}\label{4.4}
\langle A_{k|x}\otimes B_{l|y}\rangle=\sum_{a,b=0}^{d-1}\omega^{(ka+lb)}p(a,b|x,y)
\end{eqnarray}
for all $k,l,x,y$. Let us now compute the classical bound $\beta_L$ of the steering functional in Eq. \eqref{SteeringFunctional21}.

\subsection{Classical bound}

As discussed in Chapter \ref{chapter_2}, the classical bound of the steering functional \eqnref{SteeringFunctional21} can be calculated by assuming that the assemblage $\{\sigma_{b|y}\}$ admits a local hidden state model, that is,
\begin{equation}\label{LHS}
    \sigma_{b|y}=\sum_{\lambda}p(\lambda)p(b|y,\lambda)\rho_{\lambda},
\end{equation}
where $\lambda$ represents the hidden variables that are distributed with probability  $p(\lambda)$,
$p(b|y,\lambda)$ denotes the probability of obtaining outcome $b$ when Bob performs the measurement $y$ given the hidden variable $\lambda$ and $\rho_{\lambda}$
are hidden states that act on Alice's Hilbert space. Using \eqref{LHS}, the corresponding joint probabilities $p(a,b|x,y)$ for assemblages admitting a LHS model are given by

\begin{eqnarray}
p(a,b|x,y)=\Tr(M_{a|x}\sigma_{b|y})=\sum_{\lambda}p(\lambda)p(b|y,\lambda)\Tr(M_{a|x}\rho_{\lambda}),
\end{eqnarray}
where $M_{a|x}$ represents the projector corresponding to the outcome $a$ when Alice performs the measurement $x$. The above expression can also be stated as
\begin{eqnarray}\label{4.6}
p(a,b|x,y)=\sum_{\lambda}p(\lambda)p(a|x,\rho_\lambda)p(b|y,\lambda),
\end{eqnarray}
where $p(a|x,\rho_\lambda)=\Tr(M_{a|x}\rho_{\lambda})$.
Now, using the joint probability form of the steering functional \eqnref{SteeringFunctional23}, we have that 
\be \label{steeopprob.4}
W_{2,d,N} = d\sum^n_{x=1} \sum^{d-1}_{a,b=0}\sum_{\lambda} c_{a,b} \ p(\lambda)p(a|x,\rho_\lambda)p(b|x,\lambda)-N
\ee 
where $c_{a,b}$ is given in \eqnref{c_steer}. Thus, the steering functional simplifies to
\begin{eqnarray}
W_{2,d,N} = d \sum^N_{x=1} \sum_{a=0}^{d-1}\sum_{\lambda} p(\lambda)p(a|x,\rho_{\lambda})p(d-a|x,\lambda)-N.
\end{eqnarray}
The above term is upper bounded by
\begin{eqnarray}
\sum^N_{x=1} \sum_a\sum_{\lambda} p(\lambda)p(a|x,\rho_{\lambda})p(d-a|x,\lambda)\leq \sum^N_{x=1} \sum_{\lambda} p(\lambda)\max_a\{p(a|x,\rho_{\lambda})\}
\end{eqnarray}
where we used the normalisation condition $\sum_ap(a|x,\lambda)=1$ and then the fact that for any real-valued function $f(x)\in\mathbbm{R}$,
\begin{eqnarray}
\sum_ap(a)f(a)\leq \max_af(a)\qquad \text{such that}\qquad \sum_ap(a)=1.
\end{eqnarray}
Also, notice that
\begin{eqnarray}\label{4.11}
\sum^N_{x=1} \sum_{\lambda} p(\lambda)\max_a\{p(a|x,\rho_{\lambda})\}\leq \sum^N_{x=1}\max_{\rho} \sum_{\lambda} p(\lambda)\max_a\{p(a|x,\rho)\}.
\end{eqnarray}
Now, using the fact that $\sum_{\lambda} p(\lambda)=1$, we obtain an upper bound on the value of the steering functional as
\begin{eqnarray}
W_{2,d,N} \leq d \sum^N_{x=1}\max_{\rho} \max_a\{p(a|x,\rho)\}-N
\end{eqnarray}
Doing an inverse Fourier transform and expressing in terms of expectation values, we have
\begin{eqnarray}
W_{2,d,N} \leq\sum_{x=1}^{N}\max_{{\rho}}\max_a\sum_{k=1}^{d-1} \omega^{-ka}\left\langle\hat{A}_x^{k}\right\rangle_{\rho}\leq \sum_{x=1}^{N}\max_{{\rho}}\sum_{k=1}^{d-1} \left|\left\langle\hat{A}_x^{k}\right\rangle_{\rho}\right|.
\end{eqnarray}
Thus, we conclude that the local bound of $W_{2,d,N}$ is upper bounded by,
\begin{eqnarray}\label{4.15}
    \beta_L\leq\sum_{i=1}^{N}\max_{{\rho}}\sum_{k=1}^{d-1} \left|\left\langle\hat{A}_i^{k}\right\rangle_{\rho}\right|.
\end{eqnarray}
This bound can be explicitly calculated for different set of observables $A_x$. For instance, if $N=d=2$ and $A_1=\sigma_z$ and $A_2=\sigma_x$ \eqref{pauliz,x} then the classical bound is given by $\beta_L=\sqrt{2}$. Let us consider the special case, when $N=2$ such that $A_1=Z_d$ and $A_2=X_d$ which are the $d-$dimensional generalisation of the Pauli matrices $\sigma_z$ and $\sigma_x$ respectively whose explicit form is given below in Eq. \eqref{X,Z}. In this case, the classical bound is given by $\beta_L=\sqrt{2}(d-1)$ and was computed in \cite{Horodecki}. Let us now compute the quantum bound of the steering functional in Eq. \eqref{SteeringFunctional21}. 

\subsection{Quantum bound}
It is straightforward to find the maximal quantum bound of the steering functional  (\ref{SteeringFunctional21}). Let us first note that the algebraic bound of (\ref{SteeringFunctional21}) is $N(d-1)$ and can be achieved when each term of the functional equals one. Let us now consider the maximally entangled state of two qudits,
\begin{equation}\label{maxentstated}
    \ket{\phi_{+}^d}=\frac{1}{\sqrt{d}}\sum_{i}\ket{ii},
\end{equation}
and Bob's observables are unitary and conjugate of Alice's observables, that is, $B_y=A_y^{*}$ for all $y$. For these quantum realisations, the expectation value of every term in the functional \eqref{SteeringFunctional21} is one. Thus the quantum bound of the steering functional \eqnref{SteeringFunctional21} is same as its algebraic bound, that is,
\begin{eqnarray}
\beta_Q=N(d-1).
\end{eqnarray}

Consequently, the quantum state and the measurements that achieve the maximum quantum value $\beta_Q$ of the steering functional \eqnref{SteeringFunctional21} for fixed observables $A_i$ 
must satisfy the following relations

\begin{equation}
    \langle A_y^k\otimes B_{k|y}\rangle=1 \qquad \forall y,k.
\end{equation}
Since $A_y$ is unitary and $B_{k|y}^{\dagger}B_{k|y}\leq\Id$, for any state $\rho_{AB}$ that satisfies the above relation, we have that
\begin{equation}\label{Cabernet}
    A_y^k\otimes B_{k|y}\,\rho_{AB}=\rho_{AB}\qquad \forall y,k.
\end{equation}
This relation would be particularly useful for deriving the 1SDI certification results.

Let us now figure out the cases when the steering inequality $W_{2,d,N}\leq\beta_L$ is non-trivial, that is $\beta_L<\beta_Q$. For this, let us see when the upper bound \eqref{4.15} to the value attainable using classical strategies is equal to the quantum bound, that is,
\begin{equation}
    \sum_{i=1}^N\sum_{k=0}^{d-1}|\langle A_i^k\rangle_{\rho}|= N(d-1).
\end{equation}
This implies that each term in the above relation is $1$ or simply $|\langle A_i^k\rangle_{\rho}|=1$ for all $i$, $k$ and $\rho$ acting on $\mathbbm{C}^d$. Let us say that $\rho$ admits a decomposition in terms of pure states as $\rho=\sum_ip_i\ket{\psi_i}\!\bra{\psi_i}$. Since $A_i^k$ are unitary operators, this means that $\ket{\psi_i}$ are the eigenvectors of all $A_i$ with eigenvalues of modulus $1$. Thus, we can conclude that for the steering inequality to be non-trivial the observables $A_i$  cannot share any common eigenvector. However, the steering functional \eqref{SteeringFunctional21} can not be used to certify all such observables that do not share a common eigenvector, but a restricted set of such observables termed as "genuinely incompatible" observables. The reason for this ambiguity would be clarified in the later sections.

\section{Genuinely incompatible observables}
 Here we refer to the definition of genuine incompatible (GI) observables introduced in \cite{sarkar2}.
 
 \begin{defn}[Genuine incompatible observables]
Consider a set of $N$ $d$-outcome unitary observables $A_i$ 
acting on $\mathbbm{C}^d$ and obeying $A_i^d=\Id$. 
We call them genuinely incompatible (GI) if there is 
no subspace $V\subset \mathbbm{C}^d$
such that $\dim V<d$ and $A_iV\subseteq V$ for all $i$; 
in other words, the only common invariant space of all 
$A_i$ is the full space $\mathbbm{C}^d$.
\end{defn}

Let us recall the $d-$outcome observables introduced in Chapter \ref{chapter_2} in Eq. \eqref{X,Z}:
\begin{eqnarray}\label{X,Z1}
Z_d=\sum_{i=0}^{d-1}\omega^i\ket{i}\!\bra{i},\qquad X_d=\sum_{i=0}^{d-1}\ket{i+1}\!\bra{i}
\end{eqnarray}
that are $d-$dimensional generalisation of the Pauli matrices $\sigma_z, \sigma_x$ \eqref{pauliz,x}. Note that in the above sum $\ket{i+1}$ is modulo $d$. The eigenvectors of these observables are mutually unbiased (see Def. \ref{MU} below). These observables are genuinely incompatible.

\begin{defn}[Mutually unbiased bases]\label{MU}
Two orthonormal bases $\{\ket{e_i^0}\}$ and $\{\ket{e_i^1}\}$ in $\mathbbm{C}^d$ form mutually unbiased bases if  
\begin{eqnarray}
|\bra{e_i^0}e_j^1\rangle|^2=\frac{1}{d}
\end{eqnarray}
for every $i,j=0,1,\ldots,d-1$. 
\end{defn}
We then say that two unitary observables $A_0$ and $A_1$ are mutually unbiased if their eigenvectors are mutually unbiased.
There are a few interesting observations about GI observables.

\begin{enumerate}
    \item Two $d-$outcome observables $A_0$ and $A_1$ are not genuinely incompatible if there exist a basis in $\mathbbm{C}^d$ using which the observables can be decomposed as 
    \begin{eqnarray}
    A_0=A_0'\oplus A_0''\quad \text{and}\quad A_1=A_1'\oplus A_1''
    \end{eqnarray}
    such that $A_0'$ and $A_1'$ act on a $d'-$dimensional subspace of $\mathbbm{C}^d$ with $d'<d$. In this case, the observables share a common invariant subspace $\mathbbm{C}_{d'}$ spanned by the eigenvectors of $A_0'$ (or equivalently, the eigenvectors of $A_1'$).
    \item Genuinely incompatible observables do not share a common eigenvector. We verify this claim after Lemma \ref{lemma3} stated below. Thus, when Alice's observables are genuinely incompatible, the steering inequality \eqref{SteeringFunctional21} $W_{2,d,N}\leq\beta_L$ is non-trivial
    \item Consider a set of $N$ observables. If this set contains two observables that are GI, then the whole set is genuinely incompatible as well. This is because if two observables do not share a common invariant subspace, then any set containing these two observables can not share any invariant subspace. 
    \item The opposite implication of the above statement is not true. Set of observables might be genuinely incompatible even when pairwise they are not genuinely incompatible. 
    \end{enumerate}
   To illustrate the above statement with an example, let us consider three five-outcome observables with eigenvalues $\omega^i$ with $x=0,1,2,3,4$, such that $\omega=\exp{(\frac{2\pi \mathbbm{i}}{5})}$, written in the computational basis.
    
\begin{eqnarray}
    &A_0&=\frac{1}{2}\begin{pmatrix}
1+\omega & 1-\omega & 0 & 0 & 0\\
1-\omega & 1+\omega & 0 & 0 & 0\\0 & 0 & \omega^2 & 0 & 0\\0 &
0 & 0 & \omega^3 & 0\\0 & 0 & 0 & 0 & \omega^4
\end{pmatrix}, \nonumber\\  
&A_1&=\frac{1}{2}\begin{pmatrix}
2 & 0 & 0 & 0 & 0\\0 & \omega^2+\omega & \omega^2-\omega & 0 & 0\\0 &
\omega^2-\omega& \omega^2+\omega & 0 & 0\\
0 & 0 & 0 & 2\omega^3 & 0\\0 & 0 & 0 & 0 & 2\omega^4
\end{pmatrix}, \nonumber
\end{eqnarray}

\begin{eqnarray}\label{C19}
\text{and}\qquad
 &A_2&=\frac{1}{3}\begin{pmatrix}
1 & 0 & 0 & 0 & 0\\0 & \omega & 0 & 0 & 0\\0 & 0 & f_1(\omega,\omega_3) & f_2(\omega,\omega_3)& f_3(\omega,\omega_3)\\
0 & 0 & f_3(\omega,\omega_3) & f_1(\omega,\omega_3)& f_2(\omega,\omega_3)\\
0 & 0 & f_2(\omega,\omega_3) & f_3(\omega,\omega_3)& f_1(\omega,\omega_3) \end{pmatrix}
\end{eqnarray}
where $f_1(\omega,\omega_3)=\omega^2+\omega^3+\omega^4$, $f_2(\omega,\omega_3)=\omega^2+\omega_3^2\omega^3+\omega_3\omega^4$ and $f_3(\omega,\omega_3)=\omega^2+\omega_3\omega^3+\omega_3^2\omega^4$ such that $\omega_3=\exp{(\frac{2\pi \mathbbm{i}}{3})}$.
Let us first find the  eigenvectors of the above observables. The eigenvectors of $A_0's$ are $\{\ket{+_{01}},\ket{-_{01}},\ket{2},\ket{3}, \ket{4}\}$, second, $A_1's$ eigenvectors are $\{\ket{0},\ket{+_{12}},\ket{-_{12}},\ket{3}, \ket{4}\}$ 
and finally, $A_2's$ eigenvectors are $\{\ket{0},\ket{1},\ket{e_1},\ket{e_2}, \ket{e_3}\}$ where $|\pm_{ij}\ra = (|i\ra \pm |j\ra)/\sqrt{2}$ and 
\begin{eqnarray}
\ket{e_1}=\frac{1}{\sqrt{3}}\left(\ket{2}+\ket{3}+\ket{4}\right),\quad \ket{e_2}=\frac{1}{\sqrt{3}}\left(\ket{2}+\omega_3\ket{3}+\omega_3^2\ket{4}\right),\nonumber\\
\text{and}\quad\ket{e_3}=\frac{1}{\sqrt{3}}\left(\ket{2}+\omega_3^2\ket{3}+\omega_3\ket{4}\right).
\end{eqnarray}

From Lemma \ref{lemma3} which is stated below, we conclude that $A_0$ and $A_1$ have two non-trivial common invariant subspaces, spanned by $\{\ket{0},\ket{1},\ket{2}\}$ and $\{\ket{3},\ket{4}\}$.  Again, $A_0$ and $A_2$ have two non-trivial common invariant subspaces, spanned by $\{\ket{0},\ket{1}\}$ and $\{\ket{2},\ket{3},\ket{4}\}$. Finally,  $A_1$ and $A_2$ have two non-trivial common invariant subspaces, spanned by $\{\ket{0}\}$ and $\{\ket{1},\ket{2},\ket{3},\ket{4}\}$. Thus, if we consider the pair of the above observables, none of such pairs are genuinely incompatible. However, considering all three observables, we can readily see that there is no common invariant subspace shared between $A_0,\ A_1$ and $A_2$ as we simultaneously cannot express all the matrices using blocks of dimension less than $5\times 5$.

Now, we prove a lemma that would be useful for characterising genuinely incompatible observables.

\begin{customlemma}{4.1}\label{lemma3}
Two $d-$outcome observables $A_0$ and $A_1$ share a common non-trivial invariant subspace of dimension $d'<d$ if and only if $d'$ eigenvectors of $A_0$ can be expressed as a linear combination of $d'$ eigenvectors of $A_1$.
\end{customlemma}  

\begin{proof}
Let us first recall that two $d-$outcome observables $A_0$ and $A_1$ share a common invariant subspace if there exist a basis in $\mathbbm{C}^d$ using which the observables can be decomposed as 
    \begin{eqnarray}
    A_0=A_0'\oplus A_0''\quad \text{and}\quad A_1=A_1'\oplus A_1''
    \end{eqnarray}
    such that $A_0'$ and $A_1'$ act on $d'$ dimensional Hilbert space where $d'<d$. The common invariant subspace is $\mathbbm{C}_{d'}$ and is spanned by the $d'$ eigenvectors of $A_0'$ or the $d'$ eigenvectors of $A_1'$. Thus, eigenvectors of $A_0'$ can be expressed as linear combination of eigenvectors of $A_1'$ as they span the same Hilbert space. We showed here that if $A_0$ and $A_1$ share a common invariant subspace of dimension $d'$, then $d'$ eigenvectors of $A_0$ can be written as $d'$ eigenvectors of $d'$.
    
    It is trivial to show the other way round, that is, if $d'$ eigenvectors of $A_0$ can be written as $d'$ eigenvectors of $A_1$, then $A_0$ and $A_1$ share a common invariant subspace spanned by these eigenvectors. This ends the proof.
\end{proof}  

Note that if two observables share a common eigenvector, then they share a common invariant subspace of dimension one. Thus, genuinely incompatible observables can not even share a common eigenvector. A corollary of the above lemma is that mutually unbiased bases are genuinely incompatible.

\begin{ucor}\label{Lem2A}
Any two $d$-outcome observables whose eigenbases form mutually unbiased bases, are genuinely incompatible.
\end{ucor}
\begin{proof}
Consider two $d-$element mutually unbiased bases denoted by, $\{\ket{s_i}\}$ and $\{\ket{t_j}\}$ such that $|\langle s_i|t_j\rangle|^2=1/d$ for all $i,j\in\{0,1,\ldots,d-1\}$. One can construct $d-$outcome observables whose eigenbases are mutually unbiased in the following way, 
\begin{equation}
A_0=\sum_{i=0}^{d-1}\omega^{i}\ket{s_i}\!\bra{s_i},\qquad  A_1=\sum_{i=0}^{d-1}\omega^{i}\ket{t_i}\!\bra{t_i}.
\end{equation}
%
where $\omega=\exp{(\frac{2\pi\mathbbm{i}}{d})}$. Any eigenvector $\ket{s_i}$ of $A_0$ can only be written as a linear combination of all the $d$ eigenvectors $\ket{t_j}$ of $A_1$. Thus, from Lemma \ref{lemma3} we have that for $A_0$ and $A_1$ do not share a non-trivial common invariant subspace and therefore are genuinely incompatible. 
\end{proof}

Another important property of genuinely incompatible observables which is particularly useful for 1SDI certification is stated below.

\begin{customlemma}{4.2}\label{lemma4}
Consider a set of $N$ $d$-outcome unitary observables $A_y$ with eigenvalues $\omega^i$ for $i=0,1,\ldots,d-1$ such that $\omega=\exp{(\frac{2\pi\mathbbm{i}}{d})}$. Consider also a non-trivial normal matrix $P$ acting on $\mathbbm{C}^d$. If  $[P,A_y]=0$ for every $y=1,2,\ldots,N$ such that the set of observables $A_y$ are genuinely incompatible, then $P=\lambda\Id_d$ where $\lambda$ is an arbitrary complex number.
\end{customlemma}
\begin{proof}

The proof is by contradiction, that is, we assume that $P\ne\lambda\Id_d$. Let us first note that since $P$ is normal, there exists a unitary that transforms $P$ to a diagonal matrix. Thus, we can always express $P$, in terms of its orthogonal projections $P_i$ and the corresponding eigenvalues $\lambda_i$, in the following way
\begin{equation}\label{2.27}
    P=\sum_{i=1}^{m}\lambda_i P_i.
\end{equation}
Here $\lambda_i's$ are distinct complex numbers that might be even $0$ and $m$ denotes the number of such distinct eigenvalues.
 
Let us now assume that $[P,A_y]=0$ for all $y$. Expanding this relation, we have that
\begin{equation}\label{2.28}
    A_y\ P=P\ A_y.
\end{equation}
Let us now consider two orthogonal projections $P_m$ and $P_n$ of $P$ such that $m\leq n$. Now, we multiply $P_m$ from left hand side and $P_n$ from the right hand side of the above equation \eqref{2.28} to obtain
\begin{equation}\label{2.29}
    P_m\ A_y\ P\ P_n=P_m\ P\ A_y\ P_n.
\end{equation}
Using the fact that $P_iP_j=\delta_{ij}$ in \eqref{2.27}, we have that $PP_n=\lambda_nP_n$ and $P_mP=\lambda_m P_m$. This allows us to simplify the above equation \eqref{2.29} as
\begin{equation}
    (\lambda_m-\lambda_n)P_m\ A_y\  P_n=0.
\end{equation}
The above equation has two possible solutions:
\begin{eqnarray}
\lambda_m=\lambda_n\quad \text{or} \quad P_m\ A_y\ P_n=0.
\end{eqnarray}
For distinct $m,n$, the eigenvalues $\lambda_m,\lambda_n$ are also distinct. Thus, we conclude that 
$P_m\ A_i\ P_n=0$ whenever $m\neq n$. This means that $A_y$ decomposes into blocks that act on supp$(P_i)$. We can obtain the same conclusion for every observable $A_y$. Given that $P$ and $A_y$ satisfy the relation $[P,A_y]=0$ for all $y$, if $P$ is of the form \eqref{2.27}, then all $A_y's$ are of the block form 
\begin{equation}\label{dupa12}
 A_y=A_y^{(1)}\oplus \ldots \oplus A_y^{(m)}.
\end{equation}
This contradicts the fact that $A_y's$ are genuinely incompatible which implies that $P=\lambda\Id_d$ for some $\lambda\in\mathbbm{C}$. Notice that the trivial solution to the condition $[P,A_y]=0$ is when $P=0$. Any non-trivial solution such that $A_y$ are genuinely incompatible observables imposes that the rank of $P$ is $d$ and all its eigenvalues are equal and non-zero. This ends the proof.
\end{proof}
Finally, we have all the required tools for deriving the results concerning certification of incompatible measurements.
   
\section{1SDI certification }

\subsection{Exact certification of GI observables}

Here, we present the 1SDI certification of genuinely incompatible measurements that relies on maximal violation of the steering inequality \eqref{SteeringFunctional21} $W_{2,d,N}=\beta_Q$. Let us first recall that we can only characterise Bob's observables on the support of his local state $\rho_B$. Thus, without loss of generality we can consider that the local state $\rho_B$ is full rank. This can also be understood as
that Bob's observables and local state $\rho_B$ act on the same Hilbert space $\mathcal{H}_B$.

\renewcommand{\thetheorem}{4.1}

\setcounter{thm}{0}

\begin{theorem}\label{Theo4.1} 
Consider that Alice and Bob perform the quantum steering experiment and observe that the steering functional 
\begin{eqnarray}\label{stefn3}
W_{2,d,N}=
   \sum_{k=1}^{d-1} \sum_{y=1}^N\left\langle A_{y}^k\otimes B_{k|y}\right\rangle,
\end{eqnarray}
attains the maximal quantum value $\beta_Q=N(d-1)$ where $N$ is number of measurements performed by Alice and Bob and $d$ denotes the number of outcomes of each measurement. Alice's observables $A_y$ acting on $\mathbbm{C}^d$ are unitary with eigenvalues $\omega^i$ such that $\omega=\exp(\frac{2\pi\mathbbm{i}}{d})$ and are genuinely incompatible. Let us say that the maximal quantum bound is achieved using the state $\rho_{AB}$ acting on $\mathbbm{C}^d\otimes\mathcal{H}_B$ and Bob's generalised observables $B_i \ (i\in \{1,\ldots,N\})$ acting on $\mathcal{H}_B$. Then, the following statements hold true for any integer $d$ greater than or equal to two: 
\begin{enumerate}
    \item  Bob's measurements are projective. Equaivalently, the operators $B_{k|y}$ for all $k,y$ are unitary and $B_{k|y}=B_{1|y}^k\equiv B_y^k$.
    \item Bob's Hilbert space $\mathcal{H}_B$ admits a decomposition into a $d-$dimensional Hilbert space $\mathbbm{C}^d)_{B'}$ and an auxiliary Hilbert space of unkown but finite dimension $\mathcal{H}_{B''}$,
    \begin{eqnarray}
    \mathcal{H}_{B}=(\mathbbm{C}^d)_{B'}\otimes \mathcal{H}_{B''}.
    \end{eqnarray}
    \item A local unitary transformation $U_B:\mathcal{H}_B\rightarrow\mathcal{H}_B$ can be applied on Bob's side, such that
\begin{eqnarray}\label{lem1.2}
(\Id_A\otimes U_B)\rho_{AB}(\Id_A\otimes U_B^{\dagger})=\proj{\phi^+_d}_{AB'}\otimes \rho_{B''}.
\end{eqnarray}
where $\ket{\phi^+_d}$ is the maximally entangled state \eqref{maxentstated} and
\begin{eqnarray}\label{lem1.1}
\forall y, \quad U_B\,B_y\,U_B^{\dagger}=A_y^{*}\otimes \Id_{B''},
\end{eqnarray}
where $B''$ denotes Bob's auxiliary system.
\end{enumerate}

\end{theorem}
\begin{proof}
Let us first recall the relations \eqref{Cabernet} that stem from the fact that to saturate the quantum bound of the steering functional \eqref{SteeringFunctional21} each of the expectation values in the functional must amount to one,
\begin{equation}\label{SOSApp}
    A_y^k\otimes B_{k|y}\,\rho_{AB}=\rho_{AB}
\end{equation}
for $y=1,2,\ldots,N$ and $k=1,2,\ldots,d-1$. We begin by showing that the above relations can only be satisfied if Bob's measurements are projective. Note that an equivalent representation of the above relation \eqref{SOSApp} is
\begin{eqnarray}\label{SOSApp2}
 A_y^{d-k}\otimes B_{d-k|y}\,\rho_{AB}=\rho_{AB}\qquad \forall y,k.
\end{eqnarray}
Recall that by definition $B_{d-k|y}=B_{k|y}^{\dagger}$ [cf. \eqref{obs2}]. Now, multiplying \eqref{SOSApp} with $A_y^{d-k}\otimes B_{d-k|y}$, we have that
\begin{eqnarray}
 \left(A_y^{d-k} A_y^k\otimes B_{d-k|y}B_{k|y}\right)\,\rho_{AB}=\left(A_y^{d-k}\otimes B_{d-k|y}\right)\rho_{AB}\qquad \forall y,k.
\end{eqnarray}
Using the fact that $A_y^d=\Id_A$ and the relation \eqref{SOSApp2}, we have that
\begin{equation}
  \left(\Id_A\otimes B_{k|y}^{\dagger }B_{k|y}\right)\,\rho_{AB}= \rho_{AB}.
\end{equation}
Tracing over the subsystem $A$, we get that
\begin{equation}\label{4.42}
   \left( B_{k|y}^{\dagger }B_{k|y}\right)\rho_{B}= \rho_{B}
\end{equation}
and since $\rho_B$ is full-rank and thus invertible, we finally have that $ B_{k|y}^{\dagger }B_{k|y}= \Id_{B}$ for all $k,y$
such that $\Id_{B}$ is the identity acting on $\mathcal{H}_B$. Similarly, taking the relation \eqref{SOSApp2} and multiplying it with $A_y^k\otimes B_{k|y}$, we get that $B_{k|y}B_{k|y}^{\dagger}= \Id_{B}$ for all $k,y$
such that $\Id_{B}$. Thus, one straightforwardly concludes from the above conditions that $B_{k|y}$ are unitary for every $y$ and $k$. Now using Fact \ref{factgenobs1}, we conclude that Bob's measurements are projective.
Since, for projective measurements $B_{k|y}=B_y^{k}$, from here on we substitute $B_{k|y}=B_y^{k}$.

 Let us focus on the state  $\rho_{AB}$ that results in the quantum bound of the steering functional \eqref{stefn3}. Consider the eigendecomposition of $\rho_{AB}$ as
\begin{eqnarray}\label{mixed1}
\rho_{AB}=\sum_{s=1}^{K}p_s\proj{\psi_s}_{AB},
\end{eqnarray}
where $K$ is any integer greater than or equal to one and $p_s\geq 0$ such that $\sum_sp_s=1$. Further, the eigenstates $\ket{\psi_s}$ are pairwise orthogonal, that is,  $\langle\psi_s|\psi_{s'}\rangle=\delta_{ss'}$ for every $s,s'$.

First, we show that the rank of the local state of Alice is $d$. The proof is by contradiction. For this, we use the relations \eqref{SOSApp} and the fact that Alice's observables $A_i$ are 
genuinely incompatible. 
Let us assume that rank of $\rho_A$ is strictly less than $d$. Then, we consider the relation (\ref{Cabernet}) 
for $k=1$ and then project Alice's observables onto the support of the state $\rho_A$,  
\begin{eqnarray}\label{SOS22}
\quad  \Pi_A\, A_y\,\Pi_A\otimes B_y\ \rho_{AB} = \rho_{AB},
\end{eqnarray}
where $\Pi_A$ is the projector onto the support of Alice's local state 
$\rho_A$. Let us denote $\overline{A}_i\equiv \Pi_A\, A_i\,\Pi_A$. Also, considering the relation \eqref{SOSApp2} for $k=1$, we have that
\begin{eqnarray}\label{2.46}
\Pi_A\, A_y^{\dagger}\,\Pi_A\otimes B_y^{\dagger}\  \rho_{AB} = \rho_{AB}.
\end{eqnarray}
Note, that $\Pi_A\, A_y^{\dagger}\,\Pi_A=(\Pi_A\, A_y\,\Pi_A)^{\dagger}$. As proven above, Bob's measurements that result in the maximal violation are projective and thus $B_y$ are unitary and satisfy
$B_y^d=\Id$. 
Now, applying $\overline{A}_y\otimes B_y$ to the equation \eqref{2.46}, we have that
\begin{eqnarray}
\left(\overline{A}_y\overline{A}_y^{\dagger}\otimes B_yB_y^{\dagger}\right) \rho_{AB} = \rho_{AB}.
\end{eqnarray}
Again, using the fact that $B_y$ are unitary and then taking a trace of subsystem $B$, we have that
\begin{eqnarray}
\overline{A}_y\overline{A}_y^{\dagger}=\Pi_A
\end{eqnarray}
As proven in Fact \ref{fact1} in Chapter \ref{chapter_2}, since $\overline{A}_y$ is unitary 
, it
must be of block form 
\begin{equation}
    A_y=\overline{A}_y\oplus A'_y 
\end{equation}
for some unitary matrix $A'_y$ of dimension $[d-\mathrm{rank}(\rho_A)]\times[d-\mathrm{rank}(\rho_A)]$.
However, from Lemma \ref{lemma3} we conclude that $A_y's$ have a
common invariant subspace of dimension lower $\mathrm{rank}(\rho_A)$ which strictly lower than $d$. This contradicts the fact that $A_y$ are genuinely incompatible observables. 
As a consequence, $\rho_{A}$ is a full rank matrix, or equivalently the rank of the local state of Alice $\rho_A$ is $d$. 


Now, we have all the required tools to formulate the main part of the theorem which includes characterising the state $\rho_{AB}$ and  Bob's observables $B_y$ that result in the quantum bound of the steering functional \eqref{stefn3}. Let us first expand the relation \eqref{SOSApp} using the decomposition of the state $\rho_{AB}$ \eqref{mixed1} keeping in mind that Bob's observables are projective,
\begin{eqnarray}
 \sum_{s=1}^K p_s\ A_y^k\otimes B_{y}^k\,\ket{\psi_s}\!\bra{\psi_s}_{AB}=\sum_{s=1}^K p_s\ket{\psi_s}\!\bra{\psi_s}_{AB}\qquad \forall y,k.
\end{eqnarray}
Multiplying with $\ket{\psi_s}$ on the right hand side of the above equation, we arrive at the following condition,
\begin{eqnarray}\label{SOS253}
A_y^k\otimes B_{y}^k\,\ket{\psi_s}_{AB}=\ket{\psi_s}_{AB}\qquad \forall s,y,k
\end{eqnarray}
where we used the fact that $\langle\psi_s'|\psi_{s}\rangle=\delta_{ss'}$ for every $s,s'$. 

Now, we can characterise every $\ket{\psi_s}$ and then find the relation among them to fully identify the state $\rho_{AB}$. Let us first note from the relation \eqref{SOS253} that Bob's measurements acting on the support of the local state $\rho_{B,s}=\Tr_A(\proj{\psi_s}_{AB})$ are also projective. For this we can follow the exactly same procedure as was done in the beginning of the proof to conclude that $B_y$ acting on the support of $\rho_B$ is projective. Thus, we can conclude that
\begin{eqnarray}\label{Bobobs1}
B_y=\Pi_B^sB_y\Pi_B^s\oplus E_s
\end{eqnarray}
where $\Pi_B^s$ represents the projector onto the support of $\rho_{B,s}$ and $E_s$ is some unitary matrix. For completeness, let us briefly discuss the proof again. We begin by considering the relations (\ref{SOSApp}) and (\ref{SOSApp2}) for $k=1$ and project it onto the support of $\rho_{B,s}$, that is, 
\begin{equation}\label{Jumila}
  A_y\otimes  \overline{B}_{y,s}\ket{\psi}_{AB} = \ket{\psi_s}_{AB}\quad \text{and}\quad  A_y^{\dagger}\otimes  \overline{B}_{y,s}^{\dagger}\ket{\psi}_{AB} = \ket{\psi_s}_{AB}
\end{equation}
where $\overline{B}_{y,s}=\Pi_B^sB_y\Pi_B^s$. By applying $A_y\otimes\overline{B}_{y,s}$ to the left equation in (\ref{Jumila}) $d-1$ times, we
obtain that $\overline{B}_{y,s}^d=\Id_d$. Now, applying $A_y^{\dagger}\otimes \overline{B}_{y,s}^{\dagger}$ to the right equation in \eqref{Jumila} of the above equation, we conclude that $\overline{B}_{y,s}^{\dagger}\overline{B}_{y,s}=\Id_d$. Similarly, we can also conclude that $\overline{B}_{y,s}\overline{B}_{y,s}^{\dagger}=\Id_d$. As a result $\overline{B}_{y,s}$ are unitary and thus represent projective measurements with eigenvalues $\{1,\omega,\ldots,\omega^{d-1}\}$. Using the Fact \ref{fact1} proven in Chapter \ref{chapter_2}, we finally arrive at the desired block form of Bob's observables \eqref{Bobobs1}.

Since all the states $\ket{\psi_s}_{AB}$ follow the same relation \eqref{SOS253}, foe the moment let us drop the index $s$ and consider a simple state $\ket{\psi}_{AB}$. As concluded before in Lemma \ref{lemma4}, when the set of observables $A_y$ are genuinely incompatible, $\mathrm{rank}(\rho_A)=d$ which implies that the local dimension of the state is $d$. This allows us to consider the Schmidt decomposition of 
$\ket{\psi}_{AB}$ as
\begin{equation}\label{SchmidtApp}
    \ket{\psi}_{AB}=\sum_{i=0}^{d-1}\lambda_i\ket{e_i}\ket{f_i}.
\end{equation}
As $\rho_A$ is of full-rank, the Schmidt coefficients $\lambda_i$ for $(i=0,\ldots,d-1)$ are all strictly greater than zero. Also, the normalisation of the state $\ket{\psi}_{AB}$ relates the coefficients by the condition $\sum_{i}\lambda_i^2=1$. Moreover, the local vectors $\{\ket{e_i}\}$ and $\{\ket{f_i}\}$ form two orthonormal bases in $\mathbbm{C}^d$.

Let us now consider a unitary $U_B:\mathbbm{C}^d\rightarrow\mathbbm{C}^d$ such that 
$\ket{f_i}=U_B^{\dagger}\ket{e_i^*}$ for every $i$, where the asterisk denotes 
complex conjugation in the computational basis. Applying this unitary to the state \eqref{SchmidtApp}, we have that
\begin{eqnarray}
(\Id_A\otimes U_B)\ket{\psi}_{AB}=
\sum_{i=0}^{d-1}\lambda_i\ket{e_i}\ket{e_i^*}.
\end{eqnarray}
Now, let us consider a diagonal matrix $P_A$ with eigenvectors $\{\ket{e_i}\}$ 
and eigenvalues $\sqrt{d}\,\lambda_i$, that is,  $P_A=\sqrt{d}\lambda_i\sum_{i=0}^{d-1}\ket{e_i}\!\bra{e_i}$. Now, the state \eqref{SchmidtApp} can be written as
\begin{eqnarray}\label{state2App}
(\Id_A\otimes U_B)\ket{\psi}_{AB}=
(P_A\otimes\Id_B)\frac{1}{\sqrt{d}}\sum_{i=0}^{d-1}\ket{e_i}\ket{e_i^*}
\end{eqnarray}
Recall that all $\lambda_i's$ are positive real numbers which implies that $P_A$ is full rank, or equivalently $\mathrm{rank}(P_A)= d$. Notice that the state on the right hand side of (\ref{state2App}) is the two-qudit maximally entangled state (\ref{maxentstated}) as there exist a unitary $V:\mathbbm{C}^d\rightarrow\mathbbm{C^d}$ such that $V\ket{i}=\ket{e_i}$. Let us now perform following operation
\begin{eqnarray}
V\otimes V^*\frac{1}{\sqrt{d}}\sum_{i=0}^{d-1}\ket{i}\ket{i}=\frac{1}{\sqrt{d}}\sum_{i=0}^{d-1}\ket{e_i}\ket{e_i^*}.
\end{eqnarray}
Now using the Fact \ref{factmaxent}, we arrive at $V\otimes V^*\ket{\phi_d^{+}}=VV^{\dagger}\otimes\Id\ket{\phi_d^{+}}=\ket{\phi_d^{+}}$. Thus, the two-qudit maximally entangled state remains invariant under the action of the unitary of the form $V\otimes V^*$.
As a consequence, we finally arrive at the simplified version of the state \eqref{SchmidtApp} given by
\begin{equation}\label{PW}
(\Id_A\otimes U_B)\ket{\psi}_{AB}=(P_A\otimes \Id_B)\ket{\phi^{+}_d}.
\end{equation}
Substituting the above state \eqref{PW} to the relation \eqref{SOS253} for $k=1$, we obtain that
\begin{eqnarray}
(A_yP_A\otimes \widetilde{B}_y)\ket{\phi^{+}_d}=(P_A\otimes \Id_B)\ket{\phi^{+}_d},
\end{eqnarray}
where $\widetilde{B}_y=U_B^{\dagger}\, \overline{B}_y\, U_B$. Again, employing Fact \ref{factmaxent} stated in Appendix \ref{chapano} which states that  Thus, we finally arrive at
\begin{eqnarray}
(A_yP_A\widetilde{B}_y^T\otimes \Id_B)\ket{\phi^{+}_d}=(P_A\otimes \Id_B)\ket{\phi^{+}_d}.
\end{eqnarray}
Now, taking the partial trace over subsystem $B$, we arrive at 
\begin{equation}\label{CULO-A}
A_yP_A\widetilde{B}_y^T=P_A.
\end{equation}
Multiplying the above equation with its hermitian conjugate from the right hand side, we arrive at 
\begin{equation}\label{CULO-B}
\left(A_yP_A\widetilde{B}_y^T\right)\left(\widetilde{B}_y^*P_A A_y^{\dagger}\right)=P_A^2,
\end{equation}
where we used the fact that $P_A$ is hermitian, that is, $P_A=P_A^{\dagger}$. Recalling that
$\widetilde{B}_y$ are unitary allows us to conclude that
\begin{eqnarray}
\widetilde{B}_y^T\widetilde{B}_y^*=(\widetilde{B}_y^{\dagger}\widetilde{B}_y)^*=\Id_d
\end{eqnarray}
and thus from \eqref{CULO-B} we finally arrive at
\begin{equation}
    A_y\ P_A^2\ A_y^{\dagger}=P_A^2.
\end{equation}
Using the fact that $A_y$ are unitary, we arrive at a simple condition for $P_A$ that for every $y$ the commutator of $A_y$ and $P_A^2$ is zero, that is, $[A_y,P_A^2]=0$. Since,  
$P_A\geq0$, this condition is equivalent to
\begin{equation}
    [A_y,P_A]=0\quad \forall y.
\end{equation}
Taking into account that $A_y$
are genuinely incompatible, Lemma \ref{lemma4} implies that $P_A$
is proportional to identity or simply $P_A=\lambda\Id_A$ for some $\lambda\in \mathbbm{C}$.
Plugging this form of $P_A$ into Eq. (\ref{CULO-A}) we deduce that 
\begin{eqnarray}
\widetilde{B}_y=U_{B}\overline{B}_{y}U_{B}^{\dagger}=A_y^*
\end{eqnarray}

Let us now go back to every state $\ket{\psi_s}_{AB}$ and reconsider the condition \eqref{SOS253}. Now as concluded above for a particular $s$, there exist a local transformation $U_{B,s}:\mathbbm{C}^d\rightarrow\mathbbm{C}^d$  for every $s$ that transforms Bob's observables acting on the support of $\rho_{B,s}$ as
\begin{eqnarray}
\widetilde{B}_{y,s}=U_{B,s}\overline{B}_{y,s}U_{B,s}^{\dagger}=A_y^*.
\end{eqnarray}
Also, from \eqref{state2App} we get that up to a local transformation the state $\ket{\psi_s}_{AB}$ is the two-qudit maximally entangled state
\begin{eqnarray}\label{state_s}
(\Id_A\otimes U_{B,s})\ket{\psi_s}_{AB}=
\frac{1}{\sqrt{d}}\sum_{i=0}^{d-1}\ket{e_i}\ket{e_i^*}=\ket{\phi_+^d}.
\end{eqnarray}

%
%
Let us notice that the unitary transformation $U_{B,s}$
might be different for different $s$. Moreover, they also act on different subspaces of Bob's local Hilbert space. In the last part of the proof, we show that these subspaces are mutually orthogonal and thus we arrive at the form of state $\rho_{AB}$ as \eqref{lem1.2} and Bob's measurement as \eqref{lem1.1}. For this, let us first rewrite the state $\ket{\psi_s}_{AB}$ \eqref{state_s} as 
\begin{equation}\label{TintodeVerano2}
    \ket{\psi_s}_{AB}=[\Id\otimes (U_{B,s})^{\dagger}]\ket{\phi_+^d}=\frac{1}{\sqrt{d}}\sum_{i=0}^{d-1}\ket{e_i'}\ket{g_{i,s}},
\end{equation}
where the vectors $\ket{g_{i,s}}=(U_{B,s})^{\dagger}\ket{e_i'^*}$ form an orthonormal basis in $\mathbbm{C}^d$
for any $s$. Note that for convenience, we expressed the state \eqref{TintodeVerano2} in the eigenbasis of $A_0$ given by $\{\ket{e_i'}\}$. This is well justified as the two-qudit maximally entangled state remains invariant under application of the unitary $V\otimes V^{*}$ as shown above. The support of the local state $\rho_{B,s}$ is spanned by the vectors $\ket{g_{i,s}}$, that is,
\begin{equation}
    \mathrm{supp}(\rho_{B,s})\equiv V_s=\mathrm{span}\{\ket{g_{0,s}},\ket{g_{1,s}},\ldots,\ket{g_{d-1,s}}\}\subset \mathcal{H}_B.
\end{equation}
Now, we show that all the local subspaces $V_s$ corresponding to 
the eigenstates $\ket{\psi_{s}}_{AB}$ of $\rho_{AB}$ \eqref{mixed1} are mutually orthogonal. To this end, let us consider two arbitrary eigenstates, for simplicity denoted by $\ket{\psi_1}_{AB}$ and $\ket{\psi_2}_{AB}$ and the corresponding local subspaces on Bob's side as $V_1$ and $V_2$. Let us now express Alice's observable $A_0$ using its eigendecomposition as $A_0=\sum_{i=0}^{d-1}\omega^i\proj{e_i'}$. For simplicity, in the rest of the proof we denote $\ket{e_i'}$ as $\ket{i}$.
Plugging $A_0$ and the certified state \eqref{TintodeVerano2} to the relation \eqref{SOS253} for $y=0$, $s=1,2$ and $k=1$, we have that
\begin{eqnarray}
\sum_{i=0}^{d-1}\omega^i\proj{i}\otimes B_0\,\left(\frac{1}{\sqrt{d}}\sum_{i=0}^{d-1}\ket{i}\ket{g_{i,s}}\right)=\frac{1}{\sqrt{d}}\sum_{i=0}^{d-1}\ket{i}\ket{g_{i,s}}\qquad s=1,2.
\end{eqnarray}
Multiplying $\bra{i}$ from the left hand side of the above equation, we have that
\begin{equation}\label{dupa}
    B_0\ket{g_{i,s}}=\omega^{-i}\ket{g_{i,s}}\qquad s=1,2.
\end{equation}
Thus, we clearly observe that 
both local bases $\{\ket{g_{i,1}}\}$ and $\{\ket{g_{i,2}}\}$ are the eigenbases of $B_0$.
Recalling that $B_0$ is unitary, we find some orthogonality relations among the vectors of the two bases specifically that eigenvectors corresponding to different eigenvalues must be orthogonal
\begin{equation}\label{Ortrelations}
    \langle g_{i,1}|g_{j,2}\rangle=0 \qquad (i\neq j).
\end{equation}
As a consequence, to prove that $V_1$ is orthogonal to $V_2$, it is now enough to show that eigenvectors corresponding to same eigenvalues are also orthogonal, or equivalently
\begin{eqnarray}
\langle g_{i,1}|g_{j,2}\rangle=0 \quad \forall i,j.
\end{eqnarray}
To this end, we consider the decomposition of the vectors belonging to subspace $V_2$ in terms of vectors belonging to $V_1$ kepping in mind the condition \eqref{Ortrelations} as 
\begin{equation}\label{orthogonal}
    \ket{g_{i,2}}=\alpha_i\ket{g_{i,1}}+\beta_i\ket{h_i},
\end{equation}
where $\alpha_i,\beta_i\in\mathbbm{C}$ and $|\alpha_i|^2+|\beta_i|^2=1$
and $\ket{h_i}$ is a normalized vector orthogonal to $\ket{g_{i,1}}$ for any $i$.
Again, using the condition (\ref{Ortrelations}), we clearly observe from Eq. (\ref{orthogonal}) that 
\begin{equation}\label{kolejne}
\beta_j\langle g_{i,1}|h_j\rangle=0\qquad (i,j=0,\ldots,d-1).
\end{equation}
The above equation has two possible solutions, either $\beta_j=0$ or the vectors $\ket{h_j}$ are orthogonal to the whole subspace $V_1$.

Let us now revisit the conditions \eqref{SOS253} for $k=1$, $y=2,3,\ldots,N$ and $s=1,2$ written as
\begin{equation}
    (\Id_A\otimes B_y)\ket{\psi_s}_{AB}=(A_y^{\dagger}\otimes \Id_B)\ket{\psi_s}_{AB}\quad s=1,2.
\end{equation}
Using then the certified state $\ket{\psi_s}_{AB}$ given in Eq. (\ref{TintodeVerano2}) we have that
\begin{equation}
    \sum_{i=0}^{d-1}\ket{i}\otimes (B_y\ket{g_{i,s}})=\sum_{i=0}^{d-1}(A_y^{\dagger}\ket{i})\otimes \ket{g_{i,s}}.
\end{equation}
Multiplying with $\bra{i}$ on both sides of the above equation, we arrive at the following set of vector equations
\begin{equation}\label{Katarzynki1}
    B_y\ket{g_{i,s}}=\sum_{m=0}^{d-1}\langle i|A_y^{\dagger}|m\rangle\ket{g_{m,s}} \qquad s=1,2.
\end{equation}
for $i=0,\ldots,d-1$. Using the decomposition 
(\ref{orthogonal}) in (\ref{Katarzynki1}) for $s=2$,
leads us to 
\begin{equation}
\alpha_i B_y\ket{g_{i,1}}+\beta_iB_y\ket{h_i}=\sum_{m=0}^{d-1}\alpha_m\langle i|A_y^{\dagger}|m\rangle\ket{g_{m,1}}+\sum_{m=0}^{d-1}\beta_m\langle i|A_i^{\dagger}|m\rangle\ket{h_m}.
\end{equation}
Now, using Eq. (\ref{Katarzynki1}) for $s=1$ and substituting $B_y\ket{g_{i,1}}$ we have that
\begin{equation}
\sum_{m=0}^{d-1}(\alpha_i-\alpha_m)\langle i|A_y^{\dagger}|m\rangle\ket{g_{m,1}}
=\sum_{m=0}^{d-1}\beta_m\langle i|A_y^{\dagger}|m\rangle\ket{h_m}-\beta_i B_y\ket{h_i}.
\end{equation}
Multiplying the above equation with $\bra{g_{n,1}}$ on the left hand side, we obtain
\begin{equation}
(\alpha_i-\alpha_n)\langle i|A_y^{\dagger}|n\rangle
=-\beta_i \bra{g_{n,1}}B_y\ket{h_i},
\end{equation}
where we used that $\beta_m\bra{g_{n,1}}h_m\rangle=0$ for any $n,m$ \eqref{kolejne}. Again, using the condition (\ref{kolejne}) along with the property of $B_y$ acting invariantly on the subspace spanned by $\ket{g_{n,1}}$. This can also inferred from (\ref{Katarzynki1}) and thus the right-hand side of the above equation simply vanishes and we finally arrive at
\begin{equation}
(\alpha_i-\alpha_n)\langle i|A_y^{\dagger}|n\rangle=0\qquad (i,n=0,\ldots,d-1).
\end{equation}
Now, consider a matrix $Q=\sum_{i=0}^{d-1}\alpha_i\proj{i}$ and observe that the left hand side of the above condition can be expressed as the commutator of $A_y^{\dagger}$ and $Q$, 
\begin{equation}
    [A_y^{\dagger},Q]=0\quad \forall y
\end{equation}
However, using the fact that $A_y$ are genuinely incompatible and Lemma \ref{lemma3}, the above condition can only hold if $Q=\alpha\Id$ for some $\alpha\in\mathbbm{C}$. This means that all $\alpha_i's$ are equal.

Let us now recall that the eigenstates $\ket{\psi_1}$ and $\ket{\psi_2}$ are orthogonal and thus using \eqref{TintodeVerano2}, we arrive at
\begin{equation}
    0=\langle \psi_{1}|\psi_2\rangle=\frac{1}{d}\sum_{i}\langle g_{i,1}|g_{k,2}\rangle=\frac{1}{d}\sum_{i}\alpha_i=\alpha.
\end{equation}
As $\alpha_i\geq0$, we have that $\alpha_i=\alpha=0$ for any $i$. Plugging it back to Eq. (\ref{orthogonal}),
we can clearly observe that $\ket{g_{i,2}}=\beta_i\ket{h_i}$ such that $\beta_i=\exp{(\mathbbm{i}\theta_i)}$. 
Finally, from \eqref{orthogonal} the inner product $\langle g_{i,2}|g_{i,1}\rangle=\alpha=0$. Thus, we can finally say that the subspaces $V_1$ and $V_2$ are mutually orthogonal. 
Applying the same argument by considering every pair of subspaces $V_j$ and $V_k$, allows us to conclude that every pair of the subspaces are mutually orthogonal. Thus, Bob's Hilbert space decomposes as
\begin{equation}
    \mathcal{H}_B= V_1\oplus V_2\oplus\ldots \oplus V_K,
\end{equation}
where each subspace $V_s$ is of dimension $d$, that is, $\dim V_s=d$. Equivalently, Bob's Hilbert space can be represented as $\mathcal{H}_B=(\mathbbm{C}^d)_{B'}\otimes\mathcal{H}_{B''}$ for some Hilbert space
$\mathcal{H}_{B''}$ of unknown but finite dimension. Another consequence of the subspaces $V_s$ being mutually orthogonal is that we can construct a unitary $U_B:\mathcal{H}_B\to\mathcal{H}_B$ such that 
\begin{eqnarray}\label{finalU}
U_B\ket{g_{i,s}}=\ket{i}_{B'}\otimes\ket{s}_{B''},
\end{eqnarray}
for $i=0,\ldots,d-1$ and $s=1,\ldots,K$ such that $\ket{s}$ is the computational basis over $\mathcal{H}_{B''}$. Thus, the states $\ket{\psi_s}_{AB}$ transform as
\begin{equation}\label{koniec2}
    (\Id_A\otimes U_B)\ket{\psi_s}_{AB}=\ket{\phi_+^d}_{AB'}\otimes\ket{s}_{B''}
\end{equation}
for every $s$. Let us now look at the state $\rho_{AB}$ and use the decomposition \eqref{mixed1} to get that
\begin{equation}\label{koniec}
    (\Id_A\otimes U_B)\rho_{AB}(\Id_A\otimes U_B^{\dagger})=\proj{\phi_+^d}\otimes\rho_{B''},
\end{equation}
such that $\rho_{B''}=\sum_{s}p_s\proj{s}_{B''}$. Note that $\ket{s}_{B''}$ is the eigenbasis
of $\rho_{B''}$ This is exactly the form of the state we wanted to prove (\ref{lem1.2}). 
To find the desired form of Bob's observables (\ref{lem1.1}),
we first notice that applying $U_B$ \eqref{finalU} to Bob's observables $B_y$ gives us  
\begin{equation}\label{Biblock}
 U_B\,B_y\,U_B^{\dagger}=\sum_{s,t=1}^K B_{y,s,t}\otimes\ket{s}\!\bra{t}_{B''},
\end{equation}
where $B_{y,s,t}$ are $d\times d$ blocks acting on $(\mathbbm{C}^d)_{B'}$. Plugging Eqs. (\ref{koniec}) and (\ref{Biblock}) into Eq. (\ref{SOSApp}) for $k=1$ and also recalling that Bob's measurements are projective,
we obtain
\begin{equation}\label{blocks}
    \sum_{s,t}(A_y\otimes B_{y,s,t})\proj{\phi_+^d}\otimes p_t\ket{s}\!\bra{t}_{B''}=\proj{\phi_+^d}\otimes
    \sum_{s}p_s\proj{s}_{B''}.
\end{equation}
Sandwiching the above equation with $\bra{s}.\ket{t}$, we get that for $s\ne t$ 
\begin{eqnarray}
(A_y\otimes B_{y,s,t})\ket{\phi_+^d}=0.
\end{eqnarray}
Since, $A_y$ is unitary, we can clearly see that $B_{y,s,t}=0$ for $s\neq t$. The terms of (\ref{blocks}) for $s=t$ gives us 
\begin{equation}
    (A_y\otimes B_{y,s,s})\ket{\phi_+^d}=\ket{\phi_{+}^d}.
\end{equation}
Due to Fact \ref{factmaxent} proven in Appendix \ref{chapano}, we have that  $B_{y,s,s}=A_i^{*}$ which on 
substitution to Eq. (\ref{Biblock}), finally gives us the exact form of Bob's observables \eqref{lem1.1}
\begin{equation}
     U_B\,B_i\,U_B^{\dagger}=\sum_{s=1}^K A_{i}^*\otimes\proj{s}_{B''}=A_{i}^*\otimes\Id_{B''}.
\end{equation}
This ends the proof.
\end{proof} 
\subsection{Weaker certification}
Let us now consider the quantum steering scenario as described above when the set of Alice's observables are not genuinely incompatible observables and thus they share a common invariant subspace. In this case, the saturation of the quantum bound \eqref{SteeringFunctional21} is insufficient to certify the Bob's full observable but only the part of it which acts on these subspaces. For example, consider two four-outcome observables on Alice's side as, 
\be \label{ex}
A_1 = \sum^3_{j=0} (\mathbbm{i})^j |j\ra\!\la j| ,\qquad A_2 = \sum^1_{j=0} (-1)^j \left( |-_j\ra\!\la -_j| +  \mathbbm{i}^{j+1} |+_j\ra\!\la +_j| \right)
\ee
where $|\pm_0\ra = (|0\ra \pm |1\ra)/\sqrt{2},$ $|\pm_1\ra = (|2\ra \pm |3\ra)/\sqrt{2}$. Any diagonal matrix $P_A$ of the form
\[
 P_A=\left(\begin{array}{c|c}
   \lambda_1 \Id_2  & \mathbb{O} \\
\hline
  \mathbb{O} & \lambda_2 \Id_2
\end{array}\right)
\]
satisfies the commutation relation $[P_A,A_y]=0$ for $y=1,2$ given in \eqref{ex}. Consequently, any bipartite state of local dimension four of the form
\be \label{ex1}
\ket{\psi_{AB}}=\lambda_1(|00\ra + |11\ra ) + \lambda_2 (|22\ra + |33\ra )\ee 
such that $\lambda^2_1+\lambda^2_2 =1$ gives the quantum bound of the steering functional \eqref{SteeringFunctional21}. Bob's observables can be certified on the subspace where the coefficients $\lambda_i\ne 0$. For instance, when Alice's observables are given in \eqref{ex}, then the quantum bound of steering functional \eqref{SteeringFunctional21} can achieved by the two-qubit maximally entangled state and Bob's observbales are given by $B_1=\proj{0}-\mathbbm{i}\proj{1}$ and $B_2=\proj{-_0}-\mathbbm{i}\proj{+_0}$. Thus, neither Bob's observables nor the state shared among the parties can be exactly certified if the set of observables on the trusted side are not genuinely incompatible.

\subsection{Robust certification}

Let us now study the robustness of our certification scheme against experimental defects that might not lead to achieving the exact quantum bound but a value little lower than it. A numerical approach was suggested in \cite{Jordi}, where a general scheme to robustly certify steerable assemblage was devised using the semi-definite programming. However, this approach is not applicable in our case because we consider systems of arbitrary local dimension $d$. A more challenging task would be to find analytical methods to address the considered problem. Here we find a simple technique to find robustness bounds of certification in the quantum steering scenario, when the trusted side chooses a family of genuinely $d-$outcome incompatible observables that are mutually unbiased bases. Specifically, the proof given below works when the steering functional is given by \eqref{SteeringFunctional21} and the Alice's observables are $A_1=X_dZ_d^l$ with $l=0,\ldots,d-1$ and $A_2=Z_d$. Unlike our exact certification scheme, for simplicity we assume here that the underlying state is pure and Bob's observables
zare projective. These two assumptions are well justified in a non-cryptographic scenario where there is no Eve who has access to the untrusted lab as well as the preparation device. In this scenario, the states and the measurements can always be purified by extending the Hilbert space of the untrusted party. 

\renewcommand{\thetheorem}{4.2}

\setcounter{thm}{0}

\begin{theorem}
Consider that Alice and Bob perform the quantum steering experiment and observe that the steering functional $W_{d,2}$ [\eqref{SteeringFunctional21} for $N=2$] attains a value close to the quantum bound $\beta_Q=2(d-1)$, that is,
\begin{eqnarray}\label{InexIn}
W_{d,2}=
   \sum_{k=1}^{d-1} \sum_{y=1}^2\left\langle A_{y}^k\otimes B_{y}^k\right\rangle\geq 2(d-1) -\varepsilon,
\end{eqnarray}
such that Bob's measurements are projective and Alice's observables are given by $A_1=X_dZ_d^{l}$ with $l=0,1,\ldots,d-1$ and $A_2=Z_d$. Let us say that this value is attained by the state 
$\ket{\psi_{AB}}\in\mathbbm{C}^d\otimes\mathcal{H}_B$ and observables $B_y \ (i=1,2)$, that are unitary with eigenvalues $1,\omega,\ldots,\omega^{d-1}$, acting on $\mathcal{H}_B$. Then, for any integer $d$ greater than or equal to two, there exist a unitary $U_B:\mathcal{H}_B\rightarrow\mathcal{H}_B$ such that:
\begin{enumerate}
    \item The state $\ket{\psi_{AB}}$ is close to the ideal state $\ket{\phi_+^d}$ up to a function of $\epsilon$, that is,
    \begin{eqnarray}\label{Rob1}
        \left\|\left(\Id_A\otimes U_B\right)\ket{{\psi_{AB}}}-\ket{\phi^+_d}\right\|
    \leq \sqrt{2(3d+1)}\sqrt[4]{2\varepsilon}.
    \end{eqnarray}
    \item Bob's observables are close to the ideal Bob's observables up to a function of $\epsilon$, that is,
    \begin{eqnarray}\label{Rob2}
\left\|U_BB_1^{k}U_B^{\dagger}-(X_dZ_d^{-l})^{k}\right\|_2\leq \sqrt{d}
\left(\sqrt{2\varepsilon}+2\sqrt{2(3d+1)}\sqrt[4]{2\varepsilon}\right)
\end{eqnarray}
and,
\begin{eqnarray}\label{Rob3}
\left\|U_BB_2^{k}U_B^{\dagger}-Z_d^{-k}\right\|_2\leq \sqrt{d}
\left(\sqrt{2\varepsilon}+2\sqrt{2(3d+1)}\sqrt[4]{2\varepsilon}\right)
\end{eqnarray}
with $k=0,\ldots,d-1$ and $\|\cdot\|_2$ stands for the Hilbert-Schmidt norm.
\end{enumerate}

\end{theorem}
\begin{proof} Let us first manipulate the condition
(\ref{InexIn}) to obtain a few inequalities that are crucial for the proof. As $A_y, B_y$ are unitary, the absolute value of its expectation values are bounded by one due to which we have that $\mathrm{Re}\left(\langle A_y^k\otimes B_y^k\rangle\right)\leq\left|\langle A_y^k\otimes B_y^k\rangle\right|\leq 1$ for all $y,k$. Further as discussed before, the maximum value of the expectation values in the steering functional \eqref{InexIn} is one. Thus, for every $y=1,2$ and $k=1,2,\ldots,d-1$ we have that
\begin{eqnarray}
\mathrm{Re}\left(\langle A_y^k\otimes B_y^k\rangle\right)+2d-3\geq  2d-2-\varepsilon
\end{eqnarray}
which on simplification yields us that
\begin{eqnarray}\label{App:1}
\left|\langle\psi| A_y^k\otimes B_y^k|\psi\rangle\right|\geq \mathrm{Re}\left(\langle\psi| A_y^k\otimes B_y^k|\psi\rangle\right)\geq 1-\frac{\varepsilon}{2} \geq 1-\varepsilon.
\end{eqnarray}
Here, we also used the fact the absolute value of a complex number is always greater than or equal to its real value, that is, $|z|\geq$ Re$(z)$ for any complex number $z$.

Another observation that is important in our proof is that any bipartite state $\ket{\psi}$ of local dimension $d$ can be expressed in the computational basis as,
\begin{eqnarray}\label{stateRob}
\ket{\psi}=\sum_{i=0}^{d-1}\alpha_i\ket{i}\ket{b_i},
\end{eqnarray}
where $\{\ket{b_i}\}$ are set of $d$ normalised vectors belonging to $\mathcal{H}_B$ that might not be orthogonal. Also $\alpha_i$ are real numbers greater than or equal to zero that satisfy the normalisation condition $\sum_{i=0}^{d-1}\alpha_i^2=1$. 

The proof for bounding the distance between the actual state $\ket{\psi}$ and the ideal state $\ket{\phi_+^d}$ is divided into two major parts. We first show that $\alpha_i's$ are close to $1/\sqrt{d}$. Then, we show that the real part of the terms $\langle i|b_i\rangle$ are close to 1. The proof for finding robustness of the measurements \eqref{Rob2} and \eqref{Rob3} is quite straightforward and follows directly from the robustness of state \eqref{Rob1} using simple manipulations. 

Let us first consider the inequality (\ref{App:1}) for $y=1$, and then substitute into it $A_1=X_dZ_d^l$ and the state  \eqref{stateRob}, which gives
\begin{eqnarray}
   \mathrm{Re}\left(\sum_j\alpha_j\bra{j}\!\bra{b_j}\left[ \left(X_dZ_d^l\right)^{k}\otimes B_1^k\right]\sum_i\alpha_i\ket{i}\ket{b_i}\right)\geq 1-\varepsilon
\end{eqnarray}
for every $k$. Notice that 
$(X_dZ_d^l)^{k}\ket{i}=\omega^{kl\left(i+\frac{k-1}{2}\right)}\ket{i+k}$, where the sum $i+k$ is modulo $d$. Thus, we have that
\begin{equation}
\sum_{i}\alpha_i\alpha_{i+k}\mathrm{Re}\left(\omega^{kl\left(i+\frac{k-1}{2}\right)}\langle b_{i+k}|B_1^k|b_i\rangle\right)\geq 1-\varepsilon.
\end{equation}
Using the fact that
\begin{eqnarray}
    \mathrm{Re}\left(\omega^{kl\left(i+\frac{k-1}{2}\right)}\langle b_{i+k}|B_1^k|b_i\rangle\right)\leq\left|\omega^{kl\left(i+\frac{k-1}{2}\right)}\langle b_{i+k}|B_1^k|b_i\rangle\right|\leq\left|\langle b_{i+k}|B_1^k|b_i\rangle\right|\leq  1,
\end{eqnarray}
where we used that $\ket{b_i}$ are normalised, we have finally have that
\begin{equation}
\sum_{i=0}^{d-1}\alpha_i\alpha_{i+k}\geq 1-\varepsilon\quad \forall k.
\end{equation}
Notice that the above relation holds trivially for $k=0$ from the normalisation condition. Summing the above relation over $k$, we obtain
\begin{equation}\label{App:3}
\sum_{i,k=0}^{d-1}\alpha_i\alpha_{i+k}=\left(\sum_{i=0}^{d-1}\alpha_{i}\right)^2\geq d(1-\varepsilon),
\end{equation}
which gives 
\begin{equation}\label{App:2}
\sum_{i=0}^{d-1}\alpha_{i}\geq \sqrt{d}\sqrt{1-\varepsilon}.
\end{equation}
Let us now consider the following expression,
\begin{eqnarray}
    \sum_{i=0}^{d-1}\left(\alpha_i-\frac{1}{\sqrt{d}}\right)^2&=&\sum_{i=0}^{d-1}
    \alpha_i^2-\frac{2}{\sqrt{d}}\sum_{i=0}^{d-1}\alpha_i+1\nonumber\\
    &=&2\left(1-\frac{1}{\sqrt{d}}\sum_{i=0}^{d-1}\alpha_i\right),
\end{eqnarray}
where we used the normalisation condition $\sum_{i}\alpha_i^2=1$. The right hand side of the above equation can be upper bounded using (\ref{App:2}) and thus we can finally conclude that $\alpha_i's$ are close to $1/\sqrt{d}$ by a factor of $\sqrt{2\varepsilon}$, that is,
\begin{eqnarray}
    \sum_{i=0}^{d-1}\left(\alpha_i-\frac{1}{\sqrt{d}}\right)^2&\leq &
    2(1-\sqrt{1-\varepsilon})\nonumber\\
    &\leq& 2\varepsilon,
\end{eqnarray}
where we used the fact that $\sqrt{1-\varepsilon}\geq 1-\varepsilon$ for any $0\leq\varepsilon\leq1$ and thus,
\begin{equation}\label{App:4.1}
\frac{1}{\sqrt{d}}-\sqrt{2\varepsilon}\leq \alpha_i\leq 
\frac{1}{\sqrt{d}}+\sqrt{2\varepsilon}.
\end{equation}
Considering the expression $\sum_{i=0}^{d-1}\left(\alpha_i\alpha_j-\frac{1}{\sqrt{d}}\right)^2$ for any $i,j=0,\ldots,d-1$, in a similar manner as done above we can conclude that 
\begin{equation}\label{App:4}
\frac{1}{d}-\sqrt{2\varepsilon}\leq \alpha_i\alpha_{i+j}\leq 
\frac{1}{d}+\sqrt{2\varepsilon}.
    \end{equation}
We now consider the condition (\ref{App:1}) for $y=2$ by substituting $A_2=Z_d$ and the state \eqref{stateRob}
\begin{eqnarray}
    \mathrm{Re}\left(\sum_j\alpha_j\bra{j}\!\bra{b_j}\left[ Z_d^{k}\otimes B_2^k\right]\sum_i\alpha_i\ket{i}\ket{b_i}\right)\geq 1-\varepsilon\quad \forall k.
\end{eqnarray}
Notice that $Z_d^k\ket{i}=\omega^{ik}\ket{i}$ and thus the above equation simplifies to
\begin{equation}
    \sum_{i=0}^{d-1}\alpha_i^2\,\mathrm{Re}\left(\omega^{ik}\langle b_i|B_2^k|b_i\rangle\right)\geq 1-\varepsilon.
\end{equation}
Again, using the property that
\begin{eqnarray}\label{proprob}
    \mathrm{Re}(\omega^{ik}\langle b_i|B_2^k|b_i\rangle)\leq|\omega^{ik}\langle b_i|B_2^k|b_i\rangle|\leq|\langle b_i|B_2^k|b_i\rangle|\leq 1
\end{eqnarray}
for every $i,k$ as $\ket{b_i}$ is normalised. Thus, we can choose an index $j$ such that
\begin{equation}
    \sum_{i\neq j}\alpha_i^2+\alpha_j^2\,\mathrm{Re}\left(\omega^{jk}\langle b_j|B_2^k|b_j\rangle\right)\geq 1-\varepsilon.
\end{equation}
Using the normalisation condition $\sum_i\alpha_i^2=1$, we arrive at
\begin{equation}
    \alpha_j^2\left[1-\mathrm{Re}\left(\omega^{jk}\langle b_j|B_2^k|b_j\rangle\right)\right]\leq \varepsilon.
\end{equation}
Again using the property \eqref{App:4},
and then the inequality (\ref{App:4.1}), we arrive at
\begin{eqnarray}
    \frac{1}{d}\left[1-\mathrm{Re}\left(\omega^{jk}\langle b_j|B_2^k|b_j\rangle\right)\right] &\leq& \varepsilon+\sqrt{2\varepsilon}\left[1-\mathrm{Re}\left(\omega^{jk}\langle b_j|B_2^k|b_j\rangle\right)\right]\nonumber\\&\leq&  \varepsilon+2\sqrt{2\varepsilon} \nonumber\\&\leq&  3\sqrt{2\varepsilon}, 
\end{eqnarray}
where we used the fact that the maximum of the term $\left[1-\mathrm{Re}\left(\omega^{jk}\langle b_j|B_2^k|b_j\rangle\right)\right]$ is two. Thus, we can finally conclude that
\begin{equation}\label{App:5}
\mathrm{Re}\left(\omega^{jk}\langle b_j|B_2^k|b_j\rangle\right)\geq 1-3d\sqrt{2\varepsilon}.
\end{equation}
Recall that the eigenvalues of $B_2$ is $\{1,\omega,\ldots,\omega^{d-1}\}$ and thus we can consider the eigendecomposition of $B_2$ using orthogonal projections $P_i$ as
\begin{equation}
    B_2=\sum_{i=0}^{d-1}\omega^i P_i.
\end{equation}
Here $P_i$ in general might be of rank higher than one.
Plugging this decomposition into the condition (\ref{App:5}), we have that
\begin{equation}
    \sum_{i=0}^{d-1}\mathrm{Re}\left(\omega^{(i+j)k}\langle b_j|P_i|b_j\rangle\right)\geq 1-3d\sqrt{2\varepsilon}\quad \forall k.
\end{equation}
Notice that the above equation holds trivially for $k=0$. Taking a sum over $k$ gives us 
\begin{equation}\label{App:6}
    \langle b_j|P_{-j}|b_j\rangle\geq 1-3d\sqrt{2\varepsilon},
\end{equation}
which implies that each vector $\ket{b_j}$ is close to the projection $P_{-j}$ which signifies the subspace corresponding to the $(d-j)-th$ outcome of $B_2$. Now, let us look at the following normalised vectors
\begin{eqnarray}
    \ket{\overline{v}_j}=\frac{P_{-j}\ket{b_j}}{\|P_{-j}\ket{b_j}\|},
\end{eqnarray}
where we can infer that $\ket{\overline{v}_j}$ are mutually orthogonal as well, that is, $\bra{\overline{v}_i}\overline{v}_j\rangle=\delta_{i,j}$. Note that, from the condition (\ref{App:6}) the normalisation factor of each of the vector $\ket{\overline{v}_j}$ is close to one, that is,  $\|P_{-j}\ket{b_j}\|\geq 1-2d\sqrt{2\varepsilon}$. Further, using these vectors we can express $B_2$ as
\begin{equation}
    B_2=\sum_{i=0}^{d-1}\omega^{-i}\proj{\overline{v}_i}\oplus B_2',
\end{equation}
such that $B_2'$ is some operator acting on the support orthogonal to the subspace spanned by the vectors $\{\ket{\overline{v}_i}\}$. Now, there always exist a unitary $U_B:\mathcal{H}_B\rightarrow\mathcal{H}_B$ such that $U_B\ket{\overline{v}_i}=\ket{i}$ and thus,
\begin{equation}\label{App:65}
    U_B B_2 U_B^{\dagger}=\sum_{i=0}^{d-1}\omega^{-i}\proj{i}\oplus B_2''.
\end{equation}
Substituting the above form of $B_2$ to (\ref{App:5}) and then taking a sum over $k$, we can also deduce that
\begin{equation}\label{App:7}
    \mathrm{Re}\left(\langle b_i|U_B^{\dagger}|i\rangle\right)\geq 1-3d\sqrt{2\varepsilon}.
\end{equation}

Finally, we can compute the distance between the states \eqref{Rob1} using the above results as,
\begin{eqnarray}\label{App:10}
    \left\|\Id_A\otimes U_B\ket{{\psi}}-\ket{\phi^+_d}\right\|&=&\left\{2\left[1-\mathrm{Re}(\langle{\psi}|U_B^{\dagger}|\phi_{d}^+\rangle)\right]\right\}^{1/2}\nonumber\\
    &=&\left\{2\left[1-\frac{1}{\sqrt{d}}\sum_{i}\alpha_i\,\mathrm{Re}
    (\langle b_i|U_B^{\dagger}|i\rangle)\right]\right\}^{1/2}.
\end{eqnarray}
Now, using the conditions (\ref{App:2}) and (\ref{App:7}), we have
\begin{eqnarray}\label{2.129}
\sum_{i}\alpha_i\,\mathrm{Re}
    (\langle b_i|U_B^{\dagger}|i\rangle)
    &\geq &(1-3d\sqrt{2\varepsilon})\sum_{i}\alpha_i\nonumber\\
    &\geq & (1-3d\sqrt{2\varepsilon})\sqrt{d}\sqrt{1-
    \varepsilon}\nonumber\\
    &\geq & \sqrt{d}\left[1- (3d+1)\sqrt{2\varepsilon}\right],
\end{eqnarray}
where the inequality in the third line is a consequence of two inequalities, first, $\sqrt{1-\varepsilon}\geq 1-\varepsilon$ for any $\varepsilon\leq 1$ and second, $\varepsilon\leq \sqrt{2\varepsilon}$. Using the inequality \eqref{2.129}, we finally obtain the robustness of the state \eqref{Rob1} as
\begin{eqnarray}
    \left\|\left(\Id_A\otimes U_B\right)\ket{{\psi}}-\ket{\phi^+_d}\right\|\leq \sqrt{2(3d+1)}\sqrt[4]{2\varepsilon}.
\end{eqnarray}

Let us now prove the inequalities (\ref{Rob2}) and \eqref{Rob3}. For simplicity, we denote the ideal observables as $B_i'$ and $U_BB_iU_B^{\dagger}=\widetilde{B}_i$. We first observe that the Hilbert Schmidt norm can be written as a vector norm using $\ket{\phi_+^d}$ as
\begin{equation}\label{App:9}
  \left\|\widetilde{B}_i^{k}-(B'_i)^{k}\right\|_2=\sqrt{d}\left\|\left[\widetilde{B}_i^{k}-(B'_i)^{k}\right]\ket{\phi_d^{+}}\right\|.
\end{equation}
Then, using the triangle inequality $|A+B|\geq\left||A|-|B|\right|$ and denoting $\ket{\widetilde{\psi}}=\Id_A\otimes U_B\ket{\psi}$, we have that
\begin{eqnarray}
    \left\|\widetilde{B}_i^{k}\ket{\widetilde{\psi}}-(B'_i)^{k}\ket{\phi_d^+}\right\|&=&
    \left\|\widetilde{B}_i^{k}\ket{\widetilde{\psi}}-(B'_i)^{k}\ket{\phi_d^+}+(B'_i)^k\ket{\phi_d^+}-\widetilde{B}_i^k\ket{\phi_d^+}\right\|\nonumber\\
    &\geq&\left\|\left[\widetilde{B}_i^{k}-(B'_i)^k\right]\ket{\phi_d^+}\right\|-\left\|\widetilde{B}_i^{k}(\ket{\widetilde{\psi}}-\ket{\phi_d^+})\right\|,
\end{eqnarray}
Then using using the inequality (\ref{App:9}) and the fact that $B_i$ is unitary, leads us to
\begin{equation}
    \left\|\widetilde{B}_i^{k}-(B'_i)^{k}\right\|_2\leq  \sqrt{d}\left(\left\|\widetilde{B}_i^{k}\ket{\widetilde{\psi}}-(B'_i)^{k}\ket{\phi_d^+}\right\|+\left\|\ket{\widetilde{\psi}}-\ket{\phi_d^+}\right\|\right).
\end{equation}
The first term in the right hand side of the above inequality can be computed using (\ref{App:10}) by first noting that $A_i ^k\otimes B_i'^k\ket{\phi_+^d}=\ket{\phi_+^d}$ and then using the fact that $A_i$ is unitary, 
\begin{eqnarray}\label{App:8}
\left\|(A_i^{k}\otimes \widetilde{B}_i^{k})\ket{\widetilde{\psi}}-\ket{\widetilde{\psi}}+\ket{\widetilde{\psi}}-\ket{\phi^+_d}\right\|\leq\left\|(A_i^{k}\otimes \widetilde{B}_i^{k})\ket{\widetilde{\psi}}-\ket{\widetilde{\psi}}\right\|+\left\|\ket{\widetilde{\psi}}-\ket{\phi^+_d}\right\|.
\end{eqnarray}
Finally using the conditions (\ref{App:1}) we get
\begin{eqnarray}\label{App:85}
    \left\|(A_i^{k}\otimes \widetilde{B}_i^{k})\ket{\widetilde{\psi}}-\ket{\widetilde{\psi}}\right\|&=&\left\|(A_i^{k}\otimes B_i^{k})\ket{\psi}-\ket{\psi}\right\|\nonumber\\
    &=&\left\{2\left[1-\mathrm{Re}\left(\langle \psi|A_i^{k}\otimes B_i^{k}|\psi\rangle\right)\right]\right\}^{1/2}\nonumber\\
    &\leq &\sqrt{2\varepsilon}.
\end{eqnarray}
Using the above inequality, we get that
\begin{eqnarray}
\left\|(A_i^{k}\otimes \widetilde{B}_i^{k})\ket{\widetilde{\psi}}-\ket{\widetilde{\psi}}+\ket{\widetilde{\psi}}-\ket{\phi^+_d}\right\|\leq\sqrt{2\varepsilon}+2\left\|\ket{\widetilde{\psi}}-\ket{\phi^+_d}\right\|,
\end{eqnarray}
and then using the condition (\ref{Rob1}) we obtain 
(\ref{Rob2}) and \eqref{Rob3}.
\end{proof}
\section{Conclusions and discussions}
 We proposed a one-sided device-independent scheme for a certification of a large family of incompatible measurements with arbitrary number of outcomes termed here  genuinely incompatible. Our scheme also allows for the certification of the two-qudit maximally entangled of any arbitrary local dimension using only two measurements on each side. This is the first certification of mutually unbiased bases of any dimension using quantum steering. Unlike the previous approaches in literature, our scheme is more general in the sense that we do not assume that the state shared between the parties is pure or the measurements on the untrusted side are projective. This makes our scheme even significant in cryptographic scenarios where there can be an external Eve who can have access to the state as well as the untrusted measurements. We also find a simple technique for robust certification of a smaller family of genuinely incompatible observables including the complete mutually unbiased bases of any prime dimension and pair of them for non-prime dimensions. We also considered the scenario when the measurements are not genuinely incompatible and observed that only a part of the measurement can be certified based on the quantum state that realises the quantum bound of the steering functional.

One of the drawbacks of the 1SDI scheme as compared to fully device-independent schemes is that one has to assume that one of the parties is trusted. However, they still posses some of the essential features that would be useful in certification of quantum technologies.  First, if one possesses a well characterised quantum device, here it is the measuring device of the trusted party, our scheme provides a way to compare any other untrusted device with the trusted device using minimal resources. In other words, our scheme given a trusted measurement device allows one to verify that any other device performs the desired measurements. Second, 
our scheme is applicable to every 1SDI scenario where a client wants to verify the state supplied by an untrusted source along with the untrusted measuring device. Thus, one-sided quantum key distribution schemes \cite{Crypto1, Crypto2} and randomness generation would be two major applications of our certification scheme. Third, even from a practical point of view, implementing 1SDI protocols is much easier than the DI protocols \cite{Crypto1}. The reason being that demonstration of quantum steering is practically more robust to noise and can be observed with detectors of lower efficiencies than Bell violations \cite{steeexp1, steeexp2}. Fourth, there are a very few analytical methods dedicated to the field of quantum certification, let alone to higher dimensional quantum certification. Thus, purely from a mathematical perspective, we introduced techniques that can be relevant for other schemes that aim to characterise arbitrary dimensional quantum systems. Finally, in certain scenarios 1SDI schemes can be made device-independent if one can device-independently characterise the measuring device of the trusted party as pointed out recently in Ref. \cite{sarkar2}.

Some interesting follow-up problems arise from our work. The most important among them would be whether our 1SDI scheme can be made fully device-independent, that is, the question whether it is possible to design a scheme for certifying every set of genuinely incompatible observables in a device-independent way still remains open. Another interesting problem would be to find 1SDI scheme that can be used for certification of incompatible observables that are  not genuine incompatible. It would also be interesting to construct steering functionals whose quantum bound is saturated by non-maximally entangled state of arbitrary local dimension $d$ and thus eventually finding a scheme allowing for certification of every bipartite entangled state. This problem is tackled in the next chapter of this thesis. An ambitious problem would be extend the idea of genuine incompatibility to non-projective schemes and then find steering functionals for certification of POVM's. 
%

%% file: Sections/chapter_5.tex
\chapter{Certification of any pure bipartite entangled state and optimal randomness using quantum steering}
\label{chapter_5}
\vspace{-1cm}
\rule[0.5ex]{1.0\columnwidth}{1pt} \\[0.2\baselineskip]
\definecolor{airforceblue}{rgb}{0.36, 0.54, 0.66}
\definecolor{bleudefrance}{rgb}{0.19, 0.55, 0.91}

\section{Introduction}
Generating genuine random outputs inaccessible to hackers is one of the key steps in any key distribution protocol, be it classical or quantum. As discussed before in Chapter \ref{chapter_2}, in classical physics the security of such protocols relies on the fact that large numbers can not be efficiently factorized using classical computers. However, using quantum computers such protocols can be broken in polynomial time. As was shown in \cite{di4}, violation of Bell inequalities serve as the most secure way to certify genuine randomness. However,
from a practical point of view, performing a Bell experiment is extremely challenging as one requires very low levels of noise along with the detectors being highly efficient. As a consequence, we need to consider scenarios where one can efficiently generate secure randomness, easier to implement, require minimal resources and are robust to noise. 
We showed in Chapter \ref{chapter_3} that one can securely generate $\log_2d$ bits of randomness using a quantum system of dimension $d$. From a foundational point of view, it still remains an open and highly non-trivial problem whether one can securely generate the optimal amount of randomness using a quantum system of arbitrary dimension $d$ which is $2\log_2d$ bits.

In this chapter, we aim to solve the above problem by considering the one-sided device-independent (1SDI) scenario where the required resource is quantum steering. As a matter of fact, it was shown in \cite{Crypto1} that quantum steering can be observed using detectors with much lower efficiency and more noise-robust when compared to observing Bell non-locality. This makes 1SDI setting an ideal scenario to construct protocols for randomness generation that can be practically implemented. Randomness generation in 1SDI setting has been recently explored in \cite{steerand1,Dani2018}. Particularly in Ref. \cite{Dani2018}, the authors  construct a protocol for generating $\log_2d$ bits of randomness from a quantum system of dimension $d$ in a secure way by using $d$-outcome projective measurements.   

To this end, we first construct a family of steering inequalities maximally violated by any pure entangled state of local dimension $d$ and two $d-$outcome measurements on each side. Using the maximal violation of the steering inequalities, we then certify any pure bipartite entangled state and a pair of mutually unbiased bases of arbitrary dimension on the untrusted side. This is the first instance, where any pure bipartite entangled state can be certified using the least number of measurements required to observe quantum non-locality, that is, using only two measurements on each side. Additionally, we demonstrate that any rank-one extremal measurement can be certified using our protocol. Based on these results, we finally demonstrate the certification of $2\log_2d$ bits of randomness using the certified entangled state of local dimension $d$ and the certified $d^2-$outcome extremal measurement. We further show that for systems of dimension $d=3,4,5,6$, the optimal amount of randomness can be certified using partially entangled states. This further strengthens our scheme as one can generate highest amount of randomness by employing less resource in terms of entanglement.  


\section{Family of steering inequalities}
Let us shortly describe again the quantum steering scenario introduced in Chapter \ref{chapter_2} and also in Chapter \ref{chapter_4} in an analogous way to Bell scenario. Alice and Bob are located in spatially separated labs. Both of them receive two subsystems from a preparation device. Alice is trusted and performs two known $d-$outcome measurements on the received subsystem labelled by $x=1,2$. In our case, we consider these measurements in the observable form as $A_0=Z_d$ and $A_1=X_d$ which as described in Eq. \eqref{X,Z} of Chapter \ref{chapter_4}, constitute a pair of mutually unbiased bases. Bob also performs two $d-$outcome measurements on his subsystem labelled by $y=1,2$. They collect enough statistics to construct the joint probability distribution $\{p(a,b|x,y)\}$. The scenario is depicted in Fig.  \ref{fig3.1} of Chapter \ref{chapter_4} such that $N=2$.

We now construct a family of steering inequalities which is expressed in the observable picture and using a collection of positive non-zero numbers $\boldsymbol{\alpha}=\{\alpha_0,\alpha_1,\ldots,\alpha_{d-1}\}$ such that $\sum_{i=0}^{d-1}\alpha_i^2=1$, as
\begin{equation}\label{Stefn1}
I_{2,d,2}(\boldsymbol{\alpha})=\sum_{k=1}^{d-1}\left\langle A_0^{k}\otimes B_{k|0} + 
\gamma(\boldsymbol{\alpha}) A_1^{k}\otimes B_{k|1}+\delta_k(\boldsymbol{\alpha})A_0^{k}\right\rangle\leq \beta_L(\boldsymbol{\alpha}),
\end{equation}
%
%
where the coefficients $\gamma(\boldsymbol{\alpha})$ and $\delta_k(\boldsymbol{\alpha})$ are given by
\begin{eqnarray}\label{value1}
\gamma(\boldsymbol{\alpha})=d\left({\sum_{\substack{i,j=0\\i\ne j}}^{d-1}\frac{\alpha_i}{\alpha_j}}\right)^{-1},\qquad \delta_k(\boldsymbol{\alpha})=-\frac{\gamma(\boldsymbol{\alpha})}{d}\sum_{\substack{i,j=0\\i\ne j}}^{d-1}\frac{\alpha_i}{\alpha_j}\omega^{k(d-j)}.
\end{eqnarray}
Notice that $\gamma(\boldsymbol{\alpha})\geq0$ for any choice of $\boldsymbol{\alpha}$ and $\delta_k(\boldsymbol{\alpha})$ are in general complex. Alice is trusted (or fully characterised), and her measurements are expressed in the observable picture as
\begin{equation}
    A_0=Z_d=\sum_{i=0}^{d-1}\omega^{i}\ket{i}\!\bra{i},\qquad A_1=X_d=\sum_{i=0}^{d-1}\ket{i+1}\!\bra{i}.
\end{equation}
Recall that the set of vectors $\{\ket{i}\}_{i=0}^{d-1}$ represents the computational basis in $\mathbbm{C}^d$. It is worth noting here that the expression $I_{2,d,2}(\boldsymbol{\alpha})$ \eqref{stefn1} is real for any choice of $\boldsymbol{\alpha}$. This is because $B_{d-k|i}=(B_{k|i})^{\dagger}$ for any generalised observable (see Eq. \eqref{obs2} of Chapter \ref{chapter_2}) and also that the coefficient $\delta_{d-k}(\boldsymbol{\alpha})=\delta_k^*(\boldsymbol{\alpha})$ for any $k$. 
Let us now express the steering functional in \eqref{stefn1} using the joint probability picture as
\begin{eqnarray}\label{steeprob}
I_{2,d,2}(\boldsymbol{\alpha})=d\sum_{a,b=0}^{d-1}c_{a,b}p(a,b|0,0)\!&+&\!\gamma(\boldsymbol{\alpha})\left(d\sum_{a,b=0}^{d-1}c_{a,b}p(a,b|1,1)-\sum_{\substack{i,a=0\\i\ne a}}^{d-1}\alpha_i\frac{p(a|0)}{\alpha_a}\right)\nonumber\\&-&\!1-\gamma(\boldsymbol{\alpha})-\delta_0(\boldsymbol{\alpha}),
\end{eqnarray}
such that
\begin{eqnarray}\label{c_steer1}
c(a,b)=\begin{cases}
1\quad \text{if}\quad a\oplus_d b=0\\
0\quad \text{otherwise}\ \ \ \ \ \ ,
\end{cases}
\end{eqnarray}
where $a\oplus_d b$ represents $a+b$ modulo $d$. To arrive at the above expression we used the fourier transform to express the expectation values in terms of the joint probabilities as defined in \eqref{4.4}.
Note from Eq. \eqref{value1} that $\delta_0(\boldsymbol{\alpha})=-1$, which implies that
\begin{eqnarray}\label{5.6}
I_{2,d,2}(\boldsymbol{\alpha})=d\sum_{a,b=0}^{d-1}c_{a,b}p(a,b|0,0)+\gamma(\boldsymbol{\alpha})\left(d\sum_{a,b=0}^{d-1}c_{a,b}p(a,b|1,1)-\sum_{\substack{i=0}}^{d-1}\alpha_i\sum_{a=0}^{d-1}\frac{p(a|0)}{\alpha_a}\right).\nonumber\\
\end{eqnarray}
Using the above expression of the steering functional, let us now compute the classical bound $\beta_L(\boldsymbol{\alpha})$ of the above steering inequality.



\subsection{Classical bound}
\label{app:ClassicalBound}

As derived before in Eq. \eqref{4.6} of Chapter \ref{chapter_4}, to find the classical bound of a steering functional, we need to consider an LHS model such that the joint probability distribution can be expressed as 
\begin{equation}\label{5.7}
    p(a,b|x,y)=\sum_{\lambda}p(\lambda)p(a|x,\rho_{\lambda})p(b|y,\lambda),
\end{equation}
where $\lambda$ denotes some unknown variables the collection of which is denoted by $\lambda$. These variables occur with a probability distribution $p(\lambda)$. Here $p(a|x,\rho_\lambda)$ is the local probability of Alice obtaining an outcome $a$ given the input $x$ and some quantum state $\rho_\lambda$ acts on $\mathbbm{C}^d$ \footnote{This is due to the fact that Alice is trusted and is known to perform quantum measurments on some quantum state.} and $p(b|y,\lambda)$ is the local probability distribution depending on some unknown variable $\lambda$.
For such a probability distribution \eqref{5.7}, the steering functional from Eq. \eqref{5.6} can be expressed as, 
\begin{eqnarray}\label{steeopprob}
I_{2,d,2}(\boldsymbol{\alpha}) &=& d \sum^{d-1}_{\substack{a=0}}\sum_{\lambda} \ p(\lambda)p(a|0,\rho_\lambda)p(d-a|0,\lambda)\nonumber\\ &+&\gamma(\boldsymbol{\alpha})\left(d \sum^{d-1}_{\substack{a=0}}\sum_{\lambda}  \ p(\lambda)p(a|1,\rho_\lambda)p(d-a|1,\lambda)-\sum_i\alpha_i\sum_{a=0}^{d-1}\sum_{\lambda}\frac{p(\lambda)p(a|0,\rho_\lambda)}{\alpha_a}\right),\nonumber\\
\end{eqnarray}
where to obtain the last term, we used the no-signalling conditions given in Eq. \eqref{NS1} of Chapter \ref{chapter_2} such that
\begin{eqnarray}
p(a|0)=\sum_bp(a,b|0,y)\qquad \forall y.
\end{eqnarray}


Notice that in the above expression we used the fact that $\sum_bp(b|y,\lambda)=1$ for any $\lambda$ and then denoted $\rho_A=\sum_{\lambda}p(\lambda)\rho_{\lambda}$.
%
Let us first consider the first two terms in Eq. (\ref{steeopprob}) and find their upper bound in the following way,
\begin{eqnarray}
I_{2,d,2}(\boldsymbol{\alpha}) &\leq& d \sum_{\lambda} \ p(\lambda)\max_{a}p(a|0,\rho_\lambda)\nonumber\\ &+&\gamma(\boldsymbol{\alpha})\left(d \sum_{\lambda}  \ p(\lambda)\max_{a}p(a|1,\rho_\lambda)-\sum_i\alpha_i\sum_{a=0}^{d-1}\sum_{\lambda}\frac{p(\lambda)p(a|0,\rho_\lambda)}{\alpha_a}\right),\nonumber\\
\end{eqnarray}
where to obtain the first inequality we used the fact that $\sum_ap(d-a|y,\lambda)=1$ for any $y$ and $\lambda$ and then used the mathematical identity $\sum_is_iw_i\leq \max_{i}\{s_i\}$ whenever $w_i\geq0$ and $\sum_iw_i=1$.
Now as was done before in Chapter \ref{chapter_4} [see Eq. \eqref{4.11}] and using the fact that $\sum_\lambda p(\lambda)=1$, we get that
\begin{equation}\label{5.12}
    I_{2,d,2}(\boldsymbol{\alpha})\leq\max_{\rho}\left[ d\max_{a}\{p(a|0,\rho)\}+\gamma(\boldsymbol{\alpha}) \left(d\max_{a}\{p(a|1,\rho)\}-\sum_{i=0}^{d-1}\alpha_i\sum_{a=0}^{d-1}\frac{p(a|0,\rho)}{\alpha_a}\right)\right].
\end{equation}
Notice the above expression is convex in the state $\rho$. As a consequence, the maximisation in the above expression can be taken over pure states $\ket{\psi}\in\mathbbm{C}^d$.
Now expressing the state $\ket{\psi}$ in the computational basis of $\mathbbm{C}^d$ as 
$\ket{\psi}=\sum_{i}\eta_i\ket{i}$,
and then plugging in Alice's observables, $A_0=Z_{d}$
and $A_1=X_d$, the formula \eqref{5.12} can be rewritten as
\begin{equation}\label{Leon}
    I_{2,d,2}(\boldsymbol{\alpha})\leq \max_{\substack{|\eta_{0}|,\ldots,|\eta_{d-1}|\\|\eta_0|^2+\ldots+|\eta_{d-1}|^2=1}}\left\{d\max_{a}\{|\eta_a|^2\}+\gamma(\boldsymbol{\alpha}) \left[\left(\sum_{i=0}^{d-1}|\eta_i|\right)^2-\sum_{i=0}^{d-1}\alpha_i\sum_{a=0}^{d-1}\frac{|\eta_a|^2}{\alpha_a}\right]\right\}.
\end{equation}
Given any arbitrary collection of positive numbers $\boldsymbol{\alpha}$ such that the sum of the squares of those numbers is one, it is not straightforward to find this bound. 
However, we can show here that the right hand side of the above formula 
is strictly less than $d$ for any $\boldsymbol{\alpha}$. We prove this claim using the technique of contradiction. Let us first consider the term inside the square brackets of the above expression \eqref{Leon} and show that it is always negative. For this, let us recall the Cauchy–Schwarz inequality for positive real numbers also known as Sedrakyan's inequality \cite{LI2},
\begin{eqnarray}\label{CS}
\frac{(\sum_iu_i)^2}{\sum_iv_i}\leq \sum_i\frac{u_i^2}{v_i}.
\end{eqnarray}
Now, substituting $u_i=|\eta_i|$ and $v_i=\alpha_i$, we can rewrite the above expression as
\begin{equation}
    \left(\sum_{i=0}^{d-1}|\eta_i|\right)^2\leq \sum_{i=0}^{d-1}\alpha_i\sum_{j=0}^{d-1}\frac{|\eta_i|^2}{\alpha_i}.
\end{equation}
Thus, we can conclude from (\ref{Leon}) that the the classical bound of $I_{2,d,2}(\boldsymbol{\alpha})$ is less than or equal to $d$, that is,
\begin{eqnarray}
     I_{2,d,2}(\boldsymbol{\alpha})\leq d\max_{a}\{|\eta_a|^2\}\leq d.
\end{eqnarray}
We now show that the above inequlity can not be saturated by L.H.S. models. To this end, let us assume that $I_{2,d,2}(\boldsymbol{\alpha})=d$. This implies that the term inside the square brackets in (\ref{Leon}) vanishes and thus,
\begin{eqnarray}\label{Asturias}
\left(\sum_{i=0}^{d-1}|\eta_i|\right)^2= \sum_{i=0}^{d-1}\alpha_i\sum_{j=0}^{d-1}\frac{|\eta_i|^2}{\alpha_i}.
\end{eqnarray}
along with the first term
\begin{equation}\label{OtherCond}
 \max_a\{|\eta_a|^2\}=1.   
\end{equation}
Recall that equality holds in the Cauchy–Schwarz inequality \eqref{CS} iff $u_i=\kappa v_i$ for all $i$ where $\kappa$ is some real coefficient. Thus, in (\ref{Asturias}) $\alpha_i=\kappa |\eta_i|$ for each $i$. Using the condition $\sum_i\alpha_i^2=\sum_i|\eta_i|^2=1$ and that $\alpha_i>0$ for any $i$, imposes that $\kappa=1$, and therefore $\alpha_i=|\eta_i|$. But $\alpha_i<1$ for any $i$ which contradicts the second condition (\ref{OtherCond}). Thus, our initial assumption is wrong which implies that the maximal classical value of the steering function $I_{2,d,2}(\boldsymbol{\alpha})$ is strictly less than $d$ for arbitrary collection of positive numbers $\alpha_i$ such that $\alpha_0^2+\ldots+\alpha_{d-1}^2=1$. 
Now, we move onto finding the quantum bound of the steering functional in \eqref{stefn1}.

\subsection{Quantum bound} 
\label{app:SOS}

Here we show that the maximum value of the steering functional $I_{2,d,2}(\boldsymbol{\alpha})$ in Eq. \eqref{Stefn1} obtainable using quantum states and measurements is given by $\beta_Q(\boldsymbol{\alpha})=d$. The result is stated below as a mathematical theorem.
\renewcommand{\thetheorem}{5.1}

\setcounter{thm}{0}

\begin{theorem}
For any collection of positive real numbers $\boldsymbol{\alpha}=\{\alpha_0,\alpha_1,\ldots,\alpha_{d-1}\}$  such that $\alpha_0^2+\ldots+\alpha_{d-1}^2=1$,
the quantum bound of the steering functional $I_{2,d,2}(\boldsymbol{\alpha})$ is independent of $\boldsymbol{\alpha}$ and is given by
$\beta_Q(\boldsymbol{\alpha})=d$.
\end{theorem}
\begin{proof}
Let us begin by introducing the steering operator corresponding to 
the steering functional $I_{2,d,2}(\boldsymbol{\alpha})$ in (\ref{Stefn1}), 
\begin{equation}\label{Porto}
    \hat{I}_{2,d,2}(\boldsymbol{\alpha})=\sum_{k=1}^{d-1}\left(A_0^k\otimes B_{k|0}+\gamma(\boldsymbol{\alpha}) A_1^k\otimes B_{k|1}+\delta_k(\boldsymbol{\alpha})A_0^k\right).
\end{equation}
Recall that $A_0=Z_d$ and $A_1=X_d$ and $B_i's$ are any $d$-outcome generalised observables corresponding to the measurements of Bob. Our aim is to show that
\begin{equation}\label{Lizbon1}
    \beta_Q(\boldsymbol{\alpha})=\max_{\rho_{AB},B_i}\Tr\left[\hat{I}_{2,d,2}(\boldsymbol{\alpha})\rho_{AB}\right]= d,
\end{equation}
where $\rho_{AB}$ acting on $\mathbbm{C}^d\otimes\mathcal{H}_B$ represents some quantum state shared between Alice and Bob and $\mathcal{H}_B$ represents the Hilbert space of Bob of arbitrary but finite dimension. As the expression \eqref{Lizbon1} is linear, we can optimise this over pure states $\ket{\psi_{AB}}\in\mathbbm{C}^d\otimes\mathcal{H}_B$, that is,
\begin{equation}\label{Lizbon}
    \max_{\ket{\psi_{AB}},B_i}\langle\psi_{AB}|\hat{I}_{2,d,2}|\psi_{AB}\rangle= d.
\end{equation}
%
For simplicity, in the rest of the proof we drop the subscript $AB$ from the state. We first show that the expectation value of the steering operator for any $\ket{\psi}$ is upper bounded by $d$ and then find a quantum realisation that achieves this bound. For this purpose, let us break the steering operator $\hat{I}_{2,d,2}(\boldsymbol{\alpha})$ into two parts as
\begin{equation}\label{rozklad}
    \hat{I}_{2,d,2}(\boldsymbol{\alpha})=\sum_{k=1}^{d-1}A_0^k\otimes B_{k|0}+S(\boldsymbol{\alpha})
\end{equation}
such that 
\begin{equation}\label{S1}
    S(\boldsymbol{\alpha})=\sum_{k=1}^{d-1}\left[\gamma(\boldsymbol{\alpha}) A_1^k\otimes B_{k|1}+\delta_k(\boldsymbol{\alpha}) A_0^k\right].
\end{equation}
Notice that the above operator is hermitian which is due to the fact that $A_0^{d-k}=(A_0^k)^{\dagger}$ and $B_{d-k|0}=B_{k|0}^{\dagger}$ [see Eq. \eqref{obs2}]. It is trivial to see that the absolute values of the expectation value of each term in the first part of operator in Eq. \eqref{rozklad} is less than or equal to $1$ as $A_0$ is unitary and $B_{k|0}^{\dagger}B_{k|0}\leq \Id$ for any $k$ [see Eq. \eqref{obs3}], that is,
\begin{equation}\label{sat1}
|\langle\psi| A^k_0\otimes B_{k|0}|\psi\rangle|\leq 1,  
\end{equation}
for any $\ket{\psi}$ and $k=1,\ldots,d-1$.
This allows us to conclude from \eqref{rozklad} that 
\begin{eqnarray}\label{Guimerais}
    \langle\psi|\hat{I}_{2,d,2}(\boldsymbol{\alpha})|\psi\rangle&\leq& 
    d-1+\langle\psi|S(\boldsymbol{\alpha})|\psi\rangle.
\end{eqnarray}
Now, we demonstrate for any $\ket{\psi}$ that $\langle\psi|S(\boldsymbol{\alpha})|\psi\rangle\leq 1$. For this purpose, let us first notice that the state $\ket{\psi}$ belongs to $\mathbbm{C}^d\otimes\mathcal{H}_B$. Thus, as discussed before in Chapter \ref{chapter_2} any such state can be written using the computational basis in $\mathbbm{C}^d$ as 
\begin{equation}\label{representation}
 \ket{\psi_{ABE}}=\sum_{i=0}^{d-1}\lambda_i\ket{i}_A\ket{e_i}_{B},
\end{equation}
where $\lambda_i$ are real and non-negative numbers such that  $\lambda_0^2+\ldots+\lambda_{d-1}^2=1$, and $\ket{e_i}$ are vectors belonging to $\mathcal{H}_B$ which are not orthogonal in general. Plugging in this state, we find the expectation value of $S(\boldsymbol{\alpha})$ as
\begin{eqnarray}\label{Coimbra2}
\langle\psi|S(\boldsymbol{\alpha})|\psi\rangle&=&\sum_{k=1}^{d-1}\sum_{i,j=0}^{d-1}\left[\gamma(\boldsymbol{\alpha}) \lambda_i\lambda_j\bra{i}A_1^{k}\ket{j} \bra{e_i}B_{k|1}\ket{e_j}+\delta_k(\boldsymbol{\alpha})\lambda_i\lambda_j\bra{i}A_0^{k}\ket{j}\!\bra{e_i}e_j\rangle\right]\nonumber\\
&=&\sum_{k=1}^{d-1}\sum_{i=0}^{d-1}\left[\gamma(\boldsymbol{\alpha}) \lambda_i\lambda_{i-k} \bra{e_i}B_{k|1}\ket{e_{i-k}}+\delta_k(\boldsymbol{\alpha})\lambda_i^2\omega^{ik}\right],
\end{eqnarray}
where to arrive at the second equality, we plugged in the explicit forms of $A_i's$ and then used the fact that $Z_d^k\ket{i}=\omega^{ki}\ket{i}$ and $X_d^k\ket{i}=\ket{i+k}$.
Focusing on the second term of the above expression and plugging in
$\delta_0(\boldsymbol{\alpha})=-1$, we obtain that
\begin{equation}
    \sum_{k=1}^{d-1}\sum_{i=0}^{d-1}\delta_k(\boldsymbol{\alpha})\lambda_i^2\omega^{ik}=1+\sum_{k=0}^{d-1}\sum_{i=0}^{d-1}\delta_k(\boldsymbol{\alpha})\lambda_i^2\omega^{ik}.
\end{equation}
Using then the explicit form of $\delta_k(\boldsymbol{\alpha})$ given in Eq. (\ref{value1}), we arrive at
\begin{eqnarray}\label{Coimbra}
     \sum_{k=1}^{d-1}\sum_{i=0}^{d-1}\delta_k(\boldsymbol{\alpha})\lambda_i^2\omega^{ik}=1+\gamma(\boldsymbol{\alpha})-\gamma(\boldsymbol{\alpha})\sum_{i,j=0}^{d-1}\frac{\alpha_i}{\alpha_j}\lambda_j^2,
\end{eqnarray}
where we also used the identity 
\begin{equation}
    \sum_{k=0}^{d-1}\omega^{k(i-j)}=d\delta_{ij}.
\end{equation}
Notice that since $\gamma(\boldsymbol{\alpha})$ is real, the above expression \eqref{Coimbra} is also real.
Since, $S(\boldsymbol{\alpha})$ is a Hermitian operator we have that any expectation value of this operator is real.
Plugging the relations (\ref{Coimbra}) and remembering that $\lambda_i's$ are also real, we can rewrite Eq. (\ref{Coimbra2}) as
\begin{equation}\label{rownanie}
 \langle\psi|S(\boldsymbol{\alpha})|\psi\rangle=1+\gamma(\boldsymbol{\alpha}) \sum_{k=0}^{d-1}\sum_{i=0}^{d-1}\lambda_i\lambda_{i-k} \mathrm{Re}\left(\bra{e_i}B_{k|1}\ket{e_{i-k}}\right)-\gamma(\boldsymbol{\alpha})\sum_{i,j=0}^{d-1}\frac{\alpha_i}{\alpha_j}\lambda_j^2.
\end{equation}
Exploiting the fact that $\mathrm{Re}(z)\leq |z|$ for any $z\in\mathbbm{C}$ and that $B_{k|1}^{\dagger}B_{k|1}\leq \Id$ for any $k$, we get that $\mathrm{Re}\left(\bra{e_i}B_{k|1}\ket{e_{i-k}}\right)\leq1$, Thus, we finally arrive at
\begin{equation}
 \langle\psi|S(\boldsymbol{\alpha})|\psi\rangle\leq 1+\gamma(\boldsymbol{\alpha}) \left[\left(\sum_{i=0}^{d-1}\lambda_i\right)^2 -\sum_{i=0}^{d-1}\alpha_i\sum_{j=0}^{d-1}\frac{\lambda_j^2}{\alpha_j}\right].
\end{equation}
Now, using the Cauchy-Schwarz inequality \eqref{CS} in which we substitute $u_i=\lambda_i$ and $v_i=\alpha_i$, we can conclude that the term inside the square brackets of the above expression is less than or equal to $0$.
Thus, we can finally conclude that the expectation value of $S(\boldsymbol{\alpha})$ for any state $\ket{\psi}$ is less than or equal to $1$. that is,
\begin{equation}\label{sat2}
   \langle\psi|S(\boldsymbol{\alpha})|\psi\rangle\leq 1
\end{equation}
and hence, putting it back into (\ref{Guimerais}), we obtain
\begin{equation}
    \langle\psi|\hat{I}_{2,d,2}(\boldsymbol{\alpha})|\psi\rangle\leq d,
\end{equation}
for any $\ket{\psi}\in \mathbbm{C}^d\otimes\mathcal{H}_B$.  It is also interesting to note here that the relation \eqref{sat2} is saturated when  
\begin{eqnarray}\label{5.34}
    \left(\sum_{i=0}^{d-1}\lambda_i\right)^2 =\sum_{i=0}^{d-1}\alpha_i\sum_{j=0}^{d-1}\frac{\lambda_j^2}{\alpha_j}.
\end{eqnarray}
Now, using the fact that Sedrakyan's inequality \eqref{CS} is saturated if $u_i=\kappa v_i$ for some $\kappa\in\mathbbm{C}$. Substituting $u_i=\lambda_i$ and $v_i=\alpha_i$ in the Sedrakyan's inequality \eqref{CS}, we see that $\lambda_i=\kappa\alpha_i$ for all $i$. Now, using normalisation we get that the only solution of Eq. \eqref{5.34} is $\lambda_i=e^{i\phi}\alpha_i$ for some arbitrary phase $\phi$. As $\alpha_i>0$, we get that $|\lambda_i|>0$ for all $i$. Thus, we can simply conclude that any state saturating the inequality \eqref{sat2} is locally full-rank.

Let us now consider a family of states parametrized by the collection of positive real coefficients $\boldsymbol{\alpha}$,
%
\begin{eqnarray}\label{theState}
\ket{\psi(\boldsymbol{\alpha})}_{AB}=\sum_{i=0}^{d-1}\alpha_{i}\ket{i}_A\ket{i}_B,
\end{eqnarray}
where the local bases of \eqnref{theState} is the computational basis of $\mathbbm{C}^d$. Notice that the above state is a valid normalised quantum state $\sum_{i=0}^{d-1}\alpha_i^2=1$. Notice also that every pure bipartite entangled state of Schmidt rank $d$ can be expressed as these states \eqref{theState} up to local unitary transformations. Now, consider Bob's observables to be projective and to satisfy 
\begin{eqnarray}\label{idealmea1}
    B_0=Z_d^*,\qquad B_1=X_d.
\end{eqnarray} 
Plugging this state and the observables in the steering functional in \eqref{Stefn1}, we get that $ I_{2,d,2}(\boldsymbol{\alpha})=d$.
This completes the proof.
\end{proof}

Notice that the maximal violation of the steering inequality  
(\ref{Stefn1}) by a state $\ket{\psi_{AB}}$ and Bob's observables
$B_i$ can only be achieved iff the inequalities \eqref{SOS2} and \eqref{SOS3} are saturated. Thus, we arrive at the following conditions
\begin{equation}\label{SOSFALTU1}
    \langle\psi| A_0^k\otimes B_{k|0}|\psi\rangle=1
\end{equation}
for any $k=1,\ldots,d-1$ as well as
\begin{equation}\label{SOSFALTU2}
    \langle\psi| S(\boldsymbol{\alpha})|\psi\rangle=1, 
\end{equation}
where $S(\boldsymbol{\alpha})$ is defined in Eq. (\ref{S1}). Now, consider the Cauchy-Schwarz inequality for vectors given as
\begin{eqnarray}\label{CS2}
\mathrm{Re}(\langle u|v\rangle)\leq|\langle u|v\rangle|\leq\left|\langle{u}\ket{u}\langle{v}\ket{v}\right|.
\end{eqnarray}
From the condition \eqref{SOSFALTU1}, let us now substitute $\ket{u}=\ket{\psi}$ and $\ket{v}=A_0^k\otimes B_{k|0}|\psi\rangle$ in the above inequality \eqref{CS2}. Then using the fact that $B_{k|0}^{\dagger}B_{k|0}\leq\Id$, we get that both the L.H.S. and R.H.S. of \eqref{CS} are equal to one. This can only happen iff $\ket{u}$ and $\ket{v}$ are linearly dependent, that is, $\ket{u}=\lambda\ket{v}$ for some $\lambda\in\mathbbm{C}$. As both $\ket{u}$ and $\ket{v}$ are normalised, we get that $\ket{u}=e^{i\phi}\ket{v}$ where $\phi$ is some arbitrary phase. Putting it back into Eq. \eqref{SOSFALTU1}, we get that $\phi=0$. As a consequence, we obtain the following relation
\begin{eqnarray}\label{SOS2}
A_0^{k}\otimes B_{k|0}\,\ket{\psi_{AB}}=\ket{\psi_{AB}} \qquad (k=1,\ldots,d-1).
\end{eqnarray}
Let us now consider the condition \eqref{SOSFALTU2} and  express $S(\boldsymbol{\alpha})$ as
$S(\boldsymbol{\alpha})=S(\boldsymbol{\alpha})_+-S(\boldsymbol{\alpha})_-$ where $S(\boldsymbol{\alpha})_+$ is a positive matrix spanned by the eigenvectors corresponding to the positive eigenvalues of $S(\boldsymbol{\alpha})$ and $S(\boldsymbol{\alpha})_-$ is a positive matrix spanned by the eigenvectors corresponding to the negative eigenvalues of $S(\boldsymbol{\alpha})$. Plugging this decomposition in \eqref{SOSFALTU2}, we get that
\begin{eqnarray}\label{S+}
    \langle\psi| S(\boldsymbol{\alpha})_+|\psi\rangle=1+\langle\psi| S(\boldsymbol{\alpha})_-|\psi\rangle
\end{eqnarray}
 Notice now from Eq. \eqref{sat2} that the maximum eigenvalue of $ S(\boldsymbol{\alpha})$ is one and as a consequence $\langle\psi| S(\boldsymbol{\alpha})_+|\psi\rangle\leq1$ and also $\langle\psi| S(\boldsymbol{\alpha})_+^2|\psi\rangle\leq1$. Thus, from the above condition \eqref{S+}, we can conclude that  $\langle\psi| S(\boldsymbol{\alpha})_-|\psi\rangle=0$ which implies that $ S(\boldsymbol{\alpha})_-|\psi\rangle=0$ as $S(\boldsymbol{\alpha})_-$ is positive. As a consequence, we also have that $\langle\psi| S(\boldsymbol{\alpha})_+|\psi\rangle=1$. Going back to the Cauchy-Schwarz inequality \eqref{CS2}, we substitute $\ket{u}=\ket{\psi}$ and $\ket{v}=S(\boldsymbol{\alpha})|\psi\rangle$. Let us now compute $|\langle{v}\ket{v}|$ by using the decomposition of $S(\boldsymbol{\alpha})$ and also using the fact that it is hermitian
 \begin{eqnarray}
     \langle\psi| S(\boldsymbol{\alpha})|\psi\rangle\leq\langle\psi| S(\boldsymbol{\alpha})^2|\psi\rangle&=& \langle\psi| \left(S(\boldsymbol{\alpha})_+^2+S(\boldsymbol{\alpha})_-^2+S(\boldsymbol{\alpha})_+S(\boldsymbol{\alpha})_-+S(\boldsymbol{\alpha})_-S(\boldsymbol{\alpha})_+\right)|\psi\rangle\nonumber\\
    &=&\langle\psi| S(\boldsymbol{\alpha})_+^2|\psi\rangle\leq 1
 \end{eqnarray}
 where to get to the second line of the above expression we used that $ S(\boldsymbol{\alpha})_-|\psi\rangle=0$. Now, as concluded before, we get that $\ket{u}=e^{i\phi}\ket{v}$ where $\phi$ is some arbitrary phase. Now, again using Eq. \eqref{SOSFALTU2} thus we finally arrive at the relation

%
\begin{eqnarray}\label{SOS3}
\left(\sum_{k=1}^{d-1}\left[\gamma(\boldsymbol{\alpha}) A_1^{k}\otimes B_{k|1}+\delta_k(\boldsymbol{\alpha})A_0^{k}\right]\right)\ket{\psi_{AB}}=\ket{\psi_{AB}}.
\end{eqnarray}
The relations \eqref{SOS2} and \eqref{SOS3} would be particularly useful for certification of the quantum states and measurements that achieve the maximal violation of the steering functional $I_{2,d,2}$. Let us now show that observation of maximal violation of our steering inequalities allows us to certify the state shared between Alice and Bob and also the measurements performed by Bob.


\section{ISDI certification of all pure bipartite entangled states}
\label{App C}
Here, we present the 1SDI certification of all pure bipartite entangled states using the saturation of the quantum bound of the steering functional \eqref{Stefn1} $I_{2,d,2}=\beta_Q$. Let us first recall that we can only characterise Bob's observables on the support of his local state $\rho_B$. Thus, without loss of generality we assume it to be full rank. This can also be understood as
that Bob's observables and local state $\rho_B$ act on the same Hilbert space $\mathcal{H}_B$.

\renewcommand{\thetheorem}{5.2}

\setcounter{thm}{0}

\begin{theorem}\label{Theo1} 
Consider that Alice and Bob perform the quantum steering experiment and observe that the steering functional 
\begin{eqnarray}
I_{2,d,2}(\boldsymbol{\alpha})=\sum_{k=1}^{d-1}\left\langle A_0^{k}\otimes B_{k|0} + 
\gamma(\boldsymbol{\alpha}) A_1^{k}\otimes B_{k|1}+\delta_k(\boldsymbol{\alpha})A_0^{k}\right\rangle,
\end{eqnarray}
attains the maximal quantum value $\beta_Q=d$ where $d$ denotes the number of outcomes of each measurement. Alice is trusted and her measurements are given by $A_0=Z_d$ and $A_1=X_d$ [cf. Eq.  \eqref{X,Z}]. Let us say that the maximal quantum bound is achieved using the state $\rho_{AB}$ acting on $\mathbbm{C}^d\otimes\mathcal{H}_B$ 
and Bob's generalised observables $B_i \ (i\in \{1,2\})$ acting on $\mathcal{H}_B$. Then, the following statements hold true for any integer $d\geq 2$ : 
\begin{enumerate}
    \item  Bob's measurements are projective. Equaivalently, the operators $B_{k|i}$ for all $k,i$ are unitary and $B_{k|i}=B_{1|i}^k\equiv B_i^k$.
    \item Bob's Hilbert space $\mathcal{H}_B$ admits a decomposition into a $d-$dimensional Hilbert space $\mathbbm{C}^d)_{B'}$ and some unknown but finite dimensional auxiliary Hilbert space $\mathcal{H}_{B''}$,
    \begin{eqnarray}
    \mathcal{H}_{B}=(\mathbbm{C}^d)_{B'}\otimes \mathcal{H}_{B''}.
    \end{eqnarray}
    \item There exists a local unitary on Bob's side  $U_B:\mathcal{H}_B\rightarrow\mathcal{H}_B$  such that
\begin{eqnarray}\label{lem5.1.2}
(\Id_A\otimes U_B)\rho_{AB}(\Id_A\otimes U_B^{\dagger})=\ket{\psi(\boldsymbol{\alpha})}\!\bra{\psi(\boldsymbol{\alpha})}_{AB'}
\otimes\rho_{B''}^{aux}.
\end{eqnarray}
where $\ket{\psi(\boldsymbol{\alpha})}$ is the state given in \eqref{theState} and
\begin{eqnarray}
\forall i, \quad U_B\,B_i\,U_B^{\dagger}=A_i^{*}\otimes \Id_{B''},
\end{eqnarray}
where $B''$ denotes Bob's auxiliary system.
\end{enumerate}

\end{theorem}

\begin{proof}
The proof is divided into two major steps. In the first step, we exploit the relations \eqref{SOS2} and \eqref{SOS3} to find Bob's observables that result in the maximal violation of the steering inequality \eqref{Stefn1}. Again, using the relations  \eqref{SOS2} and \eqref{SOS3} and the derived Bob's observables, we find the family of states shared beween Alice and Bob parametrised by the collection of numbers $\boldsymbol{\alpha}$. For our proof, as was discussed before in Chapter \ref{chapter_2}, the state shared between Alice and Bob $\rho_{AB}$ is purified by adding an ancillary system $E$ possessed by some external agent, named Eve, such that $\rho_{AB}=\Tr_E(\ket{\psi_{ABE}}\!\bra{\psi_{ABE}})$ where $\ket{\psi_{ABE}}\in\mathbbm{C}^d\otimes\mathcal{H}_B\otimes\mathcal{H}_E$.

\subsubsection{Bob's observables}

Before finding the explicit forms of Bob's observables that maximally 
violate our steering inequality \eqref{Stefn1}, we show that these generalised observables must correspond to projective measurements. Let us concentrate on Bob's first observable and follow the exact same technique as was used in Chapter \ref{chapter_4} from Eqs . \eqref{SOSApp} to \eqref{4.42} as the relations \eqref{SOS2} and \eqref{SOSApp} are identical. First, we apply
$Z_d^{d-k}\otimes B_{d-k|0}$ to the relation (\ref{sat1}) and then recalling that $Z_d$ is unitary as well as that $B_{d-k|0}=B_{k|0}^{\dagger}$ from \eqref{obs2}, we obtain that
\begin{equation}
    \Id_{AE}\otimes (B_{k|0}^{\dagger}B_{k|0})\ket{\psi_{ABE}}=\ket{\psi_{ABE}}.
\end{equation}
Taking a partial trace over the subsystems $AE$, we arrive at the following condition 
\begin{equation}
    (B_{k|0}^{\dagger}B_{k|0})\rho_B=\rho_B,
\end{equation}
where $\rho_B=\Tr_{AE}[\proj{\psi_{ABE}}]$. Recall that $\rho_B$ is full-rank and thus it is non-singular and invertible. This allows us to immediately conclude that $B_{k|0}^{\dagger}B_{k|0}=\Id_B$ and consequently $B_{k|0}B_{k|0}^{\dagger}=\Id_B$, and thus $B_{k|0}$ is unitary for any $k=0,\ldots,d-1$. Now using Fact \ref{factgenobs1}, we can conclude that Bob's measurements are projective, that is, the positive semi-definite  operators representing the measurement are mutually orthogonal projectors. Further, the fact that $B_0$ is projective imposes that $B_{k|0}$ are powers of $B_{1|0}$ [see Chapter \ref{chapter_2}]. As a consequence, from now on we can simply denote $B_{k|0}=B_0^k$, where $B_0\equiv B_{1|0}$.

Let us now focus on Bob's second observable and show that it corresponds to a projective measurement too. 
For this purpose, we refer to the second condition (\ref{sat2}) and then consider the general representation of any state $\ket{\psi_{ABE}}\in\mathbbm{C}^d\otimes\mathcal{H}_B\otimes\mathcal{H}_E$ as in Eq. (\ref{representation})
\begin{equation}\label{genstate5.3}
    \ket{\psi_{ABE}}=\sum_{i}\lambda_i\ket{i}_A\ket{e_i}_{BE},
\end{equation}
where $\lambda_i\geq0$ and $\ket{e_i}_{BE}$ are vectors that are in general not orthogonal. Now recall that, we showed in the previous subsection we showed that any state satisfying the condition \eqref{sat2} must be locally full-rank. Thus, in the state \eqref{genstate5.3}, $\lambda_i>0$ for all $i$. For simplicity, from here on we drop all the subscripts in the notation of the state. Now, by plugging this state in the condition $\bra{\psi_{ABE}}S(\boldsymbol{\alpha})\ket{\psi_{ABE}}=1$, we obtain [cf. Eq. (\ref{rownanie})]
\begin{equation}
\sum_{k=0}^{d-1}\sum_{i=0}^{d-1}\lambda_i\lambda_{i-k} \mathrm{Re}\left(\bra{e_i}[B_{k|1}\otimes\Id_E]\ket{e_{i-k}}\right)=\sum_{i,j=0}^{d-1}\frac{\alpha_i}{\alpha_j}\lambda_j^2.
\end{equation}
Dropping the identity acting on Eve, $\Id_E$, for the time being and then using
the inequality (\ref{CS}) with in $u_i=\lambda_i$ and $v_i=\alpha_i$, we arrive at
\begin{equation}\label{ineq1979}
    \sum_{k=0}^{d-1}\sum_{i=0}^{d-1}\lambda_i\lambda_{i-k} \mathrm{Re}\left(\bra{e_i}B_{k|1}\ket{e_{i-k}}\right)\geq \left(\sum_{i=0}^{d-1}\lambda_i\right)^2.
\end{equation}
Notice that the term on the right hand side of the above inequality can be expanded as $\left(\sum_{i=0}^{d-1}\lambda_i\right)^2=\sum_{k=0}^{d-1}\sum_{i=0}^{d-1}\lambda_i\lambda_{i-k}$. As a consequence, we immediately obtain that
\begin{eqnarray}
     \sum_{k=0}^{d-1}\sum_{i=0}^{d-1}\lambda_i\lambda_{i-k} \mathrm{Re}\left[\left(\bra{e_i}B_{k|1}\ket{e_{i-k}}\right)-1\right]\geq 0.
\end{eqnarray}
Again, recalling that $B_{k|1}^{\dagger}B_{k|1}\leq \Id_B$ for any $k$ allows us to conclude that $\mathrm{Re}\left(\bra{e_i}[B_{k|1}\ket{e_{i-k}}\right)\leq 1$ 
for any $i$ and $k$. However, $\lambda_i$ being non-negative in the inequality (\ref{ineq1979}), forces the term inside the square brackets to be $0$ for all $i,k$. Thus, we finally obtain
\begin{equation}\label{5.54}
    \mathrm{Re}\left(\bra{e_i}B_{k|1}\ket{e_{i-k}}\right)= 1.
\end{equation}
As the states $\ket{e_i}$ are normalised, the above condition is satisfied iff $B_{k|1}\otimes\Id_E\ket{e_{i-k}}=\ket{e_i}$. For this purpose, we again employ the Cauchy-Schwarz inequality \eqref{CS2} where $\ket{u}=\ket{e_i}$ and $\ket{v}=B_{k|1}\ket{e_{i-k}}$ and then the fact that $B_{k|1}^{\dagger}B_{k|1}\leq\Id$ from which we obtain $B_{k|1}\otimes\Id_E\ket{e_{i-k}}=e^{i\phi}\ket{e_i}$ for some arbitrary phase $\phi$. Again using Eq. \eqref{5.54}, we get that $\phi=0$.
Now, we multiply this equation with its conjugate transpose, to observe that 
\begin{equation}
    \bra{e_{i-k}}[B_{k|1}^{\dagger}B_{k|1}\otimes \Id_E]\ket{e_{i-k}}=1.
\end{equation}
for any $i,k$. This implies that for any $k$ the above condition is satisfied for any $i$. As a consequence, we arrive at a simple relation for any $k$
\begin{equation}
    \bra{e_{i}}[B_{k|1}^{\dagger}B_{k|1}\otimes \Id_E]\ket{e_{i}}=1 \qquad (i=0,\ldots,d-1).
\end{equation}
Tracing out Eve's subsystem, further implies that 
$\Tr[B_{k|1}^{\dagger}B_{k|1}\rho_B^i]=1$ for all $i$, where $\rho_{B}^i=\Tr_E[\proj{e_i}_{BE}]$.
As $\rho_B^i$ is positive and again recalling that $B_{k|1}^{\dagger}B_{k|1}\leq \Id_B$ for any $k$ allows us to conclude that this condition holds true iff $B_{k|1}^{\dagger}B_{k|1}$ is an identity acting onto the support of $\rho_B^i$. However, notice that the support of Bob's reduced state $\rho_B=\Tr_{AE}[\proj{\psi}_{ABE}]$ is in-fact composed of the supports of $\rho_B^i$'s. To see this, we use the decomposition of the state $\ket{\psi}_{ABE}$ given in \eqref{genstate5.3} to obtain that
\begin{eqnarray}
   \rho_B&=& \Tr_{AE}\left[\sum_{i,j=0}^{d-1}\lambda_i\lambda_j\ket{i}\!\bra{j}\otimes\ket{e_i}\!\bra{e_j}\right]\nonumber\\
   &=& \Tr_{E}\left[\sum_{i=0}^{d-1}\lambda_i^2\ket{e_i}\!\bra{e_i}\right]=\sum_{i=0}^{d-1}\lambda_i^2\rho_B^i.
\end{eqnarray}
As a consequence, $B_{k|1}^{\dagger}B_{k|1}$ is an identity that acts on the entire support of Bob's reduced state $\rho_B$, and thus we finally have that $B_{k|1}^{\dagger}B_{k|1}=\Id_B$ and consequently $B_{k|1}B_{k|1}^{\dagger}=\Id_B$ for all $k$. Again, using Fact \ref{factgenobs1}, we can conclude that the second Bob's observable corresponds to projective measurements and hence from here on, we denote $B_{k|1}=B_{1}^k$, where $B_{1}\equiv B_{1|1}$. 
This completes the part of the proof to show that the maximal violation of the steering inequality \eqref{Stefn1} can only be achieved when Bob's both measurements are projective.

Now, we move onto finding the explicit form of Bob's both observables. Let us first consider the relation \eqref{SOS3} and then apply 
$\Id_{A}\otimes B_1$ to it, which after rearranging some terms gives us
\begin{eqnarray}\label{procond1.1}
\gamma(\boldsymbol{\alpha}) \sum_{k=1}^{d-1}\left( X_d^{k}\otimes B_{1}^{k+1}\right)\ket{\psi_{ABE}}=\left[\left(\Id_A-\sum_{k=1}^{d-1}
\delta_k(\boldsymbol{\alpha})Z_d^{k}\right)\otimes B_{1}\right]\ket{\psi_{ABE}}.
\end{eqnarray}
To simplify the notation, let us introduce the following operator
\begin{equation}\label{Za}
    \overline{Z}_A:=\Id_A-\sum_{k=1}^{d-1}\delta_k(\boldsymbol{\alpha})Z_d^{k}.
\end{equation}
Now, an application of $Z_d^{-1}\otimes\Id_B$ to the left hand side of Eq. \eqref{procond1.1}  gives us
\begin{eqnarray}\label{eq:number1}
\gamma(\boldsymbol{\alpha})\sum_{k=1}^{d-1}\left( Z_d^{-1}X_d^{k}\otimes B_{1}^{k+1}\right)\ket{\psi_{ABE}}=(\overline{Z}_AZ_d^{-1}\otimes B_{1})\ket{\psi_{ABE}},
\end{eqnarray}
where we can interchange the positioning of $Z_d$ and $\overline{Z}_A$ as they commute. Then, by using the commutation relation $Z_dX_d=\omega X_dZ_d$, we can rewrite Eq. \eqref{eq:number1}  as
\begin{eqnarray}\label{eq:comutationrelation}
\gamma(\boldsymbol{\alpha})\sum_{k=1}^{d-1}\left( \omega^{-k}X_d^{k}\otimes B_{1}^{k+1}\right)\left(Z_d^{-1}\otimes \Id_B\right)\ket{\psi_{ABE}}=(\overline{Z}_A\otimes B_{1})\left(Z_d^{-1}\otimes \Id_B\right)\ket{\psi_{ABE}},
\end{eqnarray}
where we again used the fact that $Z_d^{-1}\otimes\Id_B$ and $\Id_A\otimes B_1$ commute. Now, using Eq. \eqref{SOS2} for $k=1$, that is, $\left(Z_d^{-1}\otimes \Id_B\right)\ket{\psi_{ABE}}=\Id_A\otimes B_0\ket{\psi_{ABE}}$, we finally get
\begin{eqnarray}\label{procond3}
\gamma(\boldsymbol{\alpha}) \sum_{k=1}^{d-1}
\left( \omega^{-k}X_d^{k}\otimes B_{1}^{k+1}B_0\right)\ket{\psi_{ABE}}=(\overline{Z}_A\otimes B_{1}B_0)\ket{\psi_{ABE}}.
\end{eqnarray}
Applying $B_{0}$ from the left hand side of the expression \eqref{procond1.1}, we obtain 
\begin{eqnarray}\label{procond2}
\gamma(\boldsymbol{\alpha})\sum_{k=1}^{d-1}\left( X_d^{k}\otimes B_0B_{1}^{k+1}\right)\ket{\psi_{ABE}}=(\overline{Z}_A\otimes B_0B_{1})\ket{\psi_{ABE}}.
\end{eqnarray}
In the next step, we multiply $\omega^{-1}$ to Eq. \eqref{procond3} and then subtract it from Eq. $\eqref{procond2}$, which immediately gives us
\begin{equation}\label{procond4}
\gamma(\boldsymbol{\alpha})\sum_{k=1}^{d-1}\left[ X_d^{k}\otimes \left(B_{0}B_{1}^{k+1}-\omega^{-(k+1)}B_{1}^{k+1}B_{0}\right) \right]\ket{\psi_{ABE}}=[\overline{Z}_A\otimes (B_{0}B_{1}-\omega^{-1} B_{1}B_{0})]\ket{\psi_{ABE}}.
\end{equation}

 Let us again consider the relation \eqref{SOS3} and multiply it by $X_d^{-1}\otimes \Id_B$ from the left hand side, which gives us
\begin{eqnarray}\label{procond2.1}
\sum_{k=1}^{d-1}\left(\gamma(\boldsymbol{\alpha}) X_d^{k-1}\otimes B_{1}^{k}\right)\ket{\psi_{ABE}}=(X_d^{-1}\overline{Z}_A\otimes\Id_B)\ket{\psi_{ABE}}.
\end{eqnarray}
Then, after multiplying $\Id_A\otimes B_{0}$ to the above equation and then taking into account that it commutes with $X_d\otimes\Id_B$, it not difficult to see that
\begin{eqnarray}\label{procond5}
\sum_{k=1}^{d-1}\left(\gamma(\boldsymbol{\alpha}) X_d^{k-1}\otimes B_0B_{1}^{k}\right)\ket{\psi_{ABE}}=(X_d^{-1}\overline{Z}_A\otimes B_0)\ket{\psi_{ABE}}.
\end{eqnarray}
Now, let us exploit the relation \eqref{SOS2} for $k=1$, that is, $\left(Z_d^{-1}\otimes \Id_B\right)\ket{\psi_{ABE}}=\Id_A\otimes B_0\ket{\psi_{ABE}}$ and then using the fact that $\overline{Z}_A$ and $Z_d$ commutes, we finally get,
\begin{eqnarray}
\sum_{k=1}^{d-1}\left(\gamma(\boldsymbol{\alpha}) X_d^{k-1}\otimes B_0B_{1}^{k}\right)\ket{\psi_{ABE}}=(X_d^{-1}Z_d^{-1}\overline{Z}_A\otimes \Id_B)\ket{\psi_{ABE}}.
\end{eqnarray}
Next, we apply $Z_d^{-1}\otimes\Id_B$ to Eq. \eqref{procond2.1} from the left hand side to obtain, 
\begin{eqnarray}\label{eq:equation2}
\sum_{k=1}^{d-1}\left(\gamma(\boldsymbol{\alpha}) Z_d^{-1}X_d^{k-1}\otimes B_{1}^{k}\right)\ket{\psi_{ABE}}=(Z_d^{-1}X_d^{-1}\overline{Z}_A\otimes\Id_B)\ket{\psi_{ABE}}.
\end{eqnarray}
Again, by employing the relation $Z_dX_d=\omega X_dZ_d$, the above equation \eqref{eq:equation2} can be rewriten as
\begin{eqnarray}\label{procond6}
\sum_{k=1}^{d-1}\left(\gamma(\boldsymbol{\alpha}) \omega^{-(k-1)}X_d^{k-1}\otimes B_{1}^{k}B_0\right)\ket{\psi_{ABE}}=\left(\omega X_d^{-1}Z_d^{-1}\overline{Z}_A\otimes \Id_B\right)\ket{\psi_{ABE}}.
\end{eqnarray}
Notice that in the left hand side of the above equation, we again exploited the relation \eqref{SOS2} for $k=1$, that is, $\left(Z_d^{-1}\otimes \Id_B\right)\ket{\psi_{ABE}}=\Id_A\otimes B_0\ket{\psi_{ABE}}$. Now we apply $\omega^{-1}$ to the above equation \eqref{procond6} and then subtract it from Eq. \eqref{procond5} to get
\begin{eqnarray}
\gamma(\boldsymbol{\alpha})\sum_{k=1}^{d-1}\left[ X_d^{k-1}\otimes \left(B_{0}B_{1}^{k}-\omega^{-k}B_{1}^{k}B_{0}\right)\right]\ket{\psi_{ABE}}= 0,
\end{eqnarray}
which can be divided up into two parts by separating the term corresponding to $k=1$, as follows
\begin{eqnarray}\label{procond7}
\gamma(\boldsymbol{\alpha})\sum_{k=2}^{d-1}\left[ X_d^{k-1}\otimes \left(B_{0}B_{1}^{k}-\omega^{-k}B_{1}^{k}B_{0}\right)\right]\ket{\psi_{ABE}}=- \gamma(\boldsymbol{\alpha})\left(B_{0}B_{1}-\omega^{-1}B_{1}B_{0}\right)\ket{\psi_{ABE}}.
\end{eqnarray}
Notice that the left hand side of the expressions \eqref{procond4} and \eqref{procond7} are identical, which immediately allows us to get that
\begin{eqnarray}
\overline{Z}_A\otimes (B_{0}B_{1}-\omega^{-1} B_{1}B_{0})\ket{\psi_{ABE}}=-\gamma(\boldsymbol{\alpha})\left( B_{0}B_{1}-\omega^{-1} B_{1}B_{0}\right)\ket{\psi_{ABE}},
\end{eqnarray}
which after a simple rearrangement of the terms and expanding $\overline{Z}_A$ from \eqref{Za} yields,
\begin{eqnarray}\label{STF1}
\left[(1+\gamma(\boldsymbol{\alpha}))\Id_A-\sum_{k=1}^{d-1}\delta_k(\boldsymbol{\alpha})Z_d^{k}\right]\otimes \left(B_{0}B_{1}-\omega^{-1} B_{1}B_{0}\right)\ket{\psi_{ABE}}=0.
\end{eqnarray}
As proven in Observation \ref{obsST1} stated in Appendix \ref{chap5}, the operator $[1+\gamma(\boldsymbol{\alpha})]\Id-\sum_{k=1}^{d-1}\delta_k(\boldsymbol{\alpha})Z_d^{k}$ is invertible. Thus, taking trace over the subsystems $A,E$ allows us to finally conclude that
\begin{equation}\label{eq:self-testing}
    (B_0B_1-\omega^{-1}B_1B_0)\rho_B=0,
\end{equation}
%
%
%
where $\rho_B=\Tr_{AE}\proj{\psi_{ABE}}$. Recalling that $\rho_B$ is full-rank and thus invertible, the above expression \eqref{eq:self-testing} implies the following 
commutation relation between Bob's both observables
\begin{equation}
B_{0}B_{1}=\omega^{-1} B_{1}B_{0}.
\end{equation}
As stated in Fact \ref{fact2} which was proven in Ref. \cite{Jed1}, the above relation along with the fact that $B_0^d=B_1^d=\Id_B$ imposes that Bob's Hilbert space decomposes into a tensor product $\mathcal{H}_B=(\mathbbm{C}^d)_{B'}\otimes\mathcal{H}_{B''}$
where $\mathcal{H}_{B''}$ is some Hilbert space of unknown but finite dimension. Along with it, there also exists a unitary transformation
$U_B:\mathcal{H}_B\to\mathcal{H}_B$ such that
\begin{eqnarray}\label{BobsMeas}
U_B\,B_{0}\,U^{\dagger}_B= Z^{*}_{d}\otimes\Id_{B''},\quad U_B\,B_{1}\,U_B^{\dagger}=X_d\otimes\Id_{B''},
\end{eqnarray}
where $\Id_{B''}$ is the identity acting on $\mathcal{H}_{B''}$. This completes the characterisation of Bob's observables that maximally violate the steering inequality \eqref{Stefn1}.

\subsubsection{The state}
 
 We finally have all the tools required to find the state that maximally violates the steering inequality \eqref{Stefn1}. As was derived in the previous part, up to a local unitary Bob's both observables are the ideal ones (\ref{BobsMeas}). Thus, we can rewrite the relation (\ref{sat1}) and (\ref{sat2}) by plugging in Bob's derived observables as
\begin{equation}\label{condZ}
    (Z_d\otimes Z_d^{\dagger}\otimes\Id_{B''E})\ket{\tilde{\psi}_{ABE}}=\ket{\tilde{\psi}_{ABE}},
\end{equation}
and
\begin{eqnarray}\label{STstate11}
\sum_{k=1}^{d-1}\left[\gamma(\boldsymbol{\alpha}) X_d^{k}\otimes X_d^{k}\otimes\Id_{B''E}+\delta_k(\boldsymbol{\alpha})Z_d^{k}\otimes\Id_{BE}\right]\ket{\tilde{\psi}_{ABE}}=\ket{\tilde{\psi}_{ABE}},
\end{eqnarray}
where $\ket{\tilde{\psi}_{ABE}}=U_B\otimes\Id_{AE}\ket{\psi_{ABE}}$. 
From here on, for convenience we drop the all the identities from the above relations. As concluded in the previous part of the proof that Bob's Hilbert space is of dimension $(\mathbbm{C}^d)_{B'}\otimes\mathcal{H}_{B''}$ due to which the state $\ket{\tilde{\psi}_{ABE}}$ belongs to $(\mathbbm{C}^d)_A\otimes(\mathbbm{C}^d)_{B'}\otimes\mathcal{H}_{B''}\otimes\mathcal{H}_E$. As a consequence, any such state can be written using the computational basis in $\mathbbm{C}^d$ as,
\begin{equation}\label{5.73}
    \ket{\tilde{\psi}_{ABE}}=\sum_{i,j=0}^{d-1}\ket{i}_A\ket{j}_{B'}\ket{\psi_{ij}}_{B''E},
\end{equation}
where $\ket{\psi_{ij}}_{B''E}$ is some unnormalised state belonging to $\mathcal{H}_{B''}\otimes \mathcal{H}_E$. 
After plugging this state to the condition (\ref{condZ}) for $k=1$, we arrive at
\begin{eqnarray}
\sum_{i,j=0}^{d-1}\omega^{i-j}\ket{ij}\ket{\psi_{ij}}=\sum_{i,j=0}^{d-1}\ket{ij}\ket{\psi_{ij}},
\end{eqnarray}
which holds true if and only if 
$\ket{\psi_{ij}}=0$ for any $i\neq j$. As a consequence, the only terms in the state \eqref{5.73} remains when $i=j$, and thus we have the simplified form of the state given by 
\be \label{sf1}
\ket{\tilde{\psi}_{ABE}}=\sum_{i=0}^{d-1}\ket{ii}\ket{\psi_{ii}}.
\ee 
Let us now consider the condition \eqref{STstate11} where we can extend the range of the sum to $k=0$ by recalling that the $0-th$ power of an observable is identity and also that $\delta_0(\boldsymbol{\alpha})=-1$. Thus, after some rearrangement of the terms we get the following expression   
\begin{eqnarray}\label{STstate21}
\sum_{k=0}^{d-1}\left[\gamma(\boldsymbol{\alpha}) X_d^{k}\otimes X_d^{k}+\delta_k(\boldsymbol{\alpha})Z_d^{k}\otimes\Id_{B}\right]
\ket{\tilde{\psi}_{ABE}}
=\gamma(\boldsymbol{\alpha})\ket{\tilde{\psi}_{ABE}}.
\end{eqnarray}
Plugging in the simplified form of $\ket{\tilde{\psi}_{ABE}}$ as derived in (\ref{sf1}), the above expression turns out to be
\begin{eqnarray}\label{eq:secondcondition}
\gamma(\boldsymbol{\alpha})\sum_{k=0}^{d-1}\sum_{i=0}^{d-1}\ket{i+k}\ket{i+k}\ket{\psi_{ii}}+\sum_{k=0}^{d-1}\sum_{i=0}^{d-1}\omega^{ki}\delta_k(\boldsymbol{\alpha})\ket{ii}\ket{\psi_{ii}}=\gamma(\boldsymbol{\alpha})\sum_{i=0}^{d-1}\ket{ii}\ket{\psi_{ii}}.
\end{eqnarray}
Mulitplying the above expression with $\bra{ss}$ from the lest hand side, we obtain that
\begin{equation}
    \sum_{k=0}^{d-1}\gamma(\boldsymbol{\alpha})\ket{\psi_{s\ominus k,s\ominus k}}+\sum_{k=0}^{d-1}\omega^{ks}\delta_{k}(\boldsymbol{\alpha})\ket{\psi_{ss}}=\gamma(\boldsymbol{\alpha})\ket{\psi_{ss}}
\end{equation}
where $s\ominus k$ represents $s-k$ modulo $d$. We can simplify the above expression by substituting the explicit form of $\delta(\boldsymbol{\alpha})$ to obtain
\begin{eqnarray}
\sum_{k=0}^{d-1}\ket{\psi_{s\ominus k,s\ominus k}}-\sum_{k=0}^{d-1}\frac{\omega^{k(d-j+s)}}{d}\sum_{\substack{i,j=0\\i\ne j}}^{d-1}\frac{\alpha_i}{\alpha_j}\ket{\psi_{ss}}=\ket{\psi_{ss}}
\end{eqnarray}
Now, using the identity $\sum_{k=0}^{d-1}\omega^{k(j-s)}=d\delta_{j,s}$, we can simplify the above expression to find the explicit form of the state $ \ket{\psi_{ss}}$ given by
\begin{equation}
    \ket{\psi_{ss}}=\frac{\alpha_s}{\alpha_0+\ldots+\alpha_{d-1}}\ket{\Psi},
\end{equation}
where we denoted $\ket{\Psi}=\sum_{k=0}^{d-1}\ket{\psi_{s\ominus k,s\ominus k}}\equiv \sum_{k=0}^{d-1}\ket{\psi_{kk}}$ for any $s$. As a consequence, Eq. (\ref{sf1}) can be rewritten as 
\begin{eqnarray}
    (U_B\otimes\Id_{AE})\ket{\psi_{ABE}}&=&\left(\sum_{m=0}^{d-1}\alpha_{i}\ket{ii}_{AB'}\right)\otimes\ket{\xi}_{B''E}
    =\ket{\psi(\boldsymbol{\alpha})}_{AB'} \otimes\ket{\xi}_{B''E},
\end{eqnarray}
where 
\begin{equation}
   \ket{\xi}_{B''E}=\frac{1}{\alpha_0+\ldots+\alpha_{d-1}}\ket{\Psi}. 
\end{equation}
This finally completes the proof of certification of all pure bipartite entangled states along with a pair of arbitrary outcome mutually unbiased bases in the 1SDI scenario.
\end{proof}
%

Let us now proceed towards another important result of this chapter that utilises the above theorem \eqref{Theo1} involving certification of every rank-one extremal POVM's.

\section{Certification of all rank-one extremal POVM}
\label{App D}
As discussed before in Chapter \ref{chapter_2}, let us consider a rank-one extremal measurement denoted as $\mathcal{I}=\{\mathcal{I}_b\}$, such that $b$ represents its outcomes and $\mathcal{I}_b$ represents the measurement operator corresponding to the $b-th$ outcome. These measurement operators are positive semi-definite and sum up to one. Additionally, it was shown in \cite{APP05} that measurement operators of an extremal rank-one POVM are projectors scaled down by some non-negative real number, that is,  $\mathcal{I}_b=\lambda_b\ket{\mu}\!\bra{\mu}$ with $\ket{\mu}\in\mathbbm{C}^d$ and $0\leq\lambda_b\leq1$. 

\begin{figure}
    \centering
    \includegraphics[scale=.45]{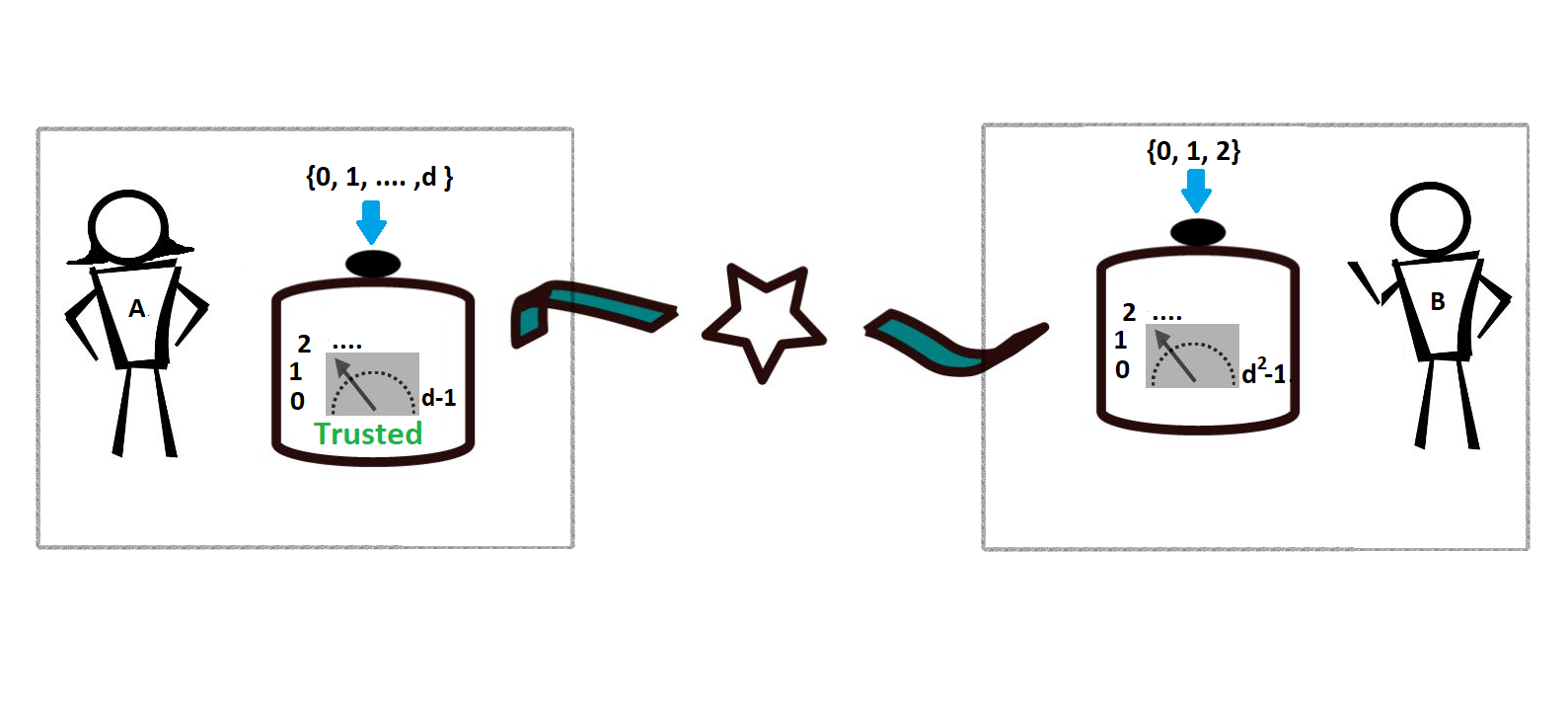}
    \caption{1SDI scenario to certify any rank-one extremal POVM: Alice and Bob receive subsystems from the preparation device on which they perform $d+1$ and $3$ measurements respectively such that Alice is trusted. All the measurements are $d-$outcome except Bob's third measurement  which is of $d^2-$outcome.}
    \label{fig5.1}
\end{figure}

It turns our that the observation of the maximal violation of our steering inequality (\ref{Stefn1}) plus an additional set of conditions enables us to design a simple method that can be used for certification of any extremal rank-one POVM. For this purpose, we again consider the 1SDI setting such that Alice is again trusted but now performs $d+1$ measurements corresponding to the observables $A_{0}=Z_d$ and $A_{i+1}=X_dZ_d^i$ for $i=0,1,\ldots,d-1$. Bob is untrusted and performs three measurements, where the first two measurements are $d-$outcome and the third one has $d^2-$outcomes. 
The scenario is depicted in Fig.  \ref{fig5.1}. 

Notice that the statistics corresponding to both Alice and Bob choosing the input $0,1$, allow us to employ the steering inequality \eqref{Stefn1} and certify any pure bipartite entangled state using Theorem \ref{Theo1}. Without loss of generality, Bob's third measurement, which is a $d^2-$outcome POVM, is denoted as $\{R_b\}$. Further, notice that the operators $X_dZ_d^i$ for $i=1,2,\ldots,d-1$ are not proper observables based on the definition introduced in Chapter \ref{chapter_2} as $(X_dZ_d^i)^d=\omega^{id(d-1)}\Id$. Thus, dividing the matrices with the scalar $\omega^{-i(d-1)}$ yields proper observables, but for simplicity we would drop this factor from further considerations. It is also worth noting that the statistics obtained from the $d+1$ observables, $Z_d$ and $X_dZ_d^i$, are enough to simulate the statistics corresponding to the operators $X_d^iZ_d^j$ for $i,j=0,1,\ldots,d-1$ that represent the Heisenberg-Weyl (HW) basis \cite{HW} \footnote{The HW basis is a collection of operators that forms a basis for operators that act on $d-$dimensional Hilbert space with $d$ being any positive integer.}. The reason for this fact is that every element in the HW basis can be generated by considering the powers of $Z_d$ and $X_dZ_d^i$ for all $i$ and then multiply them with appropriate powers of $\omega$. For instance, $X_d^2Z_d^2$ can be generated by taking $X_dZ_d$ two times and then multiplying it with $\omega^{-1}$, that is, $X_d^2Z_d^2=\omega^{-1}(X_dZ_d)^2$. The result is stated below as a simple theorem.
\renewcommand{\thetheorem}{5.3}

\setcounter{thm}{0}
\begin{theorem}\label{Theo3App} Assume that Alice and Bob perform the quantum steering experiment and are able to certify that the state shared among them as well as the measurements along with the Hilbert space of Bob as given in Theorem-\ref{Theo1}. Consider then a POVM $R=\{R_b\}$ acting on $\mathcal{H}_{B}=(\mathbbm{C}^d)_{B'}\otimes\mathcal{H}_{B''}$. If for some extremal 
POVM $\mathcal{I}=\{\mathcal{I}_b\}$ acting on $\mathbbm{C}^d$ the following identities 
\begin{eqnarray}
\langle X^iZ^j\otimes R_b\otimes\Id_E\rangle_{\ket{\psi_{ABE}}}=\langle X^iZ^j\otimes \mathcal{I}_b\rangle_{\ket{\psi(\boldsymbol{\alpha})}}
\end{eqnarray}
hold true for any $i,j=0,\ldots,d-1$, where $\ket{\psi_{ABE}}=(\Id_A\otimes U_B^{\dagger})\ket{\psi(\boldsymbol{\alpha})}_{AB'}
\otimes\ket{\xi_{B''E}}$ from \eqref{lem5.1.2} where $\ket{\psi(\boldsymbol{\alpha})}_{AB'}$ is the ideal state defined in Eq. (\ref{theState}). Then, there exist a unitary transformation $U_B:\mathcal{H}_B\rightarrow\mathcal{H}_B$ such that the measurement operators of the POVM $R$ are equivalent to the measurement operators of the ideal POVM $\mathcal{I}$ as \begin{eqnarray}\label{POVMcert}
U_B\ R_b\ U_B^{\dagger}=\mathcal{I}_b\otimes\Id_{B''}\qquad \forall b.
\end{eqnarray}
\end{theorem}

\begin{proof}
%
Our proof takes inspiration from the technique introduced in Ref. \cite{random1}, where extremal POVM's acting on two-dimensional Hilbert space were certified up to certain equivalences. Here, we generalise that approach to POVM's that act on arbitrary dimensional Hilbert space in the scenario where Alice is trusted. Let us first observe that the statistics one observes from the actual experiment must be equivalent to one observed in the ideal experiment, that is,
\begin{eqnarray}\label{POVMST1}
\forall\ b\quad \forall i,j\qquad\bra{\psi_{ABE}}X^iZ^j\otimes R_b\otimes\Id_E\ket{\psi_{ABE}}=\bra{\psi(\boldsymbol{\alpha})}X^iZ^j\otimes \mathcal{I}_b\ket{\psi(\boldsymbol{\alpha})}.
\end{eqnarray}

Solving the above condition is enough to certify the POVM $R$. For this purpose, as was done in Chapter \ref{chapter_4}, we first rewrite the state $\ket{\psi(\boldsymbol{\alpha})}$ in
terms of the maximally entangled state of two qudits $\ket{\phi_+^d}$ [see Eq. \eqref{maxentstated}] as
\begin{equation}\label{reprezentacja}
    \ket{\psi(\boldsymbol{\alpha})}=[\Id_A\otimes P(\boldsymbol{\alpha})]\ket{\phi_+^d},
\end{equation}
where 
\begin{equation}
    P(\boldsymbol{\alpha})=\sum_{i=0}^{d-1}\alpha_i\proj{i}.
\end{equation}
Recall that $\alpha_i> 0$ for all $i$ and $\sum_i\alpha_i^2=1$. Let us also introduce another set of $d^2$ number of operators, derived from HW basis
\begin{equation}\label{basis}
    W_{i,j}:=P(\boldsymbol{\alpha})^{-1}\left(X^{i}Z^{j}\right)^*P(\boldsymbol{\alpha})^{-1}.
\end{equation}
Let us observe that the above operators are linearly independent as $P(\boldsymbol{\alpha})$ is invertible and $X^iZ^j$ are orthogonal in the Hilbert-Schmidt scalar product, that is, $\Tr[X^iZ^j(X^{i'}Z^{j'})^{\dagger}]=d\delta_{i,i'}\delta_{j,j'}$. As a consequence, the set $\{W_{i,j}\}$ forms a complete basis for operators acting on $d-$dimensional Hilbert space. Recalling that the measurement operators $\mathcal{I}_b$ of the ideal POVM  act on $d-$dimensional Hilbert space and thus, using the newly defined operator basis \eqref{basis}, we can express them as 
\begin{equation}\label{POVMST6}
\mathcal{I}_b=\sum_{i,j=0}^{d-1}l^b_{i,j} W_{i,j}\qquad \forall b,
\end{equation}
where $l_{i,j}^b$ are in general complex coefficients. Let us now compute the right hand side of the expression \eqref{POVMST1} by plugging in the above representation of the POVM $\mathcal{I}$, 
\begin{eqnarray}\label{5.91}
\bra{\psi(\boldsymbol{\alpha})}X^iZ^j\otimes \mathcal{I}_b\ket{\psi(\boldsymbol{\alpha})}&=&\sum_{m,n}l^b_{m,n}\!\bra{\psi(\boldsymbol{\alpha})}X^{i}Z^{j}\otimes P(\boldsymbol{\alpha})^{-1}\left(X^{m}Z^{n}\right)^*P(\boldsymbol{\alpha})^{-1}\ket{\psi(\boldsymbol{\alpha})}\nonumber\\
&=&\sum_{m,n}l^b_{m,n}\!\bra{\phi_+^d}X^{i}Z^{j}\otimes \left(X^{m}Z^{n}\right)^*\ket{\phi_+^d},
\end{eqnarray}
where we exploited the form of the state $\ket{\psi(\boldsymbol{\alpha})}$ given in \eqref{reprezentacja}. Now, exploiting the identity $(R\otimes Q)\,\ket{\phi_+^d}=( RQ^T\otimes\Id)\ket{\phi_+^d}$ that is satisfied for any two matrices $Q$ and $R$ acting on $d-$dimensional Hilbert space  [see Fact \ref{factmaxent} in Appendix \ref{chapano}] and also the fact that $X^iZ^j$ form an orthogonal basis as mentioned above, we finally obtain that
\begin{equation}\label{5.92}
    \bra{\psi(\boldsymbol{\alpha})}X^iZ^j\otimes \mathcal{I}_b\ket{\psi(\boldsymbol{\alpha})}=l^b_{i,j} \qquad\forall b.
\end{equation}

Next, our aim is to compute the left hand side of the expression \eqref{POVMST1}. We use the fact that according to Theorem \ref{Theo1}, Bob's Hilbert space decomposes as $\mathcal{H}_{B}=\mathbbm{C}^d\otimes\mathcal{H}_{B''}$ and there exist a unitary $U_B:\mathcal{H}_B\rightarrow\mathcal{H}_B$ that transforms that transforms the state $\ket{\psi_{ABE}}$ as
\begin{eqnarray}
(\Id_{AE}\otimes U_B)\ket{\psi_{ABE}}=\ket{\psi(\boldsymbol{\alpha})}_{AB'}\otimes\ket{\xi_{B''E}}.
\end{eqnarray}
Now, any measurement operator of the POVM $R$ acting on $\mathcal{H}_{B}$ can be expressed using the basis \eqref{basis} as,
\begin{eqnarray}\label{eq77}
U_BR_bU_B^{\dagger}=\sum_{i,j=0}^{d-1}W_{i,j}\otimes \widetilde{R}_{i,j}^{b},
\end{eqnarray}
where $\widetilde{R}_{i,j}^b$ are general operators acting on $\mathcal{H}_{B''}$. 
Now, computing the left hand side of \eqref{POVMST1} by plugging in it the above mentioned form of $R_b$ \eqref{eq77} and the state $\ket{\psi_{ABE}}$, we have
\begin{equation}
\bra{\psi_{ABE}}X^iZ^j\otimes R_b\otimes\Id_E\ket{\psi_{ABE}}=\left[\sum_{m,n}\!\bra{\psi(\boldsymbol{\alpha})}X^{i}Z^{j}\otimes W_{m,n}\ket{\psi(\boldsymbol{\alpha})}\right]\bra{\xi_{B''E}}\widetilde{R}_{i,j}^{b}\otimes\Id_E\ket{\xi_{B''E}}.
\end{equation}
As was computed above to get \eqref{5.92} from \eqref{5.91}, the term inside the square bracket in the above expression is just $1$. Thus, we finally arrive at
\begin{equation}
\bra{\psi_{ABE}}X^iZ^j\otimes R_b\otimes\Id_E\ket{\psi_{ABE}}=\bra{\xi_{B''E}}\widetilde{R}_{i,j}^{b}\otimes\Id_E\ket{\xi_{B''E}}=\Tr\left(\widetilde{R}_{i,j}^{b}\sigma_{B''}\right)\qquad\forall b,
\end{equation}
where $\sigma_{B''}=\Tr_E(\proj{\xi_{B''E}})$. Let us now decompose $\sigma_{B''}$ using its eigenvectors denoted by $\ket{k}$ as $\sigma_{B''}=\sum_kp_k\proj{k}$. Plugging this into the above expression, we get that
\begin{eqnarray}
    \Tr\left(\widetilde{R}_{i,j}^{b}\sigma_{B''}\right)=\sum_kp_k\bra{k}\widetilde{R}_{i,j}^{b}\ket{k}.
\end{eqnarray}
Recalling again the identity \eqref{POVMST1} and then using Eq. \eqref{5.92},  we finally arrive at
\begin{eqnarray}\label{POVMST5}
\sum_kp_k\bra{k}\widetilde{R}_{i,j}^{b}\ket{k}=l_{i,j}^b.
\end{eqnarray}

Next, we introduce a family of POVM's $k$ as,  $\mathcal{I}_k=\{\mathcal{I}_{b,k}\}$, whose measurement operators are given by 
\begin{eqnarray}\label{5.98}
    \mathcal{I}_{b,k}&=&\Tr_{B''}\left[ \left(\Id_{B'}\otimes\ket{k}\!\bra{k}_{B''}\right)R_b\right]\nonumber\\
    &=&\sum_{i,j=0}^{d-1}\!\bra{k}\widetilde{R}_{i,j}^{b}\ket{k}W_{i,j}.
\end{eqnarray}
As $R$ is a valid POVM, as discussed in Chapter \ref{chapter_2} all its measurement operators are hermitian and positive semi-definite, that is, $R_b\geq 0$ for all $b$. Consequently, from the first line of Eq. \eqref{5.98}, we can see that $\mathcal{I}_{b,k}$ is also positive semi-definite, that is,  $\mathcal{I}_{b,k}\geq 0$ for any $k$ and $b$, as product of two positive semi-definite matrices is also positive semi-definite. Further,  $\sum_bR_b=\Id_B$ is identity and then using the first line of of Eq. \eqref{5.98}, we can directly see that $\sum_b\mathcal{I}_{b,k}=\Id_{B'}$ for any $k$. As a consequence, the family of POVM's $\{\mathcal{I}_b^k\}_b$ are valid quantum measurements. 
Let us now go back to Eq. (\ref{POVMST5}), and then rewrite it using the family of POVM's \eqref{5.98} as
\begin{eqnarray}
\mathcal{I}_b=\sum_{i,j=0}^{d-1}l_{i,j}^bW_{i,j}=\sum_{i,j=0}^{d-1}\sum_kp_k\bra{k}\widetilde{R}_{i,j}^{b}\ket{k}W_{i,j}=\sum_kp_k\mathcal{I}_{b,k}.
\end{eqnarray}
However, the POVM $\mathcal{I}$ is extremal and can not be decomposed as a convex mixture of other POVM's. Thus, we can immediately conclude that
\begin{eqnarray}
\forall k\qquad \mathcal{I}_{b,k}=\mathcal{I}_{b},
\end{eqnarray}
which is equivalent to the condition,
\begin{eqnarray}\label{eq831}
\forall k \qquad \bra{k}\widetilde{R}_{i,j}^{b}\ket{k}=l_{i,j}^b.
\end{eqnarray}
Next, we consider the following vectors belonging to $\mathbbm{C}^d$:
\begin{eqnarray}\label{eq841}
\ket{\varphi_{a,s,t}}=\frac{1}{\sqrt{2}}\left(\ket{s}\pm\mathbbm{i}^a\ket{t}\right),
\end{eqnarray}
where  $a=0,1$ and $\ket{s}$ and $\ket{t}$ are two distinct vectors belonging to the eigenbasis $\{\ket{k}\}$ of $\sigma_{B''}$. We now compute the following quantity
\begin{eqnarray}
\Tr_{B''}\left[\left(\Id_{B'}\otimes|\varphi_{a,s,t}\rangle\!\langle \varphi_{a,s,t}|_{B''}\right)R_b\right]=\sum_{i,j}\Tr(|\varphi_{a,s,t}\rangle\!\langle \varphi_{a,s,t}|_{B''}\widetilde{R}_{i,j}^{b}) W_{i,j}.
\end{eqnarray}
Expanding the above quantity using the explicit form of the vectors given in \eqref{eq841}, we obtain
\begin{eqnarray}
\Tr_{B''}\left[\left(\Id_{B'}\otimes|\varphi_{a,s,t}\rangle\!\langle \varphi_{a,s,t}|_{B''}\right)R_b\right]=\mathcal{I}_b\pm
\Tr_{B''}\left[(\Id_{B'}\otimes L^a_{B''})R_b\right],
\end{eqnarray}
where
\begin{eqnarray}\label{LB}
    L_{B''}^a=(\mathbbm{i}^a/2)\left(|t\rangle\!\langle s|+(-1)^a|s\rangle\!\langle t|\right).
\end{eqnarray} 
The fact that $R_b\geq 0$, imposes that left-hand side of the above 
expression is non-negative as it is a product of two matrices that are positive semi-definite matrices. This allows us to conclude that
\begin{eqnarray}\label{eq87}
\mathcal{I}_b\geq
\pm\Tr_{B''}\left[(\Id_{B'}\otimes L^a_{B''})R_b\right].
\end{eqnarray}
As discussed above, the measurement operators of any rank-one extremal POVM acting on $d-$dimensional Hilbert space can be expressed as $\mathcal{I}_b=\lambda_b\proj{\mu_b}$, where the vectors $\ket{\mu_b}$ are normalized and belong to $\mathbbm{C}^d$. Using this fact, we show in Observation \ref{fact5.2} stated in Appendix \ref{chap5} that the operator appearing on the right-hand side of the above expression (\ref{eq87}) must be rank one as well and must admit the following form
\begin{eqnarray}\label{POVMST2}
\Tr_{B''}\left[(\Id_{B'}\otimes L^a_{B''})R_b\right]=\lambda'_b\proj{\mu_b},
\end{eqnarray}
such that $\lambda_b\geq\pm\lambda'_b$. Recalling that $\sum_b R_b=\Id_{B}$ and that 
$\Tr L_{B''}^a=0$ for any $a$, we can finally conclude that
\begin{eqnarray}
\sum_b\Tr_{B''}\left[(\Id_{B'}\otimes L^a_{B''})R_b\right]=0=\sum_b\lambda_b'\proj{\mu_b}.
\end{eqnarray}
Since $\mathcal{I}_b$ are linearly independent, we learn from the above condition that $\lambda'_b=0$ for all $b$ which in turn implies from \eqref{POVMST2} that $\Tr_{B''}\left[(\Id_{B'}\otimes L^a_{B''})R_b\right]=0$ for any $b$ and $a$. Finally expanding the left hand side of \eqref{POVMST2} by plugging in the explicit form of $L^a_{B''}$ and also taking into account that $X^iZ^j$ are linearly independent for any $i,j$ gives us two simple conditions:
\begin{eqnarray}\label{POVMST3}
\left(X^iZ^j\right)^*\left(\bra{s}\widetilde{R}_{i,j}^{b}\ket{t}+\bra{t}\widetilde{R}_{i,j}^{b}\ket{s}\right)=0,
\end{eqnarray}
for $a=0$ and 
\begin{eqnarray}\label{POVMST4}
\left(X^iZ^j\right)^*\left(\bra{t}\widetilde{R}_{i,j}^{b}\ket{s}
-\bra{s}\widetilde{R}_{i,j}^{b}\ket{t}\right)=0,
\end{eqnarray}
for $a=1$. One can immediately see that the only possible solution of the above conditions \eqref{POVMST3} and \eqref{POVMST4} is $\bra{s}\widetilde{R}_{i,j}^{b}\ket{t}=0$ for $s\ne t$. Thus, 
from Eq. \eqref{eq831} we can conclude that the POVM acting on the support of Bob's local state is given by $R_b=\mathcal{I}_b\otimes\Id_{B''}$ for all $b's$. This completes the proof.
\end{proof}
It is worth noting that the above certification scheme works for any rank-one extremal POVM with arbitrary number of outcomes. However, it was shown in \cite{APP05} that any extremal POVM with $d^2$ outcomes have to be rank-one. As a consequence, we certify every $d^2-$outcome extremal POVM. We now show that the certified state and the certified POVM can be used for optimal randomness certification. 

\section{Optimal randomness certification}
\label{App E}

\begin{figure}
    \centering
    \includegraphics[scale=.45]{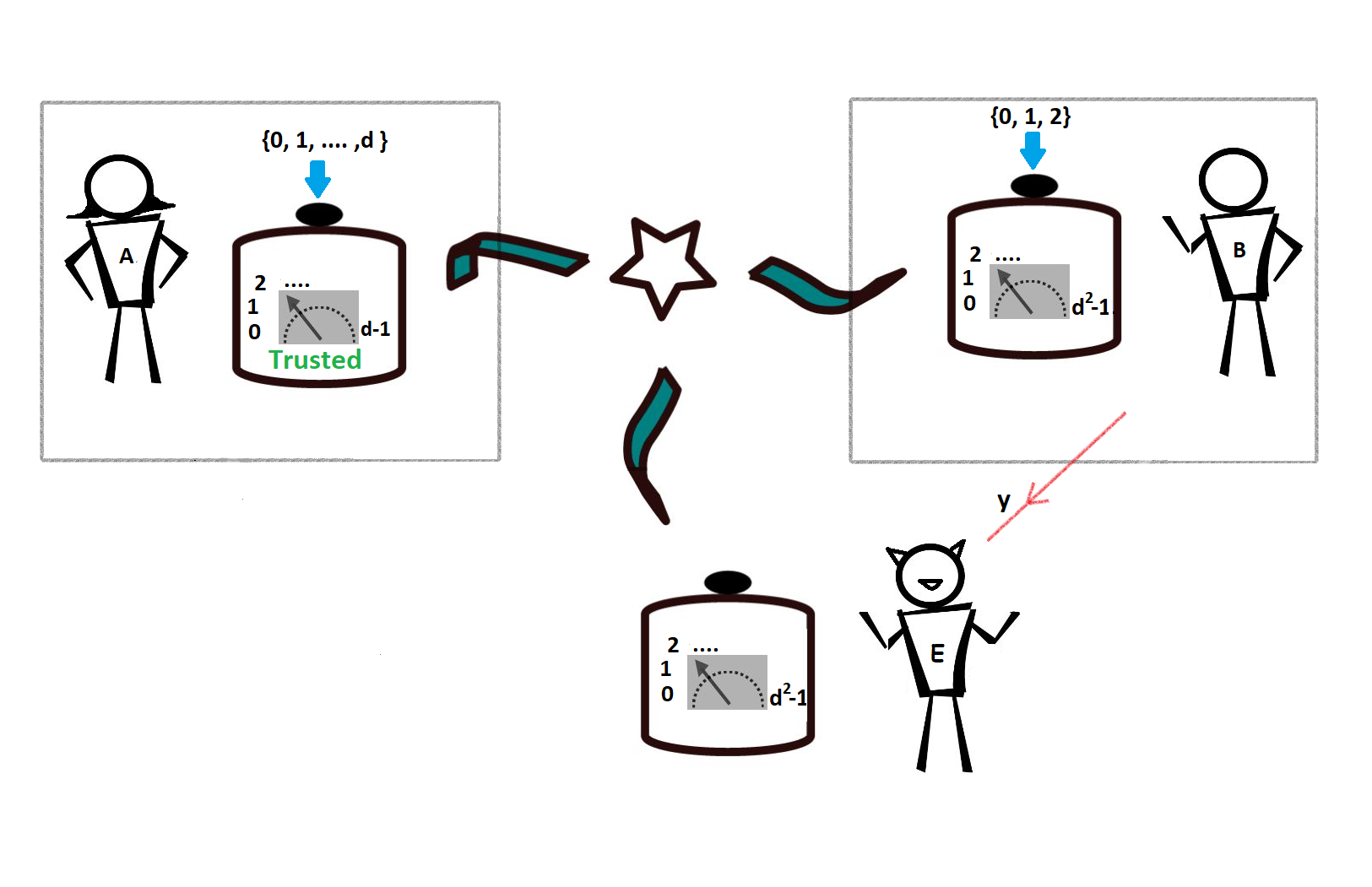}
    \caption{Optimal randomness certification in the 1SDI scenario: Alice (trusted) and Bob receive subsystems from the preparation device on which they perform $d+1$ and $3$ measurements respectively. All the measurements are $d-$outcome except Bob's third measurement which has $d^2$ outcomes. Using this measurement Bob wishes to generate $2\log_2d$ bits of randomness. Eve has knowledge about Bob's measurement choices. She also might receive an additional system from the preparation device. Using her measuring device and the received subsystem from the preparation she wants to guess Bob's outcome.}
    \label{fig5.2}
\end{figure}

Let us again go back to the previous scenario depicted in Fig. \ref{fig5.1}, but now let us assume that there is another party Eve who wants to guess Bob's outcome. As discussed in Chapter \ref{chapter_2}, Eve has full control on Bob's lab and also has access to the state sent by the preparation device. However, unlike randomness certification in the Bell scenario, in the 1SDI scenario Eve has no access to Alice's lab as it is trusted. This is depicted in Fig.  \ref{fig5.2}. 

Let us now say that Alice and Bob observe the maximal violation of the steering inequality \eqref{Stefn1} using the observables corresponding to the inputs $x,y=0,1$ where $A_0=Z_d$ and $A_1=X_d$. Now, using Theorem \ref{Theo1}, Alice and Bob can certify the quantum state shared between Alice and Bob up to local unitaries and additional degrees of freedom [see Eq. \eqref{lem5.1.2}]. Notice that in the proof we considered an external system $E$ that was used to purify the state shared between Alice and Bob. Without loss of generality, this external system in fact denotes the subsystem possessed by Eve and her Hilbert space is denoted by $\mathcal{H}_E$. Again, in the previous section, using this result we certified Bob's third measurement to be an ideal extremal measurement up to some local unitary [see Eq. \eqref{POVMcert}]. Notice that in the proof, the only condition we used apart from Theorem \ref{Theo1} was that the statistics one obtains in the actual experiment is the same as in the ideal experiment, even with the presence of the external subsystem $E$ which is held with Eve [cf. Eq. \eqref{POVMST1}]. 
%
Given the certified state and the measurements, let us now compute the probability of Eve to guess Bob's outcomes corresponding to the POVM $R$ [cf. Chapter \ref{chapter_2}], 
\begin{eqnarray}\label{G}
    G(y=2,\vec{p})&=& \sup_{S_{p}} \sum_b\bra{\psi_{ABE}}\Id_{A}\otimes R_b\otimes
E^{(b)}\ket{\psi_{ABE}}\nonumber\\
&=&\sup_{S_{p}}\!\bra{\psi(\boldsymbol{\alpha})}\Id_{A}\otimes \mathcal{I}_b\ket{\psi(\boldsymbol{\alpha})}\!\bra{\xi_{B''E}}\Id_{B''}\otimes
E^{(b)}\ket{\xi_{B''E}}.
\end{eqnarray}
Eve's strategy is composed of the states $\sigma_E=\Tr_{B''}\proj{\xi_{B''E}}$ and a measurement $Z=\{E^{(b)}\}$. Thus, simplifying the above expression we obtain
\begin{eqnarray}
 G(y=2,\vec{p})=\sup_{\sigma_E,Z}\sum_{b}\Tr[\mathcal{I}_b\rho_{B'}(\boldsymbol{\alpha})]
    \Tr[E^{(b)}\sigma_E]
\end{eqnarray}
where $\rho_{B'}(\boldsymbol{\alpha})=\Tr_A\proj{{\psi(\boldsymbol{\alpha})}}_{AB'}$. Using the fact $\sum_bE^{(b)}=\Id_E$, we can immediately observe from the above expression that
for any extremal POVM $\{\mathcal{I}_b\}$ if
\begin{equation}\label{ElCulo}
    \Tr[\mathcal{I}_b\rho_B(\boldsymbol{\alpha})]=\frac{1}{d^2} \qquad\forall b,
\end{equation}
then the guessing probability (\ref{G}) is $G(y=2,\vec{p})=1/d^2$. 
As a consequence, the maximal violation of the steering inequality (\ref{Stefn1}) along with 
with conditions (\ref{POVMST1}), can be used to certify $2\log_2d$ bits of randomness from Bob's POVM using any pure bipartite entangled state provided there exists an extremal POVM $\{\mathcal{I}_b\}$ that satisfies the condition (\ref{ElCulo}) for any $\rho_B(\boldsymbol{\alpha})$. 

Here we show an example of extremal qudit POVM that can be used to generate the optimal amount of randomness by Bob when he and Alice and share the two-qudit maximally entangled state $\ket{\phi^d_+}$.
We consider a simple construction of family of extremal $d^2-$outcome POVM's acting on arbitrary dimensional Hilbert space introduced in Ref. \cite{APP05}. It turns out that such POVM's serve as a perfect example to obtain the desired result. Consider the following $d^2$ unitary operators defined as
\begin{equation}
    U_{k,l}=X_d^kZ_d^l,
\end{equation}
where $k,l=0 \dots d-1$ and a vector $\ket{\nu}\in \mathbbm{C}^d$ such that $\Tr[U^\dag_{k,l} \ket{\nu} ] \ne 0 $ for any $k,l$. As proven in \cite{APP05}, the following $d^2-$outcome POVM given by
\begin{equation}\label{exPOVM1}
    \mathcal{I}_{k,l}:=\frac{1}{d} U_{k,l} \ket{\nu}\!\bra{\nu} U^\dag_{k,l},
\end{equation}
%
is extremal. Notice that when $\ket{\psi_{AB}}=\ket{\phi_+^d}$, then the local states of Alice and Bob are $\rho_A=\rho_B=\Id_d/d$. Now, plugging this reduced state and the above POVM \eqref{exPOVM1} into the left hand side of the condition \eqref{ElCulo}, we see that
\begin{eqnarray}
\Tr[\mathcal{I}_b\rho_B(\boldsymbol{\alpha})]=\frac{1}{d^2}\Tr(U_{k,l} \ket{\nu}\!\bra{\nu} U^\dag_{k,l})=\frac{1}{d^2}\qquad \forall k,l.
\end{eqnarray}
Thus, the example provided above can be used to obtain optimal randomness.


Now, we also demonstrate that the optimal randomness can be certified when Alice and Bob share partially entangled states. For this purpose, we find examples of extremal qudit POVMs with $d^2$ outcomes when $d=3,4,5,6$ that can be used to securely generate $2\log_2d$ amount of random bits using a particular class of partially entangled pure states $\ket{\psi(\boldsymbol{\alpha})}_{AB}$ such that $\alpha_i\geq 1/d$ for $i=0,1,\ldots,d-2$.
These extremal POVM's are given by,
\begin{equation} 
    \mathcal{I}_{b}:=\lambda_{b} \ket{\delta_b}\!\bra{\delta_b}.
\end{equation}
For $b=0, \ldots,  d-2$, the vectors $\ket{\delta_b}$ are given by
\begin{eqnarray} 
    \ket{\delta_b}=\ket{b},
\end{eqnarray}
whereas, for $b=d-1,\ldots,d^2-1$ they are defined as
\begin{eqnarray}\label{POVMele2}
    \ket{\delta_b}=\sum_{i=0}^{d-1}\mu_i \exp{\left( \frac{2\pi \mathbbm{i}\xi_i(b-d+1)}{d^2-d+1}\right)}\ket{i}
\end{eqnarray}
where
\begin{eqnarray}
    \mu_i=\sqrt{\frac{1-\lambda_i}{(d^2-d+1)\lambda_d}}\qquad (i=0,1,\ldots,d-2)
\end{eqnarray}
and
\begin{equation}
    \quad\mu_{d-1}=\sqrt{\frac{1}{(d^2-d+1)\lambda_d}}.
\end{equation}
The $\lambda_b's$ are given by
\begin{eqnarray}
    \lambda_i=\frac{1}{d^2\alpha_i^2}\qquad (i=0,1,\ldots,d-2) 
\end{eqnarray}
and
\begin{equation}
 \lambda_{d-1}=\lambda_d=\ldots=\lambda_{d^2-1}=\frac{1}{d^2-d+1}\left(d-\sum_{i=0}^{d-2}\lambda_i\right).
\end{equation}
Finally, below we provide the $\xi_i$ coefficients. \\
For $d=3$:
%
\begin{eqnarray}
    \xi_0=0,\quad\xi_1=1\quad\text{and}\quad\xi_2=3.
\end{eqnarray}
For $d=4$:
\begin{eqnarray}
    \xi_0=0,\quad\xi_1=1,\quad \xi_2=3\quad\text{and}\quad\xi_3=9.
\end{eqnarray}
For $d=5$:
\begin{eqnarray}
    \xi_0=0,\quad\xi_1=1,\quad\xi_2=4,\quad\xi_3=14\quad\text{and}\quad\xi_4=16.
\end{eqnarray}
For $d=6$:
\begin{eqnarray}
    \xi_0=0,\quad\xi_1=1,\quad \xi_2=3,\quad \xi_3=8,
    \quad\xi_4=12\quad\text{and}\quad\xi_4=18.
\end{eqnarray}
All these coefficients were found numerically such that the above constructed POVM is extremal and satisfies the condition \eqref{ElCulo}. 

\section{Conclusions and Discussions}

In this chapter, we first constructed a family of steering inequalities that are maximally violated by every pure entangled bipartite state. The only other work that provides a steering inequality that is maximally violated by any pure entangled state is Ref. \cite{Dani2018}. However, the inequality proposed in Ref. \cite{Dani2018} requires the trusted party to perform the full tomography on her subsystem. On the other hand, our scheme is the most efficient in terms of the number of measurements, as we require only two measurements to be performed by both the parties. This is the minimal number required to observe any form of quantum non-locality. We then showed that the maximal violation of our inequality allows one to certify any pure bipartite entangled state in the 1SDI scenario. A method for certification of any pure entangled bipartite state in the 1SDI scenario was also proposed in \cite{Bharti}, but their approach is a direct translation of the method of Ref. \cite{Projection} to the 1SDI scenario that relies on certification of two-qubit states as discussed before in Chapter \ref{chapter_3}. The scheme also requires both the parties to perform three and four measurements respectively. Contrary to this, our certification scheme relies only on two genuinely $d$-outcome measurements per party making it extremely useful for experimental implementation. Moreover, our scheme does not need to assume that the state shared among Alice and Bob is pure and the measurements performed by them are projective.

Improving on the results presented in the previous Chapter \ref{chapter_4}, we went on further with our self-testing result and used it to certify any rank-one extremal POVM in 1SDI scenario. Apart from it, we also showed that this task can be accomplished with quantum states that are close to separable states, that is, quantum states with a low level of entanglement. This makes our scheme resource friendly. 
Finally, we utilised both these results to devise a simple scheme for certification of the optimal amount of $2\log_2d$ bits of randomness using quantum systems of local dimension $d$ in the 1SDI scenario. Apart from its importance towards application in quantum cryptography, our result answers a long-standing question in the quantum foundations community of whether one can securely generate this optimal amount of randomness. Moreover, for a few finite dimensions, we showed that we can also accomplish this task using low levels of entanglement.

Some interesting follow-up questions arise from our work. First, and the most important one would be to explore whether our construction of the steering functionals can be used to design Bell functionals whose maximal quantum value is achieved by any pure entangled bipartite state and two measurements per site which can be later used for self-testing.
Another interesting problem would be to devise a scheme for certification of extremal measurements of arbitary rank, and thus certify any quantum measurement in the 1SDI scenario. A direct follow-up problem from our work is to find extremal POVM's satisfying the condition \eqref{ElCulo} for any $\rho_B(\boldsymbol{\alpha})$. 
A more challenging problem would be then to find a fully device-independent scheme for certification of optimal randomness using quantum systems of arbitrary local dimension. For the particular case of $d=2$ and $d=3$, schemes have been devised that can certify $2\log_22$ and $2\log_23$ bits of local randomness in \cite{random1} and \cite{sarkar5} respectively (see also Ref. \cite{Tavakoli2021}).

%% file: Sections/conclusions.tex
\chapter{Concluding remarks}
\label{chapter_6}
\vspace{-1cm}
\rule[0.5ex]{1.0\columnwidth}{1pt} \\[0.2\baselineskip]

\section{Summary of the thesis}
Let us finally summarise the major points presented throughout this thesis and their relevance to the current literature in quantum information and foundations. 

\subsubsection{From a theoretical perspective}

\begin{enumerate}
    \item All the results presented in this thesis are very general, in the sense that, they are applicable to composite quantum systems of arbitrary local dimensions. The results presented in Chapter \ref{chapter_3} are even applicable to arbitrary number of parties. As a matter of fact, the results presented in Chapter \ref{chapter_4} and Chapter \ref{chapter_5} can also be generalised to arbitrary number of parties using similar mathematical techniques.
    
    \item Most of the known certification schemes in the device-independent regime deal with states that are locally qubits and thus can utilise the Jordan's lemma \cite{Pironio1} which simplifies the mathematical considerations significantly. Our work is thus interesting from a mathematical point of view, as we develop new mathematical techniques that are applicable to arbitrary finite dimensional quantum systems.
    
    \item The result presented in Chapter \ref{chapter_3} provide an interesting insight into the structure of quantum sets. For instance, it was exploited in Ref. \cite{qs5} to show that the set of quantum correlations in a certain Bell scenario is not closed.
    
    \item An open question in quantum information has been whether one can securely generate the optimal randomness using quantum systems of arbitrary local dimension. In Chapter \ref{chapter_5}, we show  that it is possible in the 1SDI scenario. 
\end{enumerate}

\subsubsection{From a practical perspective}

\begin{enumerate}
    \item We showed in Chapter \ref{chapter_3} that one can certify the generalised GHZ state using just two genuinely $d-$outcome measurements per party in a fully device-independent way. This is the minimum number of measurements required to observe quantum non-locality and thus more efficient as compared to the existing schemes with regards to implementing it in experiments as one needs to observe minimum number of correlations to certify the state.
    
    \item Mutually unbiased bases are essential for quantum cryptographic tasks. In Chapter \ref{chapter_4}, we presented robust certification of mutually unbiased bases in the 1SDI scenario. This is the first instance where a method for certification of a general family of measurements, termed genuinely incompatible, have been introduced, which is based on quantum steering.
    
    \item In Chapter \ref{chapter_5}, we demonstrated a method to certify any pure bipartite entangled state using just two genuinely $d-$outcome measurements per party in the 1SDI scenario. This is again the most efficient protocol till date that can be used to certify such states. We then showed that any rank-one extremal measurement can be certified in the 1SDI scenario.
    
    \item  Any cryptographic task requires access to sources generating random bits. In Chapter \ref{chapter_3}, we demonstrated a protocol to generate randomness of amount $\log_2d$ bits using projective measurements in a fully device-independent way. Then, in Chapter \ref{chapter_5}, we showed that optimal randomness of amount $2\log_2d$ bits can be certified using a quantum system and a generalised measurement in the 1SDI scenario. Both of these protocols are secure against any eavesdropper who has access to quantum resources. 
\end{enumerate}
Let us list some of the interesting open questions that stem from this work.

\section{Open questions for further exploration}
\begin{enumerate}
    \item The first open question that naturally stems from this work is to find Bell inequalities inspired from the construction of our steering inequalities in Chapter \ref{chapter_5} that are maximally violated by any pure bipartite entangled state and utilises only two measurements per party. Then, building on our techniques described in Chapter \ref{chapter_3}, it would be extremely interesting to prove self-testing statements of such states in the fully DI scenario using minimal number of measurements.
    
    \item Generalising the steering inequalities in Chapter \ref{chapter_5} to the multipartite scenario in order for 1SDI certification of any pure multipartite entangled state. The author of this thesis is currently investigating the possibility to certify certain class of multipartite states in the 1SDI scenario such as graph states and Schmidt states by extending the approach presented in Chapter \ref{chapter_4} and Chapter \ref{chapter_5}. 
    
    \item Generalising the certification method of rank-one extremal measurements in the 1SDI scenario to measurements of arbitrary rank. Further, it would be interesting to explore whether one can reduce the number of measurements performed by each party, as in our scheme, the trusted party needs to perform $d+1$ measurements.
    
    \item Finding ideal POVM's that can be used to locally generate the optimal amount of randomness using any pure partially entangled bipartite state, or putting it simply, finding extremal POVM's $\{\mathcal{I}_b\}$ that satisfies the condition \eqref{POVMST1} for any non-singular local state $\rho_B$. 
    
    \item Extending all the 1SDI schemes presented in this thesis to the fully device-independent scenario. In particular, it would be extremely interesting to provide a way to certify the optimal amount of randomness in the fully device-independent scenario. Further, certifying the mutually unbiased bases in a fully device-independent way has been a long sought after question in quantum information community. For instance, if one can fully characterise the trusted side in the scheme presented in Chapter \ref{chapter_4}, then this scheme becomes fully device-independent.
\end{enumerate}

%% file: Sections/appendices.tex


\begin{appendices}

\chapter{Some general mathematical facts}\label{chapano}

We prove here some identities that were extensively used throughout this work.

\begin{fakt}\label{fact1} Consider a matrix $U$ acting on a Hilbert space $\mathcal{H}$ and $\overline{U}=\Pi U\Pi$, where $\Pi$ is a projection onto some subspace $\mathcal{K}$ of $\mathcal{H}$. Then the following properties hold true:
\begin{enumerate}
    \item If $U$ is unitary, then $\overline{U}\overline{U}^{\dagger}\leq\Id_\mathcal{K}$ where $\Id_\mathcal{K}$ is identity acting on the subspace $\mathcal{K}$. 
    \item If $U$ is hermitian, then $\overline{U}$ is also hermitian.
    \item If both $U$ and $\overline{U}$ are unitary, then $U=\overline{U}\oplus C$ such that $C$ is also unitary. 
\end{enumerate}
\end{fakt}
\begin{proof}
Given that $\overline{U}$ is a projection of $U$ onto some subspace $\mathcal{K}$, then $U$ can be written in the matrix form as
\begin{eqnarray}\label{2.128}
      U=\begin{pmatrix}
\overline{U} & A\\
B& C
\end{pmatrix},
\end{eqnarray}
where $A=\Pi U \Pi^{\perp},\ B=\Pi^{\perp} U \Pi,\ C=\Pi^{\perp} U \Pi^{\perp}$. Here, $\Pi^{\perp}$ is the projection onto the subspace $\mathcal{K}^{\perp}$ of $\mathcal{H}$ that is orthogonal to $\mathcal{K}$. Since, $U$ is unitary
\begin{eqnarray}
    U U^{\dagger} =\begin{pmatrix}
\overline{U}\overline{U}^{\dagger}+ AA^{\dagger}& \overline{U}B^{\dagger}+AC^{\dagger}\\
B\overline{U}^{\dagger}+CA^{\dagger}& BB^{\dagger}+CC^{\dagger}
\end{pmatrix}=\begin{pmatrix}
\Id_\mathcal{K}& 0\\
0& \Id_\mathcal{K^{\perp}}
\end{pmatrix}.
\end{eqnarray}
Here $\Id_\mathcal{K^{\perp}}$ is identity acting on the subspace $\mathcal{K}^{\perp}$. Since $AA^{\dagger}$ is positive, we can conclude that
\begin{eqnarray}
    \overline{U}\overline{U}^{\dagger}\leq\Id.
\end{eqnarray}
Now, if $U$ is hermitian, that is, $U^{\dagger}=U$ then for $\overline{U}$ we have that
\begin{eqnarray}
     \overline{U}^{\dagger}=(\Pi U\Pi)^{\dagger}=\Pi U^{\dagger}\Pi=\Pi U\Pi=\overline{U}
\end{eqnarray}
and thus $\overline{U}$ is also hermitian if $U$ is hermitian. One can also realise this by observing \eqref{2.128}.

Let us now also assume that $\overline{U}$ is unitary, that is, $\overline{U}\overline{U}^{\dagger}=\Id$. Since $U$ is unitary, from the diagonal element in \eqref{2.129} we have that $AA^{\dagger}=0$ which implies that $A=0$. Again, from the off-diagonal element, we have that $\overline{U}B^{\dagger}=0$ which imposes that $B=0$ as $\overline{U}$ is unitary and thus invertible. Similarly, one can prove that $C=0$. Thus, the matrix $U$ reduces to 
\begin{eqnarray}
      U=\begin{pmatrix}
\overline{U} & 0\\
0& C
\end{pmatrix}
\end{eqnarray}
which is equivalent to saying that $U=\overline{U}\oplus C$.
\end{proof}

\begin{fakt}\label{factmaxent} Consider two matrices $R, Q$ acting on $d-$dimensional Hilbert space. Then, the following relation holds true when they act on the two-qudit maximally entangled state,
\begin{eqnarray}\label{4.34}
R\otimes Q\ket{\phi_d^{+}}=RQ^T\otimes\mathbbm{1}\ket{\phi_d^{+}}
\end{eqnarray}
such that $Q^T$ represents the transpose of $Q$.
\end{fakt}
\begin{proof}
Let us first expand the matrices $R, Q$ using the $d-$dimensional computational basis as
\begin{eqnarray}
R=\sum_{i,j=0}^{d-1}r_{i,j}\ket{i}\!\bra{j},\qquad Q=\sum_{i,j=0}^{d-1}q_{i,j}\ket{i}\!\bra{j}.
\end{eqnarray}
Let us now evaluate the right hand side of \eqref{4.34} by employing the form of the maximally entangled state $\ket{\phi_d^{+}}$ given in \eqref{maxentstated},
\begin{eqnarray}\label{abcd}
R\otimes Q\ket{\phi_d^{+}}=\frac{1}{\sqrt{d}}\sum_{i,k,j=0}^{d-1}r_{i,j}q_{k,j}\ket{i}\ket{k}.
\end{eqnarray}
 Notice that $Q^T=\sum_{i,j=0}^{d-1}q_{j,i}\ket{i}\!\bra{j}$ using which we get 
 \begin{eqnarray}
 RQ^T=\sum_{i,k,j=0}^{d-1}r_{i,j}q_{k,j}\ket{i}\bra{k}
 \end{eqnarray}
 Now, using the above expression let us evaluate the left hand side of the equation \eqref{4.34}
\begin{eqnarray}
RQ^T\otimes\mathbbm{1}\ket{\phi_d^{+}}=\frac{1}{\sqrt{d}}\sum_{i,k,j=0}^{d-1}r_{i,j}q_{k,j}\ket{i}\ket{k}
\end{eqnarray}
which is exactly same as the left hand side \eqref{abcd}.
\end{proof}

\begin{fakt}\label{iden} The following identities hold true:
\begin{eqnarray} \label{FijIden1}
  \sum_{\substack{j=0\\j\ne i}}^{d-1}\frac{1-\omega^{k(j-i)}}{1-\omega^{i-j}}=k, \qquad k=1,\ldots,d-1,\qquad i=0,\ldots,d-1,
\end{eqnarray}
and
\begin{eqnarray} \label{FijIden2} 
  \sum_{k=0}^{d-1}k\omega^{kn}=\frac{d}{\omega^n-1}, \qquad n=1,\ldots,d-1.
\end{eqnarray}
\end{fakt}
\begin{proof}
Let us begin by proving the first identity \eqref{FijIden1}. For this, we express the left-hand side of (\ref{FijIden1}) as
\begin{equation}
      \sum_{\substack{j=0\\j\ne i}}^{d-1}\frac{1-\omega^{k(j-i)}}{1-\omega^{i-j}}=-\sum_{\substack{j=0\\j\ne i}}^{d-1}\omega^{j-i}
      \left(\frac{1-\omega^{k(j-i)}}{1-\omega^{j-i}}\right).
\end{equation}
Notice that the term appearing inside the bracket on the right-hand side of the above expression is a sum of a geometric sequence $\sum_{k=0}^{n-1}w^k=(1-w^n)/(1-w)$. Thus, the above expression can be rewritten as
\begin{eqnarray}
      \sum_{\substack{j=0\\j\ne i}}^{d-1}\frac{1-\omega^{k(j-i)}}{1-\omega^{i-j}}&=&-\sum_{\substack{j=0\\j\ne i}}^{d-1}\left(\omega^{j-i}+\omega^{2(j-i)}+\ldots+ \omega^{k(j-i)}\right)\nonumber\\&=&-\sum_{n=1}^k\omega^{-ni}\left(\sum_{\substack{j=0\\j\ne i}}^{d-1}\omega^{nj}\right).
\end{eqnarray}
Evaluating the term inside the brackets of the above expression over $x$ for any $n=1,\ldots,d-1$, we obtain that
\begin{equation}\label{formulaOne}
  \sum_{\substack{j=0\\j\ne i}}^{d-1}\omega^{nj}=\sum_{j=0}^{d-1}\omega^{nj}-\omega^{ni}=-\omega^{ni},
\end{equation}
where we used the fact that $\sum_{j=0}^{d-1}\omega^{nj}=\delta_{n,0}$. Plugging this last formula into Eq. (\ref{formulaOne}) we finally arrive
at Eq. (\ref{FijIden1}). 

Let us now prove the second identity \eqref{FijIden2} for which we again consider a geometric sum given by,
\begin{eqnarray}
 \sum_{k=0}^{d-1}x^k=\frac{1-x^d}{1-x} .
\end{eqnarray}
Now, we take the derivative of the above expression and then multiply the resultant equation with $x$ on both sides, to finally obtain
\begin{eqnarray}
\sum_{k=0}^{d-1}kx^k=x\frac{\mathrm{d}}{\mathrm{d} x}\left(\frac{1-x^d}{1-x}\right)=\frac{-dx^{d}}{1-x}+\frac{x(1-x^d)}{(1-x)^2} \ .
\end{eqnarray}
Substituting $x=\omega^n$ and using the fact that $\omega^d=1$ we obtain \eqref{FijIden2}.
\end{proof}


\chapter{Proofs of some observations relevant to Chapter 3}\label{chap3}

\begin{customobs}{3.1} \label{fact:4sA} For any two unitary observables $A_{1,2}$ and $A_{1,3}$ related by the condition  \eqref{Obs22} which is given by
\begin{eqnarray}\label{SOSrelaApp}
\omega^{\frac{2k-d}{2m}}A_{1,2}^{k}A_{1,3}^{-k}+\omega^{-\frac{2k-d}{2m}}A_{1,3}^{k}A_{1,2}^{-k}=
2\cos\left(\frac{\pi}{m}\right)\mathbbm{1},
\end{eqnarray}
for $k=1,2,\ldots,d-1$. Then, the traces of $A_{1,2}$ and $A_{1,3}$ are related in the following way: 
\begin{equation}\label{NewId1A}
    \Tr(A_{1,2}^x)=\omega^{\frac{2tx}{m}}\,\Tr\left(A_{1,2}^{(2t+1)x}A_{1,3}^{-2tx}\right).
\end{equation}
for any non-negative integer $t\in \mathbb{N} \cup \{0\}$ and $x=1,\ldots,\lfloor d/2\rfloor$.
\end{customobs}
\begin{proof}
To prove the above claim we use the technique of mathematical induction. For this purpose, one can immediately observe that the condition \eqref{NewId1A} holds trivially for $t=0$. Now, let us  assume that the condition \eqref{NewId1A} is satisfied for $t=s-1$, that is,
\begin{equation}\label{Obs31}
    \Tr(A_{1,2}^x)=\omega^{\frac{2(s-1)x}{m}}\,\Tr\left(A_{1,2}^{(2s-1)x}A_{1,3}^{-2(s-1)x}\right)\qquad x = 1,\dots, \left\lfloor{\frac{d}{2}}\right\rfloor.
\end{equation}
Let us now show that the condition \eqref{NewId1A} is also satisfied for $t=s$.
For this purpose, let us consider (\ref{Obs22}) for $k=2sx$ and multiply it with $A_{1,2}^x$ on both the sides. Now, taking the trace of the resultant expression yields,
\begin{equation}\label{Eq1}
   \omega^{\frac{4sx-d}{2m}} \Tr\left(A_{1,2}^{(2s+1)x}A_{1,3}^{-2sx}\right)+\omega^{\frac{d-4sx}{2m}}\Tr\left(A_{1,3}^{2sx}A_{1,2}^{(-2s+1)x}\right)=\cos\left(\frac{\pi}{m}\right)\Tr\left(A_{1,2}^x\right).
\end{equation}
Again, considering the condition (\ref{Obs22}) for $k=(2s-1)x$ but now we  
multiply it with $A_{1,3}^x$ on both the sides. Taking the trace of the resulting expression gives us,
\begin{equation}
   \omega^{\frac{2(2s-1)x-d}{2m}} \Tr\left(A_{1,2}^{(2s-1)x}A_{1,3}^{-2(s-1)x}\right)+\omega^{\frac{d-2(2s-1)x}{2m}}\Tr\left(A_{1,3}^{2sx}A_{1,2}^{(-2s+1)x}\right)=\cos\left(\frac{\pi}{m}\right)\Tr\left(A_{1,3}^x\right).
\end{equation}
Utilising the condition \eqref{Obs26} for $k=x$, the above expression simplifies to
\begin{eqnarray}\label{Eq2}
 \omega^{\frac{4(s-1)x-d}{2m}} \Tr\left(A_{1,2}^{(2s-1)x}A_{1,3}^{-2(s-1)x}\right)+\omega^{\frac{d-4sx}{2m}}\Tr\left(A_{1,3}^{2sx}A_{1,2}^{(-2s+1)x}\right)=\cos\left(\frac{\pi}{m}\right)\Tr\left(A_{1,2}^x\right),
\end{eqnarray}
for $x = 1,\dots, \lfloor d/2\rfloor$ where $\lfloor d/2\rfloor$ denotes the largest integer smaller than $d/2$. Finally, subtracting Eq. (\ref{Eq2}) from Eq. (\ref{Eq1}) we obtain
\begin{equation}\label{Eq3}
    \Tr\left(A_{1,2}^{(2s+1)x}A_{1,3}^{-2sx}\right)=\omega^{-\frac{2x}{m}}\left(A_{1,2}^{(2s-1)x}A_{1,3}^{-2(s-1)x}\right),
\end{equation}
which along with Eq. \eqref{Obs31} gives us the desired result,
\begin{equation}
    \Tr\left(A_{1,2}^{(2s+1)x}A_{1,3}^{-2sx}\right)=\omega^{-\frac{2sx}{m}}\Tr(A_{1,2}^x).
\end{equation}
\end{proof}

%

\begin{customobs}{3.2}\label{obs3.1}
 Consider two unitary observables $A_{1,2}$ and $A_{1,3}$ related by the condition  \eqref{FijEq1} which is given by
\begin{eqnarray}
A_{1,3}^k=-(k-1)\omega^{\frac{k}{m}}A_{1,2}^k+\omega^{\frac{k-1}{m}}\sum_{t=0}^{k-1}A_{1,2}^tA_{1,3}A_{1,2}^{k-1-t}
\end{eqnarray}
 for any $k=1,\ldots,d-1$ and $m\geq 2$. If $A_{1,2}=Z_d\otimes\Id$ and $A_{1,3}=\sum_{i,j=0}^{d-1}\ket{i}\!\bra{j}\otimes F_{ij}$, then the matrices $F_{ij}$ are related as:
 \begin{eqnarray} \label{FijObs3A}
  -(k-1)\sum_{i,j=0}^{d-1}\omega^{ki}\ket{i}\!\bra{j}\otimes F_{ij}+\omega^{-\frac{1}{m}}\sum_{i,j=0}^{d-1}\ket{i}\!\bra{j}\otimes\left[\sum_{\substack{l=0\\l\ne i}}^{d-1}\left(\frac{\omega^{ki}-\omega^{kl}}{\omega^{i}-\omega^{l}}\right) F_{il}F_{lj}+k\omega^{(k-1)i}F_{ii}F_{ij}\right] \nonumber\\
  = -k\omega^{\frac{1}{m}}\sum_{i=0}^{d-1}\omega^{(k+1)i}\ket{i}\!\bra{i}\otimes \Id+\sum_{i,j=0}^{d-1}\ket{i}\!\bra{j}\otimes\sum_{t=0}^{k}\omega^{k j+t(i-j)} F_{ij}.\nonumber\\
\end{eqnarray}
for any $k=1,\ldots,d-1$ and $m\geq 2$.
\end{customobs}
\begin{proof}
To begin the proof, let us first notice that $A_{1,3}^{k+1}=A^k_{1,3}A_{1,3}$. Now, plugging in
$A_{1,3}^{k+1}$ and $A_{1,3}^k$ using Eq. (\ref{FijEq1}) gives us,
\begin{equation} \label{FijEq11a}
-k\w^{\frac{1}{m}}A_{1,2}^{k+1}+\sum_{t=0}^{k}A_{1,2}^tA_{1,3}A_{1,2}^{k-t}=-(k-1)A_{1,2}^kA_{1,3}+\omega^{-\frac{1}{m}}\sum_{t=0}^{k-1}A_{1,2}^tA_{1,3}A_{1,2}^{k-1-t}A_{1,3}.
\end{equation}
Substituting the explicit forms of the observables $A_{1,2}=Z_d\otimes\Id$ and $A_{1,3}=\sum_{i,j=0}^{d-1}\ket{i}\!\bra{j}\otimes F_{ij}$, we obtain that
\begin{equation}
\sum_{t=0}^{k-1}A_{1,2}^tA_{1,3}A_{1,2}^{k-1-t}A_{1,3} 
= \sum_{i,j=0}^{d-1}\ket{i}\!\bra{j}\otimes\sum_{l=0}^{d-1}\sum_{t=0}^{k-1}\omega^{l(k-1)}\omega^{t(i-l)}F_{il}F_{lj}.
\end{equation}
Splitting the sum  over $l$ appearing on the right hand side into two parts: $l=i$ and $l\neq i$, 
and then using the sum computed using the geometric series
\begin{equation}
    \sum_{t=0}^{k-1}\w^{t(i-l)}=\frac{1-\omega^{k(i-l)}}{1-\omega^{i-l}},
\end{equation}
we obtain that
\begin{equation} \label{eq83}
\sum_{t=0}^{k-1}A_{1,2}^tA_{1,3}A_{1,2}^{k-1-t}A_{1,3}  
= \sum_{i,j=0}^{d-1}\ket{i}\!\bra{j}\otimes\left[\sum_{\substack{l=0\\l\ne i}}^{d-1}\left(\frac{\omega^{ki}-\omega^{kl}}{\omega^{i}-\omega^{l}}\right) F_{il}F_{lj}+k\omega^{(k-1)i}F_{ii}F_{ij}\right].
\end{equation}
Using exactly the same technique, the sum on the left-hand side of Eq. (\ref{FijEq11a}) can be rewritten as,
\begin{equation}  \label{eq84}
\sum_{t=0}^{k}A_{1,2}^tA_{1,3}A_{1,2}^{k-t} = \sum_{i,j=0}^{d-1}\omega^{kj}\sum_{t=0}^{k}\omega^{t(i-j)}\ket{i}\!\bra{j}\otimes F_{ij}.
\end{equation} 
Finally substituting Eqs. (\ref{eq83}) and (\ref{eq84}) into Eq. (\ref{FijEq11a}), we obtain
(\ref{FijObs3}).
\end{proof}

\begin{customobs}{3.3}
\label{fact:1}
Consider the following unitary operators $W_1,W_2, W_{\mathrm{odd}}, W_{\mathrm{ev}}:\mathbbm{C}^d\rightarrow \mathbbm{C}^d$ given by
\begin{eqnarray}\label{Unitaries}
W_1&=&\frac{1}{\sqrt{d}}\sum_{i,j=0}^{d-1}(-1)^{\delta_{j,0}}\omega^{-\frac{3i}{2m}+ij+\frac{j}{2}}\ket{i}\!\bra{j},\nonumber\\
W_2&=&\frac{1}{\sqrt{d}}\sum_{i,j=0}^{d-1}(-1)^{\delta_{j,0}}\omega^{-\frac{2i}{m}+ij+\frac{j}{2}}\ket{d-1-i}\!\bra{j}, \nonumber\\
W_{\mathrm{odd}}&=&\frac{1}{\sqrt{d}}\sum_{i,j=0}^{d-1}(-1)^{\delta_{j,0}}\omega^{-\frac{i}{m}+ij+\frac{j}{2}}\ket{i}\!\bra{j},\nonumber\\
W_{\mathrm{ev}}&=&\frac{1}{\sqrt{d}}\sum_{i,j=0}^{d-1}(-1)^{\delta_{j,0}}\omega^{-\frac{i}{m}+ij+\frac{j}{2}}\ket{d-1-i}\!\bra{j},
\end{eqnarray}
where $\{\ket{i}\}$ is the standard basis on $\mathbbm{C}^d$. These unitaries transform $Z_d,T_{d,m}$ as defined in Eq. \eqref{ZdTd} to the ideal measurements given in Eq. \eqref{measurements},\eqref{measurements2} and \eqref{measurements3} in the following way: $\mathcal{O}_{i,2}=W_i\,Z_d\,W_i^{\dagger}$
and $\mathcal{O}_{i,3}=W_i\,T_{d,m}\,W_i^{\dagger}$
%
%
for the parties $i=1,2$, and
$\mathcal{O}_{n_{\mathrm{odd}},2}=W_{\mathrm{odd}}\,Z_d\,W_{\mathrm{odd}}^{\dagger}, \mathcal{O}_{n_{\mathrm{odd}},3}=W_{\mathrm{odd}}T_{d,m}W_{\mathrm{odd}}^{\dagger}$, and
$\mathcal{O}_{n_{\mathrm{ev}},2}=W_{\mathrm{ev}}Z_dW_{\mathrm{ev}}^{\dagger}, \mathcal{O}_{n_{\mathrm{ev}},3}=W_{\mathrm{ev}}T_{d,m}W_{\mathrm{ev}}^{\dagger}$ for remaining parties. Here, the subscripts odd and ev refer to parties that are numbered by odd and even numbers respectively.

\end{customobs}
%
%
\begin{proof} Before proceeding, let us recall the measurements $Z_d,T_{d,m}$ as defined in Eq. \eqref{ZdTd} 
\begin{eqnarray}\label{ZdA}
    Z_d&=&\sum_{i=0}^{d-1}\omega^i\proj{i}\nonumber\\
 T_{d,m}&=&\sum_{i=0}^{d-1}\omega^{i+\frac{1}{m}}\proj{i}-\frac{2\mathbbm{i}}{d}\sin\left(\frac{\pi}{m}\right)\sum_{i,j=0}^{d-1}(-1)^{\delta_{i,0}+\delta_{j,0}}\omega^{\frac{i+j}{2}-\frac{d-2}{2m}}|i\rangle\!\langle j|.
\end{eqnarray}
Let us also recall the ideal observables given in Eq. \eqref{measurements},\eqref{measurements2} and \eqref{measurements3} which we express here in their matrix form as
\begin{eqnarray}\label{Obs12A}
\mathcal{O}_{1,x}&=&\sum_{i=0}^{d-2}\omega^{\gamma_m(\alpha)}\ket{i}\!\bra{i+1}+\omega^{(1-d)\gamma_m(\alpha)}\ket{d-1}\!\bra{0},\nonumber\\  \mathcal{O}_{2,x}&=&\sum_{i=0}^{d-2}\omega^{\zeta_m(\alpha)}\ket{i+1}\!\bra{i}+\omega^{(1-d)\zeta_m(\alpha)}\ket{0}\!\bra{d-1}
\end{eqnarray}
for the first two parties, and 
\begin{eqnarray}\label{ObsoddevenA}
\mathcal{O}_{\mathrm{odd},x}&=&\sum_{i=0}^{d-2}\omega^{\theta_m(\alpha)}\ket{i}\!\bra{i+1}+\omega^{(1-d)\theta_m(\alpha)}\ket{d-1}\!\bra{0}, \nonumber\\ \mathcal{O}_{\mathrm{ev},x}&=&\sum_{i=0}^{d-2}\omega^{\theta_m(\alpha)}\ket{i+1}\!\bra{i}+\omega^{(1-d)\theta_m(\alpha)}\ket{0}\!\bra{d-1}
\end{eqnarray}
for the remaining parties.
Let us begin by finding the eigen-decomposition of the ideal measurements \eqref{measurements},\eqref{measurements2} and \eqref{measurements3} for the second and third measurements for each party as,
\begin{equation}
 \mathcal{O}_{n,x}=\sum^{d-1}_{r=0} \w^r \ket{r}\!\bra{r}_{n,x}
\end{equation}
with $x=2,3$ and $n=1,2,\ldots,N$, where the eigenvectors are defined as
\begin{eqnarray}\label{idealeigen}
 \ket{r}_{1,x}&=&\frac{1}{\sqrt{d}}\sum_{q=0}^{d-1} \omega^{(r-\gamma_m(x))q}\ket{q}, \nonumber\\
 \ket{r}_{2,x}&=&\frac{1}{\sqrt{d}}\sum_{q=0}^{d-1} \omega^{-(r-\zeta_m(x))q}\ket{q},\nonumber\\
 \ket{r}_{n_{\mathrm{odd}},x}&=&\frac{1}{\sqrt{d}}\sum_{q=0}^{d-1} \omega^{(r-\theta_m(x))q}\ket{q},\nonumber\\
 \ket{r}_{n_{\mathrm{ev}},x}&=&\frac{1}{\sqrt{d}}\sum_{q=0}^{d-1} \omega^{-(r-\theta_m(x))q}\ket{q}
 \end{eqnarray}
where the coefficients $\gamma_m(x)$, $\zeta_m(x)$ and $\theta_m(x)$ are given in Eq. \eqref{measupara} and $\{\ket{q}\}$ denotes the computational basis of $\mathbbm{C}^d$. Notice that the set of vectors $\{\ket{r}_{i,x}\}$ for any particular $i$ and $x$ are mutually orthogonal.

Let us now consider the eigen-decompositions of $Z_d$ and $T_{d,m}$,
\begin{equation}
    Z_d = \sum^{d-1}_{q=0}\w^q \ket{q}\!\bra{q},\qquad T_{d,m}= \sum^{d-1}_{r=0} \w^r \ket{r}\!\bra{r}_{T}. 
\end{equation}

Here again 
$\{\ket{q}\}$ is the computational basis of $\mathbbm{C}^d$ and $\ket{r}_{T}$
denote the eigenvectors of $T_{d,m}$ given by
\begin{eqnarray} \label{rTd}
  \ket{r}_{T}= \frac{2\mathbbm{i}}{d}\sin\left(\frac{\pi}{m}\right)\w^{-\frac{d}{2m}}\sum_{q=0}^{d-1}(-1)^{\delta_{q,0}}\frac{\omega^{-\frac{q}{2}}}{1-\omega^{r-q-\frac{1}{m}}}\ket{q}.
  \end{eqnarray}

Let us now go back to the main proof and show that the unitaries $W_1, W_2, W_{\mathrm{odd}}, W_{\mathrm{ev}}$ \eqref{Unitaries} transform $Z_d$ and $T_{d,m}$ to the ideal measurements $\mathcal{O}_{i,2}$ and $\mathcal{O}_{i,3}$ for any $i=1,2,\ldots,N$.
For this purpose, we show that under the action of the unitaries the eigenvectors of one observable transform to the eigenvectors of another observable up to a complex number.

Let us begin by considering the first party. The action of $W_1^{\dagger}$ on 
the eigenvectors of $\mathcal{O}_{1,2}$, $\ket{r}_{1,2}$, given explicitly in Eq. (\ref{idealeigen}) can be expressed as
%
 %
 %
 %
 \begin{eqnarray}
 W_1^{\dagger}\ket{r}_{1,2}
 &=& \frac{1}{d}\sum_{j,q=0}^{d-1}(-1)^{\delta_{j,0}}\omega^{(r-j)q}\omega^{-\frac{j}{2}}\ket{j}.
 \end{eqnarray}
Using the fact that $ \sum_{q=0}^{d-1}\w^{(r-j)q}=d\delta_{r,j}$,
the above expression simplifies to
\begin{eqnarray}\label{A1toZd}
W_1^{\dagger}\ket{r}_{1,2}
= \w^{\delta_{j,0}-\frac{j}{2}}\ket{j}.
\end{eqnarray}
Recall that $\ket{j}$ are the eigenvectors of $Z_d$ and thus we obtain that $W_1^{\dagger}\mathcal{O}_{1,2}\, W_1=Z_d$. 
Now, the action of $W_1^{\dagger}$ on 
the eigenvectors of $\mathcal{O}_{1,3}$, $\ket{r}_{1,3}$, given explicitly in Eq. (\ref{idealeigen}) can be expressed as
 \begin{eqnarray}\label{B.16}
W_1^{\dagger}\ket{r}_{1,3}
 &=& \frac{1}{d}\sum_{j,q=0}^{d-1}(-1)^{\delta_{j,0}}\omega^{(r-j-\frac{1}{m})q}\omega^{-\frac{j}{2}}\ket{j} .
 \end{eqnarray}
Using the following relation derived from the sum of a geometric series
\be \label{sumimpW}
\sum_{l=0}^{d-1}\omega^{(r-k-\frac{1}{m})l}=\frac{1-\w^{-\frac{d}{m}}}{1-\omega^{r-k-\frac{1}{m}}}=2\mathbbm{i}\sin\left(\frac{\pi}{m}\right)\frac{\w^{-\frac{d}{2m}}}{1-\omega^{r-k-\frac{1}{m}}},
\ee 
and (\ref{rTd}), the expression \eqref{B.16} simplifies to
 \begin{eqnarray} \label{A2toTd}
  W_1^{\dagger}\ket{r}_{1,3}
 &=&\frac{2\mathbbm{i}}{d}\sin\left(\frac{\pi}{m}\right)\w^{-\frac{d}{2m}}\sum_{j=0}^{d-1}(-1)^{\delta_{j,0}}\frac{\omega^{-\frac{j}{2}}}{1-\omega^{r-j-\frac{1}{m}}}\ket{j}\nonumber\\
  &=&\ket{r}_{T_{d,m}}.
 \end{eqnarray}
As a consequence, $W_1^{\dagger}\,\mathcal{O}_{1,3}\,W_{1}=T_{d,m}$.

Let us now consider the second party. The action of $W_2^{\dagger}$ on 
the eigenvectors of $\mathcal{O}_{2,2}$, $\ket{r}_{2,2}$, given explicitly in Eq. (\ref{idealeigen}) can be expressed as

 %
 %
%
 \begin{eqnarray}
 W_2^{\dagger}\ket{r}_{2,2}
 &=& \frac{1}{d}\sum_{j,q=0}^{d-1}(-1)^{\delta_{j,0}}\omega^{(r-j)q+(d-1)\left(\frac{2}{m}-r\right)-\frac{j}{2}}\ket{j},
 \end{eqnarray}
and, after employing the fact that $ \sum_{q=0}^{d-1}\w^{(r-j)q}=d\delta_{r,j}$,
the above expression simplifies to
 \begin{eqnarray}\label{B1toZd}
  W_2^{\dagger}\ket{r}_{2,2}
  = (-1)^{\delta_{j,0}}\w^{(d-1)\left(\frac{2}{m}-j\right)-\frac{j}{2}} \ket{j}.
 \end{eqnarray}
Recall that $\ket{j}$ are the eigenvectors of $Z_d$ and thus we obtain that $W_2^{\dagger}\mathcal{O}_{2,2}\, W_2=Z_d$.
Now, the action of $W_2^{\dagger}$ on 
the eigenvectors of $\mathcal{O}_{2,3}$, $\ket{r}_{2,3}$, given explicitly in Eq. (\ref{idealeigen}) can be expressed as
 \begin{eqnarray}
 W_2^{\dagger}\ket{r}_{2,3}
 &=& \frac{1}{d}\sum_{j,q=0}^{d-1}(-1)^{\delta_{j,0}}\omega^{(r-j-\frac{1}{m})q + (d-1)\left(\frac{2}{m}-r\right)-\frac{j}{2}}\ket{j} ,
 \end{eqnarray}
which can be simplified using Eq. \eqref{sumimpW} to
 \begin{eqnarray}\label{B2toTd}
  W_2^{\dagger}\ket{r}_{2,3}
  =\omega^{(d-1)\left(\frac{2}{m}-r\right)}\ket{r}_{T}.
 \end{eqnarray}
As a consequence, $W_2^{\dagger}\,\mathcal{O}_{2,3}\,W_{1}=T_{d,m}$.
Let us now consider the parties indexed by odd numbers. The action of $W_{\mathrm{odd}}^{\dagger}$ on 
the eigenvectors of $\mathcal{O}_{n_{odd},2}$, $\ket{r}_{n_{odd},2}$, given explicitly in Eq. (\ref{idealeigen}) can be expressed as
%
%
 \begin{eqnarray}
 W_{\mathrm{odd}}^{\dagger}\ket{r}_{n_{\mathrm{odd}},2}
 &=& \frac{1}{d}\sum_{j,q=0}^{d-1}(-1)^{\delta_{j,0}}\omega^{(r-j)q}\omega^{-\frac{j}{2}}\ket{j},
 \end{eqnarray}
which by again employing the fact that $ \sum_{q=0}^{d-1}\w^{(r-j)q}=d\delta_{r,j}$, simplifies to
 \begin{eqnarray}\label{AoddtoZd}
  W_{\mathrm{odd}}^{\dagger}\ket{r}_{\mathrm{odd},2}
  = (-1)^{\delta_{j,0}}\w^{-j/2}\ket{j} .
 \end{eqnarray}
Analogously for $\mathcal{O}_{\mathrm{odd},3}$, we have that
 \begin{eqnarray}
W_{\mathrm{odd}}^{\dagger}\ket{r}_{\mathrm{odd},3}
 &=& \frac{1}{d}\sum_{j,q=0}^{d-1}(-1)^{\delta_{j,0}}\omega^{(r-j-\frac{1}{m})q}\omega^{-\frac{j}{2}}\ket{j}.
 \end{eqnarray}
 which by again utilising the sum (\ref{sumimpW}) simplifies to 
 \begin{eqnarray} \label{AoddtoTd}
  W_{\mathrm{odd}}^{\dagger}\ket{r}_{\mathrm{odd},3}
  =\ket{r}_{T}.
 \end{eqnarray}
As a consequence, $W_{\mathrm{odd}}^{\dagger}\,\mathcal{O}_{\mathrm{odd},3}\,W_{\mathrm{odd}}=T_{d,m}$. Let us finally consider the parties indexed by even numbers. The action of $W_{\mathrm{ev}}^{\dagger}$ on 
the eigenvectors of $\mathcal{O}_{n_{ev},2}$, $\ket{r}_{n_{ev},2}$, given explicitly in Eq. (\ref{idealeigen}) can be expressed as
%
 %
 %
 \begin{eqnarray}
 W_{\mathrm{ev}}^{\dagger}\ket{r}_{\mathrm{ev},2}
 &=& \frac{1}{d}\sum_{j,q=0}^{d-1}(-1)^{\delta_{j,0}}\omega^{(r-j)q+(d-1)\left(\frac{2}{m}-r\right)-\frac{j}{2}}\ket{j},
 \end{eqnarray}
which by again employing the fact that $ \sum_{q=0}^{d-1}\w^{(r-j)q}=d\delta_{r,j}$, simplifies to
 \begin{eqnarray}\label{BeventoZd}
  W_{\mathrm{ev}}^{\dagger}\ket{r}_{\mathrm{ev},2}
  =(-1)^{\delta_{r,0}} \w^{(d-1)\left(\frac{2}{m}-r\right)-\frac{r}{2}} \ket{r}.
 \end{eqnarray}
As a consequence, $W_{\mathrm{ev}}^{\dagger}\,\mathcal{O}_{\mathrm{ev},2}\,W_{\mathrm{ev}}=Z_d$. Analogously for $\mathcal{O}_{\mathrm{ev},3}$, we have that
 \begin{eqnarray}
 W_{\mathrm{ev}}^{\dagger}\ket{r}_{\mathrm{ev},3}
 &=& \frac{1}{d}\sum_{j,q=0}^{d-1}(-1)^{\delta_{j,0}}\omega^{(r-j-\frac{1}{m})q + (d-1)\left(\frac{2}{m}-r\right)-\frac{j}{2}}\ket{j},
 \end{eqnarray}
which by employing the sum \eqref{sumimpW}, simplifies to
 \begin{eqnarray}\label{BeventoTd}
  W_{\mathrm{ev}}^{\dagger}\ket{r}_{\mathrm{ev},3}
  =\omega^{(d-1)\left(\frac{2}{m}-r\right)}\ket{r}_{T}.
 \end{eqnarray}
As a consequence, $W_{\mathrm{ev}}^{\dagger}\,\mathcal{O}_{\mathrm{ev},3}\,W_{\mathrm{ev}}=T_{d,m}$.
%
\end{proof}

\chapter{Proofs of some observations relevant to Chapter 5}\label{chap5}

\begin{customobs}{5.1}\label{obsST1}
The matrix 
\begin{equation}\label{matrix}
\Tilde{Z}=\overline{Z}_A+\gamma(\boldsymbol{\alpha})\Id=[1+\gamma(\boldsymbol{\alpha})]\Id-\sum_{k=1}^{d-1}\delta_k(\boldsymbol{\alpha})Z_d^{k}
\end{equation}
with $\gamma$ and $\delta_k$ defined in Eq. (\ref{value1}) is positive and invertible.
\end{customobs}
\begin{proof}
Let us first observe that that the above matrix \eqref{matrix} is diagonal in the computational basis as well as hermitian which stems from
the fact that  $\delta_{k}^*=\delta_{d-k}$ and that $Z_d^{\dagger}=Z^{d-k}_d$. To show that this matrix is positive and invertible, we compute its eigenvalues and show that they are strictly positive. For this purpose, we plug in the explicit form of $\delta_k$ and evaluate its diagonal elements in the computational basis as
%
%
%
\begin{eqnarray}
\lambda_l=1+\gamma(\boldsymbol{\alpha})+\frac{\gamma(\boldsymbol{\alpha})}{d}\sum_{k=1}^{d-1}\sum_{\substack{i,j=0\\i\ne j}}^{d-1}\frac{\alpha_i}{\alpha_j}\omega^{k(l-j)} \quad \text{for}\quad l=0,\ldots,d-1,
\end{eqnarray}
such $\Tilde{Z}=\sum_l\lambda_l\ket{l}\!\bra{l}$.  After simple manipulations and rearrangements, the above expression simplifies to
\begin{equation}
\lambda_l=\frac{\gamma(\boldsymbol{\alpha})}{\alpha_l}\sum_{\substack{i=0}}^{d-1}\alpha_i.
\end{equation}
Let us note that $\gamma(\boldsymbol{\alpha})$ and  $\alpha_i$'s  are strictly positive for all $i$'s. Thus, we can conclude that all the eigenvalues of the matrix (\ref{matrix}) are positive and as a consequence the matrix (\ref{matrix}) is invertible. 
\end{proof}

\newpage
\begin{customobs}{5.2}\label{fact5.2}
Let us consider that the elements of two POVM's $\{\mathcal{I}_b\}$ and $\{R_b\}$ are related as
\begin{eqnarray}\label{eq87A}
\mathcal{I}_b\geq
\pm\Tr_{B''}\left[(\mathbbm{1}_{B'}\otimes L^a_{B''})R_b\right],
\end{eqnarray}
such that $L^a_{B''}$ is a matrix given in \eqref{LB} and $\{\mathcal{I}_b\}$ is an extremal rank-one POVM. Then, 
\begin{eqnarray}\label{POVMST2A}
\Tr_{B''}\left[(\mathbbm{1}_{B'}\otimes L^a_{B''})R_b\right]=\lambda'_b\proj{\mu_b}.
\end{eqnarray}
\end{customobs}
\begin{proof}
As the POVM $\{\mathcal{I}_b\}$ is extremal and of rank-one, we can express each of its elements as $\mathcal{I}_b=\lambda_b\ket{\mu_b}\bra{\mu_b}$ for all $b$. Choosing a particular $b$ and the corresponding vector $\ket{\mu_b}$, we can construct an orthonormal basis 
$\{\ket{\phi_i}\}$ with $i=0,\ldots,d-1$ such that $\ket{\phi_0}=\ket{\mu_b}$.
Now, multiplying the above inequality \eqref{eq87A} with $\bra{\phi_j}$ from left and $\ket{\phi_i}$ from right hand side, we arrive at
\begin{eqnarray}\label{VinoTinto}
\forall_{i,j}\qquad \bra{\phi_j}\mathcal{I}_b\ket{\phi_i}\geq
\pm\bra{\phi_j}\Tr_{B''}\left[(\mathbbm{1}_{B'}\otimes L^a_{B''})R_b\right]\ket{\phi_i}.
\end{eqnarray}
For $i\neq 0$ or $j\neq 0$ the above expression yields
\begin{eqnarray}
0\geq\pm\bra{\phi_j}\Tr_{B''}\left[(\mathbbm{1}_{B'}\otimes L^a_{B''})R_b\right]\ket{\phi_i},
\end{eqnarray}
which is only possible if $\bra{\phi_j}\left[(\mathbbm{1}_{B'}\otimes L^a_{B''})R_b\right]\ket{\phi_i}=0$ for any $i\neq 0$ or $j\neq 0$ and thus we can straightforwardly conclude Eq. (\ref{POVMST2A}). Moreover, for $i=j=0$, Eq. (\ref{VinoTinto}) gives 
\begin{equation}
    \lambda_b\geq \pm \bra{\mu_b}\Tr_{B''}\left[(\mathbbm{1}_{B'}\otimes L^a_{B''})R_b\right]\ket{\mu_b},
\end{equation}
which additionally imposes that $\lambda_b\geq \pm\lambda_b'$. 
\end{proof}

\end{appendices} 